%% file: thesis.tex
\documentclass[12pt,oneside]{book}
\usepackage{thesis}

\begin{document}
\pagenumbering{roman}

\include{Chapters/cover}
\include{Chapters/acknowledgements}
\include{Chapters/abstract}

\tableofcontents
\listoffigures
\printglossary[title=Acronyms, toctitle=Acronyms]

\cleardoublepage
\pagenumbering{arabic}

\include{Chapters/chapter1}
\include{Chapters/chapter2}
\include{Chapters/chapter3}
\include{Chapters/chapter4}
\include{Chapters/chapter5}
\include{Chapters/chapter6}
\include{Chapters/chapter7}

\bibliographystyle{unsrt}
\bibliography{bibliography.bib}

\appendix

\include{Chapters/appendix1}
\include{Chapters/appendix2}
\include{Chapters/appendix3}

\end{document}

%% file: Chapters/cover.tex
\begin{titlepage}
    \begin{center}
        
        \vspace*{-1.5in}
        \begin{figure}[htb]
        \centering
        \includegraphics[width=0.15\textwidth]{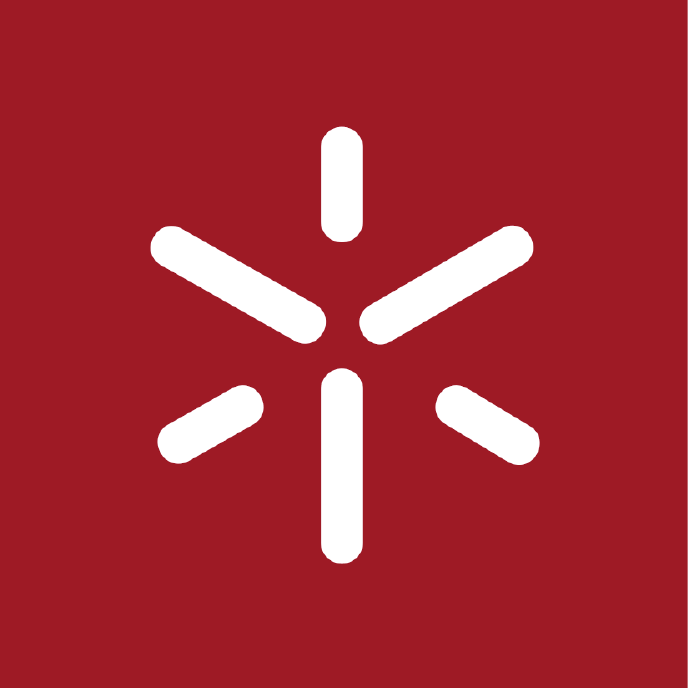}
        \hspace{-6.875pt}
        \includegraphics[width=0.15\textwidth]{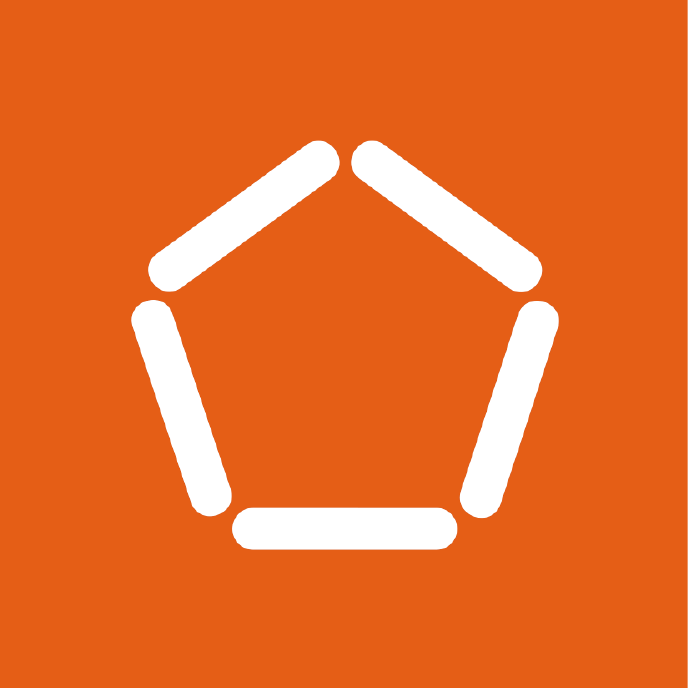}
        \end{figure}
        
        {\color{darkgray} \Large University of Minho}\\
        \vspace{0.2cm}
        {\color{gray}School of Engineering}\\
          
        \vspace*{5cm}
        
        {\color{darkgray}\LARGE
        {Mafalda Francisco Ramôa da Costa Alves}}\\
        
        \begin{doublespace}
        \vspace{1cm}
        \setstretch{3}
        \textbf{\Huge Ansätze for Noisy Variational Quantum Eigensolvers}
        \end{doublespace}
        
        \vspace{2.5cm}
            
        \large{\color{gray} Master Thesis\\
        Engineering Physics}
        
        \vspace{1.5cm}
        
        \textbf{Supervisors}\\
        \vspace{0.3cm}
        {\color{darkgray}
        Ernesto Galvão\\
        Mikhail Vasilevskiy\\}
        
        \vfill
            
        {\Large \color{gray}November, 2021}
        
        \date{\emph{November, 2021}}
            
   \end{center}
\end{titlepage}

%% file: Chapters/acknowledgements.tex
\chapter*{Acknowledgements}
\pagestyle{plain}
\addtocounter{page}{1}

I would like to offer my special thanks to Raffaele Santagati for suggesting this project, and for his guidance and support throughout the course of it. I thank also my co-supervisor Ernesto Galvão, for his assistance at every stage of the project, as well as for his invaluable advice; and Mikhail Vasilevskiy, for his support.

I thank my family and friends for their encouragement. In particular, I'm deeply thankful to my sister, colleague and friend Alexandra. This year would not have been the same without her continuous support both inside and outside the academic sphere.

I want to express my gratitude also to all the teachers and professors that, throughout the years, have taught me so many different things in so many different ways. The genuine concern for students and dedication for teaching I have encountered throughout my life have certainly impacted it along the years.

Additionally, I wish to thank the Programme New Talents in Quantum Technologies of the Gulbenkian Foundation (Portugal) for the financial support, and for the opportunity of being involved in this experience. The network of students under the support of the Gulbenkian Foundation is unmatched in the enthusiasm for learning and growing.

Finally, I wish to acknowledge the use of IBMQ's systems \cite{IBMQ} and Google Colaboratory's cloud service \cite{colab}.

\clearpage

%% file: Chapters/abstract.tex
\chapter*{Abstract}
\pagestyle{plain}

{\Large
\color{darkgray}
\textbf{Ansätze for Noisy Variational Quantum Eigensolvers}}

\vspace{1cm}

Simulating quantum mechanical systems is one of the main applications envisioned for quantum computers. In contrast with the first algorithms created for this purpose, that were devised to be implemented in a fault-tolerant quantum computer, the \gls{VQE} aims to adjust to the constraints of \gls{NISQ} devices. There is hope that the class of \glspl{VQA}, to which \gls{VQE} belongs, will be the first to achieve quantum advantage.

The choice of ansatz can dictate the success (or lack thereof) of a \gls{VQA}: too deep ansätze can hinder near-term viability, or lead to trainability issues that render the algorithm inefficient. In this context, this dissertation aimed to analyse different ansätze for quantum chemistry, examining their noise-resilience and viability in state-of-the-art quantum computers. In particular, dynamic ansätze were explored, and their performance compared against predetermined ansätze, with a focus on susceptibility to noise.

Multiple variants of \gls{VQE}, namely \gls{UCCSD}-\gls{VQE} (predetermined) and \gls{ADAPT}-\gls{VQE} (dynamic), were implemented both in simulators and cloud quantum computers. Using noise models, the impact of several noise sources on convergence was assessed. Additionally, the importance of the operator pool in \gls{ADAPT}-\gls{VQE} was analysed, and strategies to manipulate the ansatz beyond the \gls{ADAPT}-\gls{VQE} algorithm were explored.

Several conclusions could be drawn from this work. Adapting the ansatz to the problem and system was concluded to be fundamental in avoiding trainability issues, decreasing the circuit depth required for a given accuracy, and improving noise-resilience (against circuit depth dependent and independent sources alike). Dynamic ansätze were shown to be capable of enduring significantly larger error rates than predetermined alternatives, and were thus proved to be better suited for \gls{NISQ} devices. For $H_2$, as far as ground state energy calculations are concerned, \gls{ADAPT}-\gls{VQE} was shown to tolerate a 20 times lower shot count, 150 times larger error rates in state preparations and measurements, and 850 times lower coherence times than \gls{UCCSD}-\gls{VQE}. The difference is expected to increase with the size of the system.

Additionally, it was observed that there is still a margin for improving upon \gls{ADAPT}-\gls{VQE}. Further manipulation of the ansatz was shown to be capable of producing yet shallower circuits for the same accuracy. Using an idea previously proposed in the literature, a more conservative selection criterion was tested. Additionally, removing operators on the fly based on available data was attempted as a new possibility. Both approaches were shown to be capable of improving upon the \gls{ADAPT}-\gls{VQE} ansatz, resulting in an up to 35-fold decrease in the error for a similar circuit depth within the first 10 iterations of \gls{ADAPT}-\gls{VQE}.

\clearpage

%% file: Chapters/chapter1.tex
\pagestyle{plain}
\chapter{Introduction}
\section{Context and Motivation}

Richard Feynman raised the question,

\vspace{12pt}

\textit{What kind of computer are we going to use to simulate physics?}

\vspace{12pt}

, calling attention to the fact that the physical world is quantum mechanical, so \textit{quantum} physics is what we are concerned with \cite{feynman}. So arose the answer: a \textit{new} kind of computer. 

The simulation of a quantum system being a difficult task to perform classically is but a symptom of the fact that bits, as basic units of information in classical computing, are unfit to describe nature under the laws of quantum mechanics. Quantum computing was thus born with the purpose of using quantum mechanics itself to perform computation, replacing bits by their quantum counterparts: \textit{qubits}. Making use of quantum phenomena (namely entanglement; superposition; interference) empowers quantum computers with the ability to do operations without a classical analogue.

The scaling of the description of highly entangled states using classical data inspired hope that quantum computers would be able to perform tasks that are intractable even to classical supercomputers. The memory required for storing a general state of a quantum system in a classical computer scales exponentially on the size of the system; in contrast, the scaling is linear on a quantum computer, as seems intuitive in view of it being such a system itself. In fact, an interesting reason to believe that a quantum computer may bring advantage is that there is no known classical algorithm that can simulate such a computer efficiently; representing the state of a few hundred qubits may already require more bits than the number of atoms in the visible universe \cite{Preskill_2018}.

However, this fact does not constitute in itself proof that quantum computers are more powerful than the classical ones. It is a misconception that quantum computers are \textit{exponentially parallel}: even though superposition implies that qubits can be in a linear superposition of states that are exponentially many on the number of qubits, the information held by this superposition isn't necessarily accessible. Measuring the state of the qubits will cause the collapse of the wave function: transforming quantum information into classical information destroys it irreversibly, unless the measurement reveals no information about the quantum state it is inflicted upon \cite{NielsenChuang}.

Famously, by Holevo's theorem, the accessible classical information in \textit{n} qubits is bounded above by \textit{n} bits. The fact that the qubits carry more information is to no avail, unless one devises a clever way of extracting information in an efficient manner. As such, the design of techniques that allow employing quantum resources advantageously is all but straightforward; and the quest for quantum algorithms has developed into a large field of research.

In 1980, Benioff published the first article on quantum computation, a pioneering work describing a quantum mechanical model of Turing machines \cite{Benioff1980}. Since then, there has been significant progress. Notable quantum algorithms were introduced in the following decade, such as Deutsch–Jozsa in 1992 \cite{deutschjozsa} and Simon's in 1994 \cite{Simon}. Albeit not having much practical value, they hold their merit as some of the first examples of how a quantum computer could offer an exponential speedup over classical computers. Advantage of practical interest ensued when Shor, inspired by Simon's proposal, introduced his famous factoring algorithm \cite{Shor}.

Shor's algorithm enables a (hypothetical) quantum computer to factor integers in a time polynomial on their size - an exponential improvement as compared to the best-known classical algorithm for the task. While this remarkable speedup was not unprecedented, Shor's algorithm owes its distinct success to the possible applications of the problem it was created to solve. An efficient factoring algorithm defies the security of the vastly used RSA cryptosystem, which brought attention to the possible implications of quantum computers. If resource requirements in what concerns quantum hardware were met, this algorithm would allow breaking internet security: it manages to solve in a matter of minutes a problem that would take years for the best-known classical algorithm to solve, and upon the difficulty of which the RSA algorithm relies \cite{resch2019}. This was a significant landmark, as well as a stronger motivation for developing the hardware that would pave the way for physically realising these theoretical proposals. 

Albeit offering an unarguably less astounding quadratic speedup, Grover's algorithm \cite{Grover} for searching in an unstructured database has also drawn formidable attention. Aside from covering a wide array of applications, its strength is in offering a provable speedup: evidently, the number of evaluations required to solve the problem classically has to scale linearly (half of elements in the search space will have to be checked on average). By offering the uncanny possibility of finding a marked element after a number of operations that is sub-linear on the size of the search space, Grover's algorithm also contributed to a build up of the public reputation of quantum computing as a curious and strange field. 

While they seem compelling, these theoretical results are yet to be proven by experimental realization. Current hardware limitations hinder the implementation of large-scale instances of algorithms such as the aforementioned ones. 

There has been notable progress since the concept of quantum computing was introduced. The first implementation of an algorithm on a physical quantum computer took place in 1998 \cite{Chuang1998}, and several proof-of-concept demonstrations of small-scale instances of different algorithms followed. In 2012, Shor's algorithm was used to factor N=21 for the first time \cite{ShorExp2012}.

However, practical experiments up to now have been limited to toy problems and small, classically tractable demonstrations. Building a quantum computer capable of outperforming a classical one in tasks that may be of practical use seems to be strenuous at best, impossible at worse. 

The difficulty of building the physical hardware is indissociable from the very same nature on top of which the concept of quantum computing rests. Quantum mechanical states are very sensitive, causing qubits to require close to perfect isolation from the world to store and process information reliably. Building a quantum computer thus implies granting the proper conditions, i.e. conditions that reduce the coupling with the surrounding environment to the point that useful computations can be carried. Often, this involves temperatures near absolute zero and shielding from radiation; this is aggravated by the fact that the difficulty greatly increases with the system size, preventing large-scale quantum computers to have been built so far \cite{resch2019}.

There is a dichotomy between the scaling of the most compelling quantum algorithms, and the scaling of the difficulty of building a quantum computer capable of actually bringing them to fruition. In the words of two prominent French physicists vocally skeptical of quantum computing \cite{HarocheRaimond}, 

\vspace{12pt}

\textit{...the large-scale quantum machine, though it may be the computer scientist's dream, is the experimenter's nightmare}.

\vspace{12pt}

After Shor's algorithm was discovered, concerns were raised regarding the viability of quantum computing: algorithms built under the assumption of an ideal quantum computer might be rendered useless by decoherence and other types of noise \cite{Landauer1996,Unruh1995}. In light of the \textit{no-cloning} theorem \cite{Dieks1982,Park1970,Wootters1982}, that forbids quantum data from being copied, error correction didn't seem as trivial as in the classical case.

In 1995, Peter Shor himself recovered some hope with a viable way of realising error correction in quantum computing \cite{Shor1995}. Quantum error correcting codes are of the greatest importance: if error correction is important for computing in general, it is pivotal for quantum computing in particular. Unfortunately, the overhead in terms of qubit requirements and number of necessary gates is severe \cite{gottesman2009}.

An error corrected, fault-tolerant quantum computer does not seem to belong to a near future. John Preskill coined the term \glsfirst{NISQ} to describe the current era \cite{Preskill_2018}. The name encompasses qubit limitations as well as noise and decoherence.

After the promising algorithms of the 1990s shed some light on the potential of a fault-tolerant quantum computer, there has been a growing effort to build such a device - but that possibility seems at best distant. In an attempt to make as good use of \gls{NISQ} computers as possible, the attention shifted in part from the asymptotic scaling of algorithms to their near-term viability. The focus has widened, and thus, in addition to creating conceptually advantageous algorithms lacking certainty that they are ever to be implemented, we now observe a growing dedication to devising algorithms that seem implementable but aren't necessarily known to be advantageous. Variational quantum algorithms have emerged.

\section{State of the Art}

\glspl{VQA} are hybrid quantum-classical algorithms that resemble machine-learning techniques typical to classical computing. The usual outline involves using a classical computer to find the variational parameters that minimize a cost function, which is evaluated by the quantum computer. An interesting review of the structure, applications and challenges of variational quantum algorithms can be found in reference \cite{cerezo2020}.

The role of the quantum computer in variational algorithms is to prepare trial states using a state preparation circuit (called \textit{ansatz}), and measure the expectation value (and possibly the gradients) of the cost function in those states. The state preparation circuit includes a parameterized portion, the parameters of which are fed to the quantum computer each iteration by the classical optimizer, which tries to adjust them in an attempt to find the minimum.

\glspl{VQA} are currently gathering great interest, seeing that their very nature makes them resilient to noise: by outsourcing the optimization to a classical computer, they naturally require shallower circuits. The learning-based approach further aids noise mitigation.

In 2014 the first \gls{VQA} was proposed and demonstrated experimentally with photonic qubits: the \gls{VQE} \cite{Peruzzo2014}, central to this dissertation. The proposal aimed to address quantum chemistry problems using near-term quantum computers, in alternative to longer routines (such as adiabatic state preparation in conjunction with \gls{QPE}) that are beyond the capabilities of \gls{NISQ} devices.

Since it was introduced, the \gls{VQE} has become an icon of \gls{NISQ} algorithms, and a lot of research has been done to further increase its accuracy and decrease its demands on the quantum computer. There have been improvements on the efficiency of the evaluation of the cost function for chemistry problems in particular, involving a low rank factorization of the two-electron integral tensor \cite{Huggins2021}. Interestingly, making use of the specificity of the chemistry problem seems to allow for a better scaling of the required number of measurements as compared to general problems \cite{cerezo2020}. However, more general techniques can evidently be used to further increase the accuracy of the results; as an example, the readout-error mitigation technique presented in \cite{zlatko2021} can help eliminate the noise-induced bias from the expectation value of the cost function.

The qubit requirements can also be reduced: in reference  \cite{Steudtner2018} it was highlighted that it is possible to choose a more compact fermion to qubit mapping instead of the Jordan-Wigner transform employed in the original article, at the expense of more gates in the circuits. 

The optimization itself is a large field of research. While not belonging to the strictly quantum part of the algorithm, it is naturally affected by quantum noise and its particularities; the performance of classical optimizers benchmarked with noiseless function evaluations cannot simply be extrapolated to a noisy scenario. Efforts have been done to find optimizers that decrease the number of calls to the quantum computer and make convergence more likely in the presence of noise. Meta-learners were used in \cite{wilson2019} and two model-based optimization algorithms were introduced in \cite{Sung2020}. These examples are left as a sample of a much broader set of proposals for improving the optimization. 

Finally, we have the ansatz, which plays a fundamental role in the algorithm. It consists of the state preparation circuit that defines where we will be looking for in the Hilbert space. From the concept alone, one can conclude that the ansatz should satisfy two crucial aspects: contain the solution, and have a description that does not scale exponentially with the size of the system. Otherwise, it may jeopardise the accuracy and efficiency of the algorithm.

In the context of the \gls{NISQ} era, this key piece assumes an even greater importance: the ansatz corresponds to the bulk of the depth of the utilized circuits, which in turn means that the choice of ansatz might allow or prevent the \gls{VQE} from being successfully implemented on a given quantum computer.

Further, it has been shown that an improper ansatz might prevent the optimizer from finding the solution. As an example, randomly initialized parameterized quantum circuit were shown to cause the gradients of the cost function with respect to the parameters to vanish exponentially on the size of the system, an effect termed as barren plateaus \cite{McClean2018}. The effect is not limited to gradient-based optimizers, as this description might seem to suggest - the result was proven to generalize to other optimization methods \cite{Arrasmith2020}. The vanishing gradients are merely a symptom of a deeper issue regarding the optimization landscape. As the size of the Hilbert space grows exponentially on the number of qubits, it is no surprise that a generic ansatz with polynomially many parameters proves to hinder the performance of the algorithm. The minimum becomes sharper and sharper amidst a flat landscape, as a result of too much expressibility. A relationship between expressibility and trainability was derived in \cite{holmes2021}.

Multiple strategies have been used to address this. Some involve the ansatz and the formulation of the problem itself: it was shown that shallow ansätze in conjunction with local cost functions circumvent the issue. For circuits with a depth logarithmic on the number of qubits, the gradient vanishes polynomially at most \cite{Cerezo2021BPCFun,Uvarov2021}. Parameter initialization techniques that prevent the algorithm from starting in a barren plateau have been proposed. Reference \cite{Grant2019} does this by limiting the initial effective depth of the circuit, while in \cite{verdon2019} meta-learning via classical neural networks was used to find good parameter initialization heuristics. Layerwise training \cite{skolik2020} and correlated parameters \cite{Volkoff2021} were proved to aid in avoiding barren plateaus \textit{during} the optimization. As for the classical optimizer, making use of natural evolutionary strategies in the optimization as a means to avoid barren plateaus was suggested in reference \cite{Anand2021}.

Another interesing possibility is to leverage problem-specific knowledge to prevent the issue. Considering that barren plateaus are tightly related to an excess of expressibility, leveraging what is known about the problem at hand to carefully pick where we will be looking for the solution is of great help. There is a downside: the priority is placed in tailoring the ansatz \textit{to the problem} rather than to the quantum processor. The later seems preferable: in the context of \gls{NISQ} devices, the focus is usually set on making the circuit as shallow as possible. Because of this, a popular suggestion was building the ansatz from the interactions naturally available in the quantum processor - an idea that was introduced in \cite{Kandala2017} as the \textit{hardware-efficient ansatz}. However, compared to these more generic \textit{problem agnostic} ansätze, those known as \textit{problem tailored} or \textit{problem inspired} have the great benefit of usually being trainable, even when the parameters are initialized at random \cite{cerezo2020}. Further, the ansatz circuit being convenient is to no avail if the optimization process becomes intractable due to requiring exponential precision and function evaluations to overcome barren plateaus. 

The ansatz employed in the original \gls{VQE} proposal was a problem tailored one: the \gls{UCCSD} ansatz, inspired on quantum chemistry and classical variational methods. States are prepared by applying an operator, consisting of an exponentiated sum of single and double fermionic excitations, to a reference state. A  more detailed description of \gls{UCCSD} can be found in \cite{romero2018}. What is relevant here is that \gls{UCC} is the natural unitary version of the classical \gls{CC} operator, regarded as \textit{the gold standard} in classical quantum chemistry methods \cite{Barlett2007}.

Aside from its foundations providing us a motive to believe that the \gls{UCCSD} ansatz is a good choice, both the number of variational parameters and the number of gates necessary for its implementation scale polynomially with the size of the system. However, implementing it with a quantum circuit implies using a finite-order Suzuki-Trotter decomposition, which leads to ambiguities in the operator ordering. The resulting energy variations have been proved relevant at the chemical scale, meaning that the ansatz is not chemically well-defined \cite{Grimsley2020}. Additionally, it was pointed out in \cite{Wecker2015} that this ansatz has an unnecessarily large number of parameters and terms, which then reflect in an optimization more difficult and circuits deeper than required.

It is fundamental to keep the circuit as shallow as possible. In fact, while \gls{UCCSD} may avoid the barren plateaus discussed above, barren plateaus of a different kind were shown to occur as a result of incoherent noise \cite{wang2021}. In consequence of the output state being less and less pure, the minimum of the cost function becomes less and less pronounced. While with the noise-free barren plateaus the minimum value of the cost function was still represented in the optimization landscape, noise-induced barren plateaus might erase this minimum altogether, with decoherence causing the dilution of the characteristics of the cost function. 

In  light of this, efforts have been done to combine problem tailoring with hardware tailoring, in order to avoid barren plateaus while keeping circuits convenient and shallow. A few suggestions for chemistry applications follow.

Reference \cite{Gard2020} proposed efficient ansätze that respect particle number, total spin, spin projection, and time-reversal symmetries. The idea is to prevent the ansatz from covering irrelevant subsections of the Hilbert space, which allows spanning the relevant symmetry subspace with as few variational parameters as possible. As an example, we can be certain that the ground state will not be found in a region of the Hilbert space that represents a fermionic state with an incorrect number of particles. The performance of the algorithm is thus improved by excluding that region from the ansatz altogether. 

An interesting hardware-efficient, but problem-tailored ansatz was proposed in \cite{meitei2021}. The algorithm, termed ctrl-\gls{VQE} by the authors, replaces the traditional quantum circuit ansatz with a quantum control routine. Directly optimizing the pulses allows for reducing the coherence time requirements without abdicating the variational freedom.

Finally, there are the \textit{dynamically created ansätze}, which will be the main focus of this dissertation. While problem inspired ansätze \textit{leverage} knowledge on the system, the dynamically created ones \textit{collect} this knowledge themselves. In a way, they allow the problem to dictate its own ansatz, by growing it from scratch in a strongly system-adapted manner. Fermionic operators are added to the ansatz as `pieces' along the evolution of the algorithm; the chosen ones are those that, from informative measurements, are believed to be the most beneficial.

In 2019, this idea was introduced in \cite{Grimsley2019} as \glsfirst{ADAPT}-\gls{VQE}, hereby denoted \textit{Fermionic-\gls{ADAPT}-\gls{VQE}} for clarity against a more recent version. The name was attributed in light of the selection criterion being derivative-related and the circuits being reminiscent of a trotterization.

While the \gls{UCCSD} ansatz is fixed upfront and does not allow decreasing or increasing the precision of the result, or the circuit depth,
\gls{ADAPT}-\gls{VQE} grows the ansatz from zero until a convergence criterion is met. Numerical results showed that this protocol can match or surpass the accuracy of \gls{UCCSD}-\gls{VQE} with more compact ansätze. In essence, we are trading measurements and optimizations for shallower circuits, making \gls{VQE} more \gls{NISQ}-friendly and bringing us a step closer to solving real problems on quantum computers.

In 2020 a follow-up article \cite{Tang2021} was published proposing \textit{Qubit-\glsfirst{ADAPT2}-\gls{VQE}}, a modified version of the algorithm. The difference is in the operators that are available to choose from. Anti-commutation and preservation of particle number and spin are dispensed with, allowing the operators to be implemented by shallower circuits. As a consequence, they no longer have an interpretation as fermionic excitations. This typically results in an increase in the number of variational parameters and decrease of circuit depth for a given accuracy, as compared to Fermionic-\gls{ADAPT}-\gls{VQE}. Qubit-\gls{ADAPT2}-\gls{VQE} thus burdens the classical optimizer further, in exchange for shallower circuits. 

The Qubit-\gls{ADAPT2}-\gls{VQE} article also introduced some minimal complete pools whose size scaled linearly with the number of spin-orbitals (as compared to the quartic scaling of all previous pools). This reduction in pool size implies a reduction in the number of measurements per iteration. Unfortunately, because these pools do not respect fermionic symmetries, they are not adequate when \gls{ADAPT2}-\gls{VQE} is used to solve for the eigenstates of molecular Hamiltonians. 

In September 2021, the article \cite{shkolnikov2021} introduced a way of creating symmetry-adapted minimal complete pools that worked in this type of problems, while also growing linearly on the size of the system. With this, the article proved that it was possible to significantly reduce the number of measurements per iteration in \gls{ADAPT2}-\gls{VQE}, from $\mathcal{O}(N^8)$ to $\mathcal{O}(N^5)$ ($N$ the number of spin-orbitals). This brings the measurement scaling of \gls{ADAPT2}-\gls{VQE} closer to that of typical \gls{VQE} routines ($\mathcal{O}(N^4)$).

The \gls{ADAPT}-V and \gls{ADAPT}-Vx algorithms introduced in \cite{Liu2021} improve the scaling of the number of measurements using a different approach, based on \glspl{RDM}. In \gls{ADAPT}-V the measurement cost scales like $\mathcal{O}(N^4)$, the drawbacks being a larger number of variational parameters and deeper circuits. \gls{ADAPT}-Vx, in turn, leads to parameter vector lengths and circuit depths that are barely over those in \gls{ADAPT}-\gls{VQE}, at the expense of a slight measurement overhead as compared to \gls{ADAPT}-V.

\section{Objectives}

This dissertation is focused on analysing increasingly \gls{NISQ}-friendly ansatz proposals for variational quantum algorithms in chemistry applications. This encompasses shallower ansätze, as well as improved noise resilience.

Quantum chemistry and variational quantum algorithms will be reviewed, as well as relevant aspects of quantum computing.

Instances of \gls{VQE} with predetermined ansätze, both problem agnostic and problem tailored, will be studied, implemented, and simulated. The same analysis and simulation is to be done for dynamically created ansätze (fermionic-\gls{ADAPT}-\gls{VQE} and qubit-\gls{ADAPT2}-\gls{VQE}). In all cases, numerical simulation will involve the study of the impact of noise on the performance of the algorithm. This encompasses sampling noise as well as coherent, incoherent and \gls{SPAM} noise, simulated resorting to noise models, and implementation of small-scale instances in real quantum computers. 

The impact of the pool in the \gls{ADAPT}-\gls{VQE} algorithm will be analysed by comparing the one from qubit-\gls{ADAPT2}-\gls{VQE}, the one from fermionic-\gls{ADAPT}-\gls{VQE}, and other options. The impact of the choice of pool on convergence will be assessed, as well as on the conservation of expectation values of fermionic observables, which is tightly related to convergence itself.

Finally, approaches will be tested in an attempt of further reducing the ansatz depth required for a given accuracy. Heuristics for removing operators from the ansatz after their addition will be developed. Additionally, it will be attempted to grow the ansatz with a slower, more conservative method. Both strategies will be compared.

\section{Outline of the Document}

The structure of the dissertation is as follows. Chapter \ref{ch:theory} explains the fundamental theoretical framework. Section \ref{s:quantumcomputing} covers the basic concepts of quantum computing, with a focus of the aspects that are relevant for \gls{VQE} (quantum expectation estimation, quantum simulation and trotterization, and the limitations of quantum computers in the \gls{NISQ} era). Section \ref{s:quantum_chemistry} follows with the essential aspects of quantum chemistry, namely, the context in which the electronic problem arises, and the classical variational approaches that preceded (and inspired) the proposal of \gls{VQE}. Finally, section \ref{s:VQAs} explains the context, theory, and methods that are specific to variational quantum algorithms. The general outline of \glspl{VQA} is explained in this section, as well as the specific aspects concerning \gls{VQE}, such as the quantum circuit representation of fermionic operators, the possible types of ansatz, the Hamiltonian averaging, and the shot requirements necessary to achieve chemical accuracy.

Chapter \ref{cha:static_ansatze} introduces some static ansätze for \gls{VQE}, with experimental results that illustrate the viability of the procedure. The first section, \ref{s:problem_agnostic_ansatze}, covers simulations using a problem-agnostic ansatz. The bond dissociation curve of the helium hydride ion, that was obtained in the first implementation of \gls{VQE} (in a photonic chip) \cite{Peruzzo2014}, is reproduced. A circuit to span the whole state space of 2 qubits was devised, and the influence of the choice of optimizer and optimization hyperparameters studied. The implementation was tested both in simulators and real devices (IBMQ \cite{IBMQ} backends). Section \ref{s:problem_tailored_ansatze} implements VQE with the \gls{UCCSD} ansatz, a problem-tailored ansatz, and presents results for molecular hydrogen. Once again, the implementation was tested in real devices as well as simulators. The influence of sampling noise was assessed, as well as the influence of \gls{CNOT} gate errors, and of thermal relaxation and \gls{SPAM} errors.

The concept of dynamic ansätze is explained in chapter \ref{ch:adapt_VQE}, with the introduction of the \gls{ADAPT}-\gls{VQE} algorithm. Several operator pools used in the set of articles concerning this algorithm are explained, analysed, and compared both in concept and in practice. The algorithm was applied to several molecules in noise-free scenarios, as well as in real quantum computers. The impact of sampling noise, thermal relaxation and \gls{SPAM} errors were once again assessed, and compared with the results from the previous chapter (for \gls{UCCSD}-\gls{VQE}).

Other possibilities of pools, not contemplated in the original \gls{ADAPT}-\gls{VQE} articles, are introduced in chapter \ref{ch:adapt_other_pools}. The purpose of these new pools is to assess how convergence is affected by the affinity between pool operators and fermionic excitations, in multiple aspects. The operators in the original pools, and their effect on the state of the qubits, are analysed as motivation for the definitions of the new pools. The performance of \gls{ADAPT2}-\gls{VQE} with the multiple pools is compared, and the composition of the state in terms of Slater determinants is analysed.

Chapter \ref{ch:ansatz_manipulation} introduces two additional methods for manipulating the ansatz beyond what is done in \gls{ADAPT}-\gls{VQE}, with the purpose of achieving greater accuracy with shallower circuits. Section \ref{s:removing_ops} develops and applies heuristics for removing operators from the ansatz on the fly, and applies the strategy to multiple molecules. An alternative procedure, that attempts multiple optimizations per iteration before settling for a new operator, is explained and tested in section \ref{s:conservative_growth}. The approaches are compared against each other, and against the original \gls{ADAPT}-\gls{VQE} algorithm. The chapter ends with a discussion of the merits and disadvantages of the different strategies (section \ref{s:growth_vs_removal}).

Finally, chapter \ref{ch:conclusions} concludes the dissertation with a few considerations about the multiple proposals of ansätze for \gls{VQE}. The merits and demerits of each are discussed in light of the state-of-the-art of quantum computing hardware. Prospects of future work concerning the topics that were covered throughout the dissertation are discussed.

All the code created for this dissertation and used in obtaining the results presented in this document can be found online at the GitHub software repository \cite{my_repo}.

%% file: Chapters/chapter2.tex
\pagestyle{plain}
\graphicspath{{./Chapters/Figures/Ch2/}}

\chapter{Theoretical Framework}
\label{ch:theory}

\section{Quantum Computing}
\label{s:quantumcomputing}

The purpose of this section is to offer a simple overview of the concepts in quantum computing that are necessary for the following chapters. A more in-depth explanation on these topics can be found on the book Quantum Computation and Quantum Information, reference \cite{NielsenChuang}, upon which the following was based.

\subsection{Basic Concepts}

\subsubsection{Pure Single Qubit States}

The basic unit of information for classical computing, the \textit{bit}, undertakes two possible states, usually labeled \textbf{0} and \textbf{1}. Similarly, the \textit{quantum bit}, or \textit{qubit}, is a two-level system. The difference resides in the quantum mechanical nature of the qubit: while the classical bit is either in state \textbf{0} or \textbf{1}, the quantum bit can be in a superposition of these two states. 

A qubit can be physically implemented by a two-level quantum-mechanical system. Irrespective of the specific implementation, the state of an isolated qubit can be represented abstractly in Dirac notation as

\begin{equation}
\ket{\psi} = \alpha\ket{0} + \beta\ket{1}.
\label{eq:qubitstate}
\end{equation}

In the expression above $\ket{0}$ and $\ket{1}$ are the \textit{computational basis states}; they form an orthonormal basis that spans the state space of the system, a two-dimensional Hilbert space $\hilbert$ (a complex vector space equipped with inner product). By convention,

\begin{equation}
\ket{0} = 
\begin{pmatrix}
1\\
0
\end{pmatrix},
\qquad
\ket{1} = 
\begin{pmatrix}
0\\
1
\end{pmatrix}.
\label{computationalbasisstates}
\end{equation}

The numbers $\alpha$ and $\beta$ are the respective complex amplitudes, related to the respective probability of occurrence. As discussed previously, even though they characterize the state of the qubit, it is not possible to discover the values of these numbers with access to a single copy of the state: when a measurement is done on the computational basis, the outcome will be a computational basis state.

The squared absolute values of the amplitudes correspond to the probability of, upon measurement, obtaining state $\ket{0}$ or $\ket{1}$ respectively:

$$\alpha\alpha^* = |\alpha|^2 = \text{Probability of} \ket{0},$$

$$\beta\beta^* = |\beta|^2 = \text{Probability of} \ket{1}.$$

Since the probabilities must sum to one, we have a normalization condition that reads

\begin{equation}
\braket{\psi}=|\alpha|^2 + |\beta|^2 = 1.
\label{eq:normalization}
\end{equation}

To describe the state of a perfectly isolated qubit, it thus suffices to specify the corresponding unit vector in $\hilbert$. This can be done by specifying the values of $\alpha$ in $\beta$ in equation \ref{eq:qubitstate} subject to condition \ref{eq:normalization}.

The normalization removes one degree of freedom and allows the state to be parameterized by three real numbers as

\[\ket{\psi} = e^{i\gamma}\left(\cos\left(\frac{\theta}{2}\right)\ket{0} + e^{i\varphi}\sin\left(\frac{\theta}{2}\right)\ket{1}\right).\]

The parameter $\gamma$ is an irrelevant degree of freedom that does not impact any observable. Since the global phase associated with it is not detectable and lacks physical significance, it can be promptly ignored and removed from our parameterization altogether, leaving

\begin{equation}
\ket{\psi} = \cos\left(\frac{\theta}{2}\right)\ket{0} + e^{i\varphi}\sin\left(\frac{\theta}{2}\right)\ket{1}.
\label{eq:purebloch}
\end{equation}

\begin{figure}[htbp]
    \centering
    \includegraphics[width=0.5\textwidth]{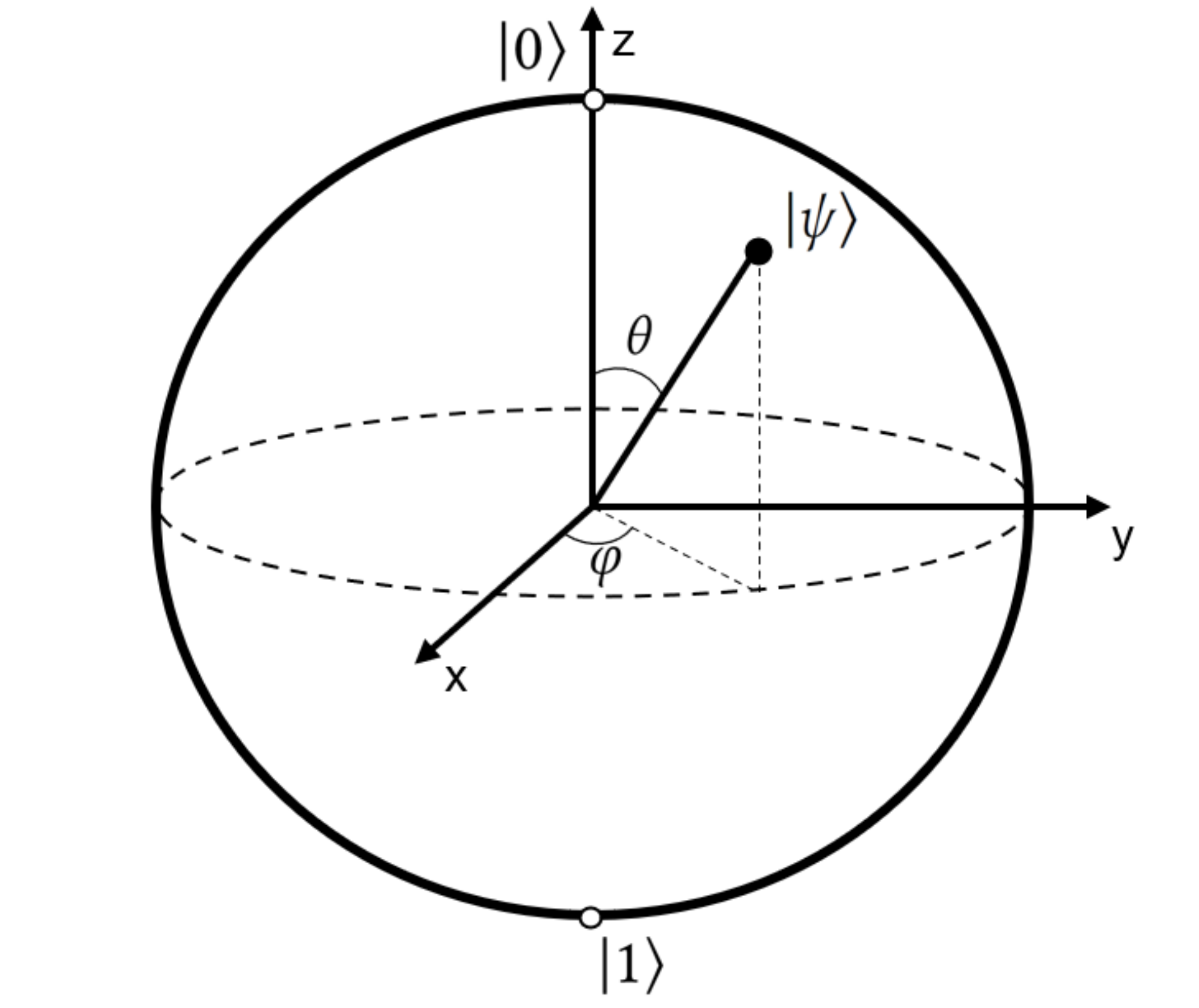}
    \caption{Representation of the state of a qubit in the Bloch sphere. The angles $\theta$ and $\varphi$ correspond to the parameters from equation \ref{eq:purebloch}.}
    \label{fig:blochsphere}
\end{figure}

With two degrees of freedom, this parameterized expression can now be visualized as lying on the surface of a sphere. The Bloch sphere depicted in figure \ref{fig:blochsphere} is a popular and convenient geometric representation of the state of a (pure) qubit.

A qubit can be acted on by \textit{operators} that are represented by two-by-two matrices. There are three matrices of particular importance, the \textit{Pauli matrices}:

\begin{align}
X\equiv\sigma_x\equiv
\begin{pmatrix}0&1\\1&0\end{pmatrix},
\quad
Y\equiv\sigma_y \equiv
\begin{pmatrix}0&-i\\i&0\end{pmatrix},
\quad
Z\equiv\sigma_z \equiv
\begin{pmatrix}1&0\\0&-1\end{pmatrix}.
\label{paulimatrices}
\end{align}

Frequently one considers the identity matrix $I$ as a Pauli matrix itself, oftentimes denoted $\sigma_0$. All Pauli matrices are unitary ($\sigma_i\sigma_i^\dagger=\sigma_i^\dagger\sigma_i=I$) and Hermitian ($\sigma_i=\sigma_i^\dagger$). 

These four matrices are orthogonal to each other, and together they span the space of the \textit{observables} of a qubit. Observables are named so because they are associated with \textit{observable physical quantities}. Each projective measurement is associated with an observable, whose eigenvalues correspond to the possible outcomes of the measurement. These outcomes are always real-valued, which is related to the fact that all observables are Hermitian.

From definitions \ref{computationalbasisstates} and \ref{paulimatrices} we can see that the computational basis states are \textit{eigenstates} of $Z$. Computational basis \textit{measurements} are projective measurements associated with the observable $Z$, and their possible outcomes correspond to the eigenvalues of the operator. These are the measurements typically available in quantum computers.

\subsubsection{Pure Multiple Qubit States}

All these ideas generalize to larger systems: the tensor product can be used to obtain a larger vector space in which the state of the multiple qubits lives. The description grows rapidly: to specify the state of \textit{n} qubits, we need $2^n$ complex amplitudes, that correspond to $2^{n+1}-2$ real-valued degrees of freedom after removing those corresponding to the normalization condition and the global phase. 

When we have more than one qubit, measurement outcomes can be correlated beyond what is possible in classical systems. This is because it is not always the case that an \textit{n} qubit state can be written in the form

\begin{equation}
     \ket{\psi_0} \otimes \ket{\psi_{1}}  \otimes ... \otimes\ket{\psi_i} \otimes...\otimes\ket{\psi_{n-2}}\otimes\ket{\psi_{n-1}},
\end{equation}

where the $\psi_i$ are the states of the individual qubits. When this is in fact possible, the state is called \textit{separable}; otherwise, we say that it is \textit{entangled}.  Entanglement plays a big role in quantum algorithms: it is an exclusively quantum-mechanical resource and a bedrock for quantum speedup.

For two qubits, the maximally entangled states are the famous \textit{Bell states}:

\begin{align}
\begin{split}
\ket{\psi^+} =
\frac{1}{\sqrt2}
\ket{01} + \ket{10},
&\quad
\ket{\psi^-} =
\frac{1}{\sqrt2}
\ket{01} - \ket{10},
\quad\\
\ket{\phi^+} =
\frac{1}{\sqrt2}
\ket{00} + \ket{11},
&\quad
\ket{\phi^-} =
\frac{1}{\sqrt2}
\ket{00} - \ket{11}.
\end{split}
\label{bellstates}
\end{align}

It can be readily seen that the outcome of a (computational basis) measurement done on one of the qubits fully determines the outcome of a measurement on the other one.

\subsubsection{Mixed States}
Most of what was discussed so far only holds for \textit{pure states} - those that are known exactly, and can be written in the form of equation \ref{eq:purebloch}. During a computation, one would ideally deal with pure states exclusively. Unfortunately, this is never the case: there are unknown errors occurring during the state preparation and the computation itself. The qubits are not fully isolated, as there is always some coupling with the environment. This results in the system being in a \textit{mixed state}, a statistical distribution over pure states. This mixture simply translates lack of knowledge about the system. 

States that are not pure cannot be written as in equation \ref{eq:purebloch} and cannot be represented in the surface of the Bloch sphere. The \textit{density matrix} $\rho$ provides a way to describe a system in a more general state as

\begin{equation}
\rho = \sum_i p_i\ket{\psi_i}\bra{\psi_i}.
\label{eq:densitymatrix}
\end{equation}

Here the $p_i$ are the probabilities of occurrence of the respective $\ket{\psi_i}$, the possible pure state configurations. The density matrix offers a way to test the purity of the state:

\begin{align}
\begin{split}
\text{tr}(\rho^2) = 1 &\longrightarrow \text{state is pure};\\
\text{tr}(\rho^2) < 1 &\longrightarrow \text{state is mixed}.
\end{split}
\label{purity}
\end{align}

There is also a way of generalizing the geometrical representation of the states: the Bloch \textit{ball}. We rewrite the density matrix representing the state as

\begin{equation}
\rho = \frac{I + \vec{r}\cdot\vec{\sigma}}{2},
\label{eq:generalbloch}
\end{equation}

where $\vec{\sigma} = (X,Y,Z)$. The three-dimensional vector $\vec{r}$, with norm lower than or equal to unity, is the Bloch vector whose coordinates are written in \ref{eq:blochcoordinates}.

\begin{equation}
\left(\sum_ip_ix_i,\sum_ip_iy_i,\sum_ip_iz_i\right)
\label{eq:blochcoordinates}
\end{equation}

The sum is over the possibilities of pure states, with $p_i$ their probabilities and $(x_i,y_i,z_i)$ their coordinates on the Bloch sphere. It is easy to verify that, for the special case of pure states, the norm of the Bloch vector $\vec{r}$ is 1 and the coordinates are as before. For mixed states, the norm is in $[0,1[$: they are \textit{inside} the Bloch sphere, in the Bloch \textit{ball}. The lower the purity, the lower the norm of  $\vec{r}$. The totally mixed state lies in the center of the Bloch sphere, representing a complete lack of knowledge about the system.

\subsection{Measuring Pauli Strings}
\label{ss:measuringpaulis}

As it was mentioned before, one can perform a $Z$ measurement on a qubit. This yields either of the operator's eigenvalues, with the respective probabilities depending on the state of the qubit. Repeated measurements allow obtaining the expectation value of this operator in this specific state, namely

\[\bra{\psi}Z\ket{\psi}.\]

It is common for problems to require measurements in bases other than the computational basis; this can be done with \textit{basis rotations}. Measuring a generic observable amounts to performing a computational basis measurement, preceded by the unitary that rotates from the eigenbasis of the desired observable to the computational basis.

Evidently, in order to be appropriate for an observable $A$, the unitary $U$ needs to obey the condition

\[\bra{\psi}U^\dagger ZU\ket{\psi} = \bra{\psi}A\ket{\psi}\]

for all states $\ket{\psi}$, so that the requirement can be written simply as

\begin{equation}
U^\dagger ZU=A.
\label{eq:basisrotation}
\end{equation}

This generalizes for observables acting on multiple qubits.

Implementing the basis rotation unitary for a generic observable may prove to hinder efficiency. Fortunately, \textit{Pauli measurements} in particular are very simple to do. 

\begin{figure}[htbp]
\centering
\begin{subfigure}[b]{0.4\textwidth}
\includegraphics[width=0.9\linewidth]{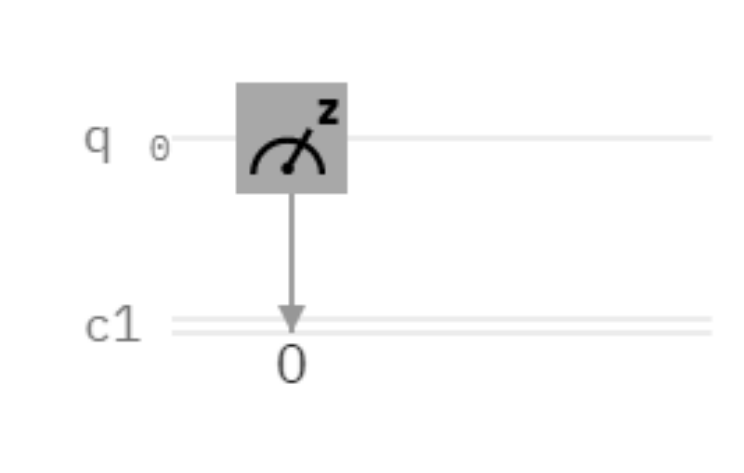} 
\caption{Circuit to measure $Z$.}
\label{fig:z_meas_circuit}
\end{subfigure}

\begin{subfigure}[b]{0.4\textwidth}
\includegraphics[width=0.9\linewidth]{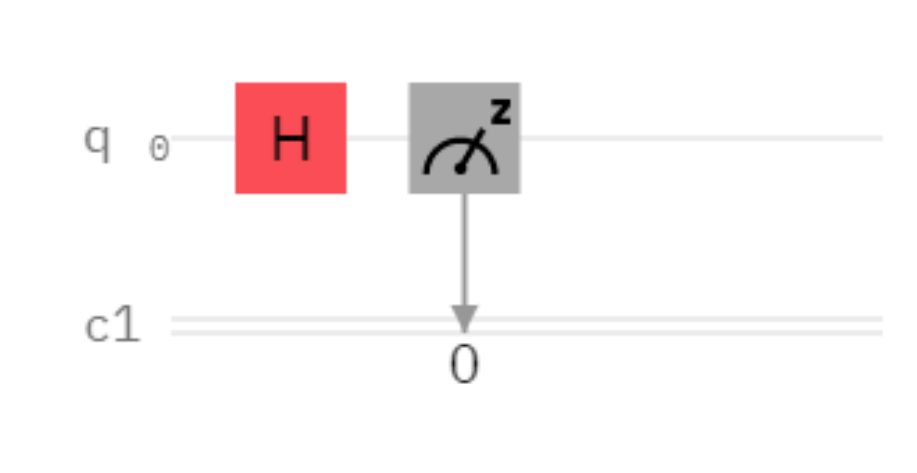} \caption{Circuit to measure $X$.}
\label{fig:x_meas_circuit}
\end{subfigure}
\hfill
\begin{subfigure}[b]{0.4\textwidth}
\includegraphics[width=0.9\linewidth]{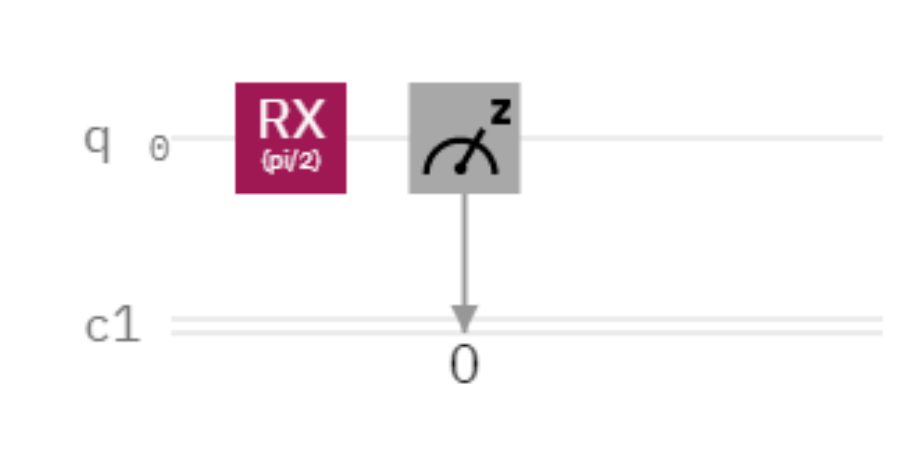} \caption{Circuit to measure $Y$.}
\label{fig:y_meas_circuit}
\end{subfigure}
\caption{Examples of circuits that perform single-qubit Pauli measurements, drawn in the IBM Quantum Composer \cite{IBMcomposer}.}
\label{fig:pauli_meas_circuits}
\end{figure}

Circuits for all three types of single-qubit Pauli measurements ($Z$,$X$,$Y$) are drawn in figure \ref{fig:pauli_meas_circuits}. The circuit in \ref{fig:z_meas_circuit} implements the typical computational basis measurement: no basis rotation is necessary. For $X$ we can use the unitary $U=H$ as the basis rotation - this is the popular Hadamard gate. The circuit is represented in figure \ref{fig:x_meas_circuit}. Finally, figure \ref{fig:y_meas_circuit} implements a measurement of $Y$, with the basis rotation being a rotation around the \textit{x} axis $U=\text{Rx}(\pi/2)$. 

These choices of unitaries are not unique. As an example, using $U=HS^\dagger$ for measuring $Y$ is also a popular option. One can choose the rotation that is the most convenient, given the operations that can natively be implemented in the hardware. 

\FloatBarrier
As well as single qubit Pauli measurements, one can easily measure \textit{Pauli strings}. Pauli strings are observables of the form

\begin{equation}
    \bigotimes\sigma_i \quad i\in{\{0,z,x,y\}}.
    \label{def:paulistring}
\end{equation}

There are multiple ways to perform such measurements. The simplest and most evident one amounts to performing the individual Pauli measurements and multiplying the outcomes. However, it is possible to include two qubit gates in addition to the single qubit rotations as a means of reducing the number of measurements that needs to be performed.

\begin{figure}[htbp]
\centering
\begin{subfigure}[b]{0.4\textwidth}
\includegraphics[width=0.9\linewidth]{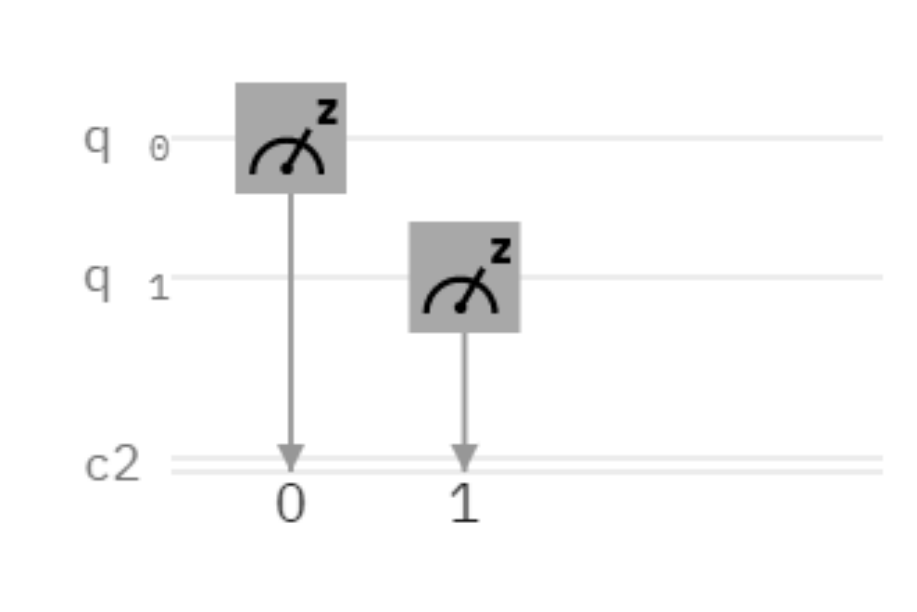} \caption{Using two single-qubit measurements.}
\label{fig:zz_2meas_circuit}
\end{subfigure}
\hfill
\begin{subfigure}[b]{0.4\textwidth}
\includegraphics[width=0.9\linewidth]{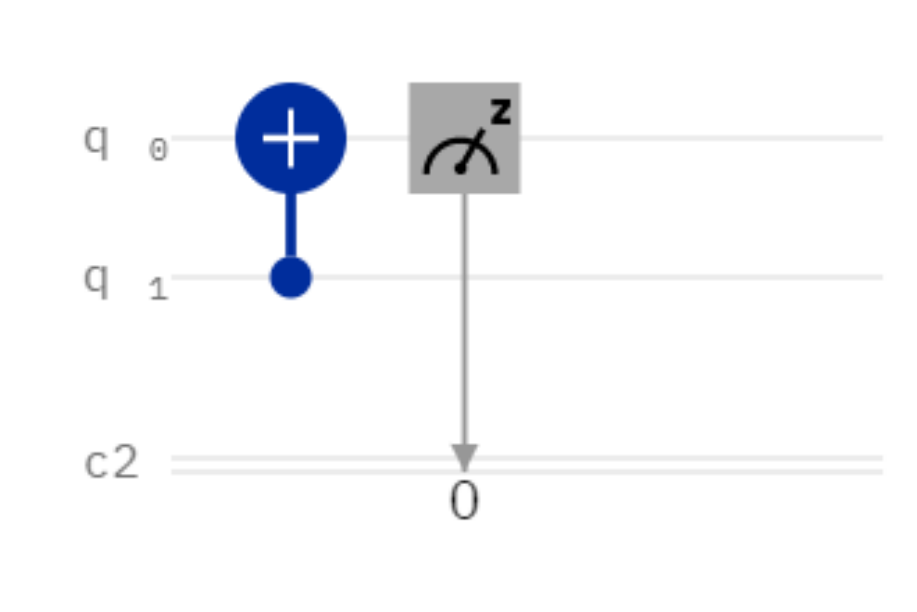} \caption{Using one single-qubit measurement.}
\label{fig:zz_1meas_circuit}
\end{subfigure}
\caption{Examples of circuits that perform a $Z\otimes Z$ measurement, drawn in the IBM Quantum Composer \cite{IBMcomposer}.}
\label{fig:zz_meas_circuits}
\end{figure}

As an example, two versions of circuits to perform a $Z\otimes Z$ measurement are illustrated in figure \ref{fig:zz_meas_circuits}.

The circuit in \ref{fig:zz_2meas_circuit} measures both qubits in the computational basis; the result is then obtained by multiplying the outcomes. The final value will be in $\{-1,1\}$. The actual value depends on the \textit{parity} of the computational basis state that was measured: we obtain 1 if the parity was even, meaning that the qubits were measured on the same state ($\ket{00}$ or $\ket{11}$). The outcome is -1 if the parity was odd, meaning that they were measured on opposite states ($\ket{01}$ or $\ket{10}$).

In contrast, the circuit in \ref{fig:zz_1meas_circuit} performs only one measurement. A \gls{CNOT} gate is used to compute and store the parity of the state of both qubits in the state of the first one alone.

Essentially, for a two-qubit observable $A$, we can use an unitary that obeys

\[
U^\dagger(Z\otimes Z)U=A
\]

and measure both qubis, or use an unitary that obeys

\[
U^\dagger(Z\otimes I)U=A
\]

and measure only one of them.

With this, it has become clear that any Pauli string, i.e. operator of the form of equation \ref{def:paulistring}, can be measured in a quantum computer. This is of great importance, because these Pauli strings span the space of Hermitian operators, and all observables are Hermitian. As a consequence, all observables can be written as a linear combination of Pauli strings; and given the linearity of quantum mechanics, all \textit{expectation values} of observables can be obtained as a weighed sum of expectation values of Pauli strings. 

If we want to measure an observable $\hat{O}$, we can start by decomposing it in the form of equation \ref{eq:pauli_decomp}. 

\begin{equation}
    \hat{O}=\sum_ih_i\hat{P}_i
    \label{eq:pauli_decomp}
\end{equation}

Here the $h_i$ are real coefficients and the $\hat{P}_i$ are Pauli strings (i.e. observables of the form of \ref{def:paulistring}). It is then straightforward to evaluate the desired expectation value: it amounts to measuring the expectation value of each Pauli string appearing in the decomposition of the observable, and plugging it into formula \ref{eq:arbitrary_expectation}.

\begin{equation}
    \langle\hat{O}\rangle=\sum_ih_i \langle\hat{P}_i\rangle
    \label{eq:arbitrary_expectation}
\end{equation}

This procedure can be followed for any observable; however, it is only tractable if the observable can be decomposed into a sum of polynomially-many Pauli strings.

\subsection{Quantum Simulation and Trotterization}
\label{ss:simulation_trotterization}

The behaviour of an isolated quantum-mechanical system is dictated by the \textit{Schrödinger equation} in \ref{eq:schrodinger}.

\begin{equation}
    i\hbar\frac{d}{dt}\ket{\psi}=H\ket{\psi}
    \label{eq:schrodinger}
\end{equation}

When the Hamiltonian operator for the system, $H$, is time-independent, this implies that the time evolution of the wave function has the form presented in \ref{eq:timeevolution}.

\begin{equation}
    \ket{\psi(t)} = e^{-iHt}\ket{\psi(0)}
    \label{eq:timeevolution}
\end{equation}

As it was mentioned before, the simulation of physical systems is one of the most promising applications of quantum computers, as it is generally a hard task for classical computers. As such, it is important to know how to simulate this time evolution on a quantum computer.

A general Hamiltonian is not easy to exponentiate - so the question becomes, \textit{how do we approximate the time evolution with sufficient accuracy?}

Typically, one deals with special classes of Hamiltonians that allow easy and efficient exponentiation. 

A Hamiltonian can be decomposed into a sum of several, simpler terms:

\begin{equation}
    H = \sum_{k=1}^{L}H_k.
    \label{eq:Hamiltoniandecomposition}
\end{equation}

And the $H_k$ may allow easier implementation on a quantum computer - for example, each one may act only on a small subsystem, or they may have a special form.

In fact, as was discussed in subsection \ref{ss:measuringpaulis}, we know that any Hermitian operator (such as the Hamiltonian) can be decomposed into a linear combination of Pauli strings. Conveniently, when the $H_k$ are Pauli strings with real coefficients, it is easy to construct the circuit that applies $e^{-iH_kt}$.

\begin{figure}[htbp]
\centering
\begin{subfigure}[b]{0.8\textwidth}
\includegraphics[width=0.9\linewidth]{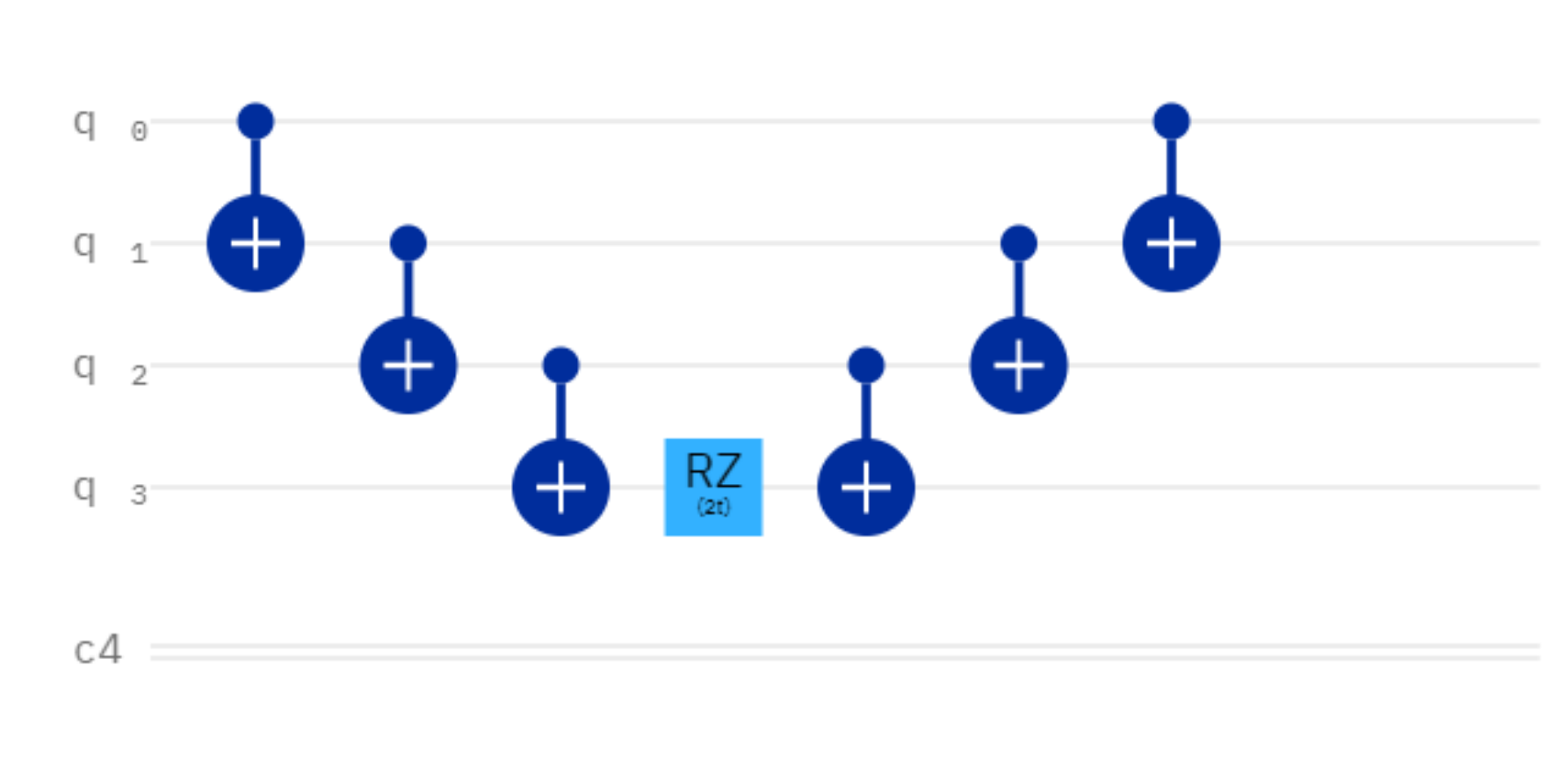} \caption{Circuit for simulating the time evolution of a system under the Hamiltonian $H_k = Z\otimes Z\otimes Z \otimes Z$.}
\label{fig:ZZZZ_exponentiation}
\end{subfigure}

\begin{subfigure}[b]{0.8\textwidth}
\includegraphics[width=0.9\linewidth]{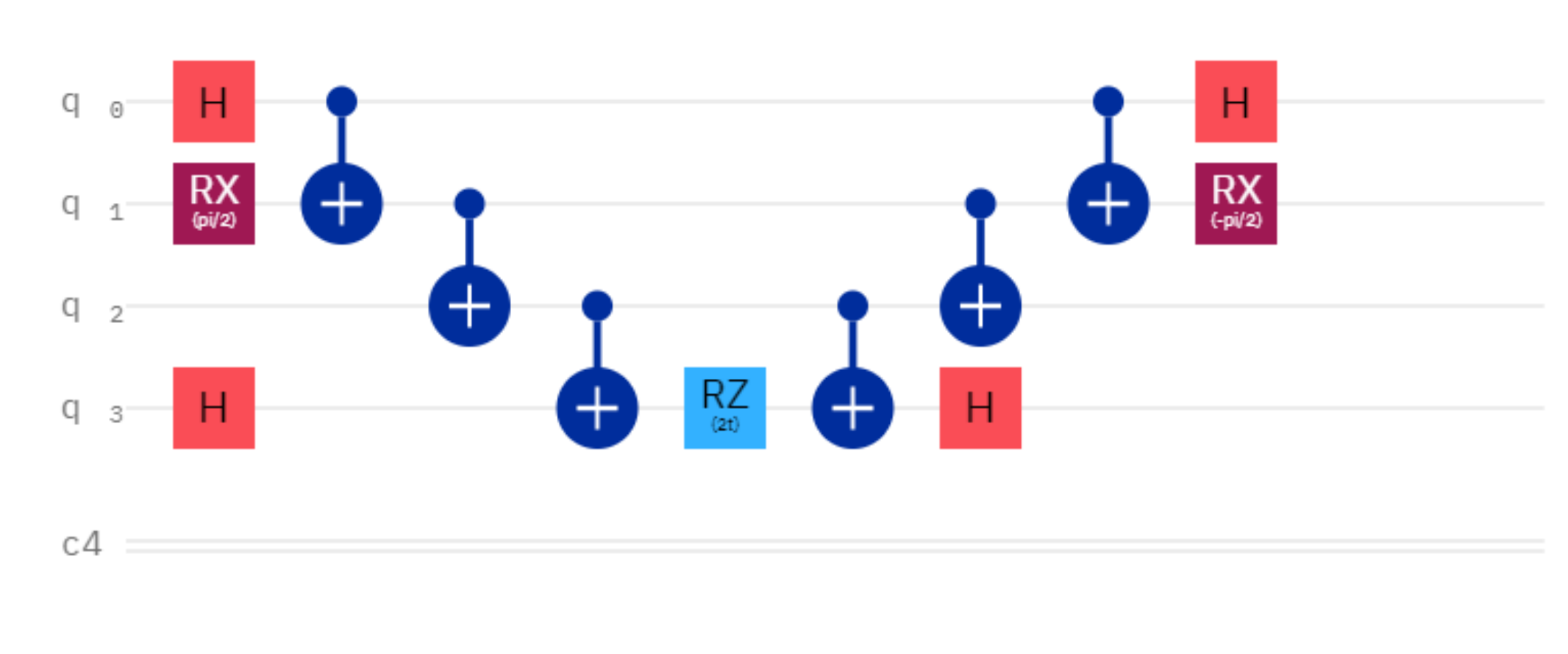} \caption{Circuit for simulating the time evolution of a system under the Hamiltonian $H_k = X\otimes Y\otimes Z\otimes X$.}
\label{fig:XYZX_exponentiation}
\end{subfigure}
\caption{Examples of circuits that apply time evolution under simple Hamiltonians $H_k$, consisting of Pauli strings. They apply $e^{-iH_kt}$ using a parameterized $Z$ rotation. The circuits were drawn in the IBM Quantum Composer \cite{IBMcomposer}.}
\label{fig:pauli_exponentiation}
\end{figure}

Figure \ref{fig:pauli_exponentiation} shows two circuits that can be used to implement the exponentiation of two specific $H_k$.

The circuit in figure \ref{fig:ZZZZ_exponentiation} is the simplest one: it implements $e^{-itZ\otimes Z\otimes Z\otimes Z}$. This amounts to applying a local phase $e^{-it}$ to the computational basis states that have an even parity, and $e^{it}$ to those that have an odd parity; the drawn circuit does precisely that. The first ladder of \gls{CNOT} gates computes the parity of the state and stores it on the last qubit. Then, all that's left is to apply $e^{-itZ}$ to the qubit that holds the parity, which can be done by rotating it around the $Z$ axis by an angle of $2t$ (i.e. applying the gate $\text{Rz}(2t)$). Finally, a ladder of \gls{CNOT} gates is applied to uncompute the parity before the computation proceeds.

In figure \ref{fig:XYZX_exponentiation}, we have the circuit that implements $e^{-itX \otimes Y\otimes Z\otimes X}$. This can be done with minor adjustments to the previous circuit.

Appendix \ref{ap:exponentiation} shows that, when $H_k$ is a Pauli string, one can write any operator of the form $e^{itH_k}$ as

\begin{equation}
e^{-i t\bigotimes_iP_i}\\
=\bigotimes_i U_i^\dagger\left( e^{-i t\bigotimes_iZ_i}\right) \bigotimes_i U_i,
\label{eq:general_exp_pauli}
\end{equation}

where the $U_i$ are single qubit unitaries obeying

\[U_i^\dagger Z_iU_i=P_i.\]

For any qubit $i$, acted on by $P_i$ in $H_k$.

This is the same condition as \ref{eq:basisrotation}: as we did to make Pauli measurements, we need an unitary that rotates from the eigenbasis of the operator in question to the computational basis. 

The form of equation \ref{eq:general_exp_pauli} is very convenient: it tells us that we can use the circuit for $e^{-i t\bigotimes_iZ_i}$ (figure \ref{fig:ZZZZ_exponentiation}) for implementing any operator of the form $e^{-i t\bigotimes_iP_i}$ (where $P_i\in\{I_i,Z_i,X_i,Y_i\}$), as long as we apply the appropriate $\bigotimes_i U_i$ before and $\bigotimes_i U_i^\dagger$ after.

The circuit in figure \ref{fig:XYZX_exponentiation} does precisely that, for the specific case of the Pauli string $\bigotimes_iP_i=X\otimes Y\otimes Z\otimes X$. The employed basis rotations were the same as those from section \ref{ss:measuringpaulis}.

So now we know how to simulate the time evolution under simple Hamiltonians $H_k$ consisting of a single Pauli string. However, the end goal is simulating the full Hamiltonian $H$ from \ref{eq:Hamiltoniandecomposition}. This amounts to applying $e^{-it\sum_{k=1}^{L}H_k}$, but

\[e^{-it\sum_{k=1}^{L}H_k}\neq \prod_{k=1}^Le^{-itH_k}\]

in general. In fact, for two operators $H_i$, $H_j$, we can calculate $e^{H_i}e^{H_j}$ as

\begin{equation}
    e^{H_i}e^{H_j}=e^{H_i + H_j + \frac{1}{2}[H_i,H_j]+\frac{1}{12}[H_i,[H_i,H_j]] - \frac{1}{12}[H_j,[H_i,H_j]]+...}.
\label{eq:BCH}
\end{equation}

Equation \ref{eq:BCH} is known as the \gls{BCH} formula. We can see that the product of the exponential series is the exponential series of the sums only in the special case of commuting operators ($[H_i,H_j]=0$).

Fortunately, we can use the \textit{Trotter formula}:

\begin{equation}
    \lim_{n\rightarrow{\infty}}(e^{iH_it/n}e^{iH_jt/n})^n = e^{i(H_i+H_j)t}.
\end{equation}

Evidently, it is not feasible to implement the circuit for $n\rightarrow{\infty}$. However, we can make $n$ finite to obtain a circuit that \textit{approximates} the desired time evolution. 

\begin{equation}
e^{i(H_i+H_j)t} \approx (e^{iH_it/n}e^{iH_jt/n})^n
\label{eq:trotterapprox}
\end{equation}

Here, $n$ is the number of \textit{repetitions} of the basic unit of our Trotter circuit, and $t/n$ is the \textit{time step}. We can interpret equation \ref{eq:trotterapprox} as approximating the evolution of a system under $(H_i+H_j)$ for a time $t$ with $n$ repetitions of the evolution under $H_j$ followed by the evolution under $H_i$, both always for a time step of $t/n$. We switch between evolving the system under the two elements in the sum repeatedly, with the duration of this evolution being always less than the actual evolution time.

The larger the number of repetitions $n$, the shorter the time step and the better the approximation: in the limit of $n\rightarrow{\infty}$, one would get the exact result. The error in the approximation will depend on the evolution time; as an example, a Trotter approximation with a single repetition will have an error $\mathcal{O}(t^2)$ \cite{NielsenChuang}.

\subsection{Current Limitations}
\label{ss:qc_limitations}

As discussed in the previous chapter, it will likely be a while before a fault-tolerant quantum computer (FTCQ) is available. Currently, one has to deal with noisy intermediate-scale quantum (\gls{NISQ}) devices. These are not error corrected and come with a multitude of limitations, namely:

\begin{itemize}
    \item Coherent noise.
    \item Incoherent noise.
    \item \gls{SPAM} errors.
\end{itemize}

A brief analysis of each follows.

\subsubsection{Coherent Noise}

Coherent errors are unitary and don't undergo fast changes (relative to the gate time); this type of error has various causes, such as systematic control noise, global external fields, cross-talk, and unwanted qubit-qubit interactions \cite{Greenbaum2017}.

For a single qubit, a coherent source of noise could be performing a rotation by an angle of $\theta+\delta\theta$ instead of just $\theta$ due to miscalibration. For two qubits, a well-known example is the unwanted $ZZ$ interaction in superconducting quantum computers. It can significantly lower the fidelity of two-qubit gates; as such, it can have a great impact on the performance of the device, and consequently pose a challenge on the scalability of the architecture \cite{Zhao2020}.

Coherent errors interfere constructively, and may take a significant toll on the computation. In the worst case scenario, the error rate of coherent noise can scale quadratically both in the number of qubits and in the length of the circuit \cite{Sheldon2016,Iverson2020}. For that reason, it can accelerate the error accumulation rate and be a challenge in the context of error correction codes. Regardless, there is hope that error correction can be used to remove undesirable coherence from the noise channel, thus avoiding the possibility that constructive interference leads to a quadratic scaling of the error rate. For the toric code and a particular coherent noise model, it was shown that the logical channel after error correction becomes increasingly incoherent with the length of the code \cite{Iverson2020}.

\subsubsection{Incoherent Noise}

Unlike coherent noise, incoherent noise can compromise the purity of the state of the qubits, causing it to become mixed. Incoherent errors may arise for example from microscopic materials defects or unwanted interactions with the environment \cite{kim2021,Lisenfeld2016}. Since it is impossible to have the qubits in perfect isolation and quantum mechanical systems are fragile, even \textit{storing} quantum information is a difficult task. 

While coherent errors are unitary and preserve the norm of the Bloch vector, incoherent errors can \textit{shrink} the Bloch vector. This is a visual translation of the fact that we've lost information about the state the qubits are in; its purity has decreased. Decoherence can destroy superposition and entanglement and cause us to lose the quantum behaviour of the system, with the information becoming classical \cite{preskill2021}.

When the errors are incoherent, the worst case scenario error rate scales linearly with the circuit size (depth and qubits), an improvement against the quadratic scaling of coherent errors \cite{Sheldon2016,Iverson2020}. 

\subsubsection{State Preparation and Measurement Errors}

State preparation and measurement errors are both classical errors, know collectively as \glsxtrshort{SPAM} errors. 

A quantum computer needs to be able to prepare a fiducial initial state, such as the all-zero state; and in order to obtain results from the computation, measurements will be necessary as well. Initialization and measurement are thus two fundamental non-unitary operations that bridge quantum and classical information, and allow quantum computing to be of use in our classical world.

However, these operations can't be done perfectly either. The prepared state will likely not be exactly the one that was intended, and the measurement operation can be faulty as well.

As \gls{SPAM} errors only occur at the beginning and end of the circuit, they evidently don't scale with the circuit depth, which makes them less disruptive. In spite of that, this is a relevant source of noise for near-term quantum computers, and efforts have been done to find efficient ways of partly mitigating it \cite{geller2020}.

\subsubsection{Other Challenges}

As well as noise, \gls{NISQ} computers come with severe limitations in qubit count and connectivity. The qubit requirements of the most promising applications are beyond what is available today; and limited connectivity demands extra routing operations or teleportation protocols, resulting in an overhead in ancillary qubits, execution time, and gate depth - each of which aggravates already critical resource requirements.

An additional challenge is that there is always some error when using a quantum computer to evaluate expectation values. This is not limited to \gls{NISQ} computers, it is an intrinsic limitation rooted in the nature of quantum mechanics. The error is associated with quantum projection noise and the always finite number of shots, and it will always exist, even if its magnitude can be decreased. In algorithms that include optimizing an expectation value, attention must be paid to this unavoidable source of noise that hampers the task of the optimizer. 

\subsubsection{Outlook}

The performance of the quantum computers available today greatly suffers from coherent and incoherent noise, \gls{SPAM} noise and the always-present shot noise.

Even if making a strong simulation of the quantum computer is hard for classical computers, producing a better approximate solution to a problem may not be so. As such, all sources of noise that contribute to corrupting the output of the computation may inhibit quantum advantage.

Errors are by no means exclusive to quantum computers; classical computations also suffer from them. Regardless, a relevant difference resides in the fact that classical information is much easier to protect and manipulate than quantum information, and error correction itself is a simpler task to perform on classical data. The viability of classical computations had been proven by as soon as 1956, with a key result by von Neumann showing that classical computers could be made robust to noise \cite{Neumann1956}.

The noise in quantum computers, along with the increased difficulty of error correction that arises in dealing with quantum data, brought concern that noise robustness was simply unattainable in quantum computers \cite{Landauer1996,Unruh1995}. Optimism was brought by the introduction of several error correction codes \cite{Shor1995,Steane1996}, followed by several proofs (comprising several error models and error correction codes) of an analogue of the result obtained by von Neumann for classical computers: the \textit{quantum threshold theorem}.

According to the quantum threshold theorem, once the physical error rate becomes smaller than a certain threshold, quantum computations can be made arbitrarily accurate by scaling up the error correction code \cite{Cai2020,Knill1998,Aharonov2008,Aliferis2006}. This provides a path to fault-tolerant quantum computing: once the hardware error rate is under the accuracy threshold for a given quantum error correction code, fault-tolerance has been achieved.

\section{Quantum Chemistry}
\label{s:quantum_chemistry}

This section aims at providing a light review of the aspects of quantum chemistry essential for the contents of the dissertation. The book used as a reference, Modern Quantum Chemistry \cite{SzaboOstlund}, offers a broader and more in-depth view of the field.

\subsection{The Electronic Problem}

We are interested in studying electronic structure and the quantum mechanical \textit{motion} of molecular systems. The objective is to understand the stable configurations of the system and its dynamics, by considering the interactions between the elements it is comprised of.

\begin{figure}[htbp]
    \centering
    \includegraphics[width=0.9\textwidth]{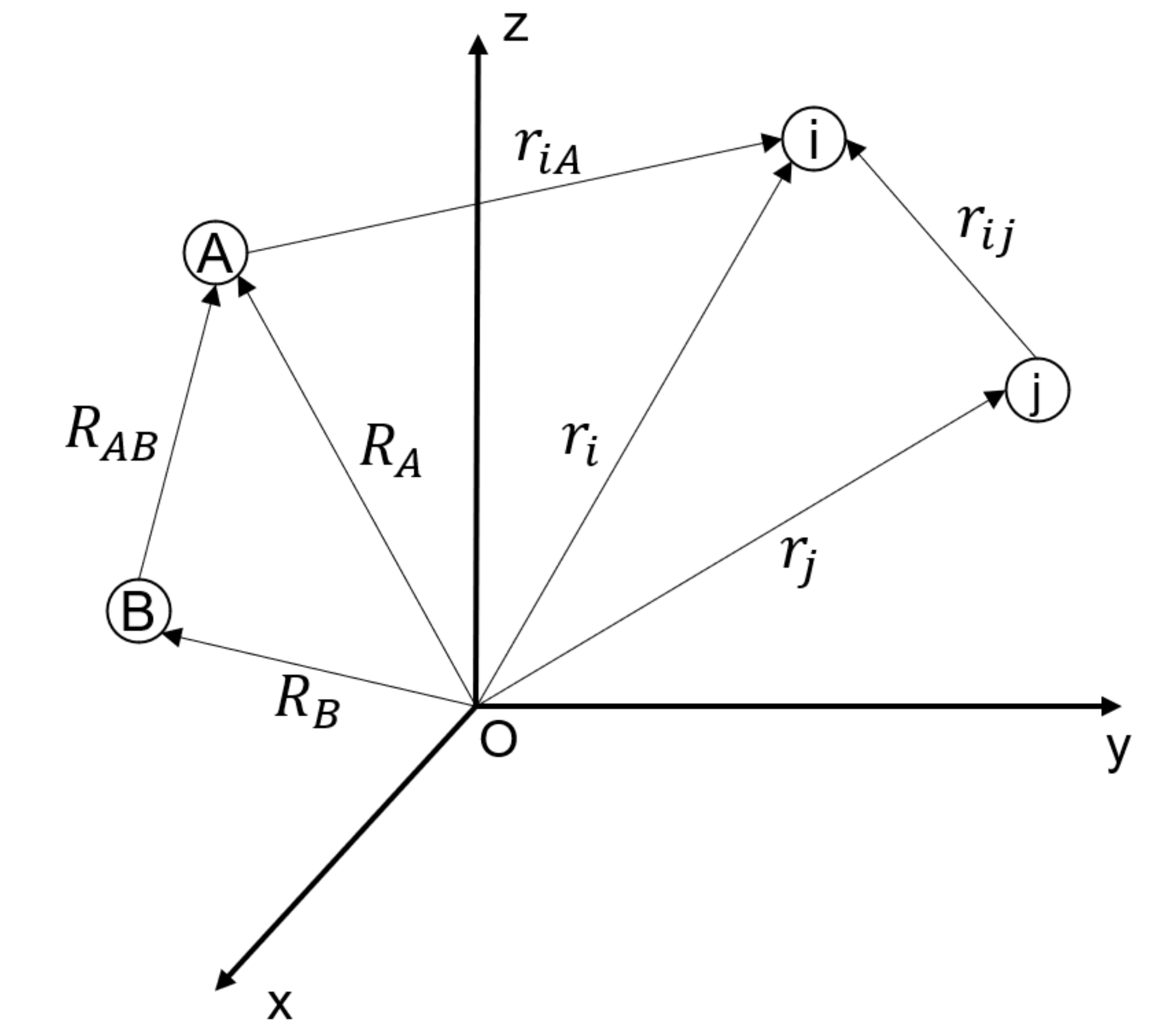}
    \caption{Representation of a molecule in a coordinate system. Upper case letters ($A$, $B$) represent nuclei, while lower case ones (i,j) represent the electrons. Similarly, upper case $R$ represent position vectors pointing to nuclei, while lower case $r$ represent those pointing to electrons. A single subscript is used on the position vector when the vectors point from the origin, while two subscripts are used to represent relative position vectors. In the latter case, the first and second subscripts indicate where the vector points to and from, respectively (e.g. $r_{ij}=r_i-r_j$).}
    \label{fig:molecular_system}
\end{figure}

A scheme of a simple molecule, at the level of electrons and nuclei, is presented in figure \ref{fig:molecular_system}. 

We are interested in finding (approximate) solutions to equation \ref{eq:schrodinger_time_ind}, the non-relativistic time-independent Schrödinger equation.

\begin{equation}
    \hat{H}\ket{\psi}=E\ket{\psi}
    \label{eq:schrodinger_time_ind}
\end{equation}

This is an \textit{eigenvalue problem}.

Here $\hat{H}$ is the Hamiltonian operator for our molecular system. It is comprised of five terms; a brief discrimination of each follows. Atomic units are to be used throughout the whole section, so that the electronic mass and the reduced Planck constant $\hbar$ are both equal to unity. The energy will then be in Hartrees (1a.u. = 27.2eV). $N$ and $M$ stand respectively for the number of electrons and  the number of nuclei. The notation regarding position vectors can be understood from figure \ref{fig:molecular_system}. 

\subsubsection{Kinetic Energy of the Electrons}

The electronic kinetic energy is the sum of the kinetic energies of the individual electrons,

\[
-\sum_{i=1}^N\frac{1}{2}\nabla_i^2.
\]

The simple multiplicative factor is due to the fact that atomic units were used. Here $\nabla_i^2$ is the \textit{Laplacian operator}, involving the second partial derivatives with respect to the coordinates of the electron labeled $i$:

\[
\nabla_i^2 = \pdv[2]{}{x}+\pdv[2]{}{y}+\pdv[2]{}{z}.
\label{def:Ekinetic}
\]

This term represents the energy associated with the motion of all the electrons in the molecule.

\subsubsection{Kinetic Energy of the Nuclei}

The term associated with the motion of the nuclei, written in \ref{def:Nkinetic}, is similar to the previous one.

\begin{equation}
-\sum_{A=1}^M\frac{1}{2M_A}\nabla_A^2
\label{def:Nkinetic}
\end{equation}

The only difference is that this term comes weighed by a factor of $\frac{1}{M_A}$, of the order of $10^4$. This factor, defined in \ref{def:M_a}, simply accounts for the fact that the mass of the nucleus is different from the mass of an electron.

\begin{equation}
M_A=\frac{\text{mass of the nucleus $A$}}{\text{mass of an electron}}
\label{def:M_a}
\end{equation}

\subsubsection{Electron-Nucleus Attraction}

What we have left now is the potential energy terms of the Hamiltonian, which include a negative term (attraction) and two positive terms (repulsion). 

The attraction term represents the interaction between molecules of opposite charge, as per Coulomb's law; it reads

\begin{equation}
-\sum_{i=1}^N\sum_{A=1}^{M}\frac{Z_A}{r_{iA}}.
\label{def:ENattraction}
\end{equation}

Here, $Z_a$ is the atomic number of the nucleus $A$. The greater the number, the greater the charge, and the greater the attraction; the larger the distance between the nucleus and the electron $r_{iA}$, the smaller the attraction.

\subsubsection{Electron-Electron Repulsion}

This is a positive term, arising from the repulsion between particles of identically negative charge. The corresponding expression is

\begin{equation}
\sum_{i=1}^{N}\sum_{j>i}^{N}\frac{1}{r_{ij}}.
\label{def:EErepulsion}
\end{equation}

Since electrons have unit charge, the only factor here is the distance between them ($r_{ij}$). The double sum runs over all distinct pairs of electrons.

\subsubsection{Nucleus-Nucleus Repulsion}

Finally, we have the positive term accounting for the repulsion between positively-charged particles:

\begin{equation}
\sum_{A=1}^M\sum_{B>A}^M\frac{Z_AZ_B}{R_{AB}}.
\label{def:NNrepulsion}
\end{equation}

\subsubsection{Complete Hamiltonian and the Born-Oppenheimer Approximation}

Contemplating all terms, the Hamiltonian of the system reads

\begin{equation}
    \hat{H} = -\sum_{i=1}^N\frac{1}{2}\nabla_i^2 - \sum_{A=1}^M\frac{1}{2M_A}\nabla_A^2 - \sum_{i=1}^N\sum_{A=1}^{M}\frac{Z_a}{r_{iA}} + \sum_{i=1}^{N}\sum_{j>i}^{N}\frac{1}{r_{ij}} + \sum_{A=1}^M\sum_{B>A}^M\frac{Z_AZ_B}{R_{AB}}.
    \label{eq:hamiltonian_full}
\end{equation}

Unfortunately, solving the Schrödinger equation \ref{eq:schrodinger_time_ind} with the Hamiltonian in \ref{eq:hamiltonian_full} is not an easy task - the computational complexity grows quickly with the size of the molecule. Obtaining an exact solution is only possible in simple cases.

To address this issue, Born and Oppenheimer proposed an approximation in 1927 \cite{Born1927}. The idea behind the proposal was that, since the mass of the nuclei is significantly larger than that of the electrons, their motion will be much slower for a comparable kinetic energy. Thus, considering the nuclei to be stationary seems like a reasonable approximation.

The Born-Oppenheimer approximation then regards the motion of the electrons as occurring in a field created by the \textit{fixed} nuclei. In this scenario, because they are not moving ($\nabla_A^2=0$), the nuclei have null kinetic energy, so that the term \ref{def:Nkinetic} in the Hamiltonian vanishes. Further, the nucleus-nucleus repulsion \ref{def:NNrepulsion} becomes constant, because their relative position $R_{AB}$ is not changing. A constant term represents only a constant shift in the eigenvalues, and does not affect the eigenstates; as such, we can ignore it as we are solving the problem, and add it back later.

Now we have

\begin{equation}
    \hat{H}_e = -\sum_{i=1}^N\frac{1}{2}\nabla_i^2 - \sum_{i=1}^N\sum_{A=1}^{M}\frac{Z_a}{r_{iA}} + \sum_{i=1}^{N}\sum_{j>i}^{N}\frac{1}{r_{ij}}
    \label{eq:hamiltonian_electronic}.
\end{equation}

We call the Hamiltonian in \ref{eq:hamiltonian_electronic} \textit{electronic} Hamiltonian for obvious reasons: it now only concerns the motion of the electrons. The nuclear coordinates may affect what the solution to the Schrödinger equation looks like, but the solution itself will be an electronic wave function, that depends explicitly \textit{only} on the electronic coordinates.

The approximation allows treating the electrons and nuclei as two decoupled systems, which greatly simplifies the problem. 

\subsection{Finding the Ground State: Classical Approaches}

There is a fundamental principle in quantum mechanics which is widely used in computational chemistry, be the resources classical or quantum: the \textit{variational principle}. This principle states that the expectation value of the Hamiltonian is lower bounded by the eigenvalue corresponding to the ground energy ($E_0$), i.e.

\begin{equation}
    \bra{\psi}\hat{H}\ket{\psi}\geq E_0,
    \label{ineq:variational_principle}
\end{equation}

assuming that the state is normalized ($\braket{\psi} = 1$). A brief sketch of the proof follows.

We consider $\{\ket{\phi_\alpha}\}$ to be the set of eigenstates of the Hamiltonian operator $\hat{H}$ and denote the eigenvalue of $\ket{\phi_\alpha}$ by $E_\alpha$, i.e.

\[\hat{H}\ket{\phi_\alpha}=E_\alpha\ket{\phi_\alpha},\hspace{10pt}\alpha=0,1,2,...\]

The eigenvalues are assumed to be ordered ($\alpha<\beta\implies E_\alpha\leq E_\beta$). Since the operator $\hat{H}$ is Hermitian, the eigenvalues are real ($E_\alpha\in\mathbb{R}\hspace{5pt}\forall\alpha$) and the eigenstates are orthonormal ($\braket{\phi_\alpha}{\phi_\beta}=\delta_{\alpha,\beta}$).

We assume that this set of eigenstates is complete, so that we can write any state $\ket{\psi}$ in the eigenbasis of $\hat{H}$:

\[\ket{\psi}=\sum_\alpha c_\alpha\ket{\phi_\alpha}\]

With this, the expectation value of $\hat{H}$ reads (using $^*$ to denote complex conjugation)

\[\bra{\psi}\hat{H}\ket{\psi}=\sum_\alpha c^*_\alpha\bra{\phi_\alpha}\hat{H}\sum_\beta c_\beta\ket{\phi_\beta} = \sum_{\alpha,\beta}c_\alpha^*c_\beta \bra{\phi_\alpha}\hat{H}\ket{\phi_\beta}\]

\[= \sum_{\alpha,\beta}c_\alpha^*c_\beta E_\beta\bra{\phi_\alpha}\ket{\phi_\beta}=\sum_{\alpha,\beta}c_\alpha^*c_\beta E_\beta\delta_{\alpha,\beta}=\sum_{\alpha}c_\alpha^*c_\alpha E_\alpha=\sum_{\alpha}|c_\alpha|^2 E_\alpha.\]

It was previously stated that the eigenvalues are ordered, so that $E_\alpha\geq E_0 \hspace{5pt}\forall\alpha$; further, assuming that the state $\ket{\psi}$ is normalized, we have $\sum_\alpha |c_\alpha|^2=1$. With this we can conclude the proof:

\[\bra{\psi}\hat{H}\ket{\psi}=\sum_{\alpha}|c_\alpha|^2 E_\alpha\geq \sum_{\alpha}|c_\alpha|^2 E_0=E_0\]

Typically, the methods for finding an approximate solution to the ground state of the electronic system aim to minimize $\bra{\psi}\hat{H}\ket{\psi}$. They differ only in the \textit{variational form} of $\ket{\psi}$, a concept identical to that of an \textit{ansatz}. It is a guess that fixes the structure of the wave function, but only specifies it up to some tunable \textit{variational parameters}. Plugging some test parameters into the variational form, we obtain a fully specified wave function in which we can calculate the expectation value of $\hat{H}$. The task is then reduced to finding the best parameters: given the variational principle, this is just a minimization problem.

A more thorough overview of the variational principle can be found in reference \cite{griffiths1995}.

\subsubsection{Self-Consistent Mean Field Methods}

As much as the Born-Oppenheimer approximation greatly simplifies our eigenproblem, solving the Schrödinger equation with the electronic Hamiltonian (equation \ref{eq:hamiltonian_electronic}) exactly is still out of reach. The problematic term is now the electron-electron repulsion (expression \ref{def:EErepulsion}), which couples the coordinates of \textit{all} the electrons. Finding the exact solution implies dealing simultaneously with the coordinates of them all; decoupling these variables would allow solving the equation. Such is made possible by the \textit{mean-field approximation}.

The essence of the mean-field approximation resides in the simplification of the effect of the electrons in each other: instead of contemplating the sum of all the individual repulsion terms, we assume that each electron experiences an \textit{average} potential created by all the other electrons. Each one of them is considered to be moving in the \textit{mean field} created by the others.

This gives rise to the operator in equation \ref{eq:scf_operator}.

\begin{equation}
    \hat{f}(i) = -\frac{1}{2}\nabla_i^2 - \sum_{A=1}^{M}\frac{Z_a}{r_{iA}} + 
    v(i)
    \label{eq:scf_operator}
\end{equation}

This operator is just the electronic Hamiltonian in equation \ref{eq:hamiltonian_electronic}, separated for each electron $i$. The impact of the remaining electrons is contained within $v(i)$, an average potential.

At this point we have managed to separate the Schrödinger equation into single fermion equations: instead of one complicated many-body problem, we face multiple simple one-body problems:

\begin{equation}
    \hat{f}(i)\chi(\vec{x}_i)=\epsilon\chi(\vec{x}_i).
    \label{eq:scf_equation}
\end{equation}

This is another eigenvalue equation, thus its form is identical to that of equation \ref{eq:schrodinger_time_ind}. But it is much simpler: we now have $\hat{f}(i)$, a one-body operator, instead of the many-body operator $\hat{H}$. Instead of the many-body wave function $\ket{\psi}$, depending on the coordinates of all of the electrons, we have single-body wave functions (spin-orbitals) $\chi(\vec{x}_i)$, each of which depends solely on the coordinates of one of them.

If we have N electrons, we have N equations of the form of \ref{eq:scf_equation}. We want to find the eigenvalues $\epsilon$ and the eigenfunctions $\chi(\vec{x}_i)$ of each operator $\hat{f}(i)$. However, this set of equations is non-linear: for the $i$th electron this operator depends (through $v(i)$) on the wave functions of the other electrons $\chi(\vec{x}_j)$, which are the solutions of all the other N-1 equations. Thus, the solution must be obtained \textit{iteratively}.

The iterative method for solving the equation \ref{eq:scf_equation} is the \gls{SCF} method. This is because the convergence condition is the self-consistency of the field.

The \textit{variational parameters} are contained within the single-particle wave functions, the $\chi(\vec{x}_i)$. We start with an \textit{initial guess} for the values of these parameters. With a concrete set of single-particle wave functions, we are then in place to calculate the average potential that each particle $i$ feels ($v(i)$). Plugging the values in the expressions of the operator $\hat{f}(i)$, we get N single-particle equations of the form of \ref{eq:scf_equation}. These are much simpler to solve than the full Schrödinger equation. The obtained solutions are new single-particle wave functions $\chi(\vec{x}_i)$. From these we can calculate new average potentials $v(i)$, and solve for the eigenfunctions of the operator $\hat{f}(i)$ once again. The procedure is repeated until self-consistency is achieved, i.e. the fields $v(i)$ stop changing  and the trial functions $\chi(\vec{x}_i)$ used to calculate those fields are also the solutions to the equations written with them. At this point, the trial functions used to construct the operator $\hat{f}(i)$ are simultaneously its eigenvalues: the field is self-consistent.

\subsubsection{Variational Form: Hartree and Hartree-Fock Methods}

The previous discussion purposefully glossed over what the many-body wave functions look like, and how they are related to the one-body wave functions. While the essence of \gls{SCF} methods is similar, they differ significantly in this one key aspect: the \textit{ansatz}, i.e. the variational form we choose for the wave functions.

It seems like the simplest approach would be to consider that the many-body wave functions are just products of one-body wave functions:

\begin{equation}
    \psi(\vec{x}_1,\vec{x}_2,...,\vec{x}_{N-1},\vec{x}_N)=\chi_\alpha(\vec{x}_1)\chi_\beta(\vec{x}_2)...\chi_\pi(\vec{x}_{N-1})\chi_\rho(\vec{x}_N).
    \label{def:hartree_product}
\end{equation}

This is what Hartree proposed upon introducing an \gls{SCF} method. However, the Hartree method does not provide satisfactory results, because it does not account for a fundamental principle: \textit{the antisymmetry principle}.

The antisymmetry principle is a crucial axiom in quantum mechanics. It states that the many-body fermionic wave function must be antisymmetric under exchange of any two particles, which we can write as

\begin{equation}
    \psi(\vec{x}_1,...,\vec{x}_i,...,\vec{x}_j,...,\vec{x}_N)=-\psi(\vec{x}_1,...,\vec{x}_j,...,\vec{x}_i,...,\vec{x}_N).
    \label{eq:antisymmetry}
\end{equation}

Evidently, the Pauli exclusion principle is enforced by the antisymmetry condition \ref{eq:antisymmetry}. The condition is rooted in the \textit{indistinguishability} of particles. Exchanging indistinguishable, or \textit{identical}, particles must result in a physically equivalent state. Such only allows variability up to a global phase. There are two main types of indistinguishable particles: fermions and bosons, with complex phase factors $e^{i\pi}=-1$ and $e^{i2\pi}=1$ respectively. These factors give rise to the antisymmetric nature of fermionic wave function, and the symmetric nature of the bosonic wave function.

In either case, the simplest possible way of combining the individual wave functions, the Hartree product (definition \ref{def:hartree_product}), is inadequate. In the scenario we are concerned with (the fermionic one), this can be dealt with by using \textit{antisymmetrized products} instead.

\begin{equation} 
\begin{split}
\psi(\vec{x}_1,\vec{x}_2,...,\vec{x}_{N-1},\vec{x}_N) &=\\
&\begin{vmatrix}
\chi_1(\vec{x}_1) & \chi_2(\vec{x}_1) & ... & \chi_{N-1}(\vec{x}_1) & \chi_N(\vec{x}_1)\\
\chi_1(\vec{x}_2) & \chi_2(\vec{x}_2) & ... & \chi_{N-1}(\vec{x}_2) & \chi_N(\vec{x}_2)\\
\vdots & \vdots & \ddots & \vdots & \vdots\\
\chi_1(\vec{x}_{N-1}) & \chi_2(\vec{x}_{N-1}) & ... & \chi_{N-1}(\vec{x}_{N-1}) & \chi_N(\vec{x}_{N-1})\\
\chi_1(\vec{x}_N) & \chi_2(\vec{x}_N) & ... & \chi_{N-1}(\vec{x}_N) & \chi_N(\vec{x}_N)
\end{vmatrix}
\end{split}
\label{def:slater_determinant}
\end{equation} 

Wave functions of the form of \ref{def:slater_determinant} are typically called \textit{Slater determinants}. Matrix determinants are antisymmetric under exchange of any two rows or columns, making them a natural choice here. For bosons, one would use matrix \textit{permanents}.

Slater determinant ansätze were used in the \textit{Hartree-Fock} method, an updated version of the Hartree method that accounted for the antisymmetry principle. 

As it was mentioned before, the choice of ansatz is what shapes a specific \gls{SCF} method. The form of the operator $\hat{f}(i)$ as written in \ref{eq:scf_operator}, and its place in the algorithm, are common to several \gls{SCF} methods. The difference is in the potential $v(i)$: it will be \textit{specific} to the assumptions we make about the wave function, i.e. it depends on how our ansatz looks like. This operator is then used to write Schrödinger-like one-particle equations \ref{eq:scf_equation}, which means that the chosen ansatz greatly impacts the formulation of our eigenproblems.

In the Hartree-Fock method, the operator of the form of equation \ref{eq:scf_operator} is called the \textit{Fock operator}:

\begin{equation}
    \hat{f}(i) = -\frac{1}{2}\nabla_i^2 - \sum_{A=1}^{M}\frac{Z_a}{r_{iA}} + 
    v^{HF}(i)
    \label{eq:hf_operator}
\end{equation}

The key difference is that the $v^{HF}(i)$ term here has a superscript, referring to the fact that the potential in this operator is the \textit{Hartree-Fock} potential. The eigenvalue equations \ref{eq:scf_equation} defined through the Fock operator are the \textit{Hartree-Fock} equations. 

As an illustration of how the potential changes depending on the ansatz, a shallow, simplified comparison of the Hartree and Hartree-Fock potentials follows.

The method proposed by Hartree used the Hartree product as the variational form. This was associated with a simple potential, consisting of a local operator that accounted solely for the electron-electron repulsion. In contrast, the potential in Hartree-Fock method includes, in addition to this local operator, an \textit{exchange} operator. This is a non-local operator that simply results from choosing Slater determinants as wave functions: it is an artifact associated with the imposition of the anti-symmetry principle.

\subsubsection{Limitations of the Hartree-Fock Method}

The Hartree-Fock method simplifies the problem in multiple ways. Demanding that the solution is a Slater determinant is in itself a restriction that precludes exactness, as is typical of variational methods: the optimal solution is only guaranteed to be the best \textit{within the variational form}. What the Hartree-Fock method does is to find the best solution of Slater determinantal form. 

As was discussed before, the motion of the nuclei is neglected (Born-Oppenheimer approximation). Further, the method also assumes the mean-field approximation, which fails to properly account for electron correlation outside of antisymmetry. 

Finally, even if none of these approximations were used, the exact solution would still be beyond reach, because for the method to be tractable the basis set must be finite. For the Slater determinants to constitute a complete N-electron basis, they would have to be built from an infinite basis set of spin-orbitals.

Regardless, the Hartree-Fock method is a cornerstone in quantum chemistry, and its approximate solution is often used as a starting point in more accurate methods, both in classical and quantum computing.

\subsubsection{Post-Hartree-Fock Methods}

Given the shortcomings of the Hartree-Fock solution, a multitude of more accurate methods has been proposed to improve upon the Hartree-Fock approximation (at the expense of extra computational cost). 

The following abbreviated explanation of post-Hartree-Fock methods is based, not only on the main reference for this section (\cite{SzaboOstlund}), but also on references \cite{romero2018} and \cite{Harsha2018}.

Several relevant simplifications assumed by the Hartree-Fock method were mentioned previously. Of these, one typically undone by posterior methods is the assumption that the solution is of the form of a Slater determinant.

To correct for the electronic correlation ignored by the Hartree-Fock method, a typical approach is to expand the wave function as a \textit{linear combination of} Slater determinants, rather than a single one of them. Since the Hartree-Fock solution is a good starting point, the wave function is often described in terms of \textit{excitation operators} that will act on this state.

It is useful to define the \textit{cluster operator} $T$:

\begin{equation}
\begin{split}
T=\sum_{i=1}^N&T_i,\\
T_1=\sum_{i,a}&t_a^ia_a^\dagger a_i\\
T_2=\sum_{i>j,a>b}&t_{ab}^{ij}a_a^\dagger a_b^\dagger a_i a_j\\
&\vdots
\end{split}
\label{def:cluster_op}
\end{equation}

The lower case $t$ are expansion coefficients. The indices $i$, $j$ and $a$, $b$ run over orbitals that are occupied and unoccupied (virtual) in the reference state, respectively. The operator $a_k^\dagger$ is a \textit{creation} operator, which creates a fermion in orbital $k$; $a_k$ is an \textit{annihilation} operator, which removes a fermion from orbital $k$. These operators belong to the second quantization formalism. 

The cluster operator consists of a sum of all possible excitation operators $T_i$: the one that generates single excitations from the reference state ($T_1$), the one that generates double excitations ($T_2$), and so forth.

With this operator, it is possible to write the \gls{FCI} variational form in \ref{def:FCI_state}.

\begin{equation}
\ket{FCI}=(1+T)\ket{HF}
\label{def:FCI_state}
\end{equation}

This is just the Hartree-Fock state acted on by an operator $(1+T)$; the expansion coefficients will be the variational parameters. The term $T\ket{HF}$ consists of a linear combination of all possible excited determinants, since the cluster operator includes all possible excitation operators. The \gls{FCI} state is then a linear combination of all Slater determinants in the N-electron space formed from the chosen set of spin-orbitals. The optimal \gls{FCI} energy \ref{def:FCI_energy} will be the exact solution within the N-electron subspace spanned by these determinants. Of course, other than the finite basis set, the solution is also only exact up to the Born-Oppenheimer approximation.

\begin{equation}
E_{FCI}=\min_{\vec{t}}\frac{\bra{FCI}H\ket{FCI}}{\braket{FCI}}
\label{def:FCI_energy}
\end{equation}

In short, the inclusion of more Slater determinants increases the variational freedom and improves the solution. The name \textit{configuration interaction} arises from the fact that each Slater determinant in the expansion is associated with a specific electronic configuration (configuration of spin-orbitals). \textit{Full} refers to the fact that all excitation operators are included.

Unfortunately, while \gls{FCI} offers accurate results, it is only tractable for small molecules: the number of determinants that need to be included in the expansion grows with the factorial of the total number of spin orbitals. Consequently, even with minimal single electron basis sets, the difficulty of calculations increases rapidly as the size of the system increases.

To make the calculations tractable, the cluster operator $T$ must be truncated.

\begin{equation}
T^{(k)}=\sum_{i=1}^kT_i
\label{def:truncated_cluster_op}
\end{equation}

The operator $T^{(k)}$ defined in \ref{def:truncated_cluster_op} is just the cluster operator from definition \ref{def:cluster_op}, truncated up to excitations of order $k$. When a truncated version of the operator is used, only a portion of the Slater determinants formed from the one-electron basis set will appear in the expansion; the associated method is simply called  \gls{CI}. The variational form is written as in \ref{def:CI_state}.

\begin{equation}
\ket{CI}=(1+T^{(k)})\ket{HF}
\label{def:CI_state}
\end{equation}

For the specific case that $k$ is two, the method is called \gls{CISD}, because it includes only single and double excitations from the reference state (variational form in \ref{def:CISD_state}).

\begin{equation}
\ket{CISD}=(1+T_1+T_2)\ket{HF}
\label{def:CISD_state}
\end{equation}

Truncating the operator to make the computation tractable results in a few issues. Namely, if we have two independent, non-interacting subsystems (e.g. two infinitely separated molecules), the \gls{CI} energy of the full system will not be the sum of the energies of the subsystems. Because it lacks this property, it is said that \gls{CI} is not \textit{size-consistent}.

A size-consistent version can be created by means of exponentiation. The corresponding method is called \gls{CC}; the variational form is written in \ref{def:CC_state}.

\begin{equation}
\ket{CC}=e^T\ket{HF}
\label{def:CC_state}
\end{equation}

Once again, the cluster operator $T$ is typically truncated at some order of excitation. For the usual choice that excitations up to order two are included, the method is called \gls{CCSD} (variational form in \ref{def:CCSD_state}).

\begin{equation}
\ket{CCSD}=e^{T_1+T_2}\ket{HF}
\label{def:CCSD_state}
\end{equation}

Even when truncated, \gls{CC} methods don't suffer from the problem of size-inconsistency. Further, they have the interesting property that even when the cluster operator $T$ is truncated up to order $k$, higher-order excitations occur in the exponential series. However, they have some problems of their own. A relevant weakness in \gls{CC} is that the similarity-transformed Hamiltonian $\bar{H}$ for conventional \gls{CC} (definition \ref{def:CC_Hamiltonian}) is not variational.

\begin{equation}
    \bar{H}=e^{-T^{(k)}}He^{T^{(k)}}
    \label{def:CC_Hamiltonian}
\end{equation}

This is due to the \gls{CC} expectation value

\begin{equation}
    E_{CC}=\bra{HF}e^{-T^{(k)}}He^{T^{(k)}}\ket{HF}
    \label{eq:CC_energy}
\end{equation}

not being symmetric, since

\[
\bra{HF}e^{-T^{(k)}}\neq(e^{T^{(k)}}\ket{HF})^\dagger,
\]

which is just a consequence of the operator $e^{T^{(k)}}$ not being unitary. This means that we can't use the variational principle \ref{ineq:variational_principle}, because the associated expectation value isn't of the proper form. The asymmetry of the expectation value impedes the \gls{CC} energy from being an upper bound to the ground energy.

To circumvent this, \gls{VCC} has been proposed. This method aims to minimize

\begin{equation}
    E_{VCC} = \frac{\bra{HF}{{e^{T^{(k)}}}^\dagger}He^{T^{(k)}}\ket{HF}}{\bra{HF}{{e^{T^{(k)}}}^\dagger}e^{T^{(k)}}\ket{HF}}.
\end{equation}

Here, the denominator is not unity because of the non-unitarity of $e^{T^{(k)}}$. The \gls{VCC} method is obviously variational, because the numerator is of the form $\bra{\psi}H\ket{\psi}$; further, much like traditional \gls{CC}, it is size-consistent. Unfortunately, the cost of the \gls{VCC} method scales exponentially with the system size, regardless of the truncation order $k$.

Another approach to make a variational version of \gls{VCC} is \gls{UCC}, that replaces the operator $e^{T}$ with another exponential operator, now unitary:

\begin{equation}
    \ket{UCC}={e^{T-T^\dagger}}\ket{HF}.
    \label{def:UCC_operator}
\end{equation}

The Hermitian conjugate of the operator $(T-T^\dagger)$ reads

\[
(T-T^\dagger)^\dagger=T^\dagger-{T^\dagger}^\dagger=T^\dagger-T = -(T-T^\dagger),
\]

which means that this operator is anti-Hermitian, and the commutators between $(T-T^\dagger)$ and $(T-T^\dagger)^\dagger$ are all zero. This grants unitarity to the \gls{UCC} operator.

\[
\left(e^{T-T^\dagger}\right)\left(e^{T-T^\dagger}\right)^\dagger = \left(e^{T-T^\dagger}\right)^\dagger \left(e^{T-T^\dagger}\right) = I
\]

Since the \gls{UCC} operator is unitary, the method is variational, and the \gls{UCC} energy is an upper bound to the ground state energy. Finding the \gls{UCC} ground state then amounts to a minimization problem (expression \ref{def:UCC_energy}).

\begin{equation}
E_{UCC}=\min_{\vec{t}}\bra{HF}e^{-(T-T^\dagger)}He^{T-T^\dagger}\ket{HF}
\label{def:UCC_energy}
\end{equation}

As usual, the cluster operator should be truncated for the computation to be tractable, so that we get an \gls{UCC} operator of the form ${e^{T^{(k)}-{T^{(k)}}^\dagger}}$. For the frequent case that only excitations up to order two are included ($k$=2), the method is called \gls{UCCSD}. This corresponds to the variational form in \ref{def:UCCSD_operator}.

\begin{equation}
    \ket{UCCSD}={e^{(T_1+T_2)-(T_1^\dagger+T_2^\dagger)}}\ket{HF}
    \label{def:UCCSD_operator}
\end{equation}

The \gls{UCCSD} method is variational, size-consistent, and has more Slater determinants contributing to the wave function than CISD, even though the truncation order of $T$ is the same.

In spite of its benefits, the \gls{UCCSD} ansatz is not used in classical computations because it can't be efficiently evaluated. Regardless of the truncation order of the cluster operator, expression \ref{def:UCC_energy} is classically intractable, because the \gls{BCH} expansion for $e^{-(T-T^\dagger)}He^{T-T^\dagger}$ is infinite.

What is interesting is that while originally the unitary variant of the \gls{CC} operator was created with the purpose of obtaining a variational version of \gls{CC} theory, this unitarity has the convenient side effect of allowing straightforward implementation of the variational form in quantum computers.

In fact, the \gls{UCCSD} ansatz was used in the article that introduced the \gls{VQE} (reference \cite{Peruzzo2014}). Further, this ansatz has inspired many other proposals of variational forms for use in chemistry applications.

\section{Variational Quantum Algorithms}
\label{s:VQAs}

This section will present the general outline of variational quantum algorithms, as well as the context that motivated their ingress into the field of quantum computing and prompted their popularity. An overview of the subject can be found in the article \cite{cerezo2020}, used as the main reference throughout the subsections to follow.

\subsection{Motivation}
The main incentives for devising variational quantum algorithms are no other than the main limitations of quantum computers. The problems mentioned in subsection \ref{ss:qc_limitations} - reduced qubit count and connectivity; coherent, incoherent and \gls{SPAM} noise -  currently inhibit quantum hardware from being large enough and good enough to implement full algorithms. The prospect of a fault-tolerant era in quantum computing seems distant, if promising.

As it was mentioned before, in 2014 the \gls{VQE} was proposed as the first algorithm inserting a quantum computer in an optimization loop and obtaining a solution via the variational principle. The problem it meant to solve was one familiar to us from the previous section: finding the eigenstates of a Hamiltonian representing a chemical system.

Variational algorithms were far from the first approach for using quantum resources to solve the Schrödinger equation. In fact, the problem is strongly tied with the idea of using quantum computers for the simulation of quantum mechanical systems as proposed by Feynman. Because of the exponential growth of the classical resources required to describe a molecule as its size grows, the task at hand is an example of something that quantum computers were created to tackle. And  by the time quantum computers were still in the conceptual realm, algorithms were certainly not developed with noise-resilience or qubit limitations in mind - rather provable speedups.

If fault-tolerant quantum computers were available to us now, the problem could be solved by means of the \gls{QPE} algorithm. \gls{QPE} allows obtaining the eigenvalue of a unitary operator corresponding to an eigenstate given as input. By applying \gls{QPE} to the (evidently unitary) time evolution operator $e^{iHt}$, it is possible to find the energy eigenstates. The only requirement is that we have a way of preparing states with high overlap with the eigenstates, which can be met by resorting to adiabatic state preparation (ASP). This procedure of using the \gls{QPE} routine in combination with ASP was detailed in \cite{Whitfield2011}.

However, this approach does not seem feasible in the \gls{NISQ} era: employing the \gls{QPE} routine to solve relevant problems would require the application of millions to billions of quantum gates \cite{Peruzzo2014}, implying a circuit depth beyond what can be endured by near-term quantum computers (given that they're not error-corrected, and coherence times are still limiting).

This is where variational hybrid quantum-classical algorithms come in. They resort to the help of a classical optimizer to spare the quantum computer from bearing the load of the algorithm in full. The goal is to provide a near-term solution for problems that are classically hard to solve but, hopefully, easy for quantum computers.

Another good example is the \gls{QAOA}, a variational algorithm aiming to solve combinatorial optimization problems introduced in \cite{farhi2014}. Among the problems suitable for \gls{QAOA} are the \gls{QUBO} problems, of which MaxCut is a famous example with several real-life applications (ranging from physics to communication networks and circuit design).

What is interesting is that an infinitely slow adiabatic evolution leading to the solution of the problem is a special case of the \gls{QAOA} circuit, in the limit of infinitely many layers. If we allow the \gls{QAOA} ansatz to be big enough, there is a performance guarantee based on the adiabatic theorem: the ansatz is sure to contain the solution.

The \gls{QAOA} both parallels and contrast with the \gls{QAA}, which provides a more solid solution by relying in full in the adiabatic theorem. \gls{QAOA} reduces the circuit depth under what would be necessary for a truly adiabatic evolution, dispenses with the assurance of the adiabatic theorem, and inserts the circuit in a classical optimization loop, converting some gate parameters to \textit{variational} parameters. The hope is that the extra variational freedom may make up for what is lost by using a shallower circuit.

While the parallel with \gls{QAA} seems like a good reason to believe the \gls{QAOA} optimization might be successful, if one employs a lower number of layers than required by the adiabatic theorem, the theorem asserts nothing. So against \gls{QAA}, an algorithm with solid foundations but demanding the unattainable in the \gls{NISQ} era, we have \gls{QAOA}: a \gls{NISQ}-friendly algorithm of tentative construction.

The priority shift from performance guarantees offered under the assumption of a fault-tolerant scenario to near-term viability is the bedrock of variational quantum algorithms.

\subsection{Outline}

\begin{figure}[htbp]
    \centering
    \includegraphics[width=1\textwidth]{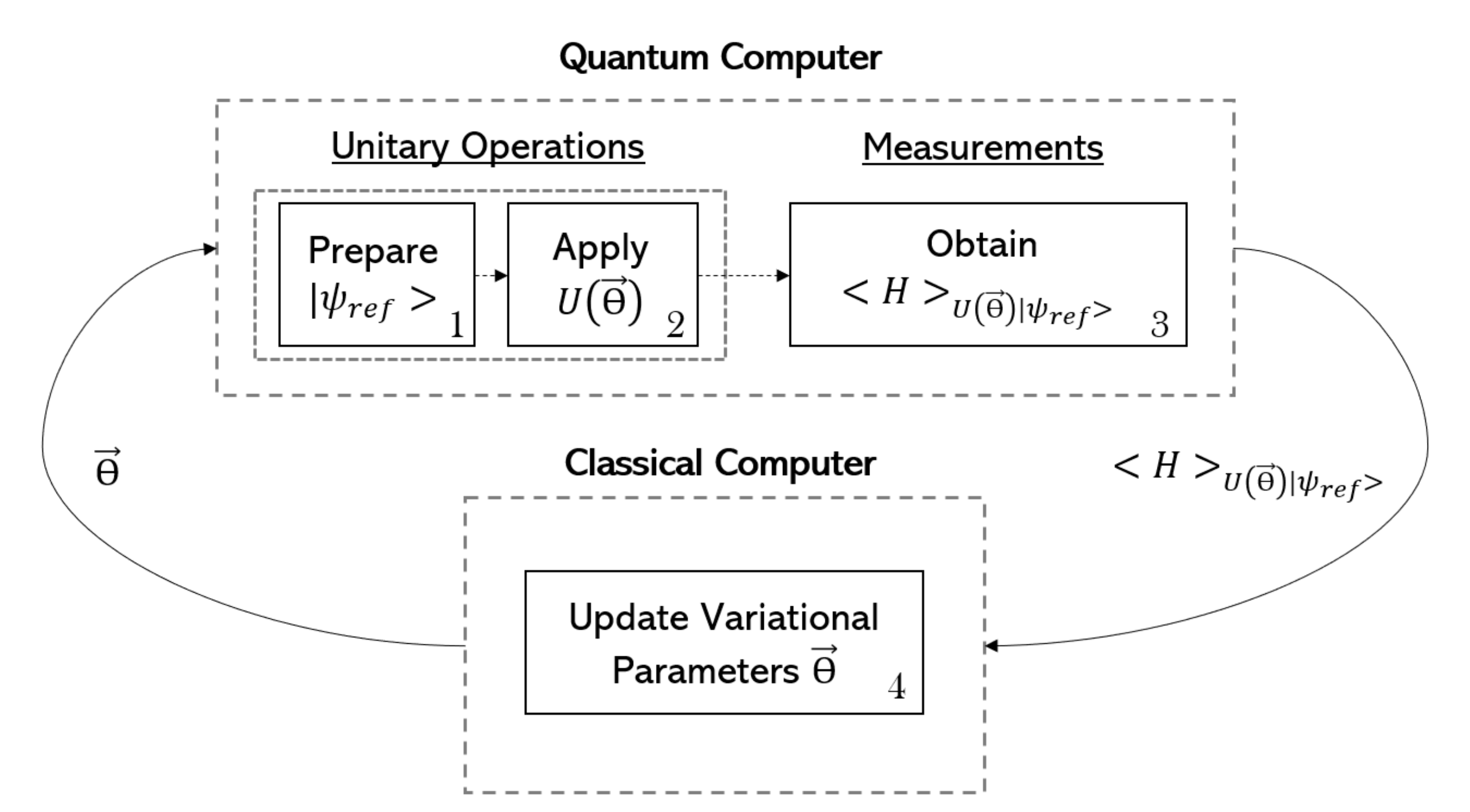}
    \caption{Typical outline of a variational quantum algorithm.}
    \label{fig:VQA_scheme}
\end{figure}

As \glspl{VQA} might differ significantly in some key aspects and cover a wide variety of applications, quantum algorithms belonging to this class exhibit great structural affinity among themselves. A high-level, greatly simplified scheme of the outline of \glspl{VQA} is presented in figure \ref{fig:VQA_scheme}.

The basic idea of \glspl{VQA} is to use a classical computer to optimize an expectation value measured in a quantum computer. In the following, the four steps labeled in the scheme will be explained in more detail, assuming the quantum circuit model.

\subsubsection*{1 - Preparation of the Reference State}

The quantum part of the algorithm starts with the preparation of a reference state $\ket{\psi_{ref}}$. This is the first portion of the \textit{ansatz}, and it will typically be a shallow (often constant-depth) circuit with fixed gates and no variational freedom.

The role of this circuit is to start the computation in a state other than the typical computational basis state $\ket{0...0}$. In the specific case of chemistry applications, this step might be of great importance in assuring that the computation starts in a state belonging to the correct symmetry subspace dictated by the problem. For example, it might be desired that the initial state has the correct particle number, and the correct total spin and $Z$ spin projection expectation values. 

Of course, it isn't always necessarily the case that a reference state different from the all-zero state is desired; when such occurs, the reference state preparation step is simply omitted.

\subsubsection*{2 - Application of the Parameterized Unitary}

After the reference state has been prepared, the parameterized portion of the ansatz circuit follows. This is the part that contains the variational freedom. 
This is represented in the scheme by the parameterized unitary $U(\vec{\theta})$. This notation simply means that, rather than a fully-fixed circuit, a parameterized circuit is applied in this step. The parameters are considered to be organized into a parameter vector $\vec{\theta}$. After application of the ansatz, we are left with the parameterized state $U(\vec{\theta})\ket{\psi_{ref}}$, which will differ among iterations (via the dependence on the parameter vector, updated each round).

The complete ansatz then consists of the combination of the circuit that prepares the reference state (step 1) with the parameterized circuit (step 2). The specific form of the ansatz greatly depends on the specifics of the problem and the approach. Ansätze can be labeled into three main categories.

\textbf{Static, problem agnostic ansätze} are those fixed upfront and completely oblivious to the problem. Ansätze of this type don't attempt to leverage any sort of previous knowledge about the problem, and don't bother spanning specific subspaces in which the solution is expected to be contained. Typically, this represents a shift of priority from problem-tailoring to hardware-tailoring. The \textit{hardware-efficient} ansätze, that attempt to reduce depth by constructing a circuit directly from gates natural to the specific hardware, belong to this category. Despite the evident advantages of strategies like this one, searching for the solution in an exponentially big space without an informed choice of variational form creates other problems. Too much expressibility and too little problem-tailoring lead to \textit{barren plateaus} in the optimization landscape, causing the gradients of the cost function to vanish exponentially on the number of qubits and rendering the full algorithm inefficient.

\textbf{Static, problem tailored ansätze} are also fixed upfront, but they do attempt to leverage information about the problem at hand. For example, in chemistry, this class includes ansätze that respect symmetries, be those symmetries generic or specific to the system in study. One can enforce time-reversal symmetry, as well as spin and particle number conservation. The \gls{UCCSD} ansatz introduced in the previous section is an example of an element of this class. While these ansätze don't typically suffer from barren plateaus (aside from the noise-induced ones) \cite{cerezo2020}, they have some issues of their own. They are not always easy to devise, because information about the problem can be limited or, if available, difficult to leverage. Additionally, they typically correspond to longer ideal circuits (since they're not developed with shallowness as the end goal) and even longer transpiled circuits (since the available gates or connectivity are also not the main concern).

\textbf{Dynamic, problem tailored ansätze} differ from the previous two in an important aspect: they are not fixed upfront, rather grown on the fly as the algorithm evolves. This implies a very special degree of tailoring: instead of human fabricated problem tailoring happening beforehand, additional measurements are used \textit{during} the computation to gather information about which parameterized gates should be added to the ansatz at each point. The idea is to optimize, as well as the parameters, the variational form itself, which is consequently not only tailored to the problem as a whole: it is tailored to each specific instance of the problem. \gls{ADAPT}-\gls{VQE} was the first such ansatz to be introduced in the literature. It grows the ansatz from scratch by selecting operators from a pool, based on their gradients. Unlike \gls{UCCSD}, which is tailored to electronic structure problems in general, \gls{ADAPT}-\gls{VQE} is tailored to each specific molecule we might want to study. The main disadvantage of dynamically created ansätze is that they result in a sizeable overhead in measurement and optimization costs; however, they have multiple advantages. Their design is not an issue, because the ansatz structure is dictated by the the problem at hand - one needn't find a way to embed information about the problem into the construction of the circuit, because the algorithm learns how to do that autonomously. The resulting circuit is typically of reduced and controllable depth, due to the higher degree of customization and the existence of a tunable convergence criterion. Further, the accuracy can surpass that of fixed-depth problem tailored ansatz circuits. 

\subsubsection*{3 - Cost Function Sampling}

The final step to occur in the quantum computer is the evaluation of the expectation value of the cost function - i.e., the function whose minimum we want to find - in the trial state that was prepared. This is the information that will be passed on to the classical computer at the end of each iteration. 

The operator whose expectation value we wish to find is typically denoted $H$, because it may correspond to a physical Hamiltonian. However, this `Hamiltonian' can also be fabricated to encode a problem that does not have any relation to a physical system. Regardless of the interpretation of the operator, we can view this step as obtaining the expectation value of $H$ measured in the state $U(\vec{\theta})\ket{\psi_{ref}}$, abbreviated to $\left\langle H\right\rangle_{U(\vec{\theta})\ket{\psi_{ref}}}$ in figure \ref{fig:VQA_scheme}. 

It was mentioned in section \ref{s:quantumcomputing} that, in general, obtaining an expectation value implies several measurements. Thus, even though this is not represented on the scheme so as to privilege terseness, the estimate of the expectation value is obtained by repeating the state preparation and measurements multiple times, and averaging over the results. Even under ideal conditions, sampling entails some noise in the result: the inevitable quantum projection noise, and shot noise, deriving from the use of a finite number of repetitions. The accuracy of the result can be increased arbitrarily by increasing the number of shots (also called repetitions) of the circuit: the error goes with the inverse of the square root of the number of shots ($\epsilon\propto\frac{1}{\sqrt{N}}$).

The diagram also disregards the evaluation of more than one expectation value in each iteration. Oftentimes, the classical optimizer will make multiple calls to the quantum processor before moving on to the next iteration (i.e. updating the parameters). For example, some optimization methods may require a few gradient evaluations per iteration. In many relevant cases, this can be done by means of measurements on the quantum computer using the parameter-shift rule \cite{Schuld2019}.

\subsubsection*{4 - Parameter Vector Update}

Finally, we have the step performed by the classical computer: the parameter update. 

Based on the expectation value of the cost function (and possibly its gradients) obtained in the state prepared by the previous parameters, a classical optimizer will decide on the parameters for the next iteration. This can be done  using the method of gradient descent, among others.

After the optimizer decides how to update the parameter vector $\vec{\theta}$, this vector will be fed to the quantum computer. The effect of this is to instantiate the ansatz, binding the specific parameter values to the gates occurring in the variational circuit. This will allow the quantum computer to obtain the expectation value of $H$ in the new trial state, which will then be passed on to the classical optimizer, and so on.

The cycle stops when some \textit{convergence criterion} is met. This is decided by the classical computer. A common stopping criterion is the absolute change in the value of the cost function between two consecutive iterations being under a certain threshold.

\subsubsection{Variability among VQAs}

While the previous succinct explanation focused mostly on the common factors between all \glspl{VQA}, this generic scheme embraces considerable diversity. As an example of this, we may compare two \glspl{VQA} that were already mentioned in this section: the \gls{VQE} and the \gls{QAOA}.

The \gls{VQE} addresses chemistry problems: its goal is to find the ground state of a system. The cost Hamiltonian is simply the electronic Hamiltonian of such system, and its expectation value is the energy. This molecular Hamiltonian only needs to be mapped onto an observable that can be measured in a quantum computer. As happens in many classical variational algorithms for chemistry, the reference state is frequently the Hartree-Fock state, and the parameterized portion of the ansatz commonly consists of circuits that mimic fermionic excitations applied to the reference determinant.

On the other hand, the \gls{QAOA} aims to solve combinatorial problems; the solution is a classical bit string, and the cost function is classical. Once again, this function must be represented by a quantum observable; in this case, the cost Hamiltonian is diagonal and consists of a sum of projectors onto computational basis states multiplied by the corresponding cost values. The reference state is the homogeneous superposition of computational basis states, which is the maximum energy eigenstate of another fabricated Hamiltonian called the \textit{mixer Hamiltonian}. The parameterized portion of the ansatz consists of \textit{p} rounds of time evolution under alternating Hamiltonians - this mixer Hamiltonian and the cost Hamiltonian. These rounds are resembling of a trotterized adiabatic process, except here the evolution times of each round correspond to independent variational parameters. In the limit that \textit{p} tends to infinity, we can make the Trotter steps very small, and the runtime very large (corresponding to a very slow evolution). In this limit we are certain that the \gls{QAOA} ansatz contains the solution, because for some variational parameters the circuit implements the slowest possible adiabatic evolution leading the maximum energy state of the mixer Hamiltonian to the maximum energy state of the cost Hamiltonian - i.e., the solution.

\gls{VQE} and \gls{QAOA} thus have distinct reference states, and the parameterized portion of their ansätze is even more distinct. Notably, while the \gls{VQE} ansatz is independent of the Hamiltonian, the \gls{QAOA} ansatz depends on it. The cost functions have distinct origins, even though both can be represented by a quantum observable: the one used in \gls{VQE} arises from a real system and the one used in \gls{QAOA} is fabricated to represent a classical problem. Consequently, there is also a different interpretation for the expectation values of these functions. The \gls{VQE} expectation value is the energy of the system in the fermionic state represented by the state of the qubits; in contrast, the \gls{QAOA} expectation value is the (weighted) average of the cost function over the several bit strings (computational basis states) that appear in the superposition state of the qubits. 

Given the classical nature of the problem, the \gls{QAOA} solution is a \textit{single} computational basis state (the one with the highest amplitude), while in \gls{VQE} the \textit{full} superposition state is the solution. This is also related to \textit{how} we expect these quantum algorithms to bring advantage as compared to classical ones. The hope for quantum advantage in \gls{VQE} is rooted in the fact that the classical description of the molecular system grows exponentially with its size, i.e. \gls{VQE} attempts to make use of quantum resources for a problem that is quantum by nature. In contrast, \gls{QAOA}  attempts to make use of \textit{quantum} superposition to simultaneously explore several solutions to a \textit{classical} problem.

\subsection{Variational Quantum Eigensolver}

As it was explained before, the variational quantum eigensolver was introduced in \cite{Peruzzo2014} with the purpose of finding the ground energy and ground state of a chemical system. The approach can be generalized to excited states by means of the folded spectrum method \cite{MacDonald1934,Wang1994}.

The essential idea behind the algorithm is to use a classical optimizer to find the minimum eigenvalue of a Hamiltonian, whose expectation value is evaluated in a quantum computer.

The purpose of this section is to explain a few important aspects that have not yet been covered, including the representation of the problem and specific details concerning the quantum expectation estimation in the context of \gls{VQE}. A deeper overview of the several components of \gls{VQE} can be found in \cite{McClean2016}.

\subsubsection{The Jordan-Wigner Transform}
\label{sss:jw_transform}

In order to be able to solve a fermionic problem in a quantum computer, we need to be able to encode the evolution of a fermionic state in the evolution of the state of the qubits through unitary operations. This is done through a \textit{fermion-to-qubit mapping}, of which there are several; the most widely used one, and the one employed in this dissertation, is the Jordan-Wigner Transform \cite{JordanWigner}.

The first thing that should be pointed out is that the state of a fermion cannot in general be described by a binary variable. For example, the state of a bound electron in an atom depends on the principal quantum number \textit{n}, the azimuthal quantum number \textit{l}, the magnetic quantum number $m_l$, and the secondary spin quantum number $m_s$. Aside from $m_s$, these quantum numbers are not even binary variables themselves - and all four are necessary for a complete description. As such, representing the state of a fermion in the state of a qubit is not feasible.

Another problem is that qubits are distinguishable, while identical fermions are indistinguishable. Even if the state of a fermion \textit{could} be fully specified by a binary variable, representing it in the state of a qubit would not work. As an example, while the two two-qubit states

\[\ket{\psi_{Q,1}}=\ket{0}\otimes\ket{1},\hspace{5pt}\ket{\psi_{Q,2}}=\ket{1}\otimes\ket{0}\]

are distinct, they would not be if we were dealing with fermions. Because fermions are identical particles, exchanging two of them cannot affect any physical process. This is enforced by the antisymmetry requirement presented in equation \ref{eq:antisymmetry}: if we exchange any two fermions, the many-body fermionic wave function simply flips the sign. Ensuring the antisymmetry of the wave function guarantees that we are not discerning between states that are not physically distinguishable. For the simple two-body example, if we have a fermion in a state labeled \textbf{0} and one in a state \textbf{1} (regardless of what those states are), we can write the wave function as 

\[\ket{\psi_F}\propto\ket{0}\otimes\ket{1}-\ket{1}\otimes\ket{0},\]

which has the required property that exchanging the two fermions (putting the fermion to the left of the tensor product in the state of the one to the right and vice-versa) results in a change of sign, but does not affect any measurable properties of the wave function. It is easy to see that any expectation value is left unchanged.

\[\ket{\psi_F}\rightarrow-\ket{\psi_F}\]

\[\bra{\psi_F}\hat{O}\ket{\psi_F}\rightarrow(-\bra{\psi_F})\hat{O}(-\ket{\psi_F})=\bra{\psi_F}\hat{O}\ket{\psi_F}\]

The two-body example is very simple, but as was discussed, we can write a generic N-body wave function through Slater determinants, formula \ref{def:slater_determinant}.

All of this trouble to express the many-body fermionic wave function is merely a symptom of the notation in use being cumbersome for our purpose. By specifying the many-bode wave function in the basis of tensor products of states of the individual particles, we are distinguishing between what isn't physically distinguishable. Because we are discriminating the state of each individual particle, it is then necessary to remove the redundancy in the notation by enforcing the antisymmetry of the wave function.

This is where the \textit{occupation number representation} of the second quantization proves useful. In dealing with identical particles, it seems more natural to specify \textit{how many} particles are in \textit{each} state, rather than \textit{which} particles are in \textit{which} state.  
Instead of having all the work antisymmetrizing the wave function, it suffices to describe it by enumerating the occupation numbers of each single-particle state (also called \textit{fermionic mode}), a formalism which naturally embraces indistinguishability and clears up the notation. In the previous example, we write

\[\ket{\psi_F} = \ket{n_0=1,n_1=1}\]

to mean that there is a fermion in the state labeled 0 (occupation number $n_0$ is 1) and another in the state labeled 1 (occupation number $n_1$ is 1), without making any reference to which fermion is in which state. In the case of a bound electron in an atom, we could specify the occupation number of each \textit{spin-orbital} (identified by the four quantum numbers \textit{n}, \textit{l}, $m_l$, $m_s$). In accordance with the Pauli exclusion principle, each occupation number must be 0 or 1, which can be readily seen from the fact that the antisymmetry principle \ref{eq:antisymmetry} causes the wave function to vanish if any two fermions occupy the same state.

The second quantization formalism clears up the notation by relaying the task of ensuring antisymmetry to the \textit{fermionic ladder operators}. For any fermionic mode $i$, there is the creation operator $a_i^\dagger$ and the annihilation operator $a_i$, which act as

\begin{align}
\begin{split}
a_i^\dagger\ket{n_i=0}=\ket{n_i=1},
&\qquad
a_i^\dagger\ket{n_i=1}=0,
\quad\\
a_i\ket{n_i=1}=\ket{n_i=0},
&\qquad
a_i\ket{n_i=0}=0.
\end{split}
\label{def:fladder_ops_action}
\end{align}

Attempting to create a fermion in an occupied state or to remove a fermion from an unoccupied state quenches the wave function.

The ladder operators have to obey the \textit{anti-commutation relations}:

\begin{align}
\begin{split}
\{a_j^\dagger,a_i^\dagger\} = 0,
&\qquad
\{a_j,a_i\} = 0,
\qquad
\{a_j,a_i^\dagger\} = \delta_{j,i},
\end{split}
\label{def:anticommutation_relations}
\end{align}

where $\{A,B\}=A+B$ is the \textit{anti-commutator} of operators $A$ and $B$, and $\delta_{j,i}$ is the Kronecker delta (which is 1 if $j=i$, and 0 if not). The anti-commutation relations in \ref{def:anticommutation_relations} enforce anti-symmetry and serve as the algebraic definition of the fermionic ladder operators.

It should be noted that there is a notion of \textit{order} at play. For example, if we use $\ket{0}$ denote the vaccuum state (no fermions) and assume $i\neq j$, the expression

\[a_j^\dagger a_i^\dagger\ket{0}=-a_i^\dagger a_j^\dagger\ket{0}\]

arising from the first anti-commutator in \ref{def:anticommutation_relations} states that if we create a fermion in state $i$, \textit{then} one in state $j$, there is a sign flip as compared to a situation in which fermions are created in the opposite order. The ladder operators are responsible for this sign flip, which ensures that the anti-symmetry principle is obeyed; thus, it is to be expected that their mapping to qubit operators will carry this notion of order.

It is now natural to introduce the \textit{Jordan-Wigner transform}, a mapping from fermions to qubits. We match qubit \textit{states} with the occupation number of fermionic \textit{modes}, and map

\begin{align}
\begin{split}
a_i^\dagger\rightarrow e^{i\pi\sum_{k=1}^{i-1}\sigma_k^+\sigma_k}\cdot\sigma^+_i,\\
\quad
a_i\rightarrow e^{i\pi\sum_{k=1}^{i-1}\sigma_k^+\sigma_k}\cdot\sigma^-_i,
\end{split}
\label{def:jw_transform}
\end{align}

where the $\sigma_i^+$, $\sigma_i^-$ are the qubit raising and lowering operators acting on qubit $i$, defined by means of Pauli operators as

\begin{align}
\begin{split}
\sigma^+_i=\frac{1}{2}(X_i-iY_i),\\
\quad
\sigma^-_i=\frac{1}{2}(X_i+iY_i).
\end{split}
\label{def:qubit_ladder_ops}
\end{align}

Evidently, the action of $\sigma_i^+$, $\sigma_i^-$ on qubit states $\ket{0}_i,\ket{1}_i$ is identical to the action of $a_i^\dagger$, $a_i$ on the fermion states $\ket{0}_i$, $\ket{1}_i$ as defined in \ref{def:fladder_ops_action}.

\begin{align}
\begin{split}
\sigma_i^+\ket{0}_i=\ket{1}_i
&\qquad
\sigma_i^+\ket{1}_i=0
\quad\\
\sigma_i^-\ket{1}_i=\ket{0}_i
&\qquad
\sigma_i^-\ket{0}_i=0
\end{split}
\label{def:qadder_ops_action}
\end{align}

For a single fermionic mode, identifying $\sigma_i^+$, $\sigma_i^-$ with $a_i^\dagger$, $a_i$ respectively would suffice. However, the antisymmetry requirement would not be satisfied, so this simpler mapping would not be fitting in a multi-mode scenario. That is the purpose of the phase that appears in \ref{def:jw_transform}.

The operator $e^{i\pi\sum_{k=1}^{i-1}\sigma_k^+\sigma_k}$ is typically called \textit{the Jordan-Wigner string operator}. It can be readily seen that it adds a phase $e^{i\pi N_{k<i}}$ to the operators acting on qubit $i$, where $N_{k<i}$ is the number of the occupied fermionic modes $k$ for which $k<i$. This is crucial for the anti-symmetry requirement, and it is where the previously mentioned notion of order comes in. We have to establish and be consistent with a sequence, i.e. choose an order $...<i<j<...$ relating all fermionic modes.

It is easy to see what is the effect of the Jordan-Wigner string by considering the successive creation of fermions in two different unoccupied modes. When the first fermion is added to the mode with the lowest index, it affects the string operator of the operator that creates the second fermion, but when the first fermion is added to the mode with the highest index, the string operator of the operator that creates the second fermion is left unchanged. This causes the wave function in these cases to differ by a phase factor $e^{i\pi}=-1$, as antisymmetry requires: whenever we swap the order of creation of fermions in two modes, we get a flipped sign.

It is convenient to write the Jordan-Wigner transform \ref{def:jw_transform} using Pauli operators. This is easily done using \ref{def:qubit_ladder_ops} and noting that the effect of the Jordan-Wigner string on operators acting on mode $i$ amounts to adding a minus sign when the parity of the occupation numbers of the modes under $i$ is odd. This parity calculation can be done through a string of Pauli $Z$ operators.

\begin{align}
\begin{split}
a_i^\dagger\rightarrow \frac{1}{2}\prod_{k=1}^{i-1}Z_k\cdot(X_i-iY_i)\\
\quad
a_i\rightarrow \frac{1}{2}\prod_{k=1}^{i-1}Z_k\cdot(X_i+iY_i)
\end{split}
\label{def:jw_transform_paulis}
\end{align}

The formulation of the Jordan-Wigner transform in \ref{def:jw_transform_paulis} is very convenient for quantum computing applications.

Finally, it is worth noting that this definition of the Jordan-Wigner mapping is not unique. Often the roles of the operators $\sigma_i^+$ and $\sigma_i^-$ in \ref{def:jw_transform} are reversed, causing the empty and occupied fermionic modes to be represented by qubit states $\ket{1}$ and $\ket{0}$ respectively (the opposite of what the definitions presented here lead to).

\subsubsection{Preparation of the Reference State}
\label{ss:prep_ref_state}

The reference state used in \gls{VQE} is typically the Hartree-Fock approximation to the ground state. Because the computational basis states usually represent the Slater determinants obtained as approximate eigenstates of the system in performing the self-consistent mean field Hartree-Fock calculations, the reference state preparation state simply amounts to a $\mathcal{O}(1)$ circuit with $N_e$ parallel Pauli $X$ gates, where $N_e$ is the number of electrons. 

Preparing the reference state amounts to applying Pauli $X$ gates on the qubits that, under the Jordan-Wigner mapping (see \ref{sss:jw_transform}), correspond to occupied spin-orbitals.

\begin{figure}
\begin{subfigure}[b]{0.4\textwidth}
\includegraphics[width=0.9\linewidth]{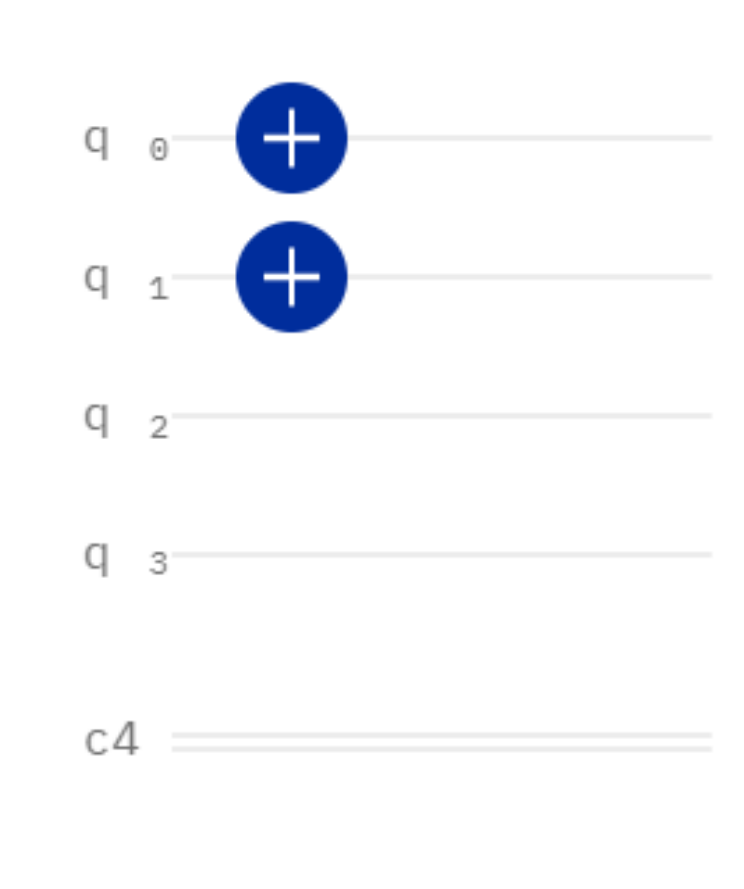} \caption{OpenFermion \cite{openfermion} orbital ordering.}
\label{fig:hf_OF}
\end{subfigure}
\hfill
\begin{subfigure}[b]{0.4\textwidth}
\includegraphics[width=0.9\linewidth]{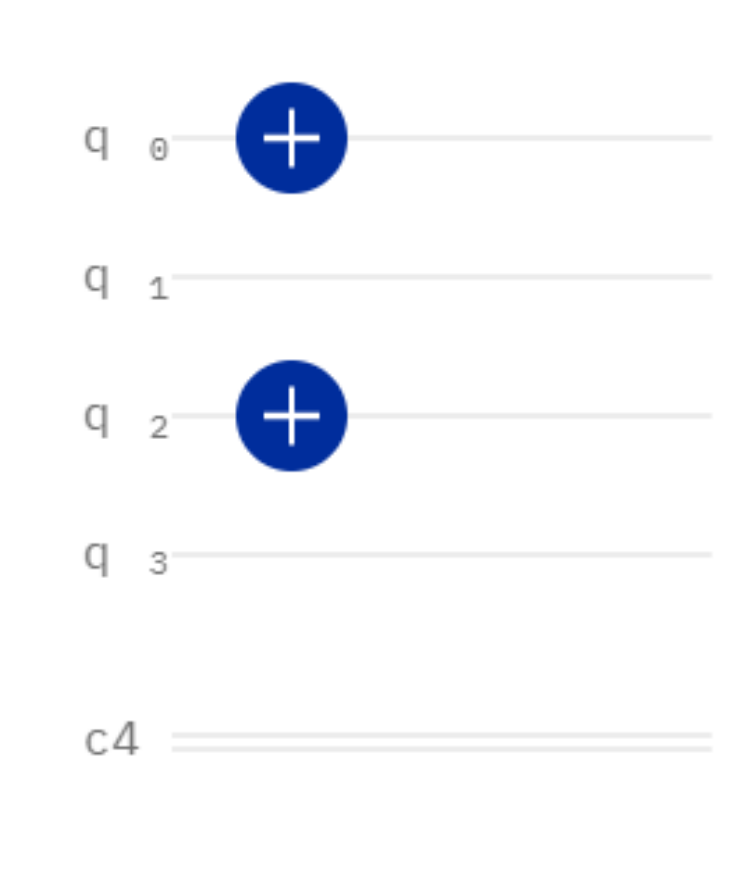} \caption{Qiskit \cite{Qiskit} orbital ordering.}
\label{fig:hf_qiskit}
\end{subfigure}
\caption{Circuit for preparing the Hartree-Fock reference state, for the $H_2$ molecule. The circuits were drawn in the IBM Quantum Composer \cite{IBMcomposer}.}
\label{fig:hf_circuits}
\end{figure}

The exact circuit depends on the convention for organizing the spin-orbitals. Figure \ref{fig:hf_circuits} contrasts the reference state preparation for the hydrogen molecule in the two software packages used for creating the circuits: OpenFermion \cite{openfermion}, used in conjunction with CIRQ \cite{CIRQ}, and Qiskit \cite{Qiskit}.

In the basis set that was always used in the simulations (the STO-3G minimal basis set), $H_2$ has four spin-orbitals, so that the electronic states are represented by four qubits.

OpenFermion alternates between $\alpha$ (up) and $\beta$ (down) type orbitals. With this ordering, each spatial orbital is represented by two qubits that are adjacent on the register. Figure \ref{fig:hf_OF} shows how to create the state $\ket{0011}$, assuming little endian ordering. Under OpenFermion's convention, this corresponds to creating two fermions on the two spin-orbitals associated with the lowest-energy spatial orbital (the orbitals are assumed to be organized in ascending energy from right to left). 

In contrast, Qiskit uses block-wise orbital ordering: all $\alpha$ orbitals come first, followed by all $\beta$ orbitals. This means that the first qubit of the register corresponds to the same spatial orbital as the one in the middle of the register, with the two representing opposite spins. Figure \ref{fig:hf_qiskit} shows how to create the state $\ket{1010}$ (assuming little endian ordering once again). In spite of the differences in convention, the states prepared by circuits \ref{fig:hf_OF} and \ref{fig:hf_qiskit} represent the same fermionic state. Evidently, the ansatz and Hamiltonian averaging must be done in accordance with the chosen convention.

\subsubsection{Choice of Ansatz}

As was discussed, the typical ansätze for \gls{VQE} are chemistry-inspired. Even though they may not lead to the shallowest circuits, the fact that they're tailored to chemistry problems in particular helps create trial states in the appropriate symmetry subspace. As a consequence, the optimization is helped by the fact that subspaces where the ground state is sure not to be are entirely avoided. Ansätze that don't leverage previous knowledge about the system typically lead to barren plateaus. These trainability issues have been shown to be a consequence of the choice of ansatz that cannot be helped through the choice of optimizer \cite{McClean2018,Arrasmith2020,holmes2021}.

\gls{UCC} theory enjoys great popularity due to its natural genesis in the context of previous computational chemistry methods. \gls{UCC} proposals are also fairly broad, and this class has been expanding as a consequence of research efforts to make the ansätze more compact and more accurate. 

The article \cite{Sokolov2020} explored singlet and pair \gls{q-UCCD} approaches (where the q stands for \textit{quantum}), which reduce the double excitations appearing in the ansatz to a smaller subset. Singlet  \gls{q-UCCD} splits the double excitations into singlet and triplet components, and makes use of symmetry / anti-symmetry to reduce the number of excitation operators. Pair  \gls{q-UCCD} limits the excitations to those which excite a \textit{pair} of electrons with opposite spins from one spacial orbital to another. Both of these ansätze were tested in several systems with results of compelling accuracy, while reducing the number of two-qubit gates in the ansatz circuit by up to $75\%$. 

A different ansatz termed k-\gls{UpCCGSD} was introduced in \cite{Lee2018}. k-\gls{UpCCGSD} consists of \textit{k} products of unitary paired coupled-cluster double excitations, as well as all generalized single excitations. Here generalized indicates that virtual-to-virtual and occupied-to-occupied excitations are also allowed (orbitals are considered virtual or occupied if they are so in the reference state). The K-\gls{UpCCGSD} ansatz increases the variational freedom as compared to the simple pair \gls{UCCGSD} (p\gls{UCCGSD}) ansatz, while still enjoying a linear scaling of the circuit depth with the system size. The k-\gls{UpCCGSD} was shown to be capable of achieving chemical accuracy with a better resource scaling than \gls{UCCSD} and \gls{UCCGSD} (the generalized version of the former), both of which require an $\mathcal{O}(N^4)$ circuit depth, where N is the number of orbitals.

Changing the form of the ansatz is not the only option. Low-rank decomposition of the basic \gls{UCC} operator was used in \cite{Motta2018} to reduce the gate complexity of each Trotter step from $\mathcal{O}(N^4)$ to $\mathcal{O}(N^3)$. By exploiting the structure and symmetries of the cluster operator and truncating small terms, the depth of the circuits can be reduced without precluding chemical accuracy.

\subsubsection{Hamiltonian Averaging}
\label{sss:hamiltonian_averaging}

The part of the algorithm that consists of obtaining the expectation value of the cost Hamiltonian (in this case a physical Hamiltonian) is typically called \textit{Hamiltonian averaging} or \textit{quantum expectation estimation}.

In subsection \ref{ss:measuringpaulis}, a procedure to measure arbitrary observables was outlined; it consisted of decomposing them as a linear combination of Pauli strings (formula \ref{eq:pauli_decomp}) and calculating the expectation value as a weighed average of the corresponding expectation values (formula \ref{eq:arbitrary_expectation}). The expectation value of the Hamiltonian can be obtained exactly this way; the question is whether this can be done efficiently.

Luckily, it is often the case that the Hamiltonian of a physical system can be decomposed as a linear combination of a polynomial number of Pauli terms. This includes the electronic structure Hamiltonian (formula \ref{eq:hamiltonian_electronic}), as well as the Hamiltonians of the Ising and Heisenberg models \cite{Peruzzo2014}. The problems addressed in this dissertation concern electronic Hamiltonians with a number of terms that scales like $\mathcal{O}(N^4)$ on the number of spin-orbitals/qubits $N$ \cite{shkolnikov2021}. An example of such a Hamiltonian, for the very simple case of the hydrogen molecule, can be found in appendix \ref{ap:h2_hamiltonian}.

While the scaling is polynomial, obtaining the expectation value can still imply a large number of measurements, especially in chemistry problems which typically require exceptional accuracy. One way this can be improved is by organizing the Pauli strings appearing in the decomposition of the operator into commuting groups, that can be measured simultaneously. 

As a simple two-qubit example, we can consider the following set of observables.

\[II, ZI, IZ, ZZ\]

The expectation values of all four terms can be obtained from the same circuit. There are multiple ways of arranging Pauli strings into commuting groups; for example, if we have the set of observables

\[ZI, IZ, ZZ, XI, IX, XX, XZ, ZX,\]

we can chose to split the Pauli strings into groups as

\[\{ZI,IZ,ZZ\}; \{XI, IX, XX\}; \{ZX\}; \{XZ\},\]

but also as

\[\{ZI,IX,ZX\};\{XI,IZ,XZ\}; \{ZZ\}; \{XX\}.\]

These divisions only exploit qubit-wise commutativity. However, two Pauli strings also commute if they have non-commuting Pauli operators in an even number of indices. This general commutativity allows us to further reduce the circuit repetitions that are required to obtain the final expectation value. In our example, it brings them down to $\frac{3}{8}$ as compared to when no grouping is used; we can organize the operators as

\[\{ZI,IX,ZX\};\{XI,IZ,XZ\}; \{ZZ,XX\}.\]

Even though this is evidently not the only possibility; we can also have

\[\{ZI,IZ,ZZ\}; \{XI, IX, XX\}; \{ZX,XZ\}.\]

In this example, both options reduce the necessary circuit repetitions by the same amount (assuming these repetitions are to be distributed equally). However, this is not always necessarily the case. What is more, the task of finding the optimal partitioning becomes harder the larger the Hamiltonian is, and taking into account sampling noise complicates the task further. Proposals of grouping strategies can be found in references \cite{McClean2016, Wecker2015, Kandala2017, Rubin2018, Izmaylov2019, jena2019, ogorman2019, Izmaylov2019snd, Yen2020, Gokhale2019, Gokhale2020, Verteletskyi2020, Crawford2021, Huggins2021}. The last article among these proposed an approach that allows for a cubic reduction in term groupings over previous proposals, with the drawback of requiring the execution of a linear-depth circuit before the measurement. A table comparing all these strategies can be found in this same reference.

\subsubsection{Estimating Shot Requirements}

The number of terms in the Hamiltonian decomposition, and the number of commuting groups, are not the only factors weighing in the shot requirements of \gls{VQE}. The impact of sampling noise in the error, which typically also increases with the size of the system, is another relevant aspect. The following estimation of shot requirements was based on reference \cite{Wecker2015}.

Assuming that the Hamiltonian operator has been decomposed in the form of formula \ref{eq:pauli_decomp} and that for each Pauli string $P_i$ we use $M_i$ shots, the error coming from the finite number of shots in the expectation estimation can be written

\[\epsilon^2 = \sum_i\frac{|h_i|^2Var(\hat{P}_i)}{M_i}.\]

The variances depend, not only on the Pauli operators appearing in the Hamiltonian, but also on the particular state. We can bound them as $Var(O_i)\leq1$, since any Pauli operator has eigenvalues $\pm1$ and so will their product.

Without further knowledge about their variance, the best choice is to distribute the shots by the operators in proportion to the norm of their coefficients: $M_i\propto|h_i|$. Assuming this distribution, one can estimate the necessary number of shots to hit a precision $\epsilon$ as

\[M\approx\frac{(\sum_i|h_i|)^2}{\epsilon^2}.\]

The term $(\sum_i|h_i|)^2$ will depend strongly on the molecule in study: it will be larger for larger molecules, requiring a larger number of shots for the same precision if the system size is increased.

Given that chemical accuracy is $1$ kcal/mol $\approx 1.59\times10^{-3}$ Hartree, achieving it requires

\[M\approx0.4\times10^6\times(\sum_i|h_i|)^2\]

shots. This is already of the order of $10^7$ to $10^8$ for small molecules like helonium and lithium hydride; $Fe_2S_2$, for instance, would require around $10^{13}$ shots.

This number of shots needs to be performed every time one needs to evaluate the energy, which happens many times throughout the optimization. The shot requirements are heavier for larger molecules, since there are more Pauli strings in the Hamiltonian, larger coefficients $|h_i|$, and more parameters to optimize (resulting in more iterations and more function evaluations per iteration being required). 

Further, this shot count doesn't guarantee proper functioning of the optimizer: it ensures that the difference between the energy estimate and the true energy \textit{in a state} are within chemical accuracy, but the final state isn't necessarily close to the ground state. The error might still be enough to confuse the optimizer into outputting a wrong guess for the ground state (e.g. by tricking it into an early convergence).

%% file: Chapters/chapter3.tex
\pagestyle{plain}
\graphicspath{{./Chapters/Figures/Ch3/}}

\chapter{Static Ansätze for VQE}
\label{cha:static_ansatze}

This chapter aims to present the results of \gls{VQE} simulations with static ansätze. In the first section, an entirely problem-agnostic ansatz will be applied in searching for the ground state of the helium hydride ion. The impact of the optimizer, optimization hyperparameters, and noise on performance will be assessed. In the second section, a problem-inspired ansatz will be applied in searching for the ground state of the hydrogen molecule. The impact of several types of noise on performance will be analysed.

For creating and executing quantum circuits, CIRQ \cite{CIRQ} and Qiskit \cite{Qiskit} were used in alternation; the CIRQ simulator was used as well as IBMQ's simulators and real quantum computers. Which package and backend was used in obtaining which results will be clear along the chapter. 

When analysing the results, the final \gls{VQE} energy and state will be compared with those obtained from performing exact diagonalization on the \gls{FCI} Hamiltonian. The \gls{FCI} solution is exact up to the Born-Oppenheimer approximation, relativistic effects, and the chosen basis set.

\section{Problem Agnostic Ansätze}
\label{s:problem_agnostic_ansatze}

As was explained before, problem agnostic ansätze come with several advantages: they don't demand previous knowledge about the problem or system, nor a way of leveraging such knowledge. Additionally, taking the focus away from the problem allows for a more hardware-friendly circuit implementation, which may decrease the depth of the circuits. The main disadvantage of this type of ansatz is that randomly initialized parameterized quantum circuits often lead to barren plateaus in the optimization landscape \cite{McClean2018}.

This section will present simple a example of a problem agnostic ansatz, applied to the helium hydride ion. An ansatz spanning the whole Hilbert space was created and employed in the simulations; such an approach is feasible given the dimensionality of the problem. The section contains results from running the algorithm on the CIRQ simulator and on IBMQ's backends (simulators or real quantum processors).

\subsection{Application to Helium Hydride}

The purpose of the following is to implement \gls{VQE} as in the article that originally introduced it \cite{Peruzzo2014}, which applied the algorithm to the helium hydride ion using a photonic quantum processor. 

In addition to the CIRQ and Qiskit implementations, an independent noise-free version of \gls{VQE}, based on matrix algebra, was implemented to verify correctness.

The Hamiltonians for the different bond distances were taken from the article (which in turn obtained them via the PSI3 computational package \cite{psi3}), as was the parameterization of an arbitrary 2-qubit state. The ansatz employed here searches through the whole Hilbert space, much like the ansatz in the original article. However, while in the article the parameterization was rewritten into a convenient form that directly dictated the values of six phase shifters in the photonic chip, here the state preparation was abstracted of the specifics of any physical implementation and instead mapped to a generic circuit.

\subsubsection{Preparing an Arbitrary 2 Qubit State}

An arbitrary (pure) two qubit state can be written as in equation \ref{def:2qubit_coordinates}. 

\begin{equation}
    \ket{\psi} = \alpha\ket{00} + \beta\ket{01} + \gamma\ket{10} + \delta\ket{11}
    \label{def:2qubit_coordinates}
\end{equation}

The computational basis coordinates $\alpha$, $\beta$, $\gamma$ and $\delta$ are complex numbers that fully determine the state. 

Since optimizers typically deal with real variables, it is convenient to rewrite the state using real parameters. Further, the degrees of freedom corresponding to the normalization and the global phase should be removed. With this, the two qubit state can be described by six real variational parameters. A possible parameterization is presented in \ref{def:2qubit_parameterization}.

\begin{align}
\begin{split}
    \ket{\psi} = &\cos{\frac{\theta_0}{2}}\cos{\frac{\theta_1}{2}}\ket{00} +
    \cos{\frac{\theta_0}{2}}\sin{\frac{\theta_1}{2}}e^{i\omega_1}\ket{01}+\\
    &\sin{\frac{\theta_0}{2}}e^{i\omega_0}\cos{\frac{\theta_2}{2}}\ket{10}+
    \sin{\frac{\theta_0}{2}}e^{i\omega_0}\sin{\frac{\theta_2}{2}}e^{i\omega_2}\ket{11}
    \end{split}
    \label{def:2qubit_parameterization}
\end{align}

Here, $\theta_i \in [0,\pi]$ and $\omega_i \in [0,2\pi]$. This was the parameterization used in the original \gls{VQE} article \cite{Peruzzo2014}, and it will be used here as well.

We wish to transform this parameterized description of the state into a parameterized circuit that we can use as our ansatz. The concept of Schmidt decomposition, presented in theorem 2.7 of reference \cite{NielsenChuang}, is useful here.

If we have a pure state of a composite system $AB$, $\ket{\psi_{A,B}}$, then the Schmidt decomposition allows us to write it as 

\begin{equation}
\ket{\psi_{A,B}}= \sum_i\lambda_i\ket{i_A}\ket{i_B},
\label{def:schmidt_decomp}
\end{equation}

where the $\ket{i_A}$, $\ket{i_B}$ are orthonormal states of the systems $A$, $B$ respectively. They form the \textit{Schmidt basis} of the respective system. The $\lambda_i$, called \textit{Schmidt coefficients}, are real non-negative numbers that satisfy $\sum_i\lambda_i^2=1$.

We can find the Schmidt decomposition of a two qubit state written in the computational basis (formula \ref{def:2qubit_coordinates}) by using \gls{SVD}. Firstly, we must organize the coordinates into a two-by-two matrix $M$.

\begin{equation}
M=
\begin{pmatrix}
\alpha & \beta\\
\gamma & \delta
\end{pmatrix}
\label{eq:coordinate_matrix}
\end{equation}

This is done by identifying the coordinate of computational basis state $\ket{ij}$ with the matrix entry $M_{i,j}$: the left qubit, $A$, is associated with row indices, and the right qubit, $B$, is associated with column indices. Calculating the scalar product of the state of the composite system with a state of system $A$ or $B$ ($\ket{\psi_A}$ or $\ket{\psi_B}$) corresponds to left- or right-multiplying matrix $M$ by the bra $\bra{\psi_A}$ or the ket $\ket{\psi_B}$, which results in a linear superposition of the rows or columns of matrix $M$, respectively. Calculating the scalar product of the state of the composite system $AB$ with a computational basis state of the same system amounts to doing both simultaneously, i.e.

\begin{equation*}
\braket{ij}{\psi_{A,B}} \equiv \bra{i}
\begin{pmatrix}
\alpha & \beta\\
\gamma & \delta
\end{pmatrix}
\ket{j}.
\end{equation*}

We can now use \gls{SVD} decomposition to write $M$ in the more convenient way of \ref{eq:svd_decomp}.

\begin{equation}
M= U
\begin{pmatrix}
\lambda_0 & 0\\
0 & \lambda_1
\end{pmatrix}
V
\label{eq:svd_decomp}
\end{equation}

The $\lambda_i$ are the \textit{singular values} of matrix $M$, non-negative real numbers that we can easily identify with the Schmidt coefficients of formula \ref{def:schmidt_decomp}. $U$ and $V^T$ (note the transpose) are two-by-two unitary matrices, whose column vectors form the Schmidt basis for $A$ and $B$ respectively.

It will be necessary to create a circuit to implement $U$ and $V^T$. A generic single-qubit unitary can be written by means of three rotations, along with a global phase. This is done in \ref{eq:1qubit_su2}.

\begin{align}
    U&=e^{i\theta_0}R_Z(\theta_1)R_Y(\theta_2)R_Z(\theta_3)
    =e^{i\theta_0}e^{\frac{-i\theta_1Z}{2}}e^{\frac{-i\theta_2Y}{2}}e^{\frac{-i\theta_3Z}{2}}\\
    &=e^{i\theta_0}
    \begin{pmatrix}
    e^{\frac{-i(\theta_1+\theta_3)}{2}}\cdot\cos(\frac{\theta_2}{2}) & -e^{\frac{-i(\theta_1-\theta_3)}{2}}\cdot\sin(\frac{\theta_2}{2}) \\
    e^{\frac{-i(-\theta_1+\theta_3)}{2}}\cdot\sin(\frac{\theta_2}{2}) &
    e^{\frac{-i(-\theta_1-\theta_3)}{2}}\cdot\cos(\frac{\theta_2}{2})
    \end{pmatrix}
    \label{eq:1qubit_su2}
\end{align}

The rotation angles can easily be found from this; for example, they can be obtained using the expressions in \ref{eq:1qubit_su2_angles}, where $\angle U_{ij}$ denotes the phase $\phi$ of the complex number $\rho e^{i\phi}$ in line \textit{i} and column \textit{j} of matrix $U$, and $|U_{ij}|$ denotes its modulus $\rho$.

\begin{align}
\begin{split}
    \theta_0=\frac{\angle U_{00} + \angle U_{11}}{2} 
    &\qquad
    \theta_1=-\angle U_{00} + \angle U_{10}\\
    \theta_2=2\arccos{|U_{00}|} 
    &\qquad
    \theta_3=-\angle U_{00} + \angle (-U_{10})
\end{split}
\label{eq:1qubit_su2_angles}
\end{align}

It should be noted that $\theta_0$ corresponds to a global phase with no relevance. The only parameters that are important and should be found are $\theta_1$, $\theta_2$ and $\theta_3$.

A procedure to create a circuit that prepares a state from its four complex coordinates in the computational basis can now be easily described. Once $U$, $V$, $\lambda_0$ and $\lambda_1$ have been found by \gls{SVD} decomposition (allowing us to rewrite the coordinate matrix as in \ref{eq:svd_decomp}), it can be divided into the three following steps:

\begin{enumerate}
    \item Create the state $\lambda_0\ket{00}+\lambda_1\ket{10}$ (from $\ket{00}$). This can be done by rotating qubit $A$ around the $Y$ axis by an angle of $2\arccos{|\lambda_0|}$. Because $\lambda_0^2+\lambda_1^2=1$, this is the same angle as $2\arcsin{|\lambda_1|}$.
    \item Create the state $\lambda_0\ket{00}+\lambda_1\ket{11}$ (from $\lambda_0\ket{00}+\lambda_1\ket{10}$). This can be done by applying a \gls{CNOT}, with qubit $A$ as the control and qubit $B$ as the target.
    \item Create the state $U\otimes V^{T} (\lambda_0\ket{00}+\lambda_1\ket{11})$ (from $\lambda_0\ket{00}+\lambda_1\ket{11}$). This can be done by decomposing the single qubit unitaries $U$ and $V^{T}$ into single qubit rotations, and applying them to qubits $A$ and $B$ respectively.
\end{enumerate}

Steps 1 and 2 create the state $\lambda_0\ket{00}+\lambda_1\ket{11}$, which is the desired state \textit{in the Schmidt basis}. Step 3 rotates each qubit from its Schmidt basis to the computational basis, creating the final state. The corresponding unitaries are those formed by the Schmidt vectors of the respective qubits.

\begin{figure}[htbp]
    \centering
    \includegraphics[width=0.8\textwidth]{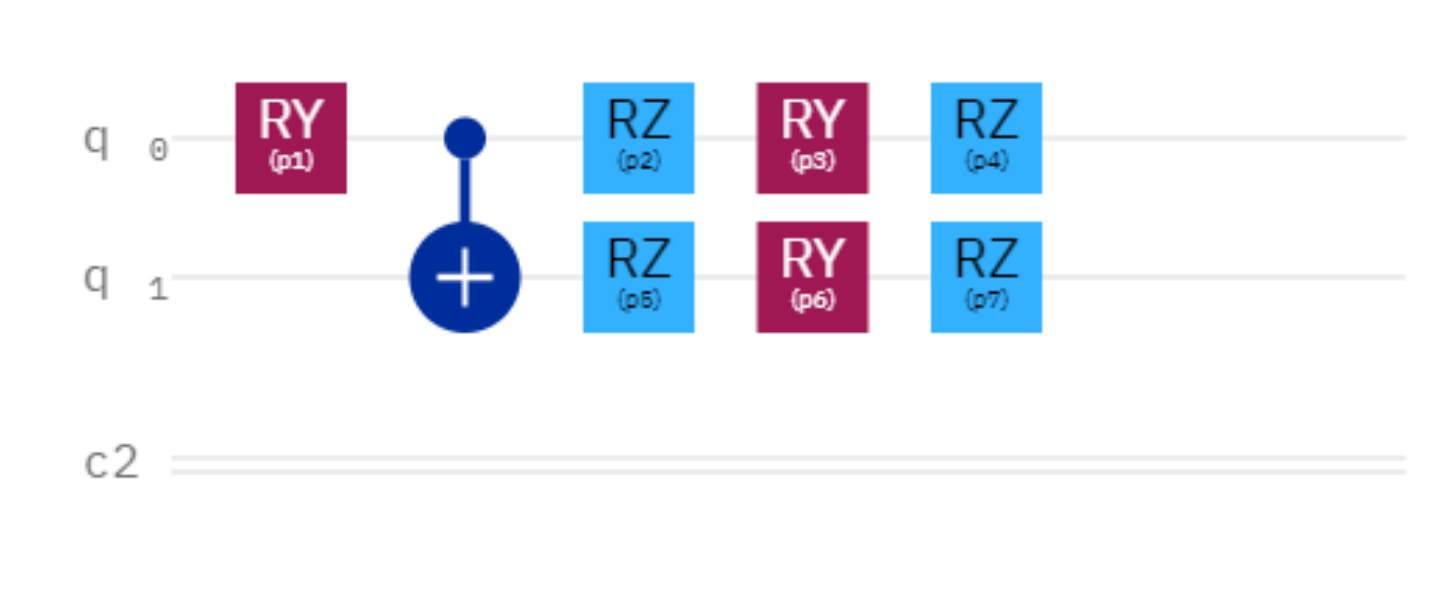}
    \caption{Circuit with 7 variational parameters for preparing an arbitrary 2-qubit state, drawn in the IBM Quantum Composer \cite{IBMcomposer}.}
    \label{fig:ansatz_7params}
\end{figure}

The outlined procedure leads to the state preparation circuit presented in figure \ref{fig:ansatz_7params}. Such an ansatz has 7 variational parameters: one from the rotation on the first qubit in step 1, and 3 from the decomposition of each of the unitaries in step 3.

One thing that stands out is that there were \textit{six} variational parameters in the parameterization \ref{def:2qubit_parameterization}. In fact, there are six degrees of freedom left in a two-qubit state once we remove those corresponding to the normalization and the global phase. This means that there is a redundant degree of freedom in our parameterized circuit.

Luckily, this can easily be solved by noting that after the two first steps (i.e. after the \gls{CNOT} gate), we have a symmetric state $\lambda_0\ket{00}+\lambda_1\ket{11}$. This means that the first two parallel Z rotations, with parameters $p2$ and $p5$ in figure \ref{fig:ansatz_7params}, can be merged into a single Z rotation with angle $p2+p5$ on either of the qubits, with exactly the same effect. 

\begin{figure}[htbp]
    \centering
    \includegraphics[width=0.8\textwidth]{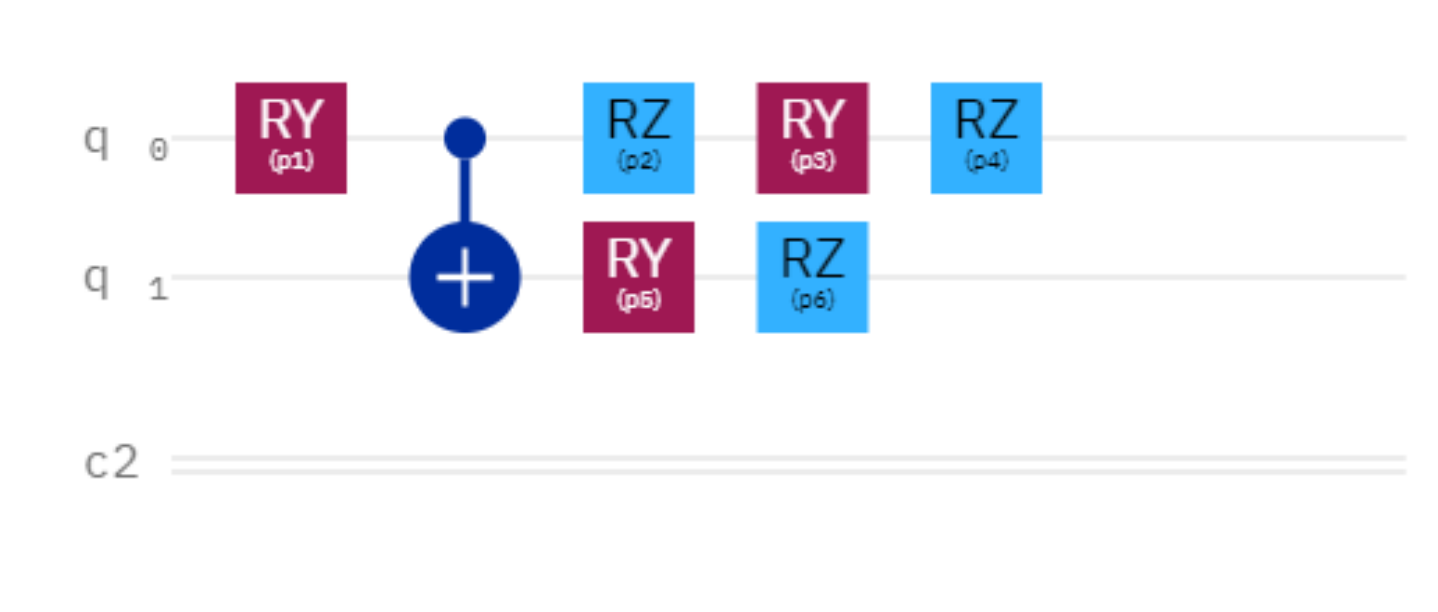}
    \caption{Circuit with 6 variational parameters for preparing an arbitrary 2-qubit state, drawn in the IBM Quantum Composer \cite{IBMcomposer}.}.
    \label{fig:ansatz_6params}
\end{figure}

The final 6-parameter ansatz is presented in figure \ref{fig:ansatz_6params}, where the parameter labels were redefined to run from $p1$ to $p6$.

In this project's \gls{VQE} implementation, the parameters chosen by the optimizer were the ones corresponding to parameterization \ref{def:2qubit_parameterization}. At each step, the parameters output by the optimizer were treated as follows:

\begin{itemize}
    \item Calculate the complex coordinates $\alpha$, $\beta$, $\gamma$, $\delta$ (formula \ref{def:2qubit_coordinates}) from the parameterization \ref{def:2qubit_parameterization}. Organize them into a two-by-two matrix $M$ as in formula \ref{eq:coordinate_matrix}.
    \item Use \gls{SVD} to obtain the $U$, $V$, $\lambda_0$, $\lambda_1$ that allow rewriting the matrix $M$ as in formula \ref{eq:svd_decomp}. This was done using the \textit{svd} function from SciPy's \cite{scipy} linear algebra module.
    \item Use the three-step procedure outlined before to create the 6-parameter circuit that prepares the desired state.
\end{itemize}

\subsubsection{Bond Dissociation Graph}

\begin{figure}[htbp]
    \centering
    \includegraphics[width=0.9\textwidth]{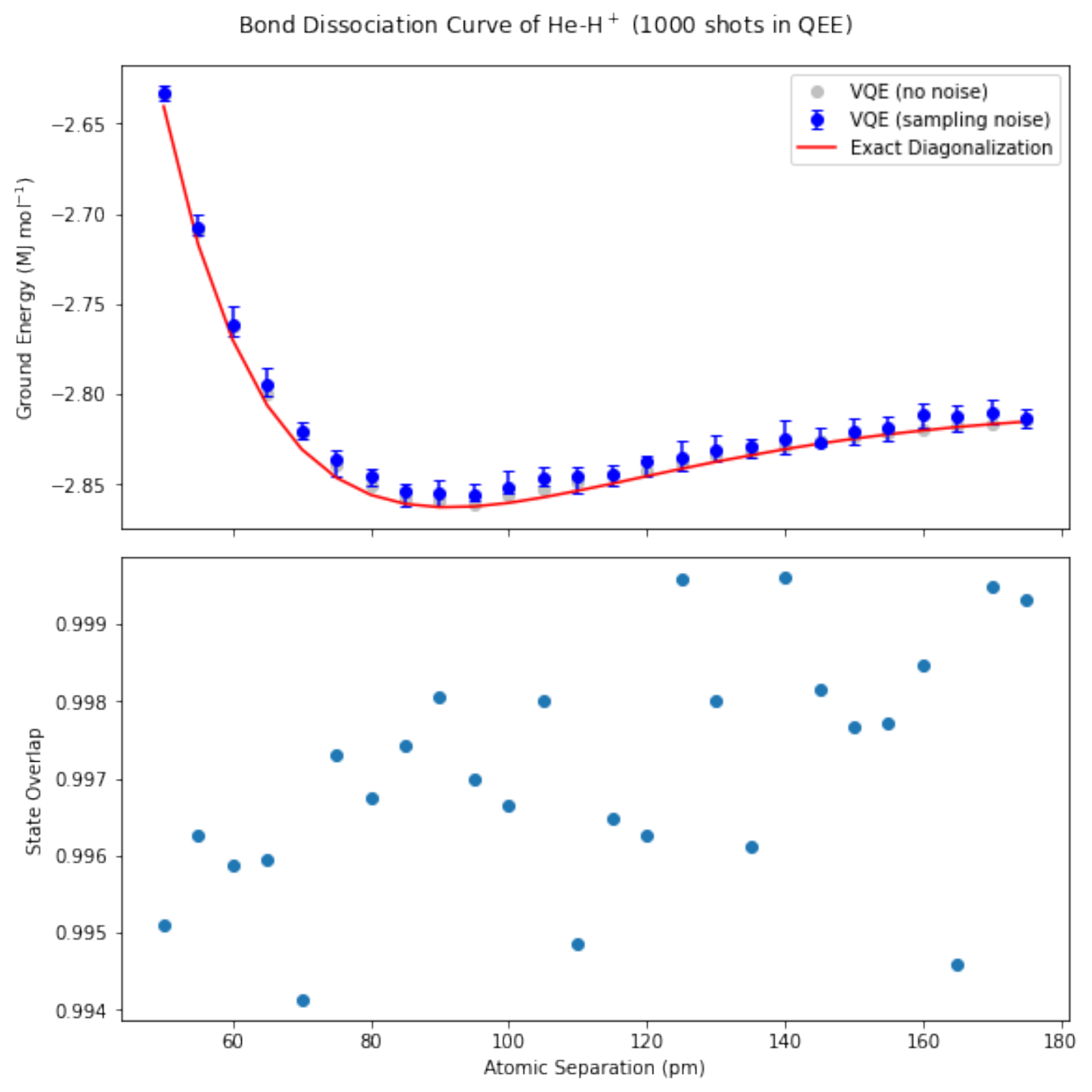}
    \caption{Performance of the \gls{VQE} algorithm along the bond dissociation graph of $HeH^+$, with and without sampling noise. The upper plot shows the \gls{VQE} energy, with and without sampling noise, plotted against the exact diagonalization energy (which is exact within the basis set and up to the Born-Oppenheimer approximation and relativistic effects). The lower plot shows the overlap of the \gls{VQE} state (with sampling noise) with the ground state obtained by exact diagonalization. In the cases with sampling noise, 1000 shots were used. 25 points were used for each interatomic distance; the median was used as the estimator, and the error bars show the interquartile ranges. The quantum circuits were run on the CIRQ simulator \cite{CIRQ}.}
    \label{fig:p_NM_25_opt}
\end{figure}

Before delving into a deeper analysis of the impact of the optimizer and of generic noise in the \gls{VQE} results, a simple illustration of the performance of the algorithm is presented in figure \ref{fig:p_NM_25_opt}. This figure shows the \gls{VQE} results for the helium hydride ion along its bond dissociation curve, for simulations with and without sampling noise (and no other types). The Nelder-Mead simplex method from SciPy \cite{scipy} was used as the optimizer, with properly adjusted hyperparameters.

It can be seen that the results are mostly consistent from run to run, and even with only 1000 shots in the expectation estimation the energy is consistently close to the ground energy obtained from exact diagonalization. 

It must be noted that the employed ansatz was overly generic (and overly expressible as a consequence), and it was used along with a completely random starting state. In most runs, the starting state did not have the correct particle number (among other symmetries) nor appreciable overlap with the ground state. Regardless, the \gls{VQE} energy points clearly trace out the bond dissociation curve. Further, at any given point, the final \gls{VQE} state has an overlap of over $99\%$ with the true ground state.

\subsubsection{Effect of the Optimizer and Hyperparameters}

The role of the classical optimizer is pivotal not only in \gls{VQE}, but in \glspl{VQA} in general: a bad optimizer might hinder the performance of the algorithm. Further, what is a \textit{good} and a \textit{bad} optimizer is specific to the quantum scenario, since noise (be it sampling or others) implies an added difficulty, and one that is not typically contemplated in benchmarking classical optimizers.

Another relevant factor in the optimization are the hyperparameters - the parameters that control the learning process. Because optimization-based algorithms are themselves computationally heavy, it is typically not easy to find the optimal hyperparameters, and they may not generalize from small-scale instances of the problem.

To illustrate the relevance of all of this, the performance of two different optimizers in obtaining \gls{VQE} results (under the same conditions as figure \ref{fig:p_NM_25_opt}) will be compared.

\begin{figure}[htbp]
    \centering
     \begin{subfigure}[b]{0.45\textwidth}
         \centering
         \includegraphics[width=\textwidth]{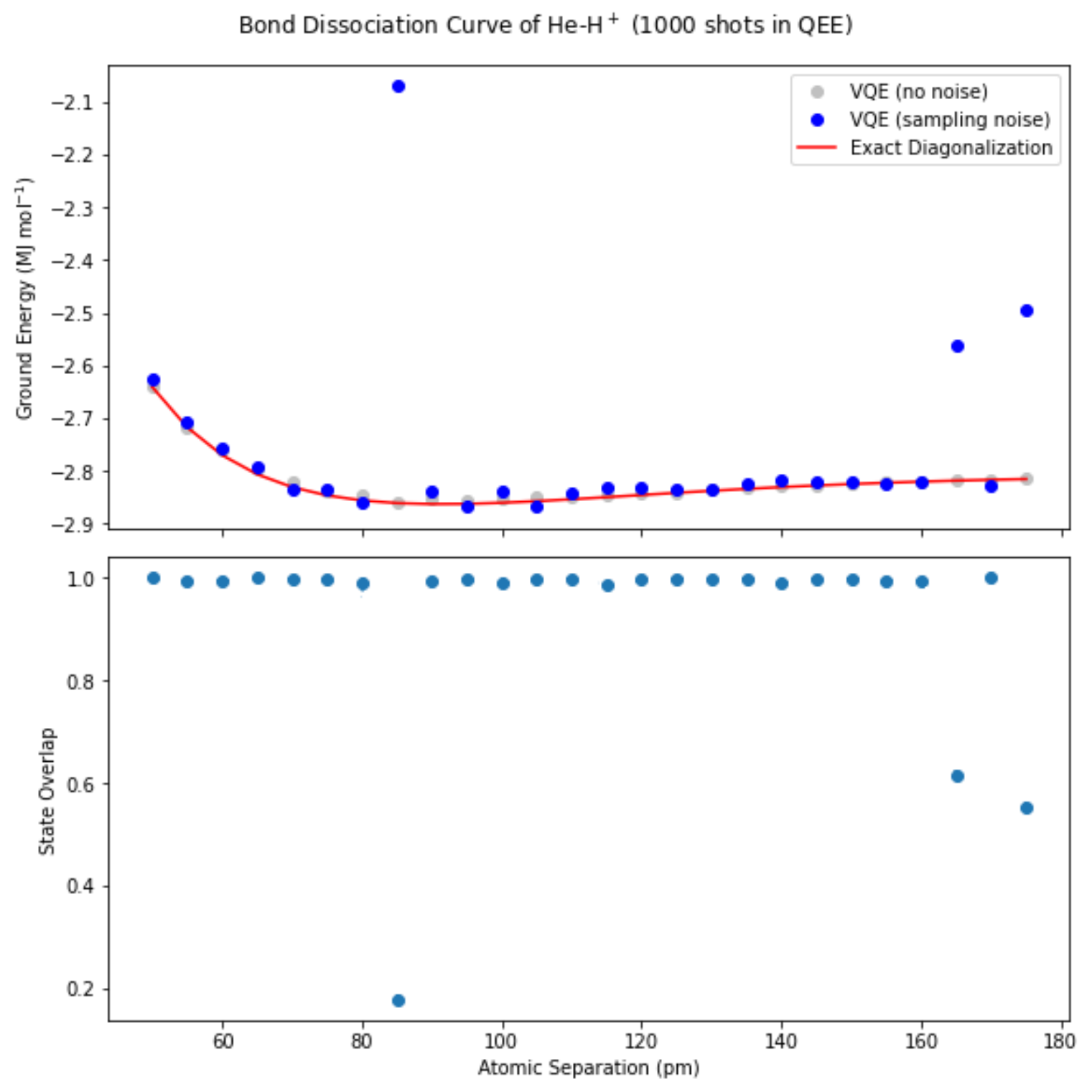}
         \caption{1 run}
         \label{fig:p_COBYLA_1}
     \end{subfigure}
     \hfill
     \begin{subfigure}[b]{0.45\textwidth}
         \centering
         \includegraphics[width=\textwidth]{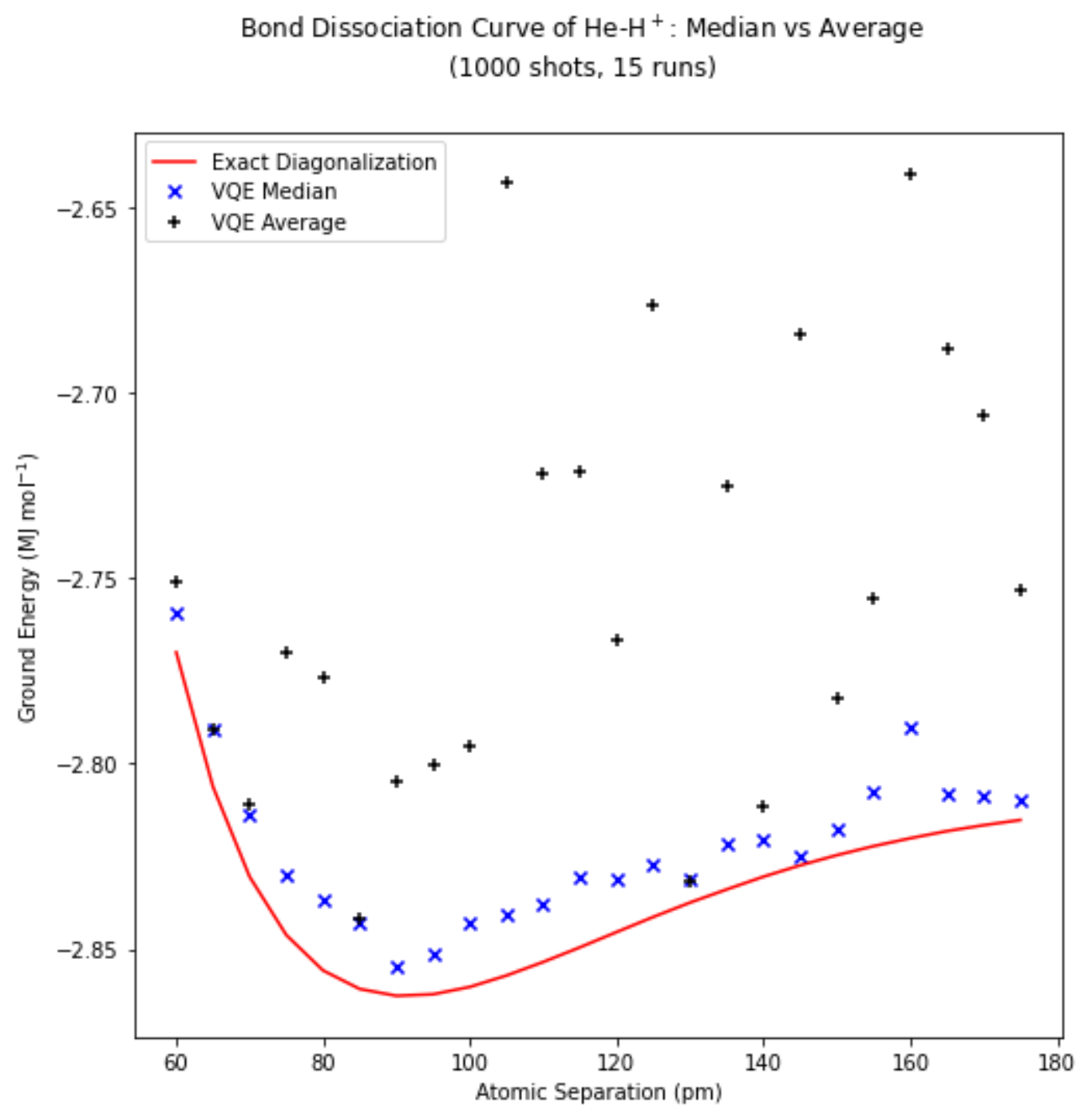}
         \caption{15 runs}
         \label{fig:p_COBYLA_15_mvsa}
     \end{subfigure}
     \caption{Performance of the \gls{VQE} algorithm with the \gls{COBYLA} optimizer along the bond dissociation graph of $HeH^+$, with default hyperparameters. The upper plot in figure \ref{fig:p_COBYLA_1} compares the single-run \gls{VQE} energy with the exact diagonalization energy, with and without sampling noise (1000 shots). For each interatomic distance, the same random starting point was used for both cases. The lower plot shows the overlap of the \gls{VQE} state, obtained in the simulations with sampling noise, with the ground state obtained from exact diagonalization. Figure \ref{fig:p_COBYLA_15_mvsa} compares the median and the average of the \gls{VQE} energy over 15 runs with the exact diagonalization value, in the case where 1000 shots are used per energy evaluation. The quantum circuits were run in the CIRQ Simulator \cite{CIRQ}.}
     \label{fig:p_COBYLA}
\end{figure}

Figure \ref{fig:p_COBYLA} shows the \gls{VQE} results with the \gls{COBYLA} (Constrained Optimization BY Linear Approximation) optimization method. Figure \ref{fig:p_COBYLA_1} shows the single-run performance of this optimizer with the default hyperparameters in the SciPy \cite{scipy} implementation. In this case, the relevant hyperparameters are the maximum number of iterations, the final accuracy required for convergence, and the magnitude of initial changes to the variables. Remarkably, even with the default hyperparameters, the optimizer converges to a reasonable approximation to the ground state ($>90$\% overlap) in all but three points. In can be seen that sampling noise represents a significant challenge to successful convergence: in the few points whose bad results stand out, the completely noise-free version of the algorithm still outputs an energy that essentially coincides with the exact value, even though the starting point was the same.

Figure \ref{fig:p_COBYLA_15_mvsa} compares the median of the \gls{VQE} energy against the average for 15 runs, in simulations with 1000 shots in the quantum expectation estimation. While in some cases they almost coincide, it is more frequent that the median is a significantly better estimator. This is caused by aberrant runs in which the optimizer doesn't successfully converge to the ground state: they are promptly excluded by the median, while they still weigh in on the average. The robustness of the median causes it to be a suitable choice here, and this is the estimator that was used in this project whenever the gathered data included more than one run per point.

Taking the median over as few as ten runs proved to be enough for the \gls{COBYLA} optimizer (with default hyperparameters) to output states with over $90\%$ overlap with the solution throughout the whole bond dissociation curve, which is a solid result considering that the whole Hilbert space is being searched through and the starting state is entirely random.

\begin{figure}[htbp]
    \centering
     \begin{subfigure}[b]{0.45\textwidth}
         \centering
         \includegraphics[width=\textwidth]{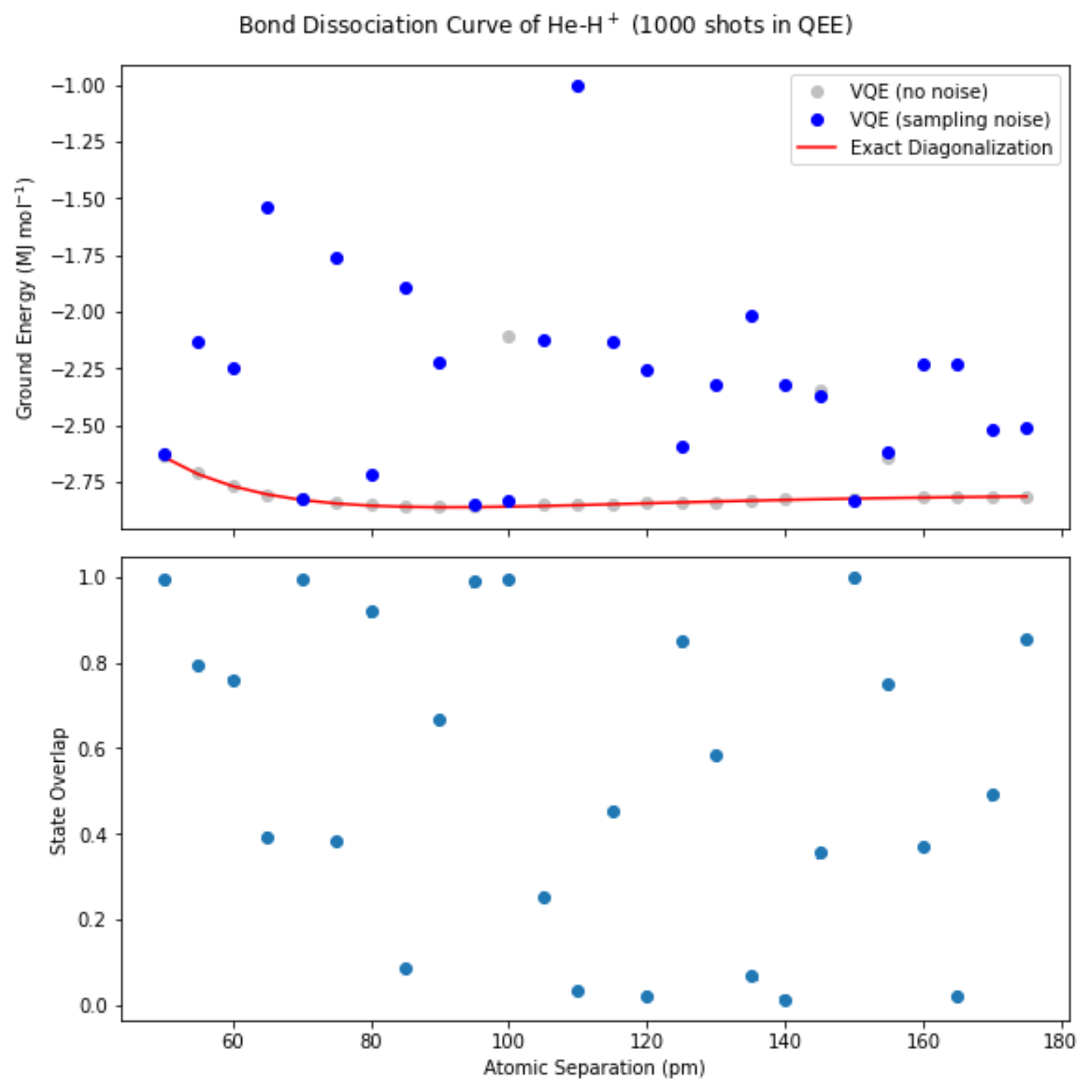}
         \caption{Single run, default hyperparameters}
         \label{fig:p_NM_1_unopt}
     \end{subfigure}
     \hfill
     \begin{subfigure}[b]{0.45\textwidth}
         \centering
         \includegraphics[width=\textwidth]{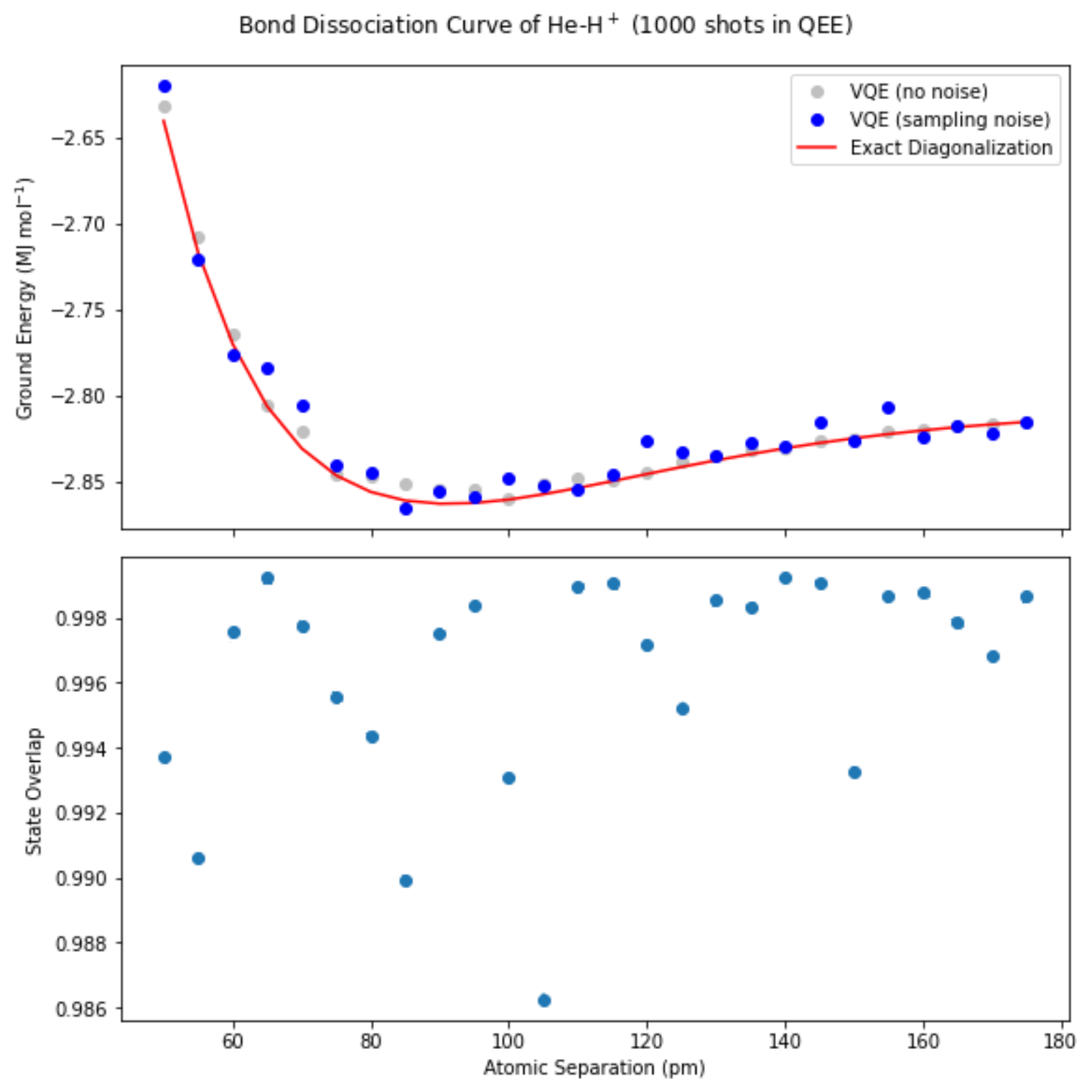}
         \caption{Single run, optimized hyperparameters}
         \label{fig:p_NM_1_opt}
     \end{subfigure}
     \\
     \begin{subfigure}[b]{0.45\textwidth}
         \centering
         \includegraphics[width=\textwidth]{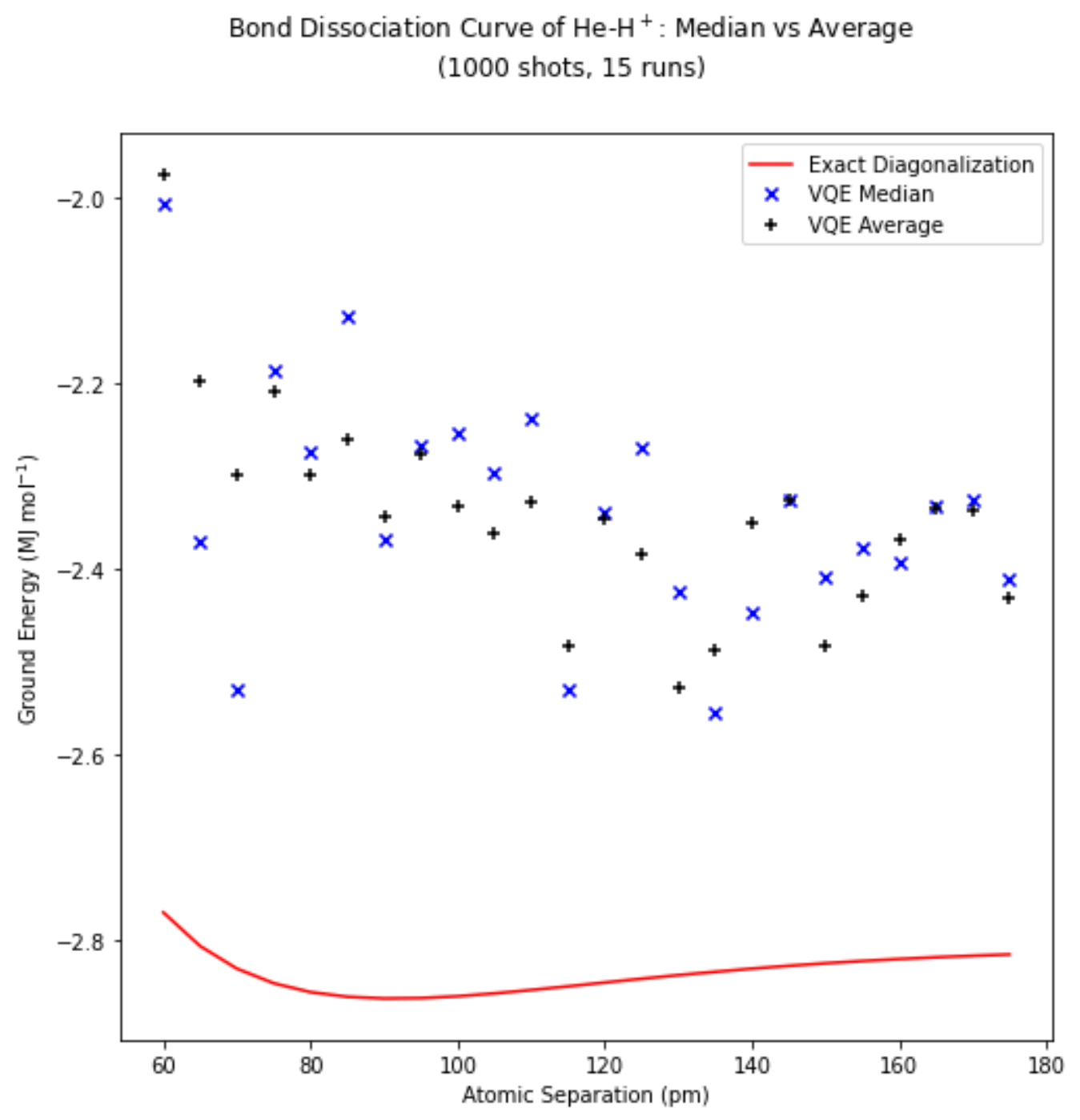}
         \caption{15 runs, default hyperparameters}
         \label{fig:p_NM_15_mvsa_unopt}
     \end{subfigure}
     \hfill
     \begin{subfigure}[b]{0.45\textwidth}
         \centering
         \includegraphics[width=\textwidth]{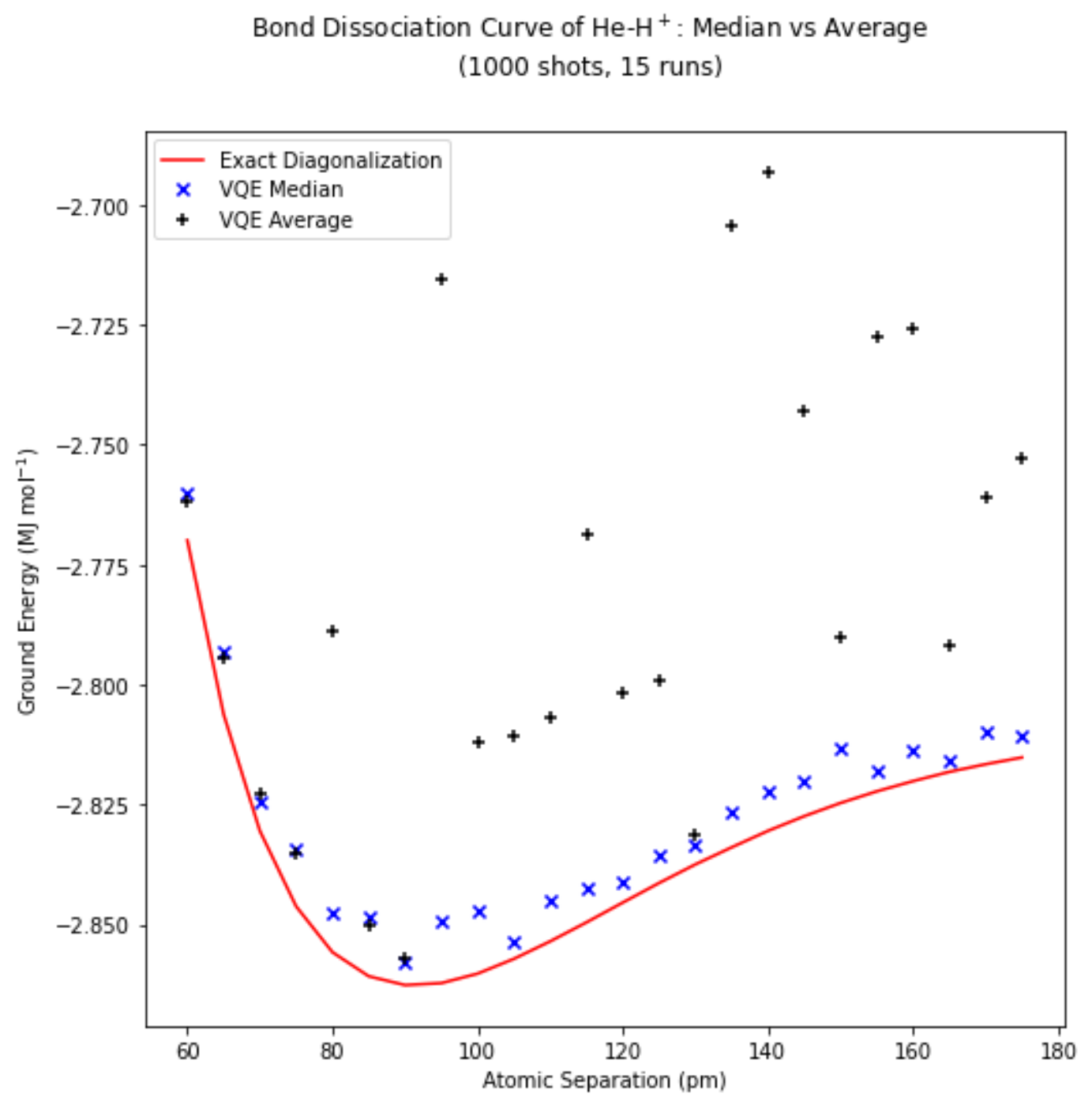}
         \caption{15 runs, optimized hyperparameters}
         \label{fig:p_NM_15_mvsa_opt}
     \end{subfigure}
    \caption{Performance of the \gls{VQE} algorithm with the Nelder-Mead optimizer along the bond dissociation graph of $HeH^+$, before (\ref{fig:p_NM_1_unopt}, \ref{fig:p_NM_15_mvsa_unopt}) and after (\ref{fig:p_NM_1_opt}, \ref{fig:p_NM_15_mvsa_opt}) hyperparameter optimization. The upper plots in figures \ref{fig:p_NM_1_unopt}, \ref{fig:p_NM_1_opt} compare the single-run \gls{VQE} energy with the exact diagonalization energy, with and without sampling noise (1000 shots). For each interatomic distance, the same random starting point was used for both cases. The lower plots show the overlap of the \gls{VQE} state, obtained in the simulations with sampling noise, with the ground state obtained by exact diagonalization. Figures \ref{fig:p_NM_15_mvsa_unopt}, \ref{fig:p_NM_15_mvsa_opt} compare the median and the average of the \gls{VQE} energy over 15 runs with the exact diagonalization value, in the case where 1000 shots are used per energy evaluation. The quantum circuits were run in the CIRQ Simulator \cite{CIRQ}.}
    \label{fig:p_NM}
\end{figure}

In figure \ref{fig:p_NM_1_unopt}, we can see that the performance of the Nelder-Mead optimization method is very different. With the default hyperparameters, Nelder-Mead optimized single-run \gls{VQE} often fails to converge to the ground state even without sampling noise. With sampling noise, this optimizer only successfully converges to the ground state ($>90$\% overlap) in six of the points. The noise in the data does not allow statistical filtering to improve upon the single-run behaviour: in figure \ref{fig:p_NM_15_mvsa_unopt} we can see that neither the average nor the median allow recovering a reasonable approximation to the ground state. In fact, the two estimators are always quite close, suggesting that nothing but noise is being recovered.

This poor performance of Nelder-Mead can greatly improve through an adjustment of the hyperparameters. In this case, the initial simplex is the most relevant. However, the error in parameters/function deemed acceptable for convergence, and the maximum number of function evaluations, may also pose too strict or too relaxed requirements for successful convergence. All of these hyperparameters of the Nelder-Mead optimizer were optimized using \gls{COBYLA}. In adjusting the initial simplex, a single parameter affecting the simplex size was used. The impact of this `meta-optimization' is reflected on figure \ref{fig:p_NM_1_opt}, which shows that each single run posterior to hyperparameter tuning resulted in an overlap with the ground state over 98.6\%. In \ref{fig:p_NM_15_mvsa_opt}, we can confirm that after hyperparameter optimization, Nelder-Mead \gls{VQE} allows for statistical treatment: the median is filtering out the aberrant runs, and for each interatomic distance we obtain an energy that is visibly close to the true ground energy. The median \gls{VQE} energies clearly trace out the bond dissociation curve.

Evidently, conclusions regarding which among these two is the best optimizer cannot be drawn from such scant results. In particular, this example has an overly expressible ansatz that might favour optimizers with greedier default initial changes to the variables. In fact, it was observed that when narrower ansätze (covering smaller regions of the Hilbert space) were used, the tendency was reversed. In that case, the \gls{COBYLA} optimizer frequently failed unless the initial changes to the variables were greatly decreased.

These examples serve as a demonstration of how the optimization process can affect the performance of \gls{VQE}, and how the choice of optimizer and hyperparameters is of the utmost importance.

\subsubsection{Running the Algorithm on Cloud Quantum Computers}

Up to this point, only \textit{sampling} noise was included in the simulations. Unfortunately, this is not realistic: real quantum computers suffer from many other sources of noise. A brief examination of the impact of generic noise on \gls{VQE} follows.

\begin{figure}[htbp]
     \centering
     \begin{subfigure}[b]{0.45\textwidth}
         \centering
         \includegraphics[width=\textwidth]{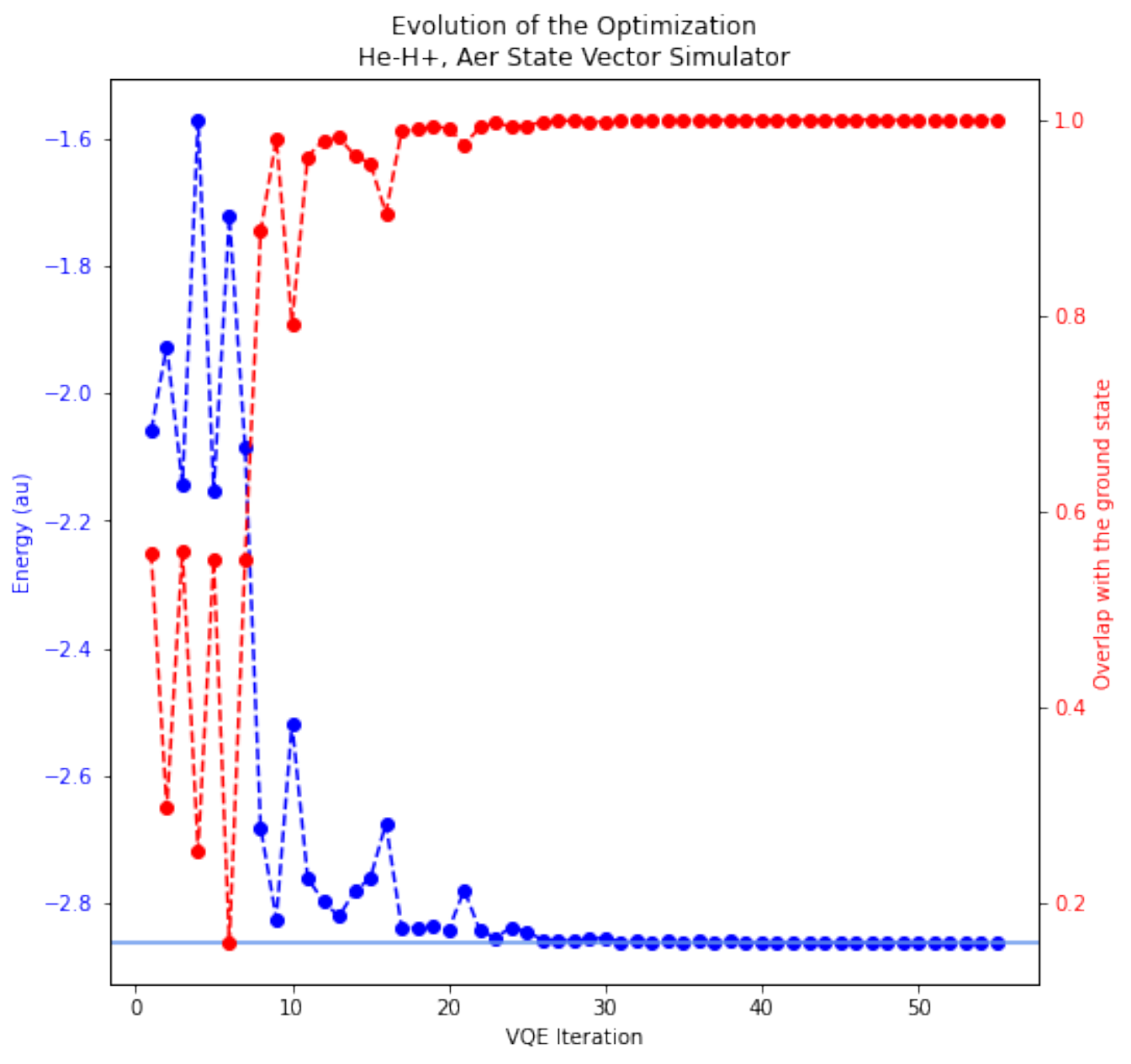}
         \caption{Statevector simulator}
         \label{fig:p_sv}
     \end{subfigure}
     \hfill
     \begin{subfigure}[b]{0.45\textwidth}
         \centering
         \includegraphics[width=\textwidth]{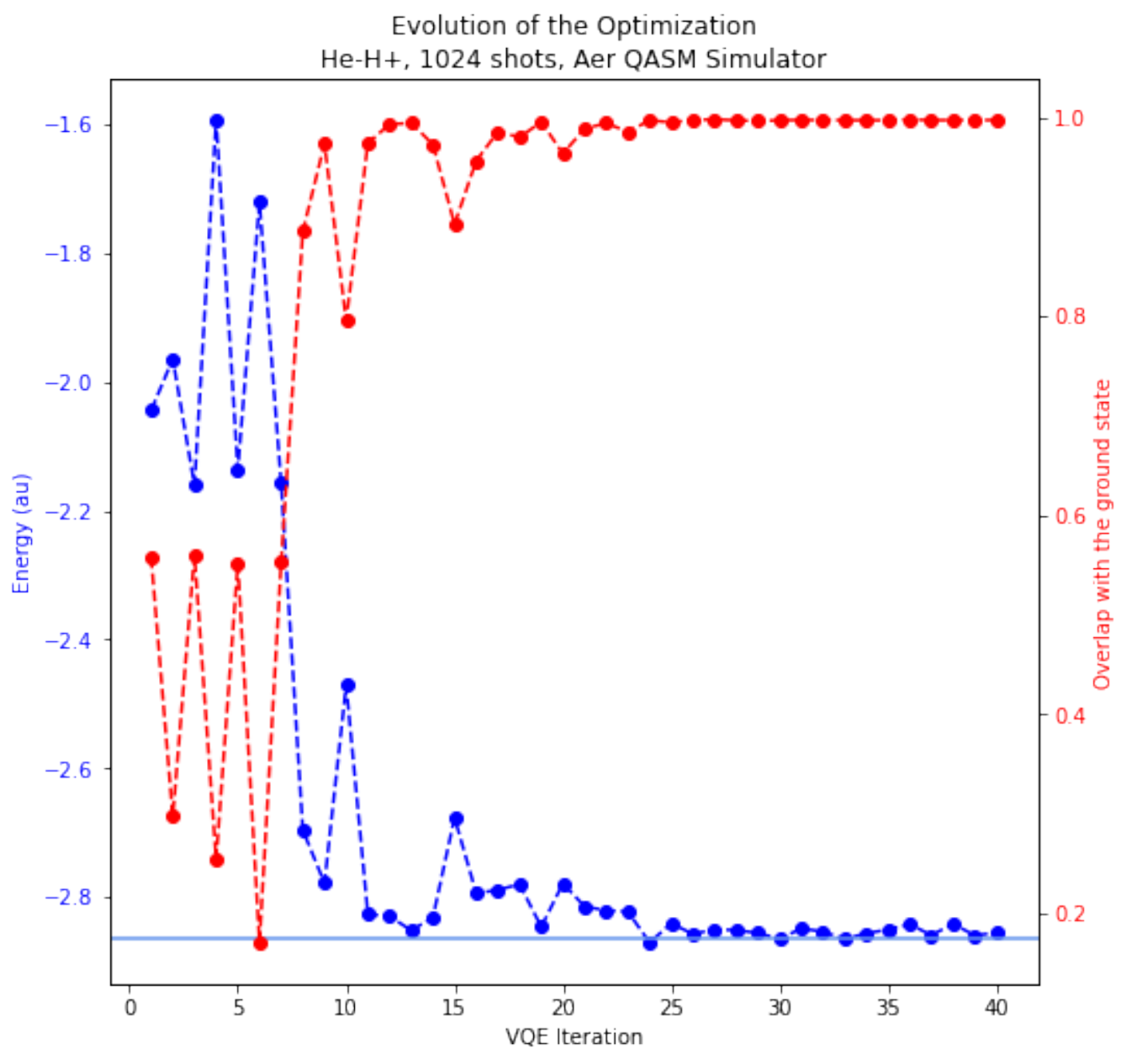}
         \caption{QASM Simulator}
         \label{fig:p_QASM_1024}
     \end{subfigure}
     \\
     \begin{subfigure}[b]{0.45\textwidth}
         \centering
         \includegraphics[width=\textwidth]{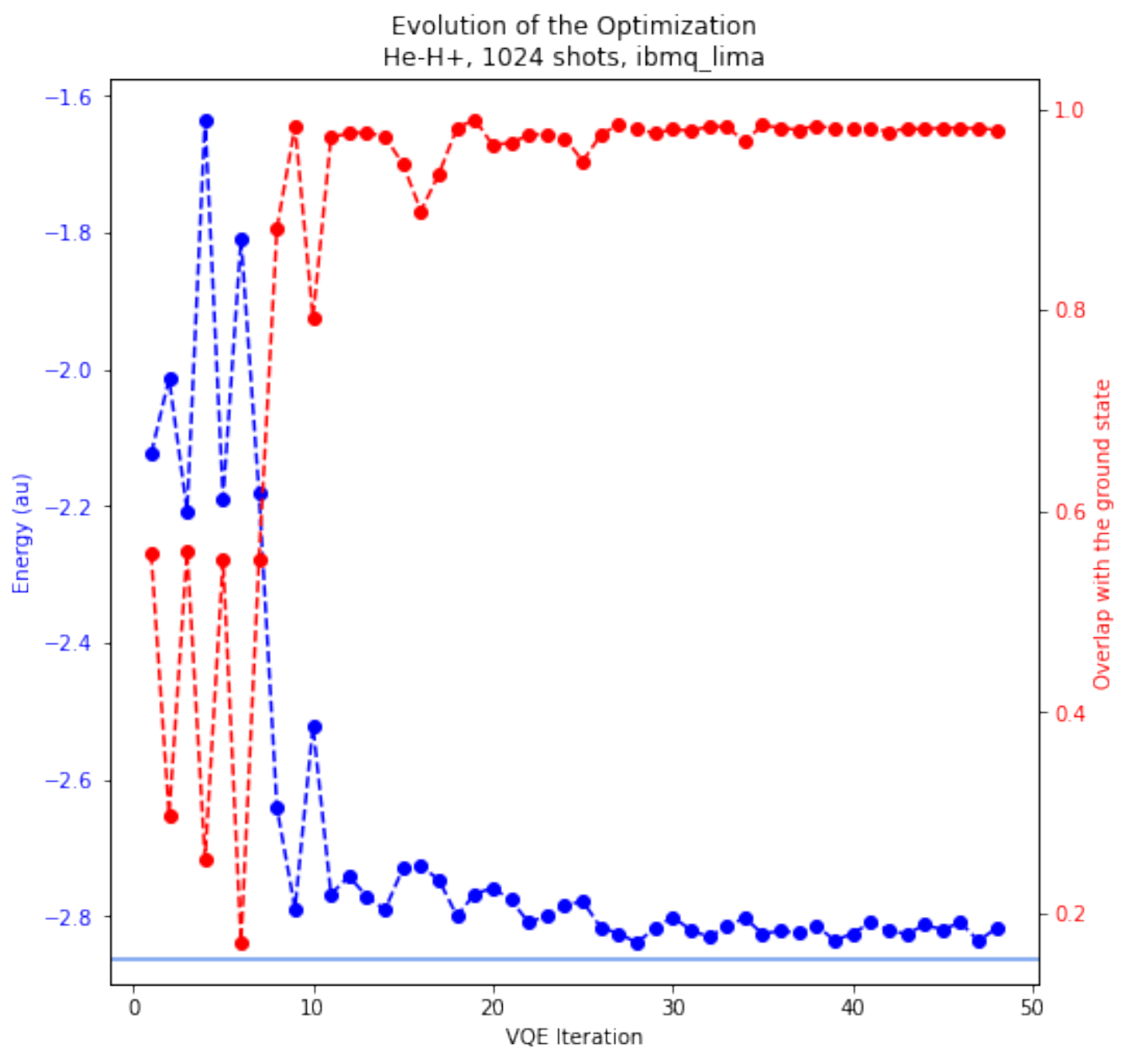}
         \caption{IBMQ Lima}
         \label{fig:p_lima_1024}
     \end{subfigure}
     \caption{Evolution of the \gls{VQE} optimization for $HeH^+$ at an interatomic distance of 90pm, using three different IBM Quantum \cite{IBMQ} backends: the statevector simulator (with no noise of any kind), the QASM simulator (with sampling noise) and the Lima device (a 5-qubit Falcon processor). In the last two cases, 1024 shots were used. The energy is plotted in blue and the overlap with the true ground state in red; the pale blue line marks the true ground energy. The same random starting point was used for all backends. The chosen optimizer was \gls{COBYLA}.}
     \label{fig:p_sv_qasm_lima}
\end{figure}

Figure \ref{fig:p_sv_qasm_lima} compares three distinct scenarios: noise-free (\ref{fig:p_sv}), sampling noise only (\ref{fig:p_QASM_1024}), and generic noise (real quantum computer; \ref{fig:p_lima_1024}). 

As expected, in the noise-free scenario, the optimizer has no issue in converging to the ground state. It stabilizes the energy around the ground energy, and after an uneventful convergence \gls{VQE} outputs a state whose overlap with the ground state is around 100\%. The effect of adding sampling noise (1024 shots) can be seen on figure \ref{fig:p_QASM_1024}. The curves show less stability, and oscillate more as convergence is reached. Regardless, the final energy still visibly nears the exact diagonalization value. When the scenario is shifted to a real quantum processor (figure \ref{fig:p_lima_1024}), all types of noise come into play (decoherence; \gls{SPAM} noise; coherent gate errors;...). 

The difference between the ideal performance of a quantum processor and a real one is the difference between figures \ref{fig:p_QASM_1024} and \ref{fig:p_lima_1024}. When \gls{VQE} is ran using IBM Quantum's Lima device, the energy does not only oscillate: even when convergence is reached, it seems to have a constant offset from the line marking the exact diagonalization energy. This is a symptom of noise-induced barren plateaus. Incoherent noise is capable of decreasing the magnitude of the minimum of the cost function. In the limit that we are left with the fully mixed state by the end of the circuit, the cost landscape is evidently flat. Before that limit is reached, noise will simply prevent the optimizer from ever finding the exact diagonalization ground energy, with a shift that will be the larger the greater the impact of noise.

\begin{figure}[htbp]
    \centering
     \begin{subfigure}[b]{0.45\textwidth}
         \centering
         \includegraphics[width=\textwidth]{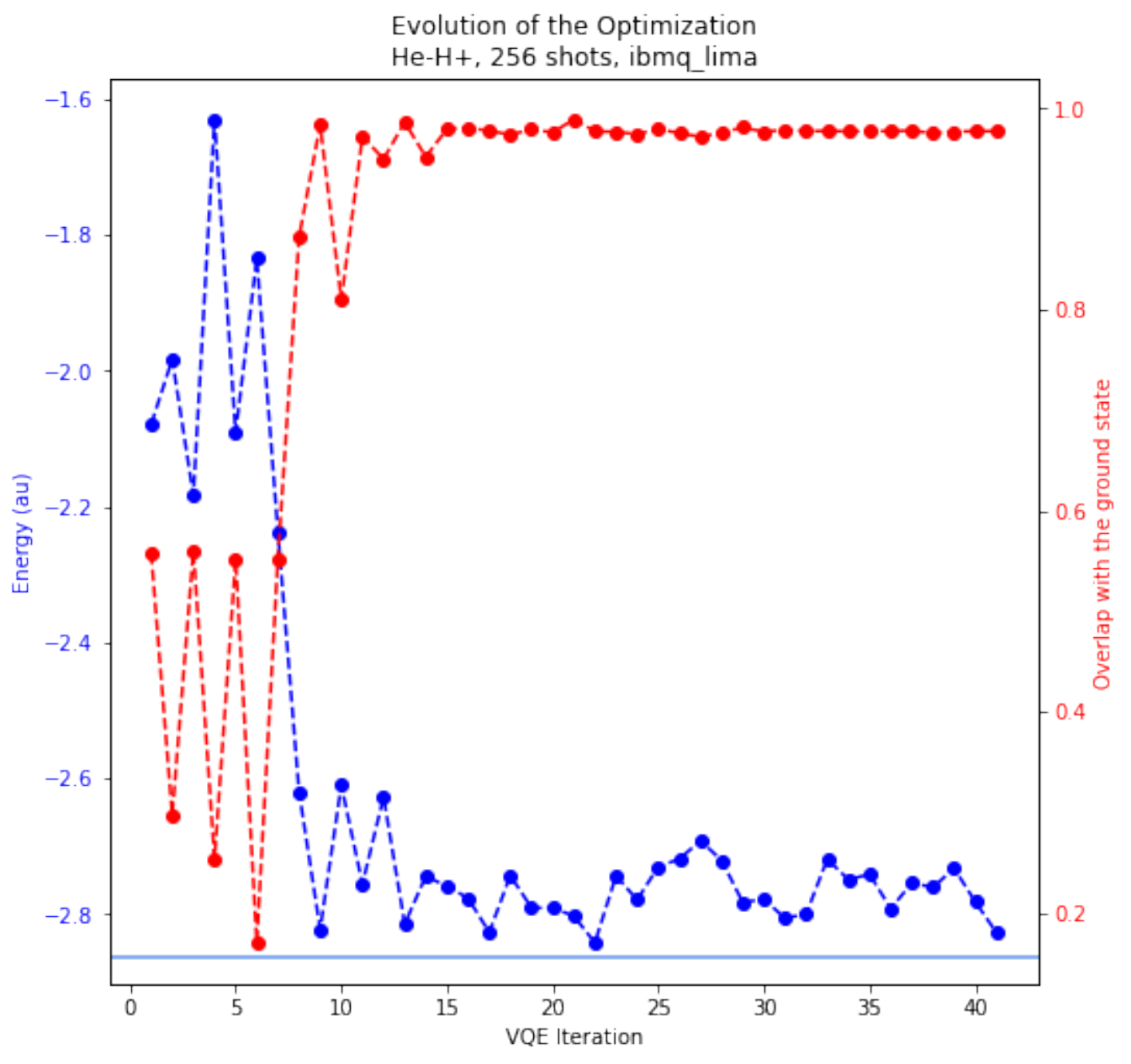}
         \caption{256 shots}
         \label{fig:p_lima_256shots}
     \end{subfigure}
     \hfill
     \begin{subfigure}[b]{0.45\textwidth}
         \centering
         \includegraphics[width=\textwidth]{p_lima_1024.pdf}
         \caption{1024 shots}
         \label{fig:p_lima_1024shots}
     \end{subfigure}
     \\
     \begin{subfigure}[b]{0.45\textwidth}
         \centering
         \includegraphics[width=\textwidth]{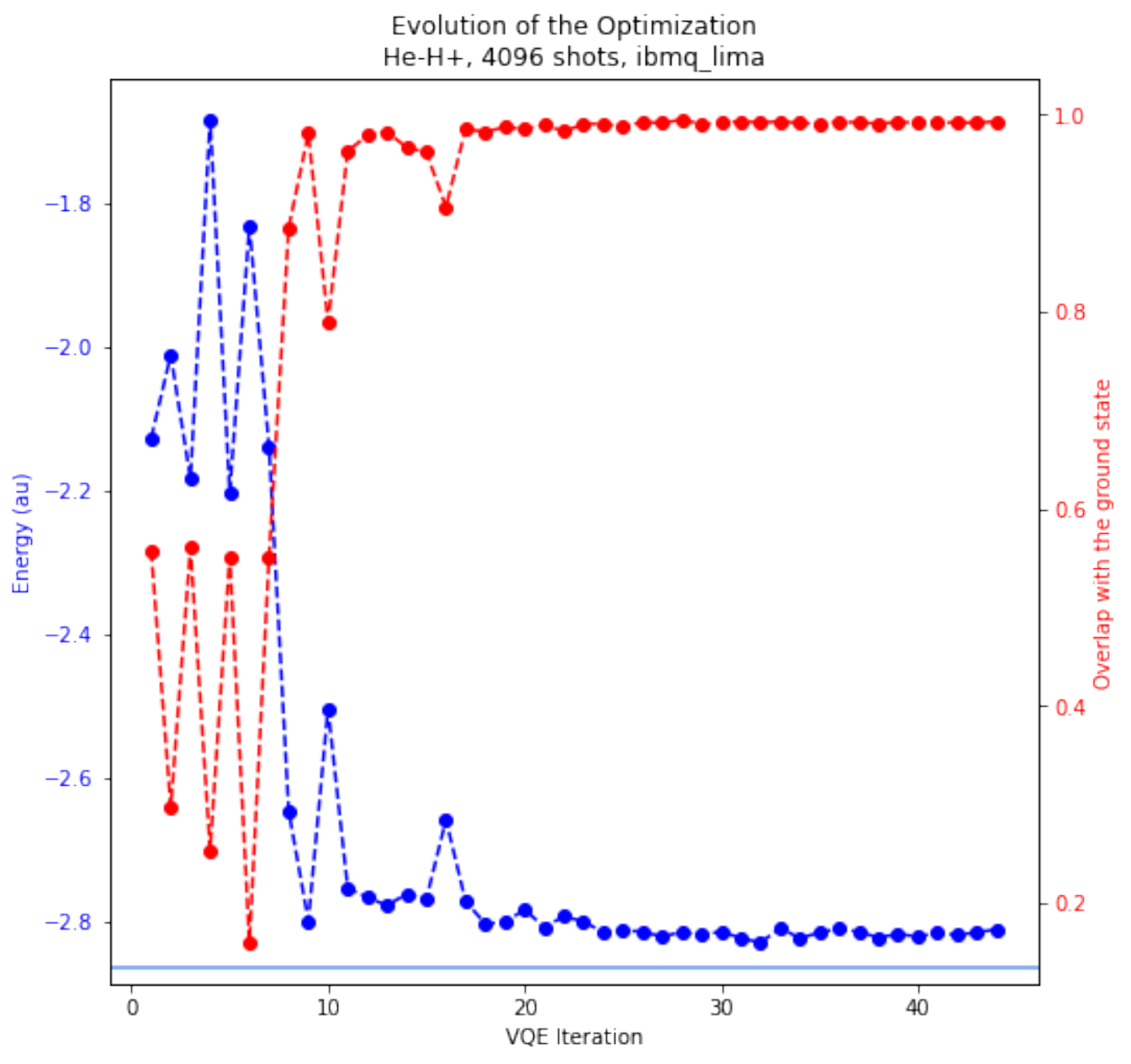}
         \caption{4096 shots}
         \label{fig:p_lima_4096shots}
     \end{subfigure}
     \hfill
     \begin{subfigure}[b]{0.45\textwidth}
         \centering
         \includegraphics[width=\textwidth]{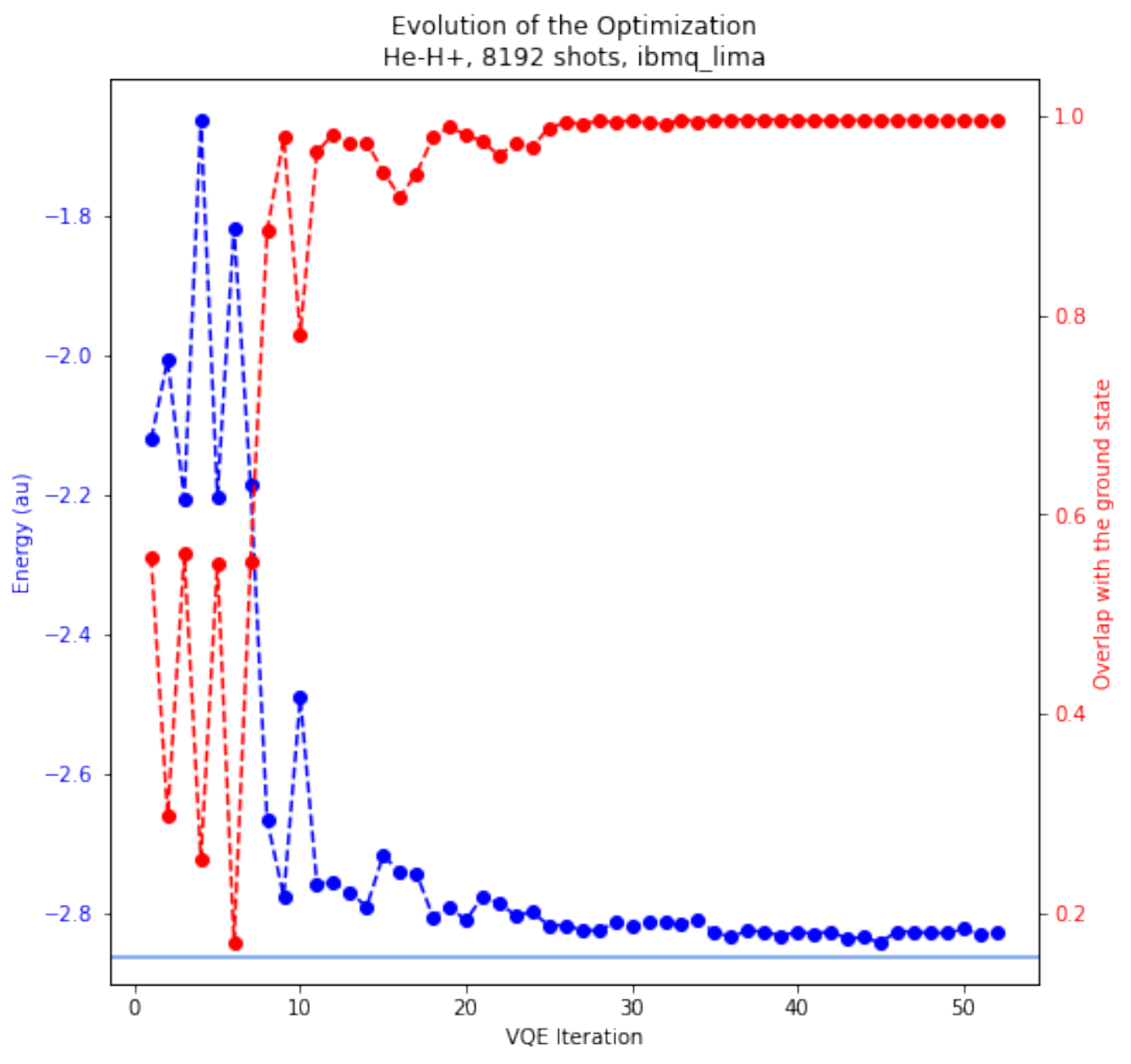}
         \caption{8192 shots}
         \label{fig:p_lima_8192shots}
     \end{subfigure}
    \caption{Evolution of the \gls{VQE} optimization for $HeH^+$ at an interatomic distance of 90pm, for different shot counts and using the Lima device. The remaining conditions (starting point, optimizer) are the same as in figure \ref{fig:p_sv_qasm_lima}.}
    \label{fig:p_lima}
\end{figure}

To better illustrate this, figure \ref{fig:p_lima} shows how the \gls{VQE} optimization changes as the number of shots is increased. The same quantum computer was used (IMBQ's Lima).

In theory, an arbitrary accuracy could be reached by increasing the number of shots. However, these plots suggest that a plateau is reached after a certain point. In figure \ref{fig:p_lima_8192shots}, we can see that even with a 4-fold increase in the number of shots, the Lima backend cannot perform as well as an ideal quantum processor would (\ref{fig:p_QASM_1024}). While with only 256 shots (\ref{fig:p_lima_256shots}) the shot count seemed to be the limiting factor in accuracy, that is no longer the case when it is increased to 8192 shots. At this point, the energy seldom oscillates from iteration to iteration, and it is evident that the issue goes beyond sampling noise. Evidently, increasing the shot count only allows us to go so far. A limit will be reached where only improving the characteristics of the quantum computer, or resorting to a quantum error correction will further improve accuracy.

\FloatBarrier
\section{Problem Tailored Ansätze}
\label{s:problem_tailored_ansatze}

Leveraging problem-specific knowledge to design the ansatz allows narrowing the search space by excluding regions that are not expected to contain the solution. Such techniques reduce the expressibility of the ansatz without abdicating precision, because the solution is still contained within the variational form. Since problem-tailoring is of great importance in preventing barren plateaus, and this trainability issue renders variational quantum algorithms inefficient, a great part of the research surrounding \glspl{VQA} is aimed at developing precise and shallow ansätze of this type.

This section contains the results obtained from applying \gls{VQE} to molecular hydrogen ($H_2$) both in simulators and real quantum computers., using a problem-tailored ansatz. The employed ansatz was \gls{UCCSD}, introduced in section \ref{s:quantum_chemistry}. As was explained then, this variational form arises in the context of conventional \gls{CC} theory as a natural modification to the non-variational wave form of the latter, which in turn is the natural size-consistent successor to the Configuration Interaction variational form. While \gls{UCCSD} is not classically tractable, the unitarity that forces variationality allows for convenient implementation in quantum circuits. Due to being unitary and having solid roots in older computational chemistry methods, the \gls{UCCSD} ansatz has become a popular choice in \glspl{VQA} addressing chemistry applications.

\subsection{Application to Molecular Hydrogen}

The \gls{UCCSD}-\gls{VQE} algorithm was implemented in this project using CIRQ \cite{CIRQ} and Qiskit \cite{Qiskit}, alternative software libraries for manipulating and running quantum circuits. Tasks such as obtaining the Hamiltonians and performing fermion-to-qubit mappings were done via OpenFermion \cite{openfermion} in the first case and Qiskit's chemistry module in the second. In both, the underlying electronic structure information was extracted from PySCF \cite{pyscf}. Additionally, an independent noise-free \gls{UCCSD}-\gls{VQE} based on matrix algebra was implemented to verify correctness.

Simulations were done to analyse the effect of sampling noise on \gls{UCCSD}-\gls{VQE}, as well as of \gls{CNOT} gate errors, and of thermal relaxation and \gls{SPAM} errors, using noise models created in Qiskit. Density matrix simulations were employed to analyse the purity of the final state, using noise models created from the specifications of IBMQ \cite{IBMQ} quantum processors to mimic their behaviour.

\subsubsection{Bond Dissociation Graph}

Much like in the previous section, we will start with the bond dissociation plots to get a general sense of performance before delving into the evolution of individual optimizations.

\begin{figure}[htbp]
    \centering
     \begin{subfigure}[b]{0.45\textwidth}
         \centering
         \includegraphics[width=\textwidth]{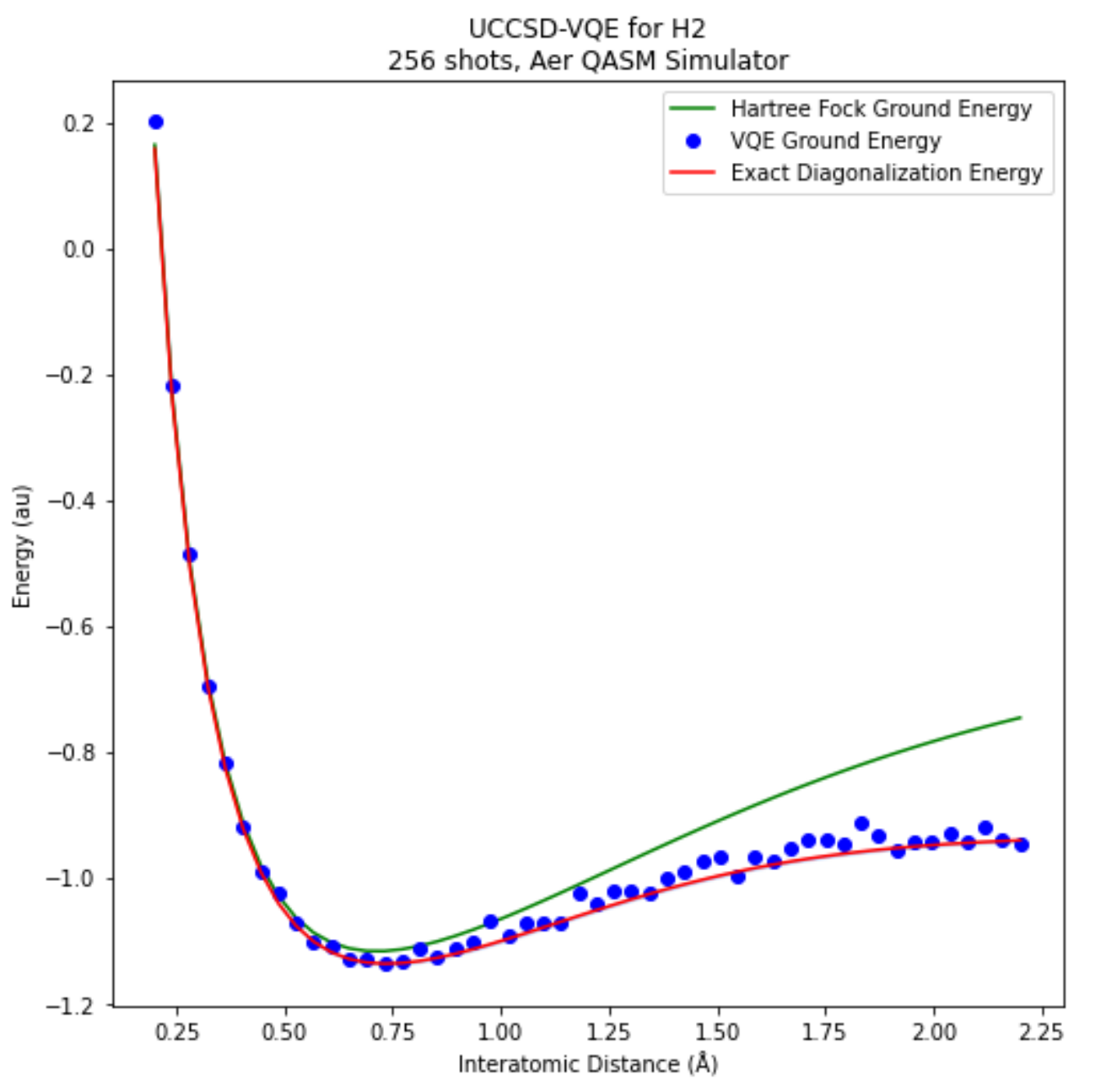}
         \caption{QASM simulator, 256 shots}
         \label{fig:h2_fullbc_uccsd_qasm_256}
     \end{subfigure}
     \hfill
     \begin{subfigure}[b]{0.45\textwidth}
         \centering
         \includegraphics[width=\textwidth]{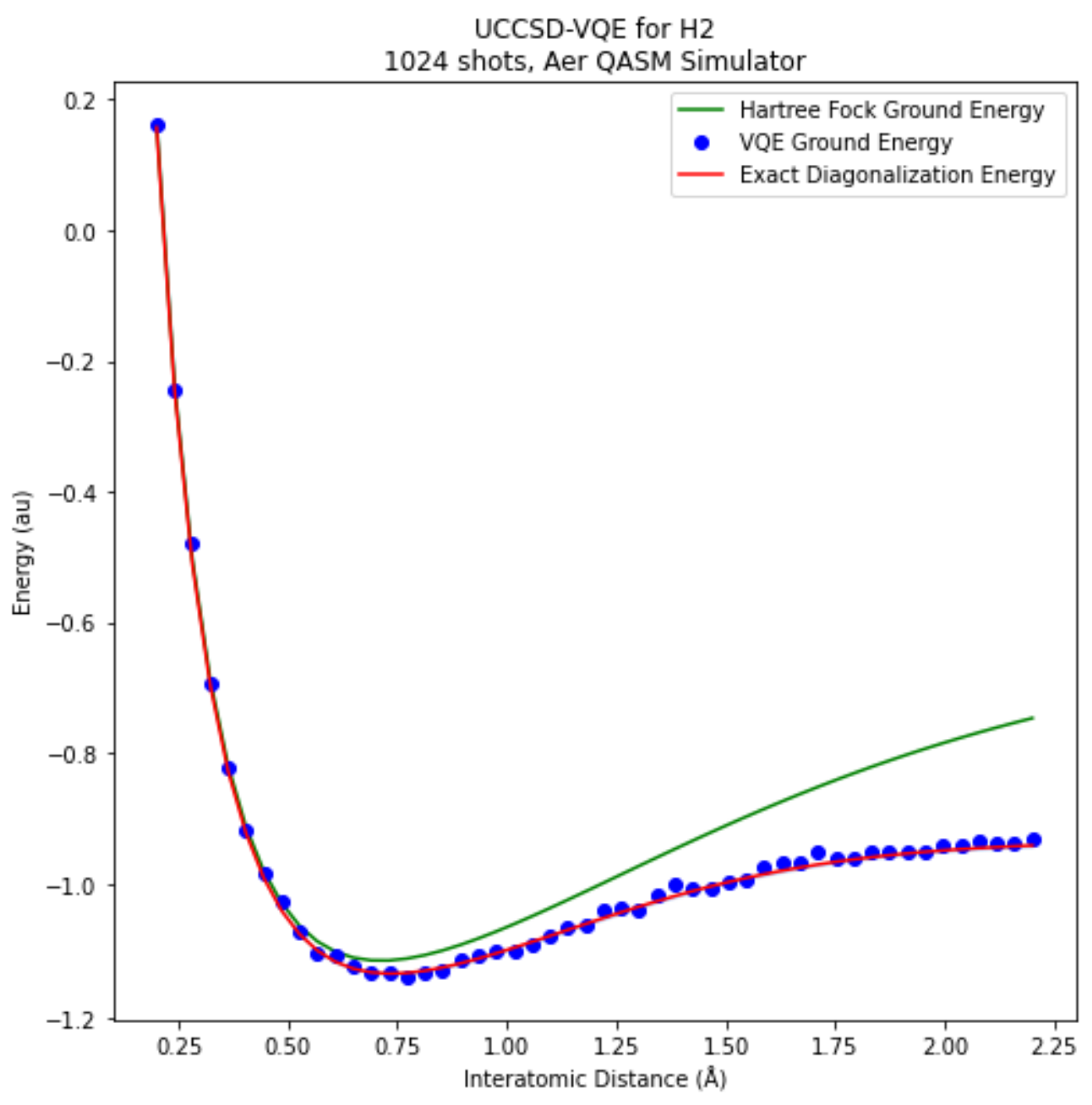}
         \caption{QASM simulator, 1024 shots}
         \label{fig:h2_fullbc_uccsd_qasm_1024}
     \end{subfigure}
     \\
     \begin{subfigure}[b]{0.45\textwidth}
         \centering
         \includegraphics[width=\textwidth]{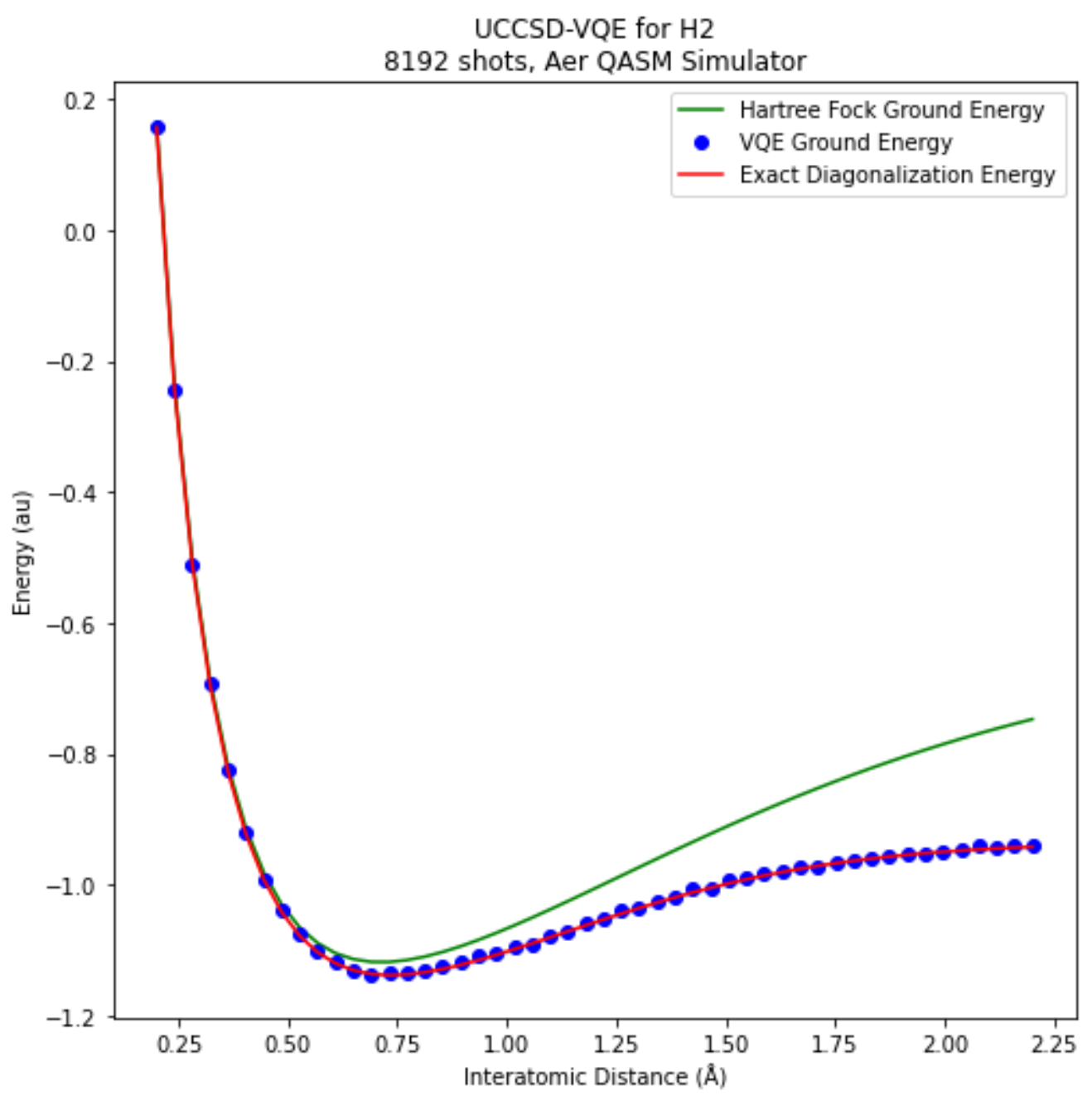}
         \caption{QASM simulator, 8192 shots}
         \label{fig:h2_fullbc_uccsd_qasm_8192}
     \end{subfigure}
     \hfill
     \begin{subfigure}[b]{0.45\textwidth}
         \centering
         \includegraphics[width=\textwidth]{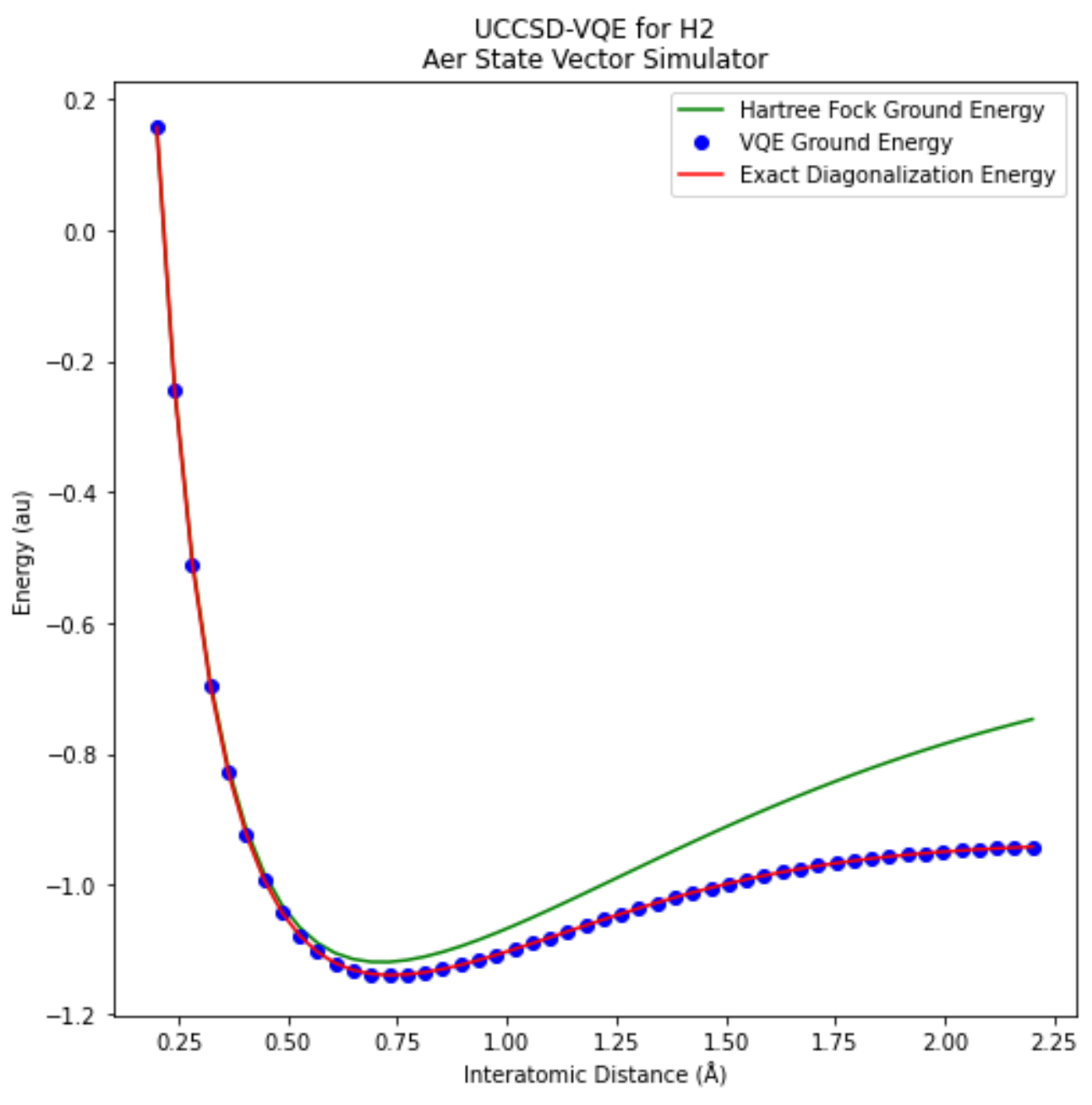}
         \caption{Statevector simulator}
         \label{fig:h2_fullbc_uccsd_statevector_simulator}
     \end{subfigure}
    \caption{Single-run \gls{UCCSD}-\gls{VQE} ground energy along the bond dissociation curve of the hydrogen molecule, obtained using IBMQ \cite{IBMQ} simulators. The results plotted in figures \ref{fig:h2_fullbc_uccsd_qasm_256}-\ref{fig:h2_fullbc_uccsd_qasm_8192}, ordered by increasing shot count, were obtained using the QASM simulator. The results plotted in figure \ref{fig:h2_fullbc_uccsd_statevector_simulator} were obtained using the state vector simulator (equivalent to infinite shots). The red and green curves represent the \gls{FCI} energy and the Hartree-Fock energy, respectively.}
    \label{fig:h2_uccsd_fullbc}
\end{figure}

In figure \ref{fig:h2_uccsd_fullbc}, we can see the \gls{VQE} energies tracing out the bond dissociation curve of $H_2$. Here, the simulations include no noise other than sampling.

The effect of the number of shots clearly shows in the results, with the \gls{VQE} energy getting closer and closer to the exact diagonalization curve from figure \ref{fig:h2_fullbc_uccsd_qasm_256} to \ref{fig:h2_fullbc_uccsd_statevector_simulator}. In the limit of infinite shots, the \gls{VQE} energy essentially matches the \gls{FCI} result (the observed error was of the order of $10^{-9}$).

It is interesting to compare the \gls{VQE} energy with that obtained by performing the Hartree-Fock self-consistent field calculations. The state resulting from these calculations is the reference state in \gls{VQE}, upon which the \gls{UCCSD} ansatz performs a `correction'. It becomes evident that, in the presence of sufficiently weak noise, optimizing the \gls{UCCSD} variational form is enough to bring the energy away from the Hartree-Fock approximation and into the \gls{FCI} value. The variational freedom takes care of electronic correlation effects that go unaccounted for in the Hartree-Fock approximation. 

\begin{figure}[htbp]
    \centering
     \begin{subfigure}[b]{0.45\textwidth}
         \centering
         \includegraphics[width=\textwidth]{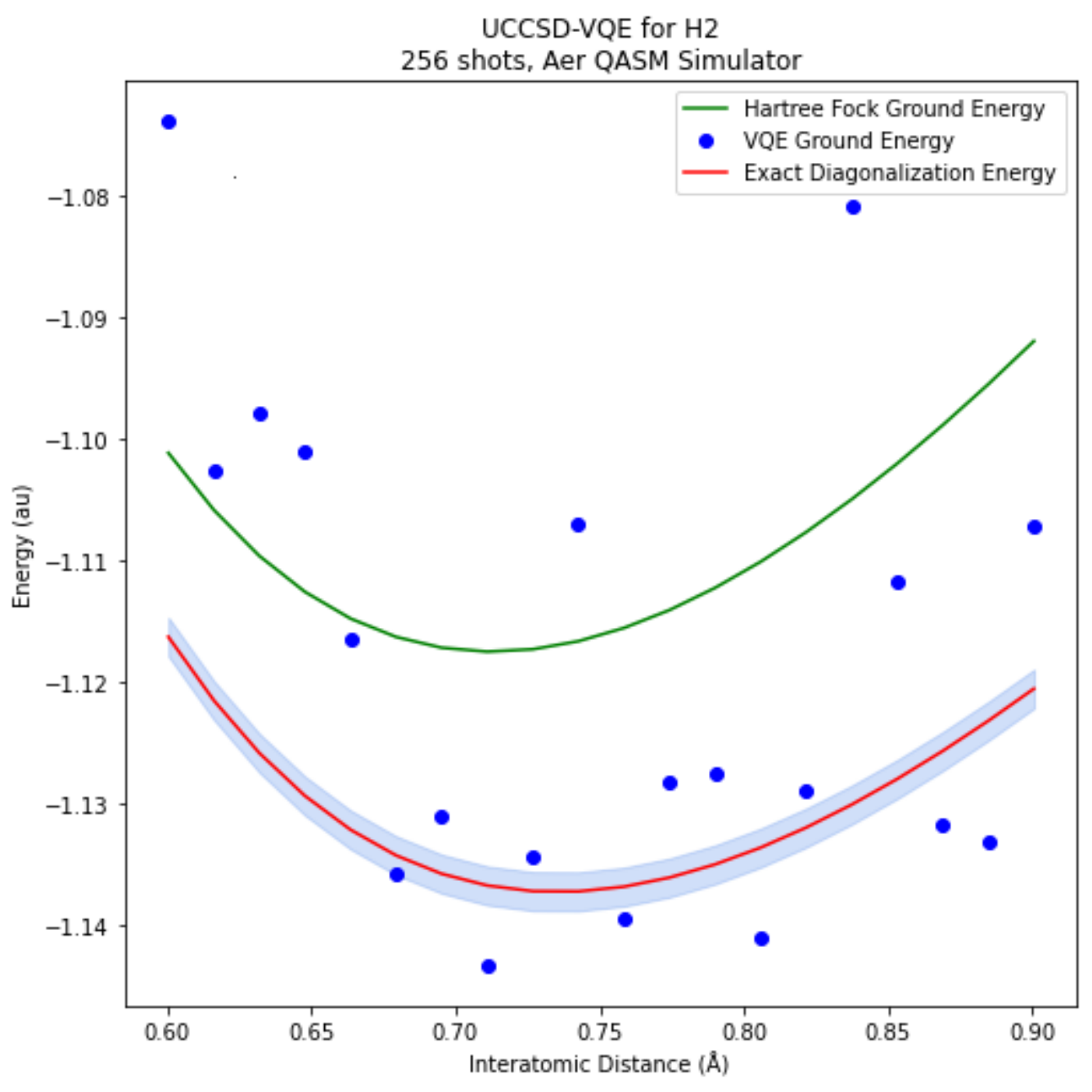}
         \caption{QASM simulator, 256 shots}
         \label{fig:h2_zoombc_uccsd_qasm_256}
     \end{subfigure}
     \hfill
     \begin{subfigure}[b]{0.45\textwidth}
         \centering
         \includegraphics[width=\textwidth]{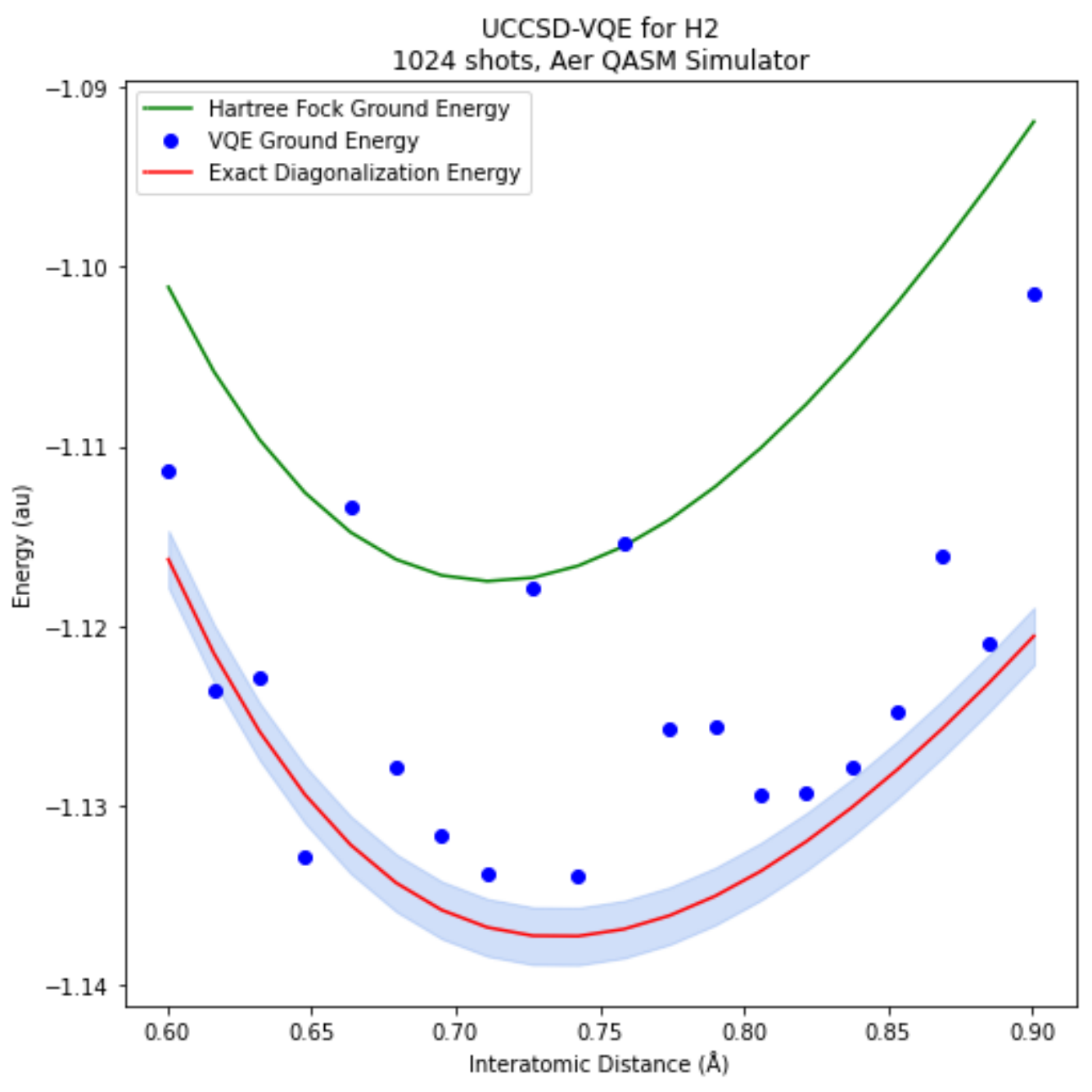}
         \caption{QASM simulator, 1024 shots}
         \label{fig:h2_zoombc_uccsd_qasm_1024}
     \end{subfigure}
     \\
     \begin{subfigure}[b]{0.45\textwidth}
         \centering
         \includegraphics[width=\textwidth]{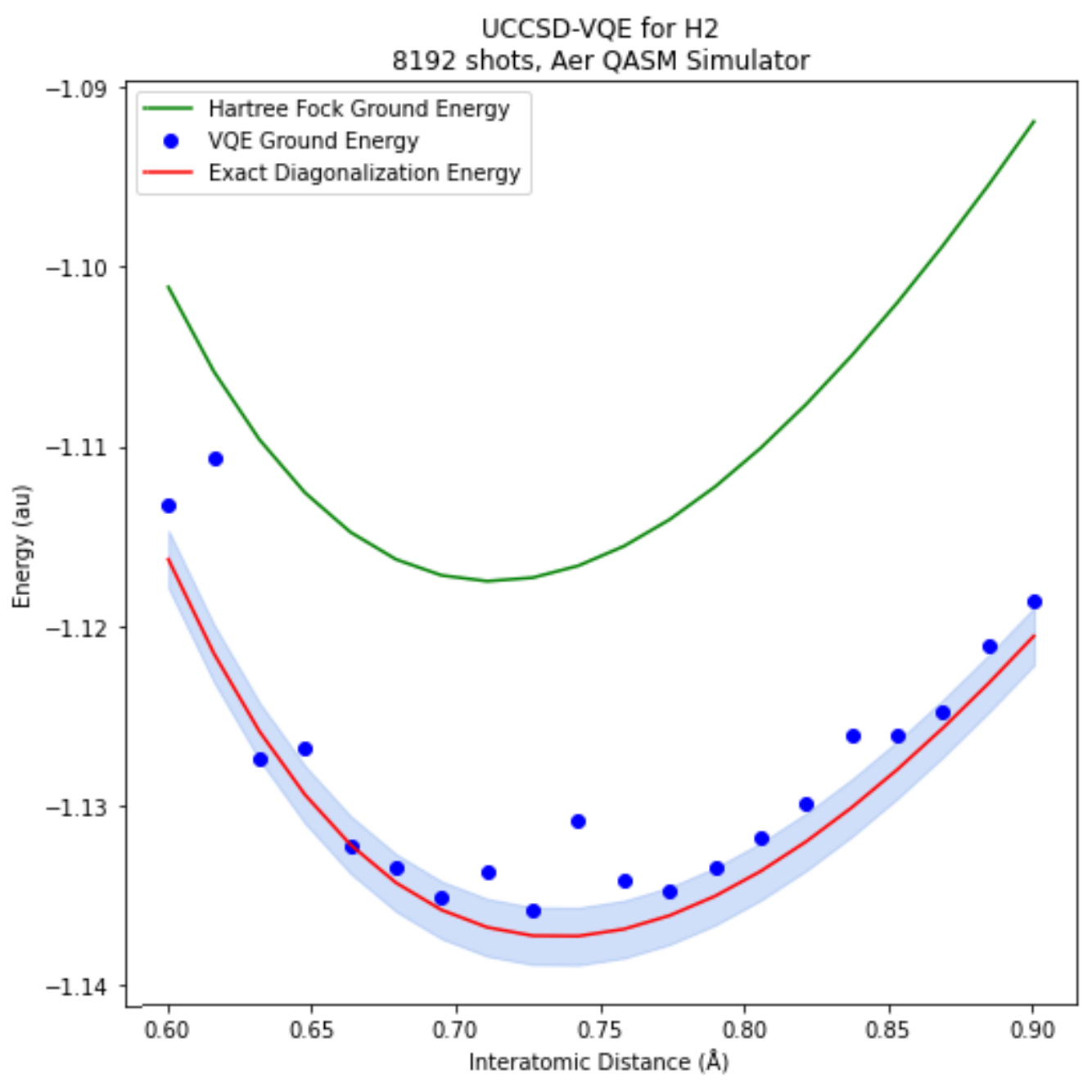}
         \caption{QASM simulator, 8192 shots}
         \label{fig:h2_zoombc_uccsd_qasm_8192}
     \end{subfigure}
     \hfill
     \begin{subfigure}[b]{0.45\textwidth}
         \centering
         \includegraphics[width=\textwidth]{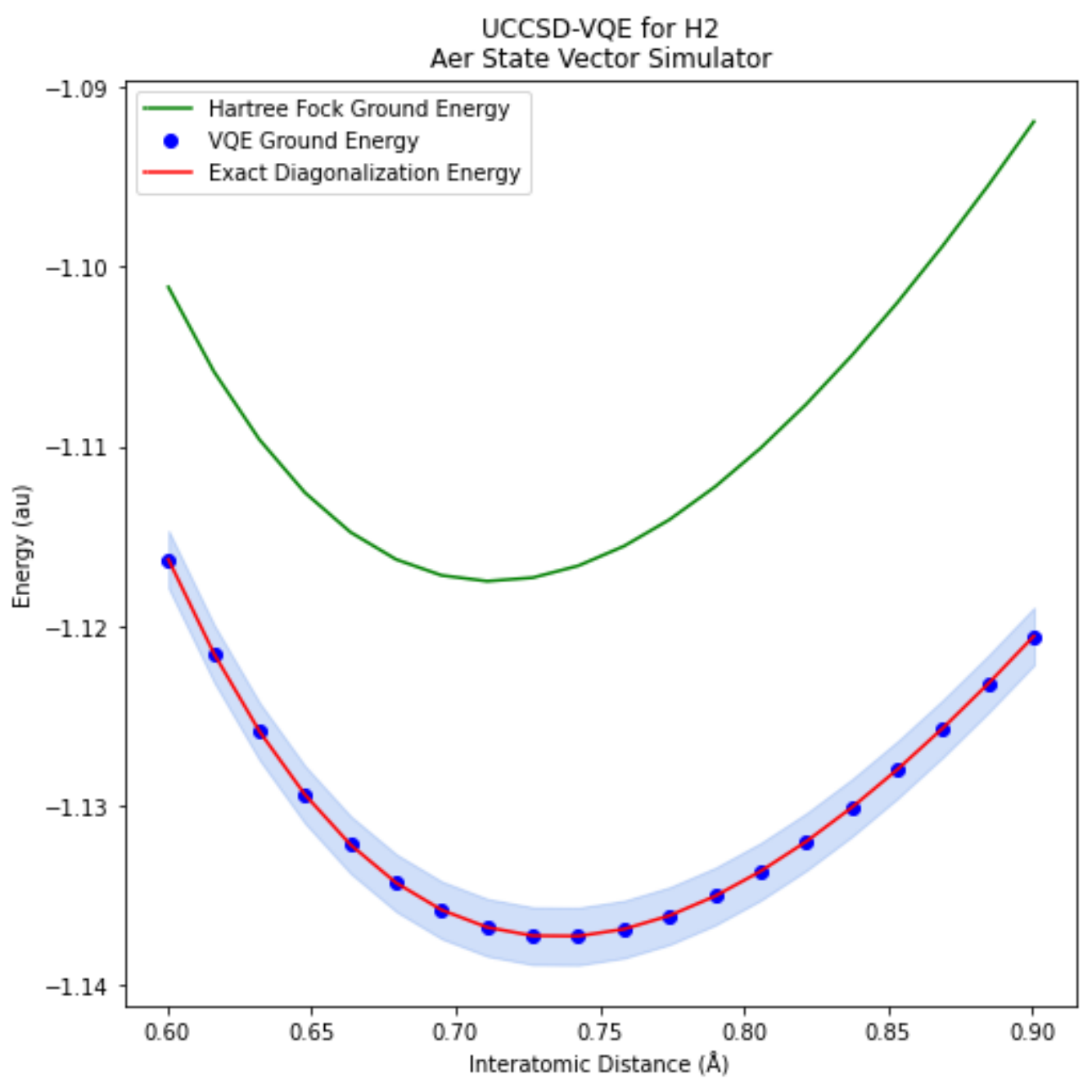}
         \caption{Statevector simulator}
         \label{fig:h2_zoombc_uccsd_statevector_simulator}
     \end{subfigure}
    \caption{Bond dissociation curves of figure \ref{fig:h2_uccsd_fullbc}, zoomed in on the region nearing the equilibrium bond length. The blue area marks the region of chemical accuracy (error of less than 1kcal/mol).}
    \label{fig:h2_uccsd_zoombc}
\end{figure}

To better analyse the results, figure \ref{fig:h2_uccsd_zoombc} shows the same plots as figure \ref{fig:h2_uccsd_fullbc}, now magnified around the equilibrium bond length (the minimum of the curve). In these plots, it is visible that reaching chemical accuracy requires a considerable number of shots. With only 256 shots (figure \ref{fig:h2_zoombc_uccsd_qasm_256}), \gls{UCCSD}-\gls{VQE} is often outperformed by the Hartree-Fock approach; when this happens, we know that the \gls{UCCSD} correction didn't at all improve upon the reference state, and the optimization was entirely in vain. 1024 shots (figure \ref{fig:h2_zoombc_uccsd_qasm_1024}) seem to be enough for \gls{UCCSD}-\gls{VQE} to consistently beat the Hartree-Fock method, but the single-run energy is still far from reaching chemical accuracy. In figure \ref{fig:h2_zoombc_uccsd_qasm_8192}, we can see that with 8192 shots, the energy is already often inside or nearing the region of chemical accuracy, even with a single run (taking the median over several runs would further refine the result). Of course, in the limit of infinite shots (figure \ref{fig:h2_zoombc_uccsd_statevector_simulator}), the performance is nearly perfect.

\subsubsection{Running the Algorithm on Cloud Quantum Computers}

So far, the simulations have included at most sampling noise - but they all assumed ideal behaviour from the quantum processor. In the previous section, we saw that the performance of the algorithm in real backends could prove significantly worse because of the other types of noise they inevitably suffer from. With \gls{UCCSD} the difference becomes even more relevant, because the circuit is deeper: the 2-qubit ansatz from the previous section had a single \gls{CNOT} and a total of only 7 gates. Even for $H_2$, the simplest molecule possible, implementing the \gls{UCCSD} operator requires significantly longer circuits.

\begin{figure}[htbp]
     \centering
     \begin{subfigure}[b]{0.45\textwidth}
         \centering
         \includegraphics[width=\textwidth]{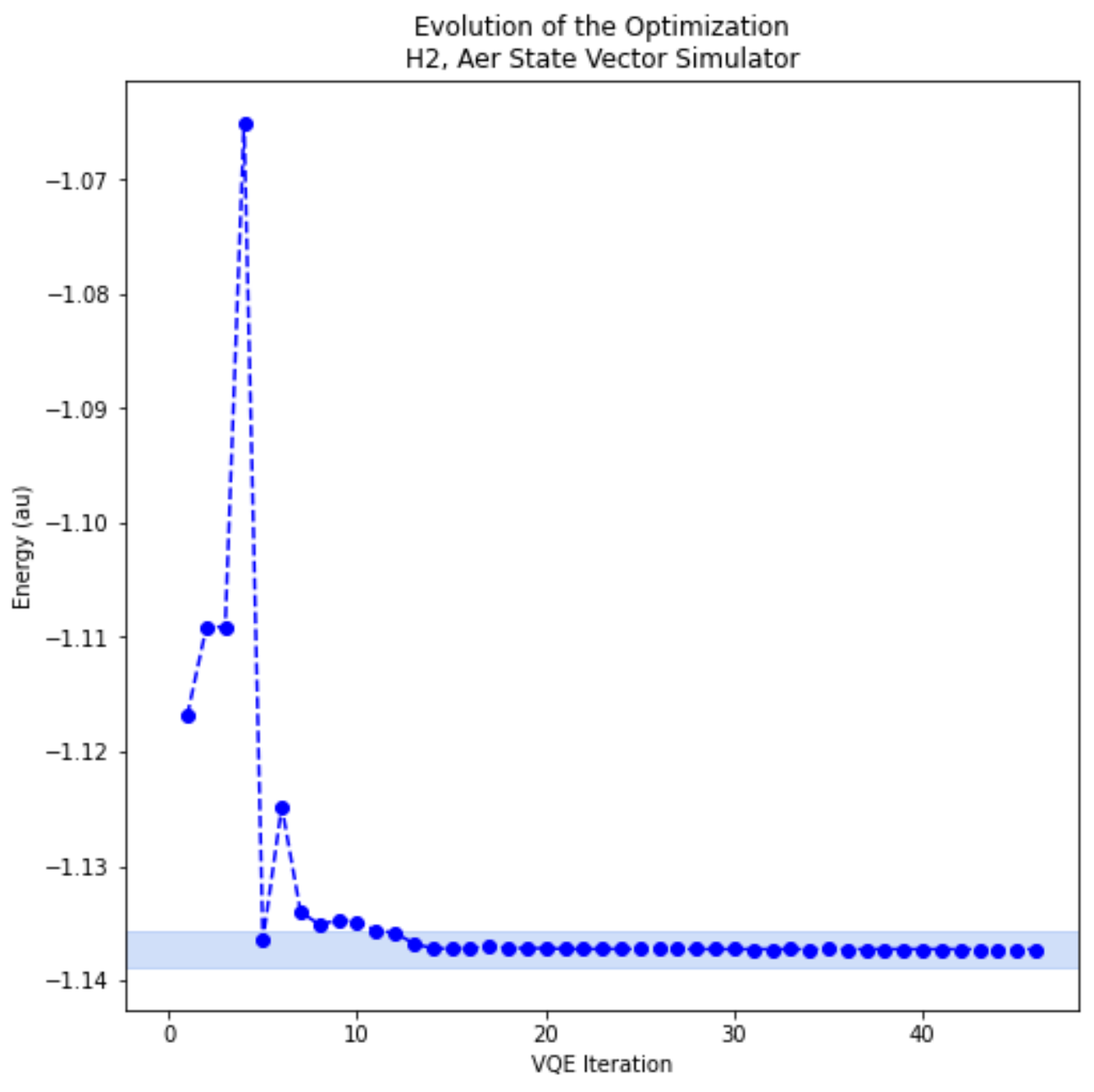}
         \caption{Statevector simulator}
         \label{fig:h2_uccsd_StateVectorSimulator}
     \end{subfigure}
     \hfill
     \begin{subfigure}[b]{0.45\textwidth}
         \centering
         \includegraphics[width=\textwidth]{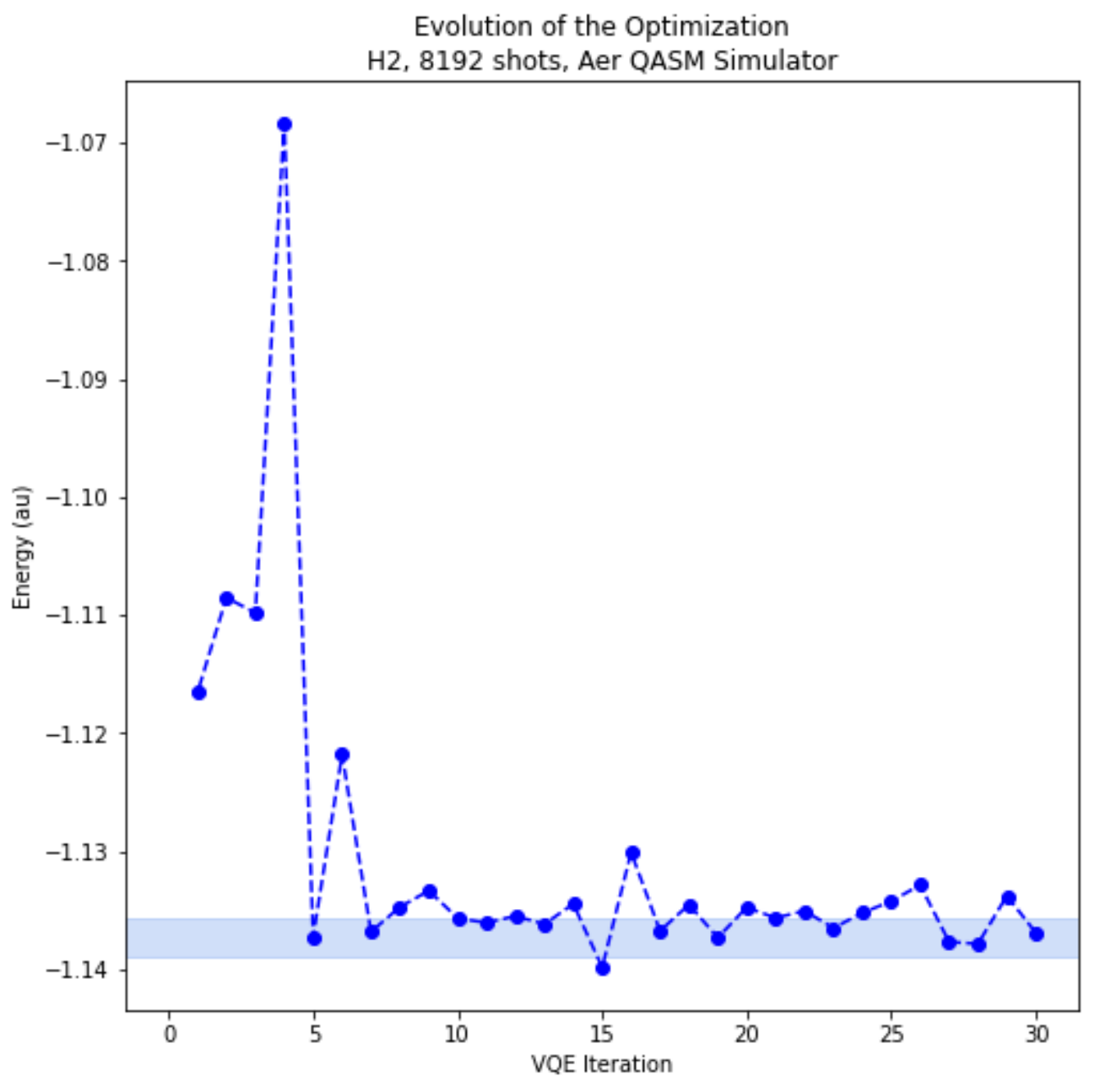}
         \caption{QASM Simulator}
         \label{fig:h2_uccsd_qasm}
     \end{subfigure}
     \\
     \begin{subfigure}[b]{0.45\textwidth}
         \centering
         \includegraphics[width=\textwidth]{h2_uccsd_belem_8192.pdf}
         \caption{IBMQ Belem}
         \label{fig:h2_uccsd_belem}
     \end{subfigure}
     \caption{Evolution of the \gls{UCCSD}-\gls{VQE} optimization for $H_2$ at an interatomic distance of 0.74Å, using three different IBM Quantum \cite{IBMQ} backends: the statevector simulator (with no noise of any kind), the QASM simulator (with sampling noise) and the Belem device (a 5-qubit Falcon  processor). In the last two cases, 8192 shots were used. The energy is plotted in blue; the pale blue region marks the area of chemical accuracy. The same starting point, the Hartree-Fock reference state, was used for all backends. The chosen optimizer was \gls{COBYLA}.}
     \label{fig:h2_uccsd_sv_qasm_belem}
\end{figure}

Figure \ref{fig:h2_uccsd_sv_qasm_belem} showcases the evolution of the \gls{VQE} optimization for a single point (interatomic distance) in three different scenarios: noise-free (figure \ref{fig:h2_uccsd_StateVectorSimulator}), sampling noise only (figure \ref{fig:h2_uccsd_qasm}), real quantum processor (figure \ref{fig:h2_uccsd_belem}). Here, the optimization of a single run was plotted instead of the final results of multiple runs (as in the bond dissociation graphs) because of the time required to run hybrid algorithms on cloud quantum computers with fair-share queuing. 

The difference between the noise-free and sampling noise only scenarios is already remarkable. While in the former the energy is fast to stabilize inside the region of chemical accuracy, in the latter the energy oscillates enough that it's in and out of the region throughout the whole optimization. Even towards the last iterations, the energy does not meet chemical accuracy consistently: it is outside the region in the second-to-last iteration.

However, if sampling noise is significant, it does not come close to the impact of other types of noise. Figure \ref{fig:h2_uccsd_belem} shows that running the algorithm in IBMQ's Belem backend, a real quantum processor, does not allow recovering any valid results. The \gls{VQE} output energy oscillates about 1 Hartree over the true ground energy, without being remotely close to reaching chemical accuracy (of the order of the millesimals of Hartree). This is a sign that quantum information has been washed out by noise. The \gls{UCCSD} ansatz for $H_2$, obtained using Qiskit's variational form and transpiled onto Belem, has 130 \glspl{CNOT}; the circuit depth is approximately proportional to the \gls{CNOT} count. The results show that the circuit in question is too deep for any meaningful data to be recovered by the end.

An approximate density matrix representing the state at the end of the noisy \gls{UCCSD} ansatz was obtained by simulating the algorithm on the QASM simulator, now with a noise model aiming to mimic the behaviour of the Belem processor. The noise model was obtained from Qiskit \cite{Qiskit} and is built from data concerning the specific backend. This procedure allowed obtaining estimates to the purity of the final state and its fidelity with the ideal noiseless output.

The fidelity of the final state with the ideal state was found to be only 0.081, in contrast with its fidelity with the fully mixed state, which was 0.40. The final purity of the state was found to be around 0.16. The STO-3G basis set was used; in a minimal basis set (as is the case), $H_2$ has four spin-orbitals, so that fermionic states of the molecule were mapped into states of four qubits. This implies a Hilbert space of dimension 16, so that the purity is lower bounded by 0.0625. 

The expectation value of the energy calculated in the fully mixed state was found to be -0.097 Hartree. In figure \ref{fig:h2_uccsd_belem}, we can see that the energy of the state at the end of the \gls{UCCSD} circuit in the Belem backend remains close to that value throughout the whole optimization. The optimizer spends 25 iterations trying to adjust the parameters that will lead to a minimum of the energy, but it is to no avail: the energy shows no more than slight deviations from the mean value, with no sign of approaching the ground energy.

\begin{figure}[htbp]
     \centering
     \begin{subfigure}[b]{0.3\textwidth}
         \centering
         \includegraphics[width=\textwidth]{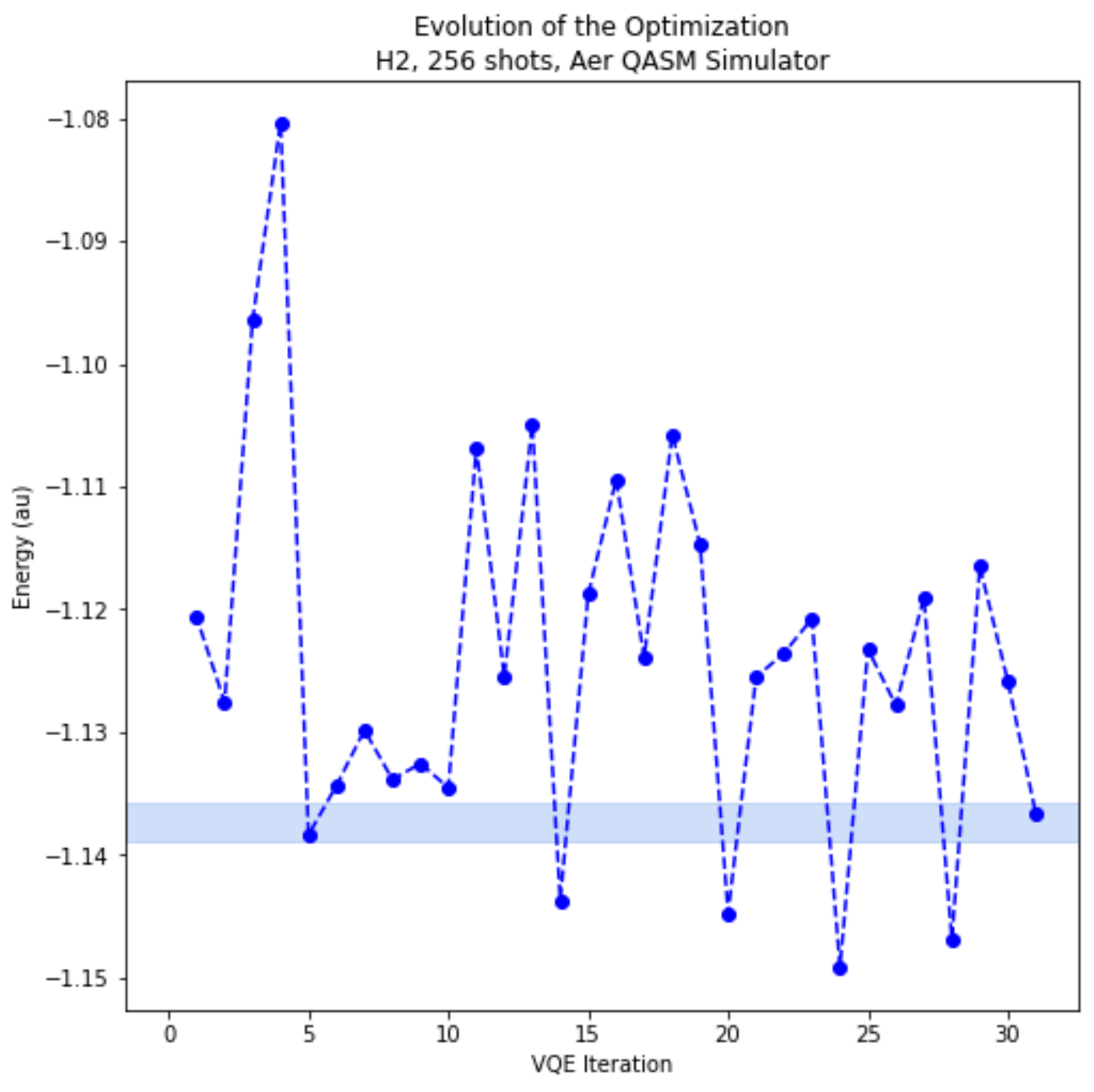}
         \caption{QASM Simulator, 256 shots}
         \label{fig:h2_uccsd_qasm_256}
     \end{subfigure}
     \hfill
     \begin{subfigure}[b]{0.3\textwidth}
         \centering
         \includegraphics[width=\textwidth]{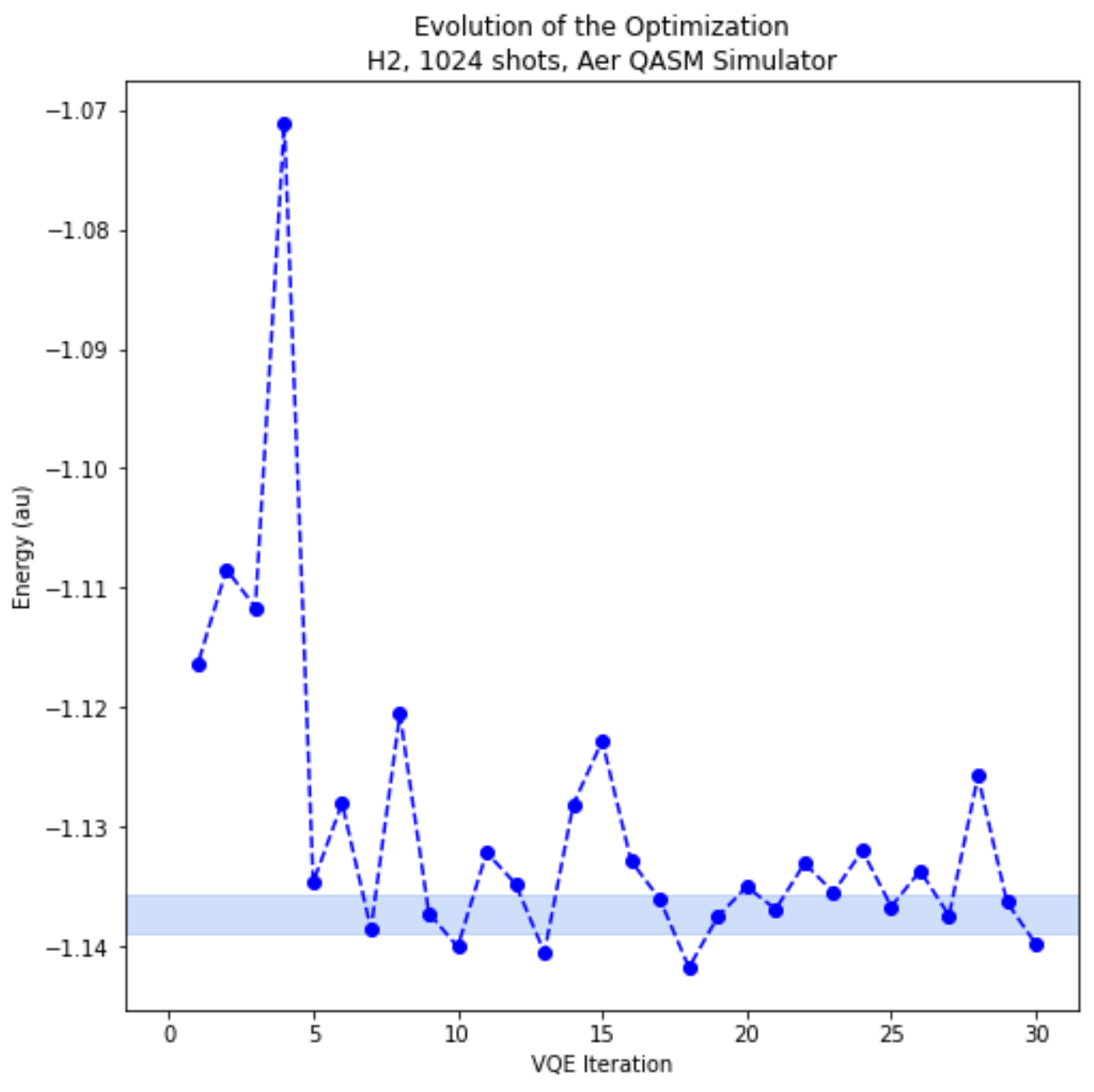}
         \caption{QASM Simulator, 1024 shots}
         \label{fig:h2_uccsd_qasm_1024}
     \end{subfigure}
     \hfill
     \begin{subfigure}[b]{0.3\textwidth}
         \centering
         \includegraphics[width=\textwidth]{h2_uccsd_qasm_8192.pdf}
         \caption{QASM Simulator, 8192 shots}
         \label{fig:h2_uccsd_qasm_8192}
     \end{subfigure}
     \\
     \centering
     \begin{subfigure}[b]{0.3\textwidth}
         \centering
         \includegraphics[width=\textwidth]{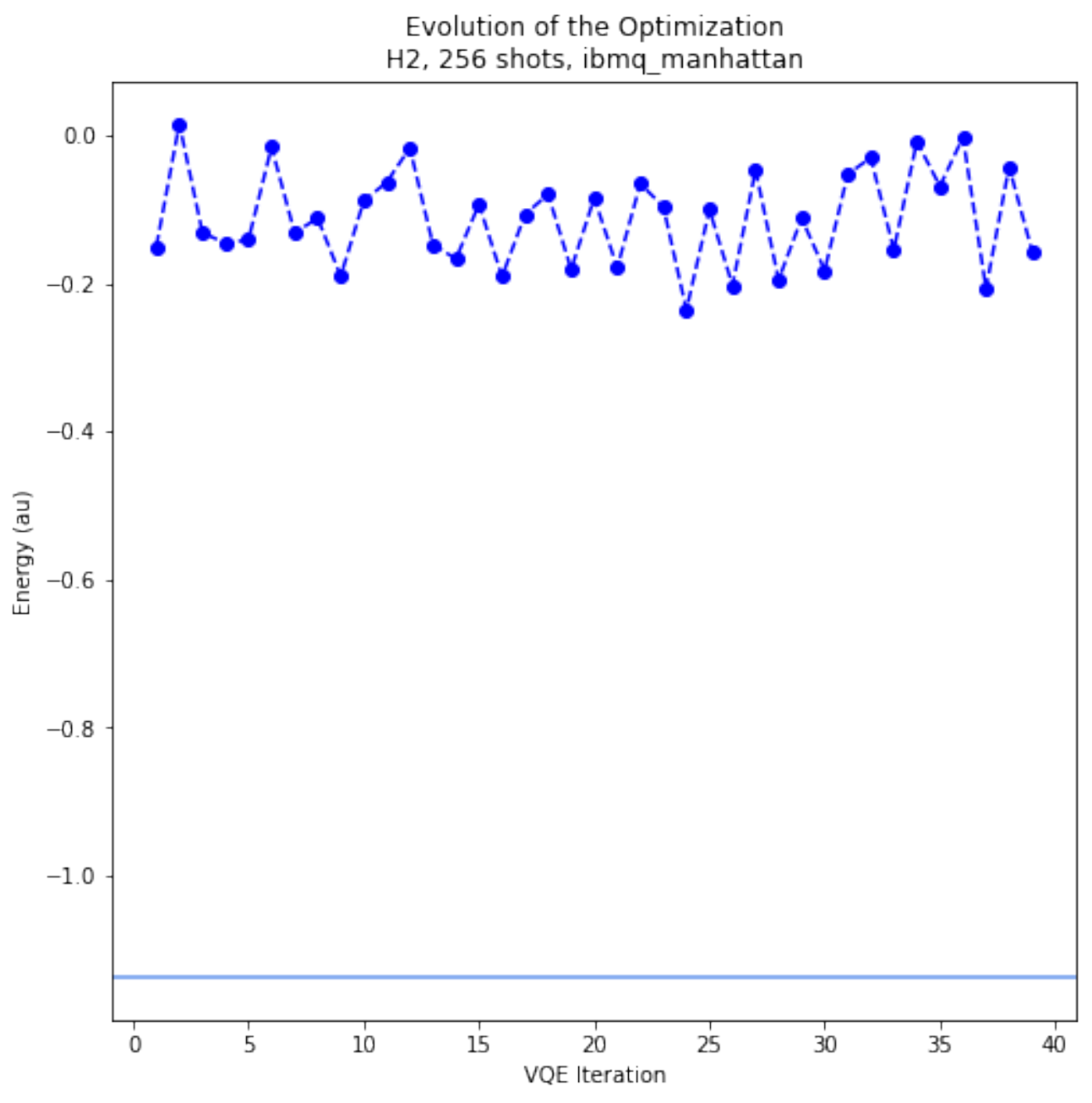}
         \caption{IBMQ Manhattan, 256 shots}
         \label{fig:h2_uccsd_manhattan_256}
     \end{subfigure}
     \hfill
     \begin{subfigure}[b]{0.3\textwidth}
         \centering
         \includegraphics[width=\textwidth]{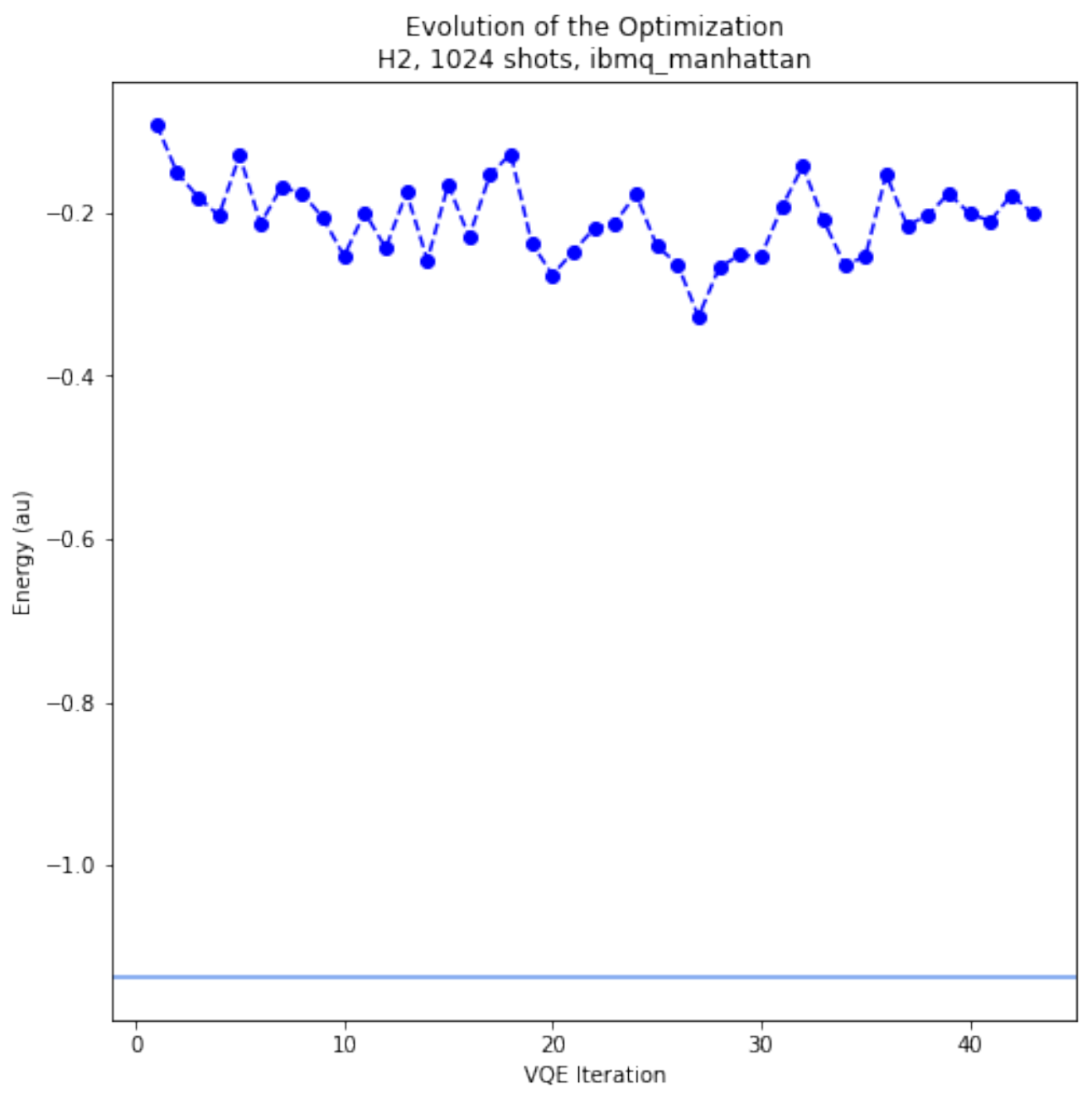}
         \caption{IBMQ Manhattan, 1024 shots}
         \label{fig:h2_uccsd_manhattan_1024}
     \end{subfigure}
     \hfill
     \begin{subfigure}[b]{0.3\textwidth}
         \centering
         \includegraphics[width=\textwidth]{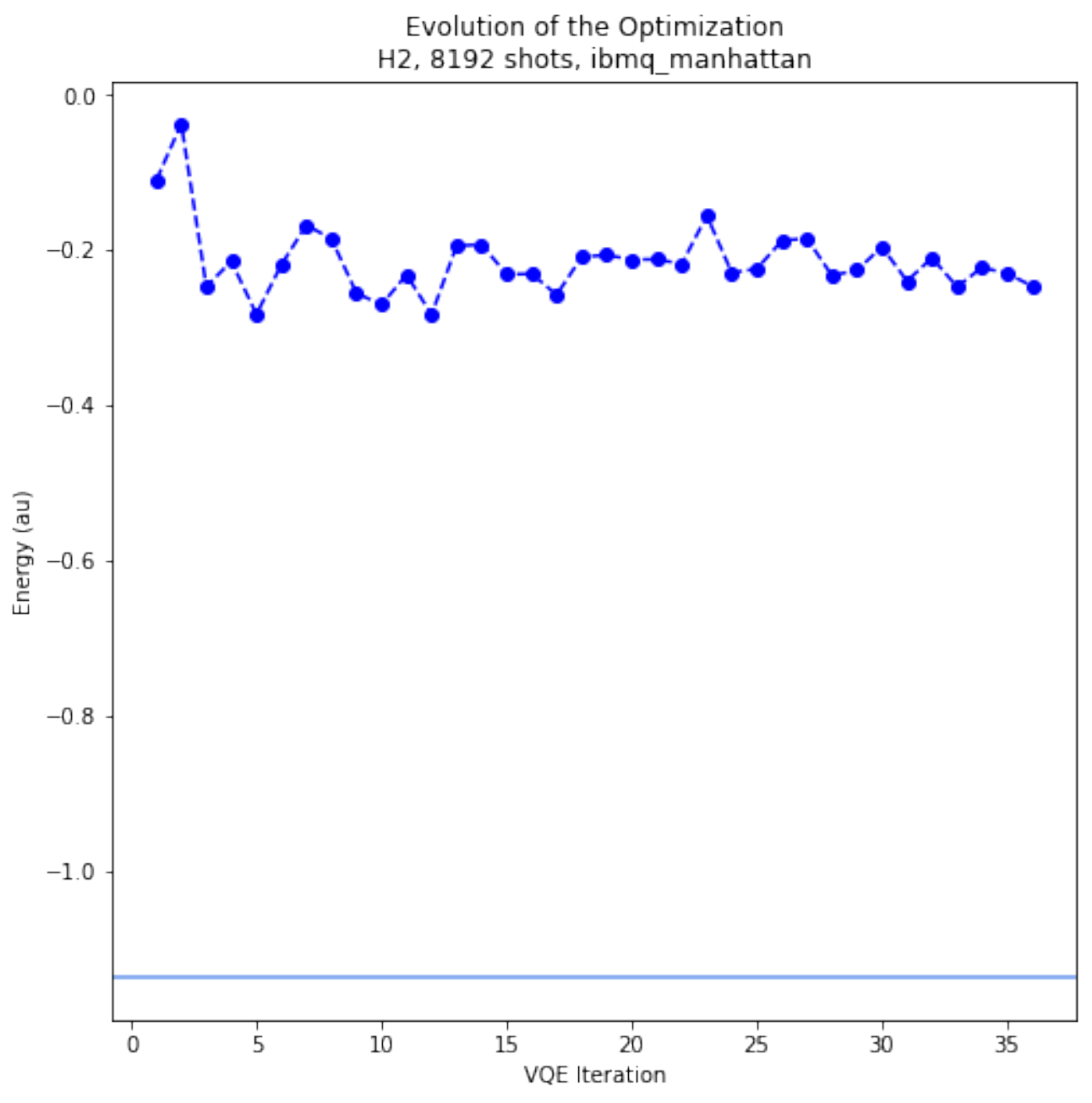}
         \caption{IBMQ Manhattan, 8192 shots}
         \label{fig:h2_uccsd_manhattan_8192}
     \end{subfigure}
     \caption{Evolution of the \gls{VQE} optimization under the same conditions of figure \ref{fig:h2_uccsd_sv_qasm_belem} (same molecule, interatomic distance and optimizer), plotted for three different shot counts and two different backends. The QASM simulator (\ref{fig:h2_uccsd_qasm_256}-\ref{fig:h2_uccsd_qasm_8192}) portraits the behaviour of an ideal quantum processor, in contrast with IBMQ Manhattan (\ref{fig:h2_uccsd_manhattan_256}-\ref{fig:h2_uccsd_manhattan_8192}), a 65-qubit Hummingbird processor.}
     \label{fig:h2_uccsd_qasm_manhattan}
\end{figure}

From figure \ref{fig:h2_uccsd_sv_qasm_belem}, we can already conclude that sampling noise is far from being the limiting factor here. To further illustrate this, figure \ref{fig:h2_uccsd_qasm_manhattan} compares \gls{VQE} optimizations using the QASM simulator (with sampling noise only) and using the Manhattan backend (another real device) for 256, 1024, and 8192 shots.

The improvement of the \gls{VQE} performance with increased shot count is evident when the QASM simulator is used: the energy is visibly faster to stabilize. This is to be expected, since here sampling is the only source of noise. With 256 shots only (figure \ref{fig:h2_uccsd_qasm_256}) the energy shows changes of the order of the centesimals of Hartree up until the last iterations. With 8192 shots (figure \ref{fig:h2_uccsd_qasm_8192}) the energy change is of the order of the millesimals of Hartree in every iteration from the seventh to the last one (30).

When the Manhattan processor is used, the energy shows the same tendency for stabilization: from 256 shots (figure \ref{fig:h2_uccsd_manhattan_256}) to 8192 shots (figure \ref{fig:h2_fullbc_uccsd_qasm_8192}), there is also a decrease in the order of magnitude of the oscillations in the energy towards the last iterations. However, here, this does not translate into any significant improvement in accuracy, because the energy is still very far from the ground energy, with an error of around 1 Hartree. Much like it happened with the Belem backend, the state at the end of the \gls{UCCSD} circuit in the Manhattan backend is too corrupted for the optimizer to converge to the ground state: the energy evaluations are too far from the exact value. The purity is so low that changing the parameters is barely reflected on the expectation value. We can conclude that the circuits are too deep to be implemented in these quantum computers.

\subsubsection{Analysing Noisy Instructions}

\begin{figure}[htbp]
    \centering
    \includegraphics[width=0.7\textwidth]{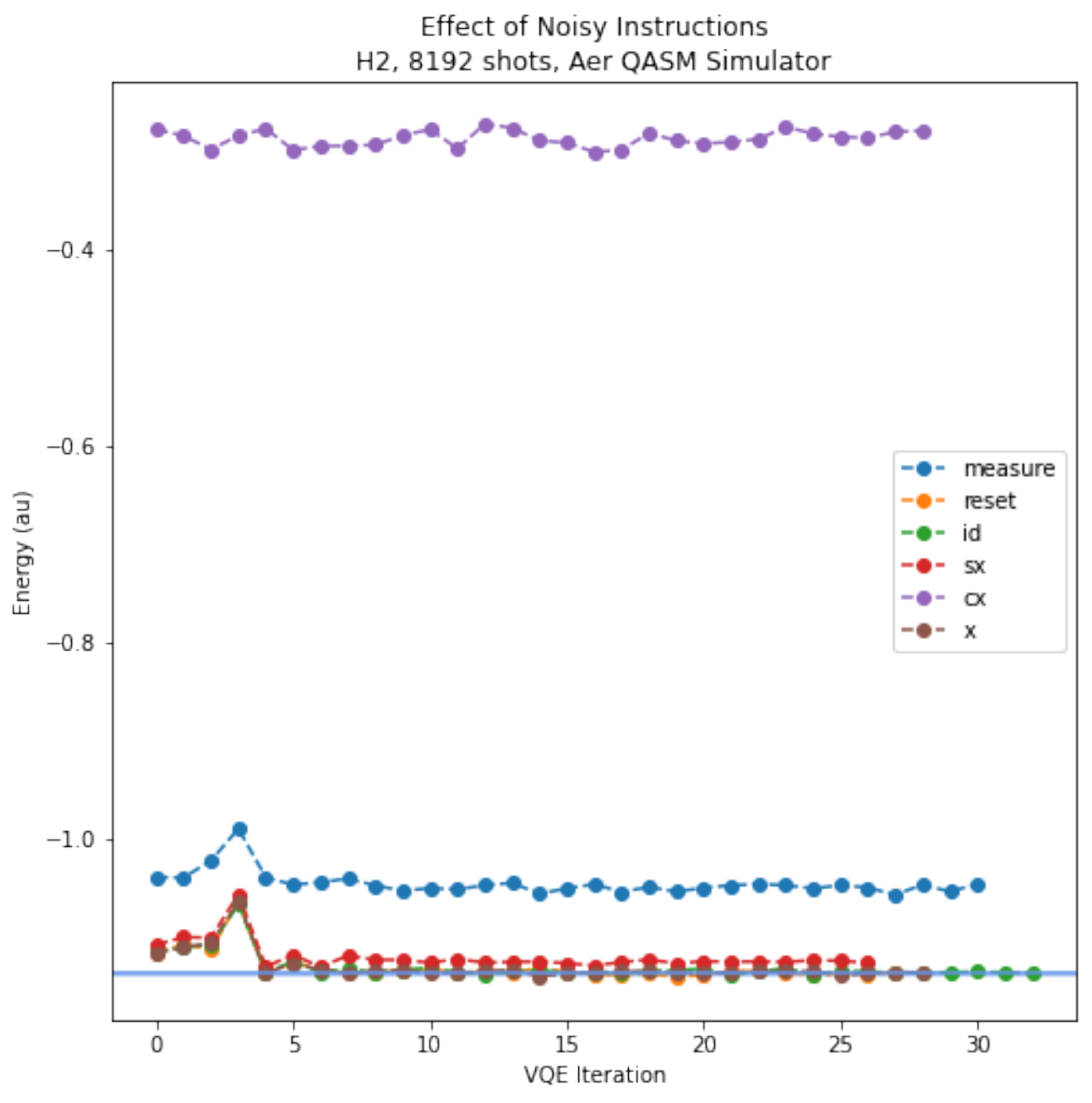}
    \caption{Impact of the noise associated with different instructions on the performance of \gls{UCCSD}-\gls{VQE}, using Belem's noise model from Qiskit's \cite{Qiskit} \textit{providers.aer.noise} module. The model is generated automatically from the properties of IBMQ's \cite{IBMQ} device. Each curve represents the \gls{UCCSD}-\gls{VQE} optimization, simulated with a model including the noise associated with the corresponding instruction exclusively. 8192 shots were used in all curves. The conditions are the same as those from figure \ref{fig:h2_uccsd_sv_qasm_belem} (same molecule, interatomic distance and optimizer).}
    \label{fig:h2_uccsd_isolated_noisy_instructions}
\end{figure}

To better understand which sources of noise were most affecting the algorithm, the noise model emulating the behaviour of the Belem backend was decomposed into several noise models, each considering only the noise associated with a specific instruction. \gls{UCCSD}-\gls{VQE} was then simulated with each of the noise models; the results are plotted in figure \ref{fig:h2_uccsd_isolated_noisy_instructions}.

The curve that stands out is the one associated with the noise model that isolated \gls{CNOT} noise. Even though this model ignored all other sources of noise, the performance is remarkably close to that of the complete Belem noise model (figure \ref{fig:h2_uccsd_belem}). In here, the error in the final energy is once again close to 1 Hartree; further, the energy barely moves away from the starting energy throughout the whole optimization. One again, the impact of noise is felt too strongly for any meaningful results to be recovered.

It is not surprising that this is the noisiest gate: it is the only two-qubit gate. In this particular backend (Belem), the average \gls{CNOT} gate time is around 550 nanoseconds, while the single-qubit gate time is typically of the order of tens of nanoseconds. Among the available gates, the \gls{CNOT} thus corresponds by far to the biggest cost in circuit depth.

\begin{figure}[htbp]
    \centering
    \includegraphics[width=0.7\textwidth]{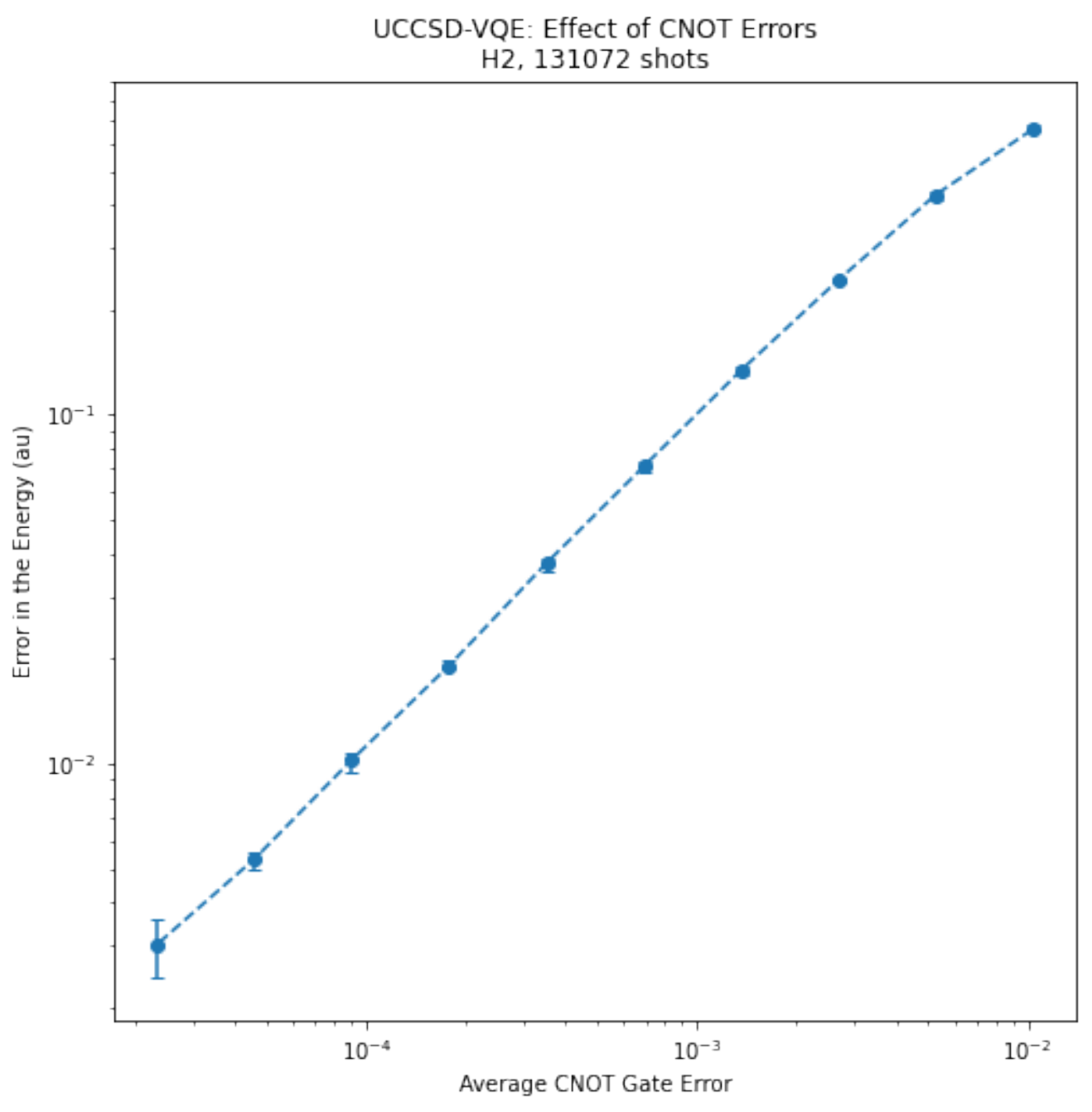}
    \caption{Impact of \gls{CNOT} gate noise on the performance of \gls{UCCSD}-\gls{VQE}. The error in the final energy is plotted as a function of the average \gls{CNOT} gate error (equal to $1-F$, where $F$ is the average gate fidelity). Twenty runs were used per point; the median was used as the estimator, and the error bars show the interquartile ranges. The number of shots was increased to ensure that sampling noise would not be the limiting factor for any gate error. The conditions are the same as those from figure \ref{fig:h2_uccsd_sv_qasm_belem} (same molecule, interatomic distance and optimizer).}
    \label{fig:h2_uccsd_cnot_errors}
\end{figure}

Since the noise associated with \gls{CNOT} gates is the one that most impacts \gls{UCCSD}-\gls{VQE}, it will be interesting to analyse the relation between the magnitude of \gls{CNOT} gate errors and the performance of the algorithm. A graphical representation of the impact of the average \gls{CNOT} gate error on the final \gls{UCCSD}-\gls{VQE} error can be found in figure \ref{fig:h2_uccsd_cnot_errors}. To obtain these results, a noise model was created to simulate \gls{CNOT} gate errors. This model consisted of a two-qubit depolarizing error, succeeded by single qubit thermal relaxation errors on the involved qubits.

This is a similar approach to the one employed in Qiskit's backend noise models, mentioned previously and used in obtaining the results presented in figure \ref{fig:h2_uccsd_isolated_noisy_instructions}. In these noise models, made available by the Qiskit package, single qubit thermal relaxation errors are created for both qubits using the relaxation parameters of the backend, and a two-qubit depolarizing error is then added so that the total gate error matches the experimental value. In our case, the purpose was precisely to vary the gate error, which prevented us from using it to dictate the characteristics of the depolarizing channel. Instead, the ratio between the gate error due to the depolarizing channel and due to thermal relaxation was assumed constant. This constant was set to the average value of this ratio for a real device, Belem (calculated from available data). For simplicity, all \gls{CNOT} gates were assumed to behave identically.

Figure \ref{fig:h2_uccsd_cnot_errors} suggests that in the case of $H_2$, the relation between the average \gls{CNOT} gate error and the error in the final energy (in atomic units) is roughly linear in the interval of values contemplated in the plot. When the magnitude of the \gls{CNOT} error is of the order of $10^{-2}$, as is the case for the quantum processors made available on the cloud by IMBQ \cite{IBMQ}, the error can reach almost 1 Hartree. For the error in the \gls{UCCSD}-\gls{VQE} energy for this molecule to be under 0.1 Hartree, the \gls{CNOT} gate error must be of the order of $10^{-4}$. 

The noise associated with each of the remaining instructions in figure \ref{fig:h2_uccsd_isolated_noisy_instructions} has a significantly lower impact on the algorithm than the noise associated with the \glspl{CNOT}. The second worst performance, already quite better, was associated with the noise model that isolated measurement errors. Such errors could be partly helped by employing measurement error mitigation techniques. The remaining instructions considered - namely, reset (preparing the qubits in state $\ket{0}$), the identity operation, and the SX and X gates - are associated with significantly less noise. The final \gls{UCCSD}-\gls{VQE} energies obtained with noise models contemplating each of these instructions are exceptionally close the \gls{FCI} value as compared to those isolating \gls{CNOT}-related noise (mostly) and measurement-related noise (to a lesser degree).

\subsubsection{Impact of Thermal Relaxation, SPAM errors, and Sampling}
\label{ss:noise_types_uccsd}

\begin{figure}[htbp]
     \centering
     \begin{subfigure}[b]{0.45\textwidth}
         \centering
         \includegraphics[width=\textwidth]{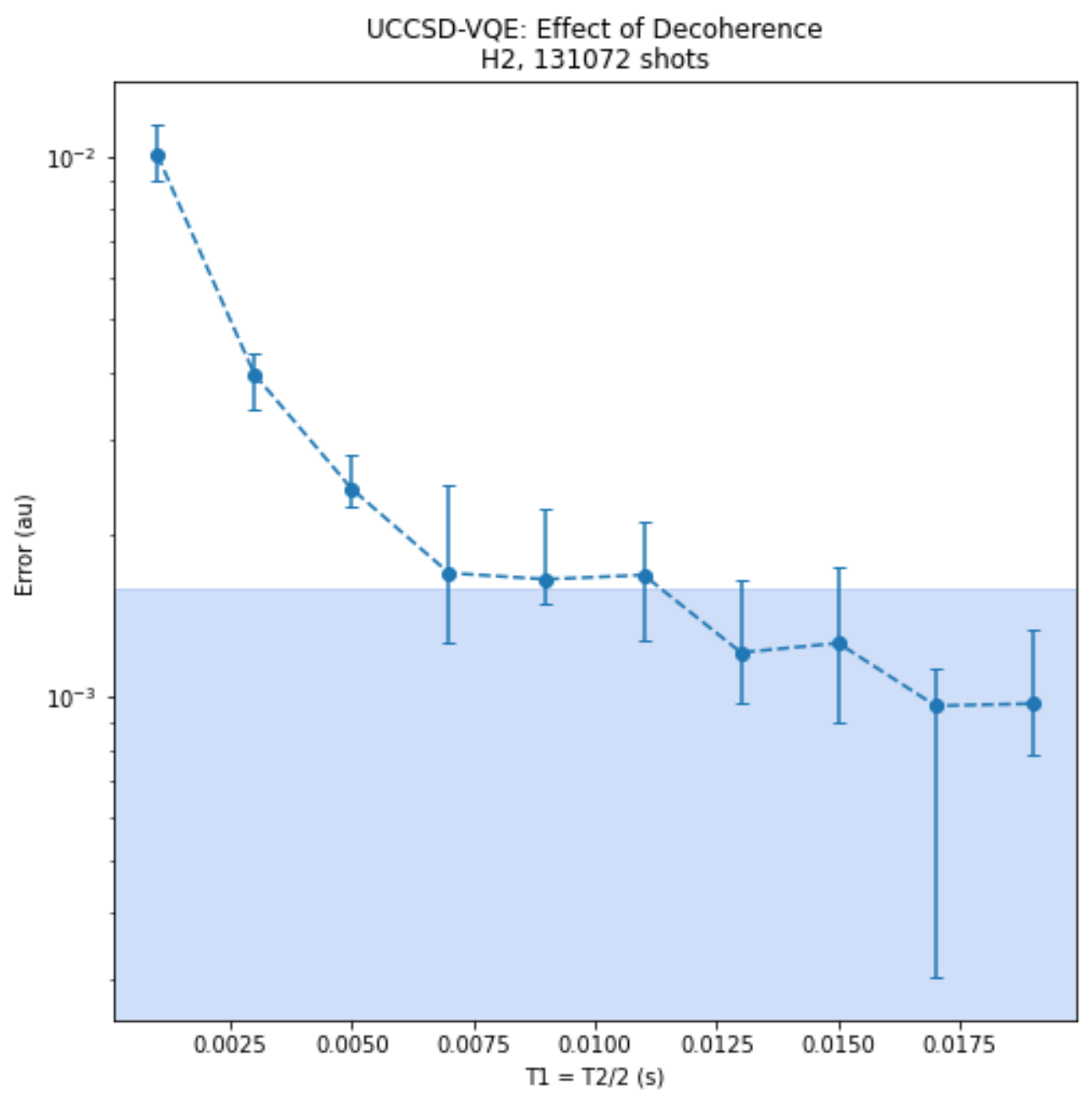}
         \caption{Thermal relaxation}
         \label{fig:h2_UCCSD_decoherence}
     \end{subfigure}
     \hfill
     \begin{subfigure}[b]{0.45\textwidth}
         \centering
         \includegraphics[width=\textwidth]{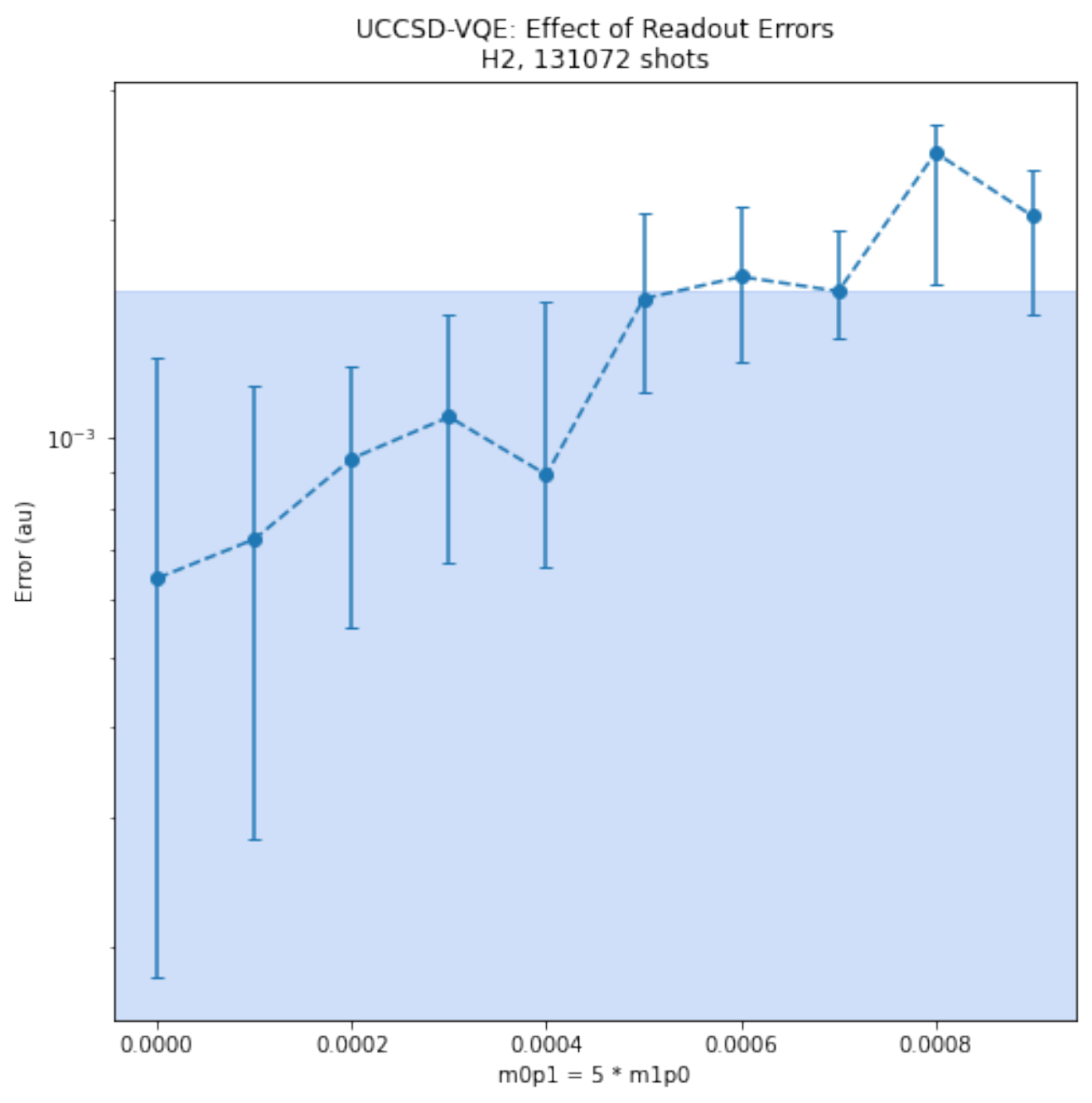}
         \caption{\gls{SPAM} errors}
         \label{fig:h2_uccsd_readout}
     \end{subfigure}
     \\
     \begin{subfigure}[b]{0.45\textwidth}
         \centering
         \includegraphics[width=\textwidth]{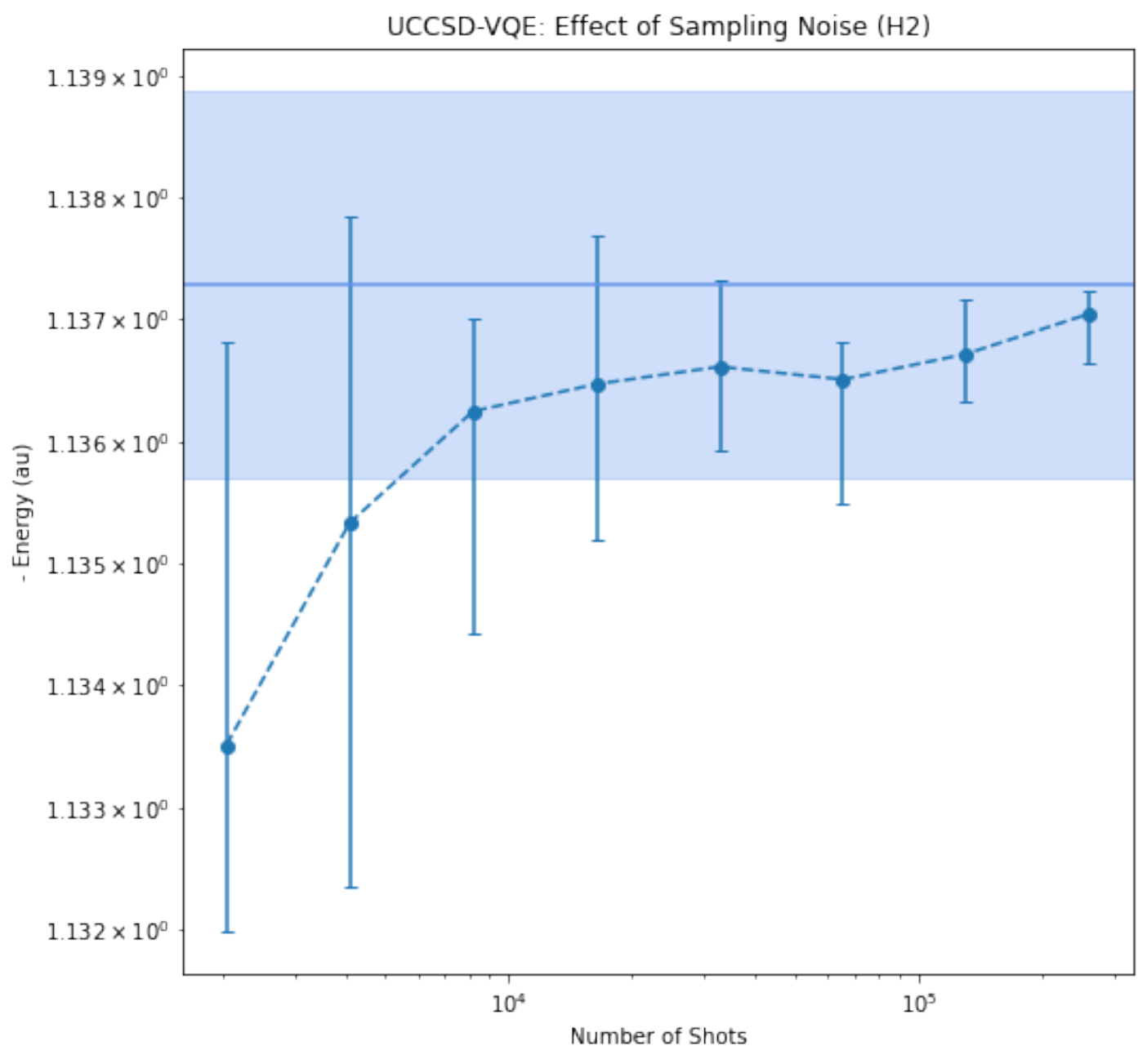}
         \caption{Sampling noise}
         \label{fig:h2_uccsd_samp_loglog}
     \end{subfigure}
     \caption{Plots showing the impact of different types of noise on the performance of \gls{UCCSD}-\gls{VQE}. For thermal relaxation, a simple model assuming average gate times and a constant ratio between the coherence times T1 and T2 was used. Figure \ref{fig:h2_UCCSD_decoherence} shows the evolution of the error in the energy as a function of these times. For \gls{SPAM} noise, the ratio between the probability of measuring 0 after preparing 1 and  the probability of measuring 1 after preparing 0 was fixed (and greater than one, for the model to be realistic). Figure \ref{fig:h2_uccsd_readout} shows the evolution of the error as a function of these probabilities. Figure \ref{fig:h2_uccsd_samp_loglog} illustrates the impact of sampling noise by plotting the error as a function of the number of shots. In all of the plots, twenty runs were used per point; the median was used as the estimator and the error bars show the interquartile ranges. All noise models were created using Qiskit \cite{Qiskit}. The conditions are the same as those from figure \ref{fig:h2_uccsd_sv_qasm_belem} (same molecule, interatomic distance and optimizer).}
     \label{fig:h2_uccsd_noisemodels}
\end{figure}

Figure \ref{fig:h2_uccsd_isolated_noisy_instructions} analysed noise by dividing it into the instructions it is associated with. However, it is also interesting to analyse it by nature - as it was briefly covered in \ref{ss:qc_limitations}, there are different types of noise affecting quantum computers. That is the intent of figure \ref{fig:h2_uccsd_noisemodels}: it shows how scaled down thermal relaxation effects (from interaction with the environment), \gls{SPAM} errors (from faulty state preparations and measurements), and sampling noise (from using a finite number of shots) should individually be for chemical accuracy to be confidently reached. In each case, all noise sources other than the one being analysed were ignored, except for sampling noise, that was included in each of the simulations due to being intrinsic to quantum computing. When fixed, the number of shots was set to $2^{17}$, chosen to be high enough not to impede chemical accuracy.

Figure \ref{fig:h2_UCCSD_decoherence} shows the impact of thermal relaxation on \gls{UCCSD}, by plotting the final energy error as a function of T1 (the relaxation time) and T2 (the dephasing time). The process of thermal relaxation leads to the loss or quantum coherence (\textit{decoherence}), causing quantum information to be irreversibly lost. Currently, on the quantum computers made available on the cloud by IBMQ, the coherence times are of the order of $10^{-4}$ seconds. In contrast, \gls{UCCSD}-\gls{VQE} seems to require them to be of the order of $10^{-2}$ seconds to reach chemical accuracy in the middle 50\% runs (the runs yielding the energies between the first and the third quartiles).

Figure \ref{fig:h2_uccsd_readout} showcases the effect of \gls{SPAM} errors. Chemical accuracy was secured in all runs within the interquartile range only when the probability of measuring $\ket{0}$ after preparing $\ket{1}$ reached $4\times10^{-4}$ (with the probability of measuring $\ket{1}$ after preparing $\ket{0}$, made proportional, at $8\times10^{-5}$). For reference, the recently used Belem backend has average error probabilities around 0.04 and 0.009, respectively - which amounts to around 100 times more than that.

Finally, figure \ref{fig:h2_uccsd_samp_loglog} shows the effect of sampling noise, plotting the error as a function of the number of shots. The simulations showed that for the 50\% center runs to fall inside the region of chemical accuracy, the number of shots would have to be roughly 80000. Currently, IBMQ limits the shot count on cloud quantum computers to 8192.

Concluding, when running \gls{UCCSD}-\gls{VQE} on IBMQ cloud quantum computers, chemical accuracy is beyond reach. Achieving it would require a significant improvement of readout and state preparation errors, thermal relaxation effects, etc. Even the shot count necessary for such an accuracy is 10-fold what is allowed; of course, this is not a relevant hardware limitation because it would not exist if a dedicated quantum processor was available. Regardless, it is a good show of how demanding chemistry problems can be.

It is also worth noting that, throughout the section, we were dealing with the smallest possible molecule. For bigger molecules, the situation is even worse: they would require more qubits, more shots for the same accuracy (see \ref{sss:hamiltonian_averaging}), deeper circuits, and more variational parameters (leading to a harder optimization). 

%% file: Chapters/chapter4.tex
\pagestyle{plain}
\graphicspath{{./Chapters/Figures/Ch4/}}

\chapter{Problem Tailored, Dynamically Created Ansätze: ADAPT-VQE}
\label{ch:adapt_VQE}

If predetermined, problem-tailored ansätze already bypass several of the issues that problem-agnostic ansätze suffer from, dynamic ones reach farther. They are not simply problem-tailored, they are \textit{system}-tailored, with all the advantages that arise from that.

This section will introduce and compare two similar dynamically created ansätze (fermionic-\gls{ADAPT}-\gls{VQE} and qubit-\gls{ADAPT2}-\gls{VQE}), and compare the noise resilience and measurement cost of \gls{ADAPT2}-\gls{VQE} against \gls{UCCSD}-\gls{VQE}.

The \gls{ADAPT2}-\gls{VQE} algorithm was implemented in CIRQ \cite{CIRQ}, in Qiskit \cite{Qiskit}, and independently using matrix algebra. In all cases, the OpenFermion \cite{openfermion} software package was used for creation and manipulation of fermionic operators, along with the corresponding plugin with PySCF \cite{pyscf}, an electronic structure package. The Qiskit implementation used the \gls{VQE} class from the Aqua module with a personalized ansatz, while in the CIRQ implementation not only the ansatz but also the Hamiltonian averaging were done from scratch, with the optimization loop having also been implemented independently up to SciPy's \cite{scipy} minimization function. Qiskit's noise module was used for the creation of noise models meant to analyse the impact of noise in the algorithms via simulations.

\section{The ADAPT-VQE Algorithm}

\subsection{Fermionic-ADAPT-VQE}

In 2019, \gls{ADAPT}-\gls{VQE} was introduced in \cite{Grimsley2019} with the purpose of creating a more accurate, compact, and problem-customized ansatz for \gls{VQE}. This version will hereby be denoted fermionic-\gls{ADAPT}-\gls{VQE} to distinguish it against a more recent version that will be covered shortly.

The idea behind the proposal is to let the molecule in study `choose' its own state preparation circuit, by creating the ansatz in a strongly system-adapted manner. 

\begin{figure*}[htbp]
    \centering
    \includegraphics[width=\textwidth]{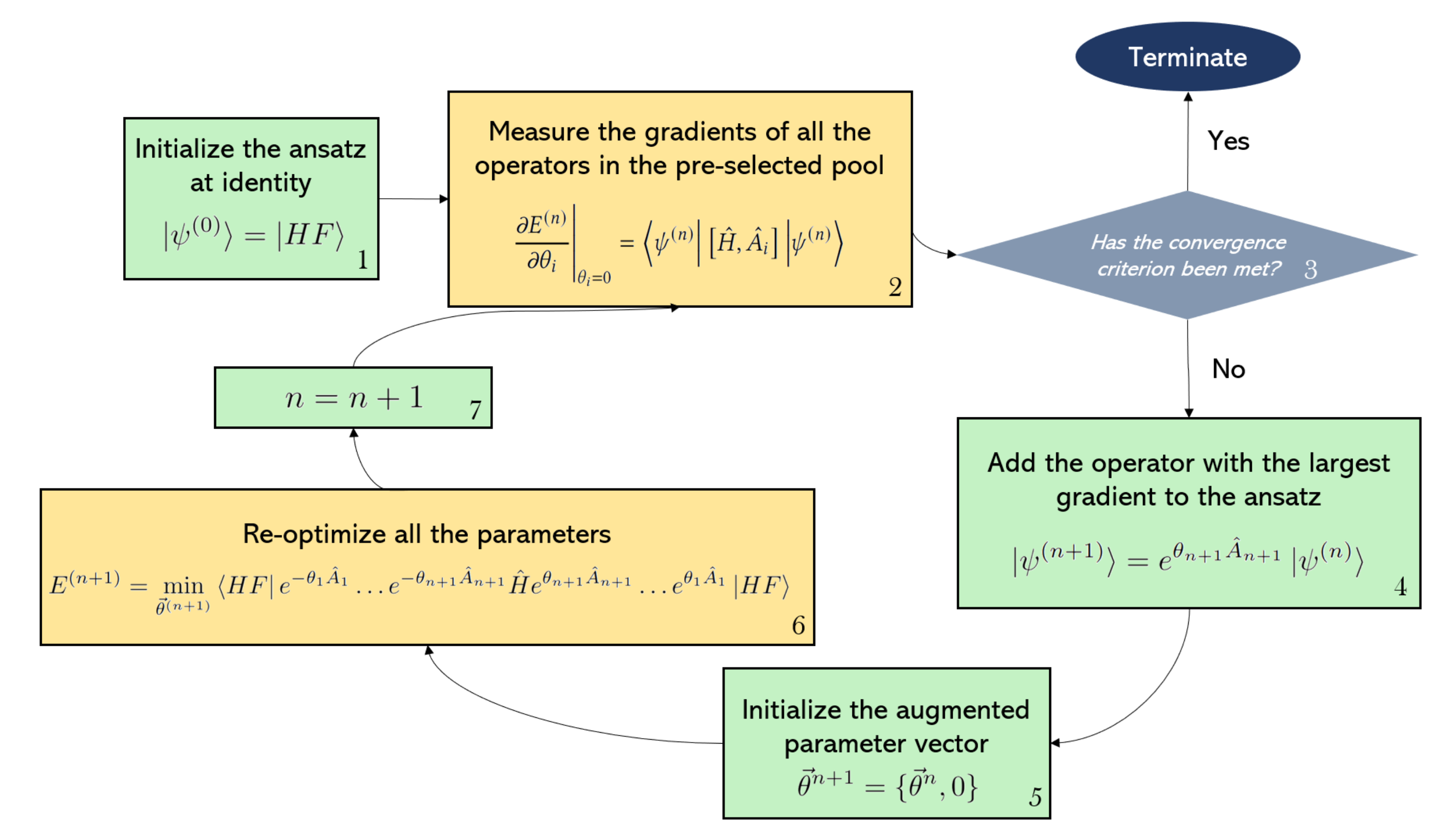}
    \caption{Scheme illustrating the \gls{ADAPT}-\gls{VQE} algorithm. The green boxes represent processes that are exclusively classical. The orange boxes represent hybrid processes (that include state preparations and measurements in the quantum computer).}
    \label{fig:adapt_scheme}
\end{figure*}

Figure \ref{fig:adapt_scheme} shows a schematic outline of the algorithm. The only human role contributing to the creation of the ansatz is the selection of an \textit{operator pool}. Any operator in this pool may or may not be a part of the final ansatz, in its exponentiated and parameterized version. In the numerical simulations presented in the original article, \gls{SCGSD} excitations were used. In this context, much like in the unitary version of \gls{CC} theory, \textit{excitations} is used to mean anti-Hermitian sums of excitation and de-excitation operators, that are mapped to parameterized unitary operators upon exponentiation. The convenience of this form goes beyond unitarity, as will be seen shortly.

The ansatz is initialized at identity: in the first iteration, the prepared state is simply the reference Hartree-Fock state. Each iteration adds an operator to the ansatz, along with the corresponding variational parameter, initialized at zero. Thus, the state preparation circuit and the parameter vector grow from iteration to iteration.

The criterion for operator selection evidently must tend to choose operators that are more beneficial for the wave function (i.e. lower the energy more) at a given point. Similarly to a wide class of optimization methods, the \textit{gradient} is used as the indicator. 

In the beginning of an iteration \textit{n}, \gls{ADAPT}-\gls{VQE} determines the instantaneous rate of change of the energy with respect to each candidate for variational parameter, associated with each candidate for operator, evaluated at the point where the parameter will be initialized (zero). The selected operator is the one which, if added along with the corresponding variational parameter $\theta_i$, has the derivative $\pdv{E^{(n)}}{\theta_i}$ of greatest magnitude.

The question is, how can these derivatives be evaluated? Luckily, the convenient form of the operators offers a solution. For each aspirant ansatz operator $e^{\theta_i\hat{A}_i}$ associated with a pool operator $\hat{A}_i$, we can calculate the energy slope at point $\theta_i=0$ as

\begin{equation}
    \pdv{E^{(n)}}{\theta_i}\Bigg|_{\theta_i=0} = \bra{\psi^{(n)}}\left[\hat{H},\hat{A}_i\right]\ket{\psi^{(n)}}.
    \label{eq:adapt_gradient}
\end{equation}

A sketch of the proof can be found on appendix \ref{ap:adaptgradient}. 

The only thing left to explain is when the algorithm terminates: in order to know when to stop adding operators and growing the ansatz, and finally output an attempted solution, a \textit{convergence criterion} is necessary. There may be different possibilities, but the one employed in the original article (and reproduced in this project) was to terminate when the total gradient norm falls below a certain threshold $\epsilon$. At the beginning of each iteration, after calculating all of the derivatives, they are used to calculate this norm. If the norm is still over the threshold, the operator with the highest derivative is added to the ansatz; if not, the algorithm terminates. Evidently, the lower $\epsilon$ is, the stricter the convergence requirement, the larger the final ansatz, and the higher the precision.

We can then summarize the \gls{ADAPT}-\gls{VQE} algorithm in 7 steps (as labeled in figure \ref{fig:adapt_scheme}). 

\begin{enumerate}
    \item Initialize the ansatz at identity, so that the initial state is simply the Hartree-Fock solution. The corresponding state preparation circuit consists of $N$ parallel Pauli $X$ gates, where $N$ is the number of electrons.
    \item Measure the gradient of each operator in the pool on the quantum computer. This can be done using formula \ref{eq:adapt_gradient}.
    \item Calculate the square norm of the gradient vector. If it is under a predetermined threshold $\epsilon$, terminate. If not, proceed to the following step.
    \item Using the information obtained in step 2, select the operator with the largest gradient and append it to the ansatz.
    \item Initialize the parameter vector with the values from the previous iteration. The newly added variational parameter should be initialized at zero.
    \item Re-optimize \textit{all} the parameters. This is a typical \gls{VQE} optimization, with the current ansatz.
    \item Move on to the next iteration, going back to step 2.
\end{enumerate}

It should be noted that before these steps take place, one must have done the necessary classical computations (common to the original \gls{VQE}) in preparation, to obtain the Hamiltonian operator and the Hartree-Fock spin-orbitals. Further, one must have chosen the constitution of the pool, and the convergence threshold $\epsilon$.

Steps 2 and 6 require preparing the \gls{ADAPT}-\gls{VQE} state multiple times, and measuring the expectation value of the Hamiltonian $\hat{H}$ as well as of the commutators $\left[\hat{H},\hat{A}_i\right]$. A brief explanation on how this can be done follows.

In iteration $n$, the \gls{ADAPT}-\gls{VQE} state is written $\ket{\psi^{(n)}}=e^{\theta_{n} \hat{A}_{n}}\dots e^{\theta_{1}\hat{A}_{1}}\ket{HF}$. Under the Jordan-Wigner transform \ref{def:jw_transform_paulis} the excitation operators $\hat{A}_i$ are mapped to linear combinations of Pauli strings with imaginary coefficients. The operators $e^{\theta_{i} \hat{A}_{i}}$ can then be implemented as explained in subsection \ref{ss:simulation_trotterization}. Along with the preparation of the Hartree-Fock state presented in \ref{ss:prep_ref_state}, this tells us how to create the state preparation circuit.

The measurements are also straightforward: the Hamiltonian $\hat{H}$ also consists of a linear combination of Pauli strings, so that its expectation value can be obtained using the procedure outlined in subsection \ref{ss:measuringpaulis} (possibly improved with the strategies for grouping commuting operators mentioned in \ref{sss:hamiltonian_averaging}). Since the product of two Pauli strings will also be a Pauli string, measuring $\left[\hat{H},\hat{A}_i\right]$ amounts to defining a new observable that can be measured in a similar way. In all cases, the different strings can be measured in parallel, if multiple quantum processors are available.

It is important to note that the pool operators $\hat{A}_i$ are basic spin-complemented one and two-body excitations, that are Jordan-Wigner transformed into a linear combination of at most 16 Pauli strings (8 from each of the two anti-Hermitian two-body excitations with opposite spins), independently of the system size. Further, the Hamiltonian itself has been decomposed into a sum of polynomially-many Pauli strings (the number being $\mathcal{O}(N^4)$ on the number of spin-orbitals $N$). The observable corresponding to the commutator $\left[\hat{H},\hat{A}_i\right]$ can then be efficiently measured in a quantum computer. Thus, the extra measurements of the \gls{ADAPT}-\gls{VQE} algorithm do not in principle hinder efficiency, as long as the scaling of the size of the operator pool is reasonable. In the case considered in the original article, including only excitations up to order two, this scaling is evidently also quartic on the number of spin-orbitals of the chosen basis set. The measurements per iteration in \gls{ADAPT}-\gls{VQE} thus scale like $\mathcal{O}(N^8)$ (instead of $\mathcal{O}(N^4)$ like in \gls{VQE} with static ansätze). It should be noted that, as mentioned previously, modified variants such as \gls{ADAPT}-V and \gls{ADAPT}-Vx manage to decrease this measurement overhead at the expense of an acceptable increase in the number of variational parameters and circuit depth \cite{Liu2021}.

While polynomial, the measurement overhead of \gls{ADAPT}-\gls{VQE} as compared to \gls{UCCSD}-\gls{VQE} or other predetermined ansätze is considerable. This is the price to pay for growing the ansatz using a procedure this customized: to choose the new operator, the gradients of all operators in the pool must be measured. \gls{ADAPT}-\gls{VQE} also implies an overhead in optimizations: instead of a single optimization with a fixed ansatz, we have multiple optimizations with a growing ansatz. The larger number of optimizations implies in itself a larger number of energy evaluations, as well as a bigger burden on the optimizer. All this sacrifice is done in favour of shallower circuits, and in some cases, greater accuracy. The \gls{ADAPT}-\gls{VQE} ansatz was designed to be a \gls{NISQ}-friendly, high-accuracy ansatz for chemistry applications.

\subsection{Qubit-ADAPT-VQE}

In 2020, Qubit-\gls{ADAPT2}-\gls{VQE} was proposed in \cite{Tang2021} with the purpose of improving upon the previous \gls{ADAPT}-\gls{VQE} algorithm. The entirety of the modification lies in the operator pool. 

Spin-complemented fermionic single and double excitations can consist of a linear combination of up to 18 Pauli strings, and spin-adapted ones of up to 48 Pauli strings. Using the ladder-of-\glspl{CNOT} implementation of Pauli exponentials \cite{NielsenChuang}, the circuit for one of such operators can require already thousands of \glspl{CNOT} when as few as a dozen spin-orbitals are used to describe a molecule, even assuming all-to-all connectivity.

Qubit-\gls{ADAPT2}-\gls{VQE} drastically reduces the circuit depth required for the implementation of each operator, by dispensing with the rigorous representation of fermionic excitations. Instead of using a linear combination of Pauli strings that respects fermionic symmetries (e.g. particle number), the operators are broken down into the individual Pauli strings. Because of this, the pool is called \textit{qubit} pool, whereas the pool with the proper excitations is called \textit{fermionic} pool. In addition to the one directly arising from the decomposition fermionic excitations, the original Qubit-\gls{ADAPT2}-\gls{VQE} article introduced a few more pools, which will be covered in the next section.

\section{Original Pool Choices}
\label{s:pools}

This section provides a more thorough description of the different pools used in the original articles: as it was stated before, they can be of \textit{fermion} or \textit{qubit} inspiration. The former take inspiration from fermionic excitations, and the latter opt for more hardware-friendly operators, that are easier to implement in quantum computers.

\subsection{Fermion Inspired Pools}
\label{ss:fermion_pools}

The first \gls{ADAPT}-\gls{VQE} article \cite{Grimsley2019} proposed pools with operators of fermionic inspiration. The ansätze resulting from pools of this type are remarkably similar to the \gls{UCCSD} operator, in that they also typically contain exponentiated single and double excitations. Of course, the ordering will usually be different, as it is dictated by the system in study; further, the number of operators in the \gls{ADAPT}-\gls{VQE} will vary (not only with respect to \gls{UCCSD}-\gls{VQE}, but also within the algorithm, depending on the molecule and on the convergence criterion).

The operators in fermionic pools respect fermionic symmetries: namely, particle number, $S_Z$ and $S^2$ preservation, and the anticommutation requirement.

A simple $N$-order fermionic excitation consists of exciting $N$ electrons from some $N$ orbitals to other $N$ orbitals. Often, \textit{generalized} excitations are used: this means that we include operators that excite fermions from occupied orbitals to occupied orbitals, for example, rather than limiting ourselves to occupied-to-virtual excitations. 

The fermionic pools used in this project were taken or adapted from the simulation code of the original \gls{ADAPT}-\gls{VQE} article \cite{Grimsley2019}, that can be found in the public GitHub repository \cite{adapt_repo}.

\subsubsection{Spin Adapted Generalized Singles and Doubles}

One option is the \gls{SGSD} pool, also called \gls{SAGSD}. This pool was compared against the qubit-inspired pool of qubit-\gls{ADAPT2}-\gls{VQE} in the article that introduced the latter \cite{Tang2021}.

In this pool, the double excitation operators are split into their singlet and triplet components.

In the first quantization formalism, spin-adapted single $\tau_1$ and double $\tau_2$ excitation operators can be written in the basis of eigenstates of the total $S_z$ and $S^2$ operators as:

\begin{align}
\begin{split}
    \hat{\tau}_1 &= \lvert s=1/2,s_z=1/2 \rangle_p \langle  s=1/2,s_z=1/2 \lvert_q\\
    &+\lvert  s=1/2,s_z=-1/2 \rangle_p \langle  s=1/2,s_z=-1/2 \lvert_q \\
    &- h.c.,\\
    \hat{\tau}_{2,A} &= -\frac{2}{\sqrt{12}} (\lvert s=1,s_z=1 \rangle_{rs} \langle s=1,s_z=1 \lvert_{pq}\\
    &+ \lvert s=1,s_z=0 \rangle_{rs} \langle s=1,s_z=0 \lvert_{pq} \\
    &+  \lvert s=1,s_z=-1 \rangle_{rs} \langle s=1,s_z=-1 \lvert_{pq})\\
    &- h.c.,\\
    \hat{\tau}_{2,B} &= -\frac{1}{2} \lvert s=0,s_z=0 \rangle_{rs} \langle s=0,s_z=0 \lvert_{pq}\\
    &- h.c.,
\end{split}
\end{align}

where $\tau_{2,A}$ is the triplet operator and $\tau_{2,B}$ is the singlet operator, and $h.c.$ denotes the Hermitian conjugate of the expression before. Of course, not all sets of four spin-orbitals allow all these excitations. If $p$, $q$ or $r$, $s$ are spin-orbitals with the same spatial part, then there is only the singlet operator, since the triplet wave function is symmetric and thus can't be the spin part of the wave function of two fermions with the same spatial part (as the antisymmetry principle would not be respected).

Rewriting these operators in the basis of tensor products of single particle $S_z$ and $S^2$ eigenstates, and using the convention that an overbar on the spatial orbital index implies a $\beta$ (down) spin (while no overbar implies $\alpha$, or up, spin), we get:

\begin{align}
\begin{split}
    &\hat{\tau}_1 = \lvert p \rangle \langle q \lvert + \lvert \bar{p} \rangle \langle \bar{q} \lvert - h.c.,\\
    &\hat{\tau}_{2,A} = -\frac{2}{\sqrt{12}} \left(\lvert rs \rangle \langle pq \lvert + \frac{1}{2}\left( \lvert r\bar{s} \rangle + \lvert \bar{r} s\rangle \right)\left( \langle p\bar{q} \lvert + \langle \bar{p} q\lvert \right) + \lvert \bar{r}\bar{s} \rangle \langle \bar{p}\bar{q} \lvert \right)-h.c.,\\
    &\hat{\tau}_{2,B} = -\frac{1}{2}\left( \lvert r\bar{s} \rangle - \lvert \bar{r} s\rangle \right)\left( \langle p\bar{q} \lvert - \langle \bar{p} q\lvert \right) - h.c.
\end{split}
\end{align}

And finally, we can transform this into the second quantization formalism to obtain the operators in \ref{def:singlet_ops_2nd_quantization}.

\begin{align}
\begin{split}
    &\hat{\tau}_1 = a_p^\dagger a_q +a_{\bar{p}}^\dagger a_{\bar{q}} - h.c.,\\   
    &\hat{\tau}_{2,A} = \frac{1}{\sqrt{12}} \Big{(}2a_r^\dagger a_p a_s^\dagger a_q  + 2a_{\bar{r}}^\dagger a_{\bar{p}} a_{\bar{s}}^\dagger a_{\bar{q}} + a_r^\dagger a_p a_{\bar{s}}^\dagger a_{\bar{q}} + a_{\bar{r}}^\dagger a_{\bar{p}} a_s^\dagger a_q + \\
    &\qquad a_r^\dagger a_{\bar{p}} a_{\bar{s}}^\dagger a_q + a_{\bar{r}}^\dagger a_p a_s^\dagger a_{\bar{q}}\Big{)} - h.c.,\\
    &\hat{\tau}_{2,B} = \frac{1}{2}\left(a_r^\dagger a_p a_{\bar{s}}^\dagger a_{\bar{q}} + a_{\bar{r}}^\dagger a_{\bar{p}} a_s^\dagger a_q - a_r^\dagger a_{\bar{p}} a_{\bar{s}}^\dagger a_q - a_{\bar{r}}
    ^\dagger a_p a_s^\dagger a_{\bar{q}} \right) - h.c.
    \label{def:singlet_ops_2nd_quantization}
\end{split}
\end{align}

For convenience, the fermionic anticommutation relations \ref{def:anticommutation_relations} were used to reorganize the order of the operators; for this purpose, they are more conveniently formulated as in \ref{def:anticommutation_relations_rewritten}.

\begin{align}
\begin{split}
a_i^\dagger a_j^\dagger = -a_j^\dagger a_i^\dagger ,
&\qquad
a_i a_j = -a_j a_i,
\qquad
a_i a_j^\dagger = \delta_{ij} - a_j^\dagger a_i
\end{split}
\label{def:anticommutation_relations_rewritten}
\end{align}

The formulation of the pool operators using creation and annihilation operators \ref{def:singlet_ops_2nd_quantization} is particularly useful, as they are directly mapped to qubit operators via the Jordan-Wigner transformation \ref{def:jw_transform_paulis}. The OpenFermion \cite{openfermion} library allows creating instances of the FermionOperator class directly from expressions like this one.

Under the Jordan-Wigner transformation, it would appear that each product of four fermionic ladder operators is mapped to a linear combination of 16 Pauli strings; however, the Pauli strings with real coefficients don't change under conjugation and are thus canceled out by the Hermitian conjugate, which is why adding it forces unitarity in the first place. The Pauli strings with imaginary coefficients simply switch sign under conjugation, so that each anti-Hermitian sum of four body operators results in a linear combination of only 8 Pauli strings. As a consequence, the operators in the \gls{SGSD} pool consist of a linear combination of at most 48 Pauli strings. Even though the order of the excitations is the same, because of the anticommutation string, these Pauli strings act on more qubits for larger systems: their length grows on average linearly with the size of the system. The number of operators in this pool scales as $\mathcal{O}(N^4)$, where $N$ is the number of spin-orbitals.

\subsubsection{Spin Complemented Generalized Singles and Doubles}

In the \gls{SCGSD} pool, each excitation is grouped along with its spin-complement (i.e. the excitation acting on the opposite spin-orbitals of the same spatial orbitals). 

In the first quantization formalism, they can be written

\begin{align}
\begin{split}
    &\hat{\tau}_1 = \lvert p \rangle \langle q \lvert + \lvert \bar{p} \rangle \langle \bar{q} \lvert - h.c.,\\
    &\hat{\tau}_{2,A} = - \left(\lvert rs \rangle \langle pq \lvert +\lvert \bar{r}\bar{s} \rangle \langle \bar{p}\bar{q} \lvert \right) - h.c.,\\
    &\hat{\tau}_{2,B} =- \left(\lvert r\bar{s} \rangle \langle p\bar{q} \lvert +\lvert \bar{r}s \rangle \langle \bar{p}q \lvert \right) - h.c., \\
    &\hat{\tau}_{2,C} =- \left(\lvert r\bar{s} \rangle \langle \bar{p}q \lvert +\lvert \bar{r}s \rangle \langle p\bar{q} \lvert \right) - h.c. \\
\end{split}
\end{align}

And if we transform them into the second quantization formalism, we obtain (using equations \ref{def:anticommutation_relations_rewritten} to reorder the expressions once again)

\begin{align}
\begin{split}
&\hat{\tau}_1 = a_p^\dagger a_q +a_{\bar{p}}^\dagger a_{\bar{q}} - h.c.,\\
&\hat{\tau}_{2,A}=a_r^\dagger a_p a_s^\dagger a_q + a_{\bar{r}}^\dagger a_{\bar{p}} a_{\bar{s}}^\dagger a_{\bar{q}} - h.c.,\\
&\hat{\tau}_{2,B} = a_r^\dagger a_p a_{\bar{s}}^\dagger a_{\bar{q}} + a_{\bar{r}}^\dagger a_{\bar{p}} a_s^\dagger a_q - h.c.,\\
&\hat{\tau}_{2,C} = a_r^\dagger a_{\bar{p}} a_{\bar{s}}^\dagger a_q + a_{\bar{r}}^\dagger a_p a_s^\dagger a_{\bar{q}} - h.c.\\
\end{split}
\end{align}

The operators in this pool consist of a linear combination of at most 16 Pauli strings, whose length also scales linearly (on average) with $N$, the number of spin-orbitals; and once again, the number of operators in the pool scales like $\mathcal{O}(N^4)$. While the \gls{SCGSD} pool has more operators consisting of less Pauli strings on average than the \gls{SGSD} pool, the asymptotic scaling with the system size remains unchanged between the two.

\subsection{Qubit Inspired Pools}

The first pools of qubit inspiration to be introduced broke the fermionic excitation operators down to their constituent Pauli strings. From that, other qubit pools were proposed with the goal of further decreasing the circuit depth per operator and the size of the pool. 

In general, operators in qubit pools don't respect fermionic symmetries. They usually abdicate of particle number, $S_Z$ and $S^2$ preservation, and may also forgo anticommutation.

The pools to follow were presented in the article introducing Qubit-\gls{ADAPT2}-\gls{VQE} \cite{Tang2021}.

\subsubsection{Pauli String Pool}

The first pool to be considered in Qubit-\gls{ADAPT2}-\gls{VQE}, then called simply Qubit Pool, was the one consisting of all individual Pauli strings appearing in the \gls{SGSD} pool (or equivalently, the \gls{SCGSD} one: they result in the same set of Pauli strings, even if doubles might differ).

As a simple example, we can consider the following (fermionic) single excitation.

\[
0.5a_0^\dagger a_4 + 0.5a_1^\dagger a_5 - 0.5a_4^\dagger a_0 - 0.5 a_5^\dagger a_1
\]

This excitation exists (when there are more than 6 spin-orbitals) both in the \gls{SCGSD} and \gls{SGSD} pool, since they don't differ in single excitations. The two last terms are the Hermitian conjugates of the former two, and OpenFermion's \cite{openfermion} orbital ordering was used, so that 0, 1 and 4, 5 are pairs of spin-complements corresponding to the same spatial orbital. Under the Jordan-Wigner transformation, this becomes

\begin{align*}
&0.25i \cdot X_0\otimes Z_1\otimes Z_2\otimes Z_3\otimes Y_4 + 0.25i \cdot X_1\otimes Z_2\otimes Z_3\otimes Z_4\otimes Y_5 \\
-& 0.25i \cdot Y_0\otimes Z_1\otimes Z_2\otimes Z_3\otimes X_4 - 0.25i \cdot Y_1 \otimes Z_2\otimes Z_3\otimes Z_4\otimes X_5.
\end{align*}

This excitation then gives rise to four operators in the Qubit Pool:

\begin{align*}
&i \cdot X_0\otimes Z_1\otimes Z_2\otimes Z_3\otimes Y_4, \qquad &i \cdot X_1\otimes Z_2\otimes Z_3\otimes Z_4\otimes Y_5, \\
&i \cdot Y_0\otimes Z_1\otimes Z_2\otimes Z_3\otimes X_4, \qquad &i \cdot Y_1 \otimes Z_2\otimes Z_3\otimes Z_4\otimes X_5.
\end{align*}

The norm of the coefficients no longer matters because now each Pauli string will have its own variational parameter.

The number of Pauli strings per fermionic excitation remains approximately the same on average as the size of the system grows. Thus, the size of the pool is still $\mathcal{O}(N^4)$, $N$ the number of spin-orbitals - albeit with a significantly larger prefactor that implies a larger number of measurements per iteration. The Qubit Pool operators also act on a number of qubits that grows like $\mathcal{O}(N)$. However, the circuit depth associated with each operator is significantly changed (in this case, improved). While \gls{SGSD} operators contain up to 48 Pauli strings, each operator in the Qubit Pool consists of a single one. This implies an up to almost 50-fold reduction in the circuit depth associated with a single operator, and is the motive behind the Qubit-\gls{ADAPT2}-\gls{VQE} proposal.

\subsubsection{Pauli String Pool Without the Jordan-Wigner String}

The Qubit-\gls{ADAPT2}-\gls{VQE} article also introduced a modified version of the Qubit Pool, consisting of the same operators, now cleared of the Jordan-Wigner string. This is the $\prod_{k=1}^{i-1}Z_k$ factor that appears in the Jordan-Wigner mapping (formula \ref{def:jw_transform_paulis}) to account for the anticommutation of fermions. The four operators introduced by the same single fermionic excitation used as an example before would simply be

\[i \cdot X_0\otimes Y_4, \qquad i \cdot X_1\otimes Y_5, \qquad i \cdot Y_0\otimes X_4, \qquad i \cdot Y_1 \otimes X_5.\]

The scaling of the total number of operators in this pool is once again $\mathcal{O}(N^4)$, with $N$ being the number of spin-orbitals. However, the scaling of the length of the Pauli strings has changed from $\mathcal{O}(N)$ to $\mathcal{O}(1)$. Without the Jordan-Wigner string, only qubits representing spin-orbitals directly involved in the original excitations will be acted on by the operators they give rise to. Since the excitations we are considering are at most doubles, the Pauli strings will have length four at most.

\subsubsection{Minimal Pools}

Finally, the article that introduced Qubit-\gls{ADAPT2}-\gls{VQE} proposed a way of creating minimal pools containing operators capable of transforming any real state into any real state. These minimal pools were proven to be complete and allow convergence when used on random real Hamiltonians. An example of one such pool is the pool G, presented in the original article and reproduced here in \ref{def:minimal_g_pool}.

\begin{align}
\begin{split}
&G_1=iZ\otimes Y \otimes I^{\otimes(N-2)},\quad
G_2=iI^{\otimes 1}\otimes Z\otimes Y \otimes I^{\otimes(N-3)},\\
&\qquad...,\quad
G_{N-2}=iI^{\otimes (N-3)}\otimes Z\otimes Y \otimes I^{\otimes 1},\quad
G_{N-1}=iI^{\otimes (N-2)}\otimes Z\otimes Y,\\
&G_N = iY\otimes I^{\otimes (N-1)},\quad
G_{N+1} = iI^{\otimes 1}\otimes Y \otimes I^{\otimes (N-2)},\\
&\qquad...,\quad
G_{2N-3} = iI^{\otimes N-3} \otimes Y \otimes I^{\otimes 2},\quad
G_{2N-2} = iI^{\otimes N-2} \otimes Y \otimes I^{\otimes 1}\quad
\end{split}
\label{def:minimal_g_pool}
\end{align}

The minimal pools proposed scale like $\mathcal{O}(N)$ on the number of qubits, with a total of only $2N-2$ operators. This is a remarkable reduction from the quartic scaling from the previous pools: the number of gradients that must be evaluated per iteration is reduced from $\mathcal{O}(N^4)$ to $\mathcal{O}(N)$.

The minimal pools allow for a significant decrease in the circuit depth per operator, and can have a very local structure (as does $G$, that only ever acts on two followed qubits).

Unfortunately, while they work for generic Hamiltonians (such as random real Hamiltonians), they are not adequate for the specific structure of molecular Hamiltonians that is imposed by fermionic symmetries. In particular, we can see that none of the operators in the G pool conserves the spin or particle number in the Hartree-Fock state, which has severe consequences.

The Hartree-Fock state can be written using OpenFermion's \cite{openfermion} orbital convention as $\ket{000...111}$. Acting on this state with a parameterized $Y$ rotation acting on qubit $k$ that is in a computational basis state $\ket{x}$, we get:

\[\ket{\psi}=e^{jY_k\theta}\ket{000...}\otimes\ket{x}_k\otimes\ket{...111}\]
\[=\cos{\theta}\ket{000...}\otimes\ket{x}_k\otimes\ket{...111} + (-1)^{1-x}\sin{\theta}\ket{000...}\otimes\ket{1-x}_k\otimes\ket{...111}\]
\[=\cos{\theta}\ket{n} + (-1)^{1-x}\sin{\theta}\ket{m}\]

Clearly, $\ket{n}$ and $\ket{m}$ represent computational basis states with a different number of electrons (different number of 1 states). Considering that the Hamiltonian matrix is real and Hermitian, so that $H_{m,n}=H_{n,m}$, the energy and gradient evaluated in this state are 

\[E_{\ket{\psi}} = \bra{\psi}H\ket{\psi} = \cos^2{\theta} H_{n,n} + \sin^2{\theta} H_{m,m} + 2(-1)^{1-x}\sin{\theta}\cos{\theta}H_{m,n},\]

\[\frac{dE_{\ket{\psi}}}{d\theta}\biggr\vert_{\theta=0} =\left(2\cos{\theta}\sin{\theta}(H_{m,m}-H_{n,n})+2(-1)^{1-x}\cos{(2\theta)}H_{m,n}\right) \biggr\vert_{\theta=0}\]
\[= 2(-1)^{1-x}H_{m,n}\]

Since $\ket{n}$, $\ket{m}$ correspond to Slater determinants different numbers of electrons, $H_{m,n}$ is zero and the gradient vanishes. Since $\ket{n}$ represents the Hartree-Fock state, the lowest-energy Slater determinant, its energy corresponds to the lowest value in the diagonal of the Hamiltonian. Thus, it is evident that a single parameterized $Y$ rotation (acting on the Hartree-Fock state) has null gradient and does not lower the energy beyond the Hartree-Fock energy. 

All operators in pool G are either $Y$ rotations or conditional $Y$ rotations, so that the the same applies. Since all gradients will be zero, the \gls{ADAPT2}-\gls{VQE} algorithm will be prevented from starting in the first place, because the selection method will fail. This is a consequence of the minimal pool operators not respecting particle number and spin symmetries.

For the \gls{ADAPT2}-\gls{VQE} algorithm to be capable of starting (by adding the first operator to the ansatz), the pool must include operators that preserve symmetries in the Hartree-Fock state. However, this still does not guarantee proper convergence; for the ground state to be reached, the pool operators must obey extra restrictions. An explicit approach for creating symmetry-adapted minimal pools was introduced in 2021 in \cite{shkolnikov2021}. Because of the added restrictions, the symmetry-adapted version of the minimal pools can actually have less operators than the minimal complete pools presented before. The size of the pool is still $\mathcal{O}(N)$, $N$ the number of qubits/spin-orbitals. This means that the measurements per iteration scale as $\mathcal{O}(N^5)$ (against $\mathcal{O}(N^8)$ in fermionic-\gls{ADAPT}-\gls{VQE} or qubit-\gls{ADAPT2}-\gls{VQE} with the original pools), a more modest increase against the $\mathcal{O}(N^4)$ scaling of static ansatz \gls{VQE}.

\section{Application to LiH}

In order to analyse the femionic- and qubit-\gls{ADAPT2}-\gls{VQE} performance, and compare \gls{ADAPT2}-\gls{VQE} with \gls{UCCSD}-\gls{VQE}, the algorithm was applied to multiple molecules. 

$LiH$ is a molecule represented by 12 qubits in a minimal basis. Its Hartree-Fock ground energy is significantly distant from the \gls{FCI} value, and the Hartree-Fock approximation can be significantly improved upon in this case. Unlike $H_2$, a smaller molecule, more than a single operator is required to reach chemical accuracy in the case of $LiH$, resulting in a more interesting evolution of the ansatz. While the simulations are heavier computationally, \gls{ADAPT}-\gls{VQE} with a threshold $\epsilon$ of 0.01 is still viable, and this suffices to reach an interesting accuracy. Because of this, $LiH$ was chosen to test the algorithm in a fully noise-free scenario.

\subsection{Evolution of the Optimization}

\begin{figure}[htbp]
    \centering
    \includegraphics[width=0.9\columnwidth]{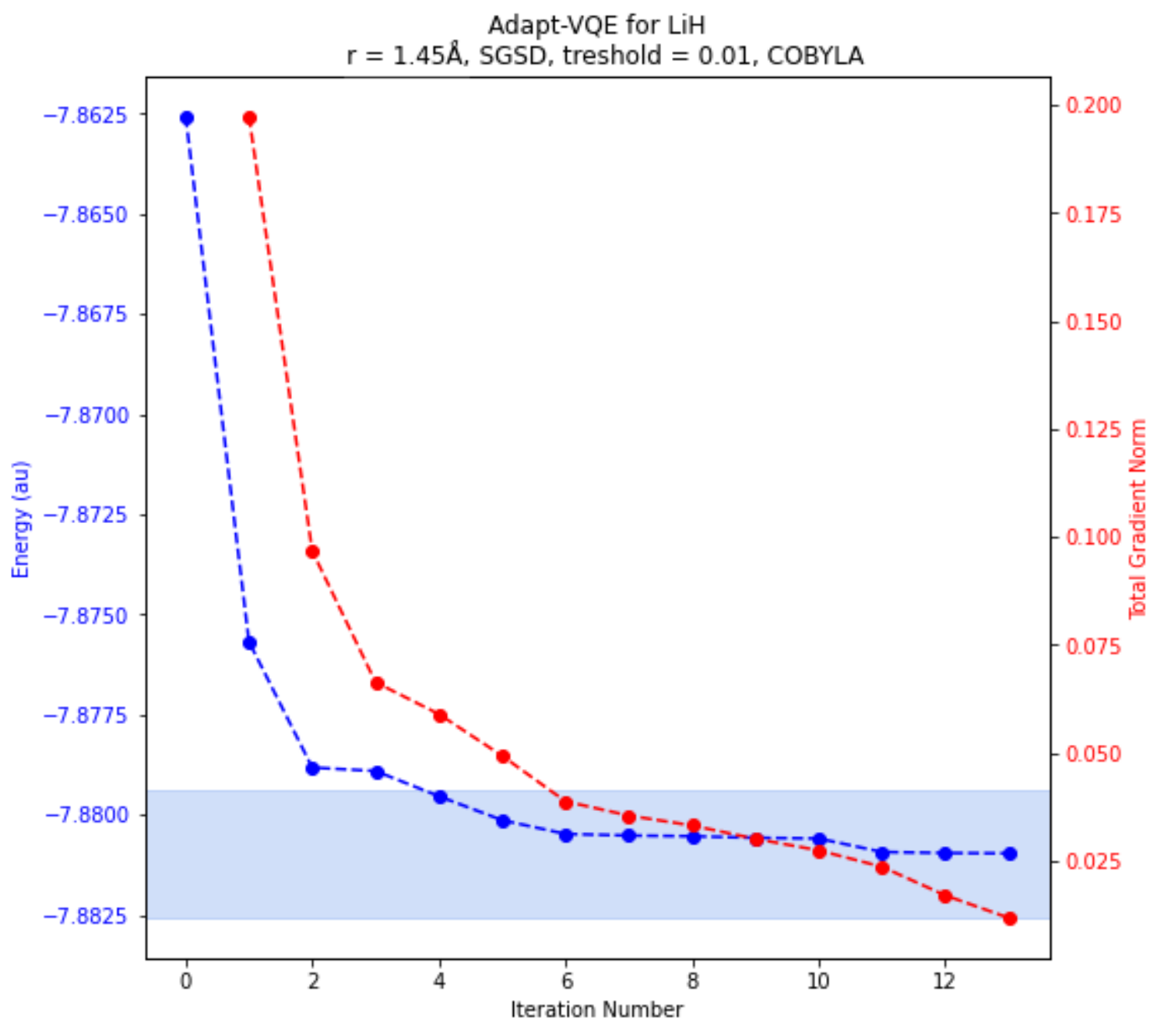}
    \caption{Evolution of the fermionic-\gls{ADAPT}-\gls{VQE} energy and gradient norm along the iterations, for the lithium hydride ($LiH$) molecule at an interatomic distance of 1.45Å. The convergence criterion was set to a gradient treshold of 0.01. The \gls{SGSD} pool and the COBYLA optimizer were used. The shaded blue area represents the region of chemical accuracy (error of less than 1kcal/mol). The simulation was done via matrix algebra, and does not include any type of noise.}
    \label{fig:LiH_singlet}
\end{figure}

Figure \ref{fig:LiH_singlet} shows the evolution of the \gls{ADAPT}-\gls{VQE} algorithm for $LiH$ at an interatomic distance of 1.45Å. 

It can be seen that the energy reaches chemical accuracy after four iterations; at this point, the ansatz has 4 spin-adapted operators and the corresponding 4 variational parameters. For reference, the \gls{UCCSD} ansatz obtained from OpenFermion \cite{openfermion} has for the same molecule 44 variational parameters and 64 (simple) excitation operators. Evidently, this also corresponds to a greater accuracy than the 4-operator ansatz of \gls{ADAPT}-\gls{VQE}; but in the plot we see that as we let the algorithm evolve further, it manages to steadily improve the energy estimate. In the original article \cite{Grimsley2019}, through simulations including several molecules, it was shown that \gls{ADAPT}-\gls{VQE} is capable of outperforming \gls{UCCSD}-\gls{VQE}.

\FloatBarrier
\section{Application to Molecular Hydrogen}

Like \gls{UCCSD}-\gls{VQE}, the Qubit-\gls{ADAPT2}-\gls{VQE} (with the Pauli string pool cleared of the Z string) was applied to the hydrogen molecule. The small size of this molecule allows it to be represented by only 4 qubits, and results in shallower ansätze. Because of this, it was chosen to run the algorithm on real quantum computers and analyse the effect of noise. Since it is enough to reach chemical accuracy in this case, a single iteration of \gls{ADAPT2}-\gls{VQE} was used to obtain the ground state. At this point, there is a single variational parameter in the ansatz, which can be implemented with a \gls{CNOT} count and depth of 6 assuming all-to-all connectivity.

In this section, \gls{ADAPT2}-\gls{VQE} results will be presented and compared against \gls{UCCSD}-\gls{VQE}, with a focus on noise-resilience and (to a lesser extent) measurement costs.

\subsection{Bond Dissociation Graph}

\begin{figure}[htbp]
    \centering
     \begin{subfigure}[b]{0.45\textwidth}
         \centering
         \includegraphics[width=\textwidth]{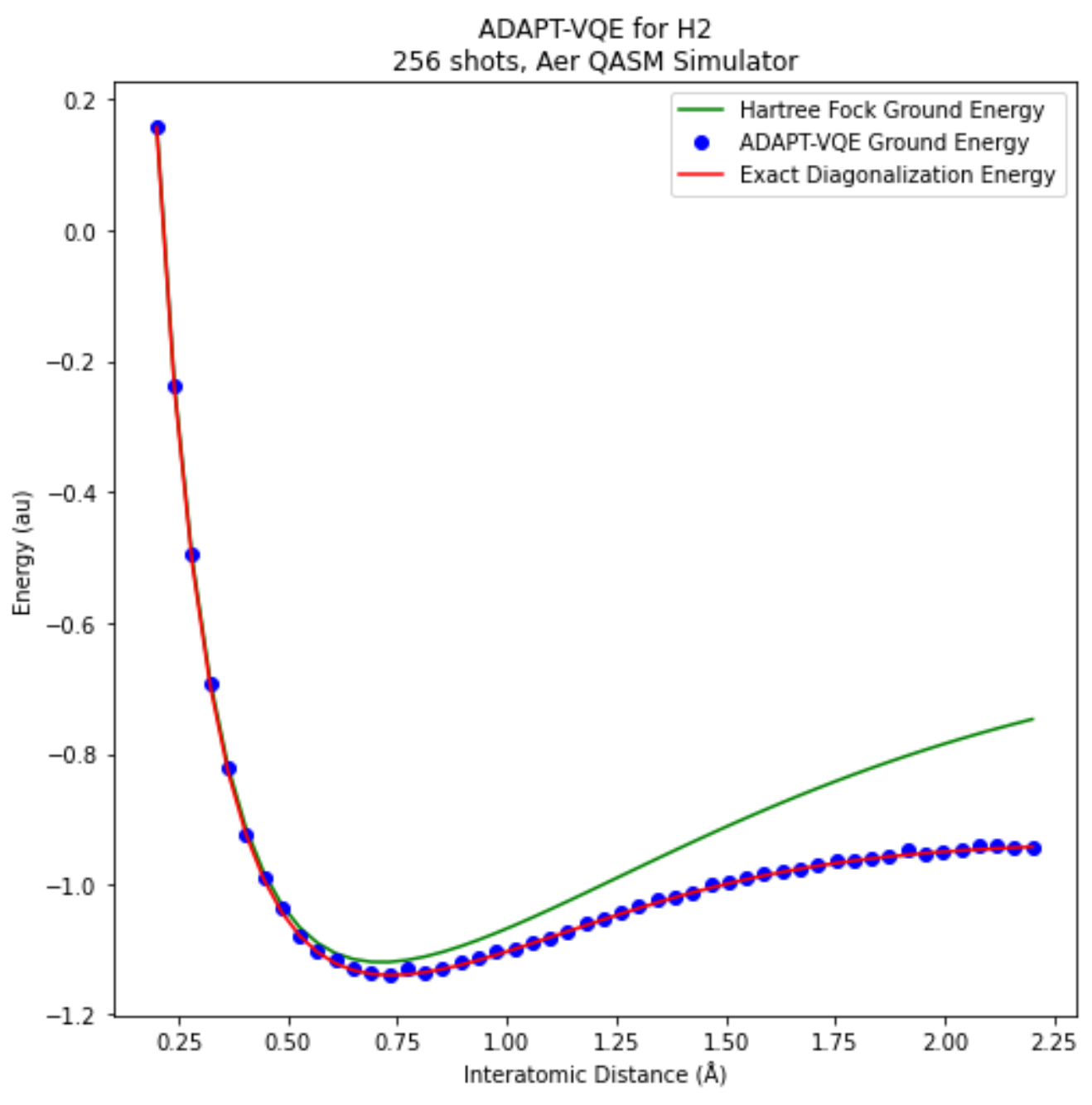}
         \caption{QASM simulator, 256 shots}
         \label{fig:h2_fullbc_adapt_qasm_256}
     \end{subfigure}
     \hfill
     \begin{subfigure}[b]{0.45\textwidth}
         \centering
         \includegraphics[width=\textwidth]{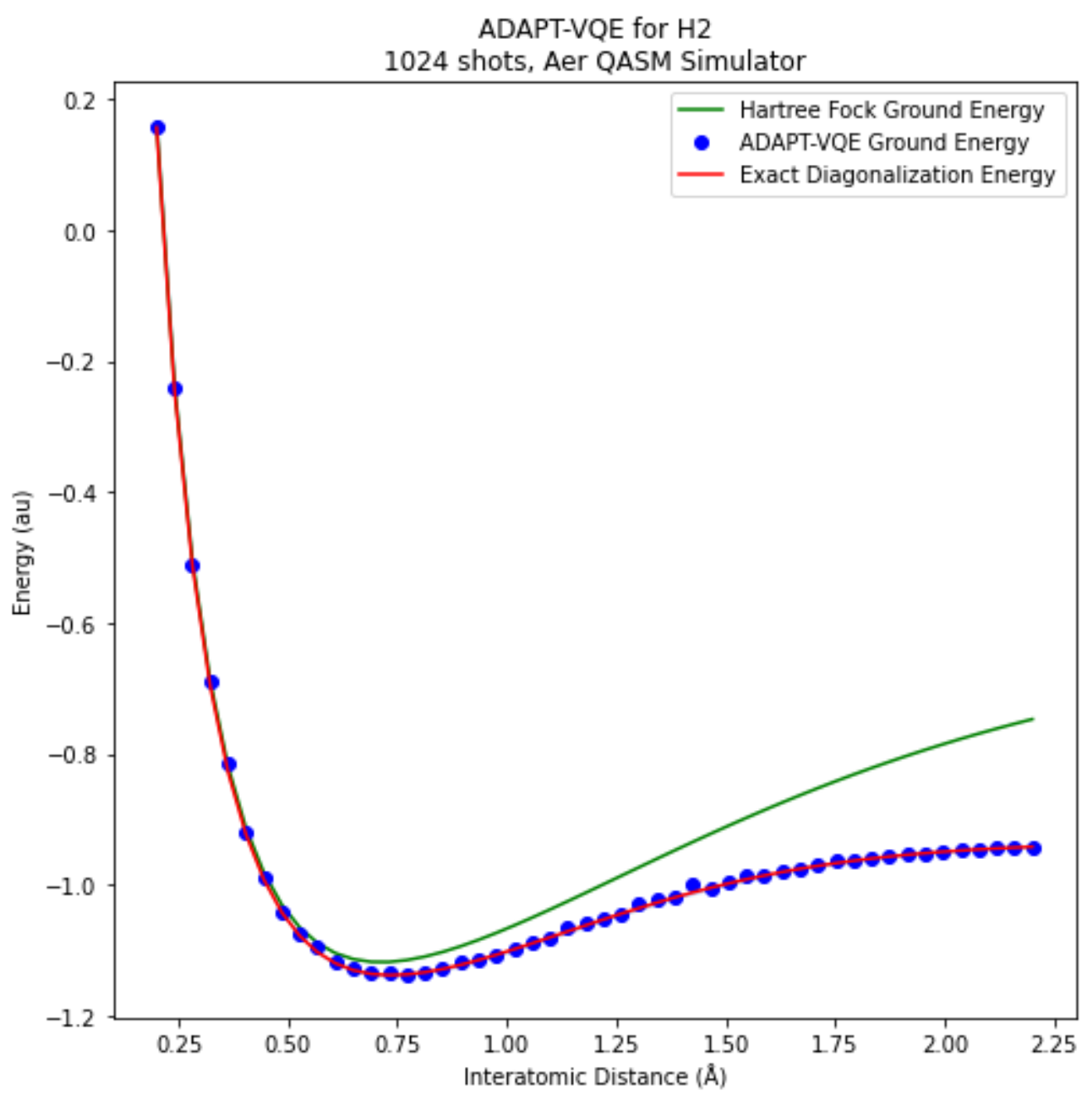}
         \caption{QASM simulator, 1024 shots}
         \label{fig:h2_fullbc_adapt_qasm_1024}
     \end{subfigure}
     \\
     \begin{subfigure}[b]{0.45\textwidth}
         \centering
         \includegraphics[width=\textwidth]{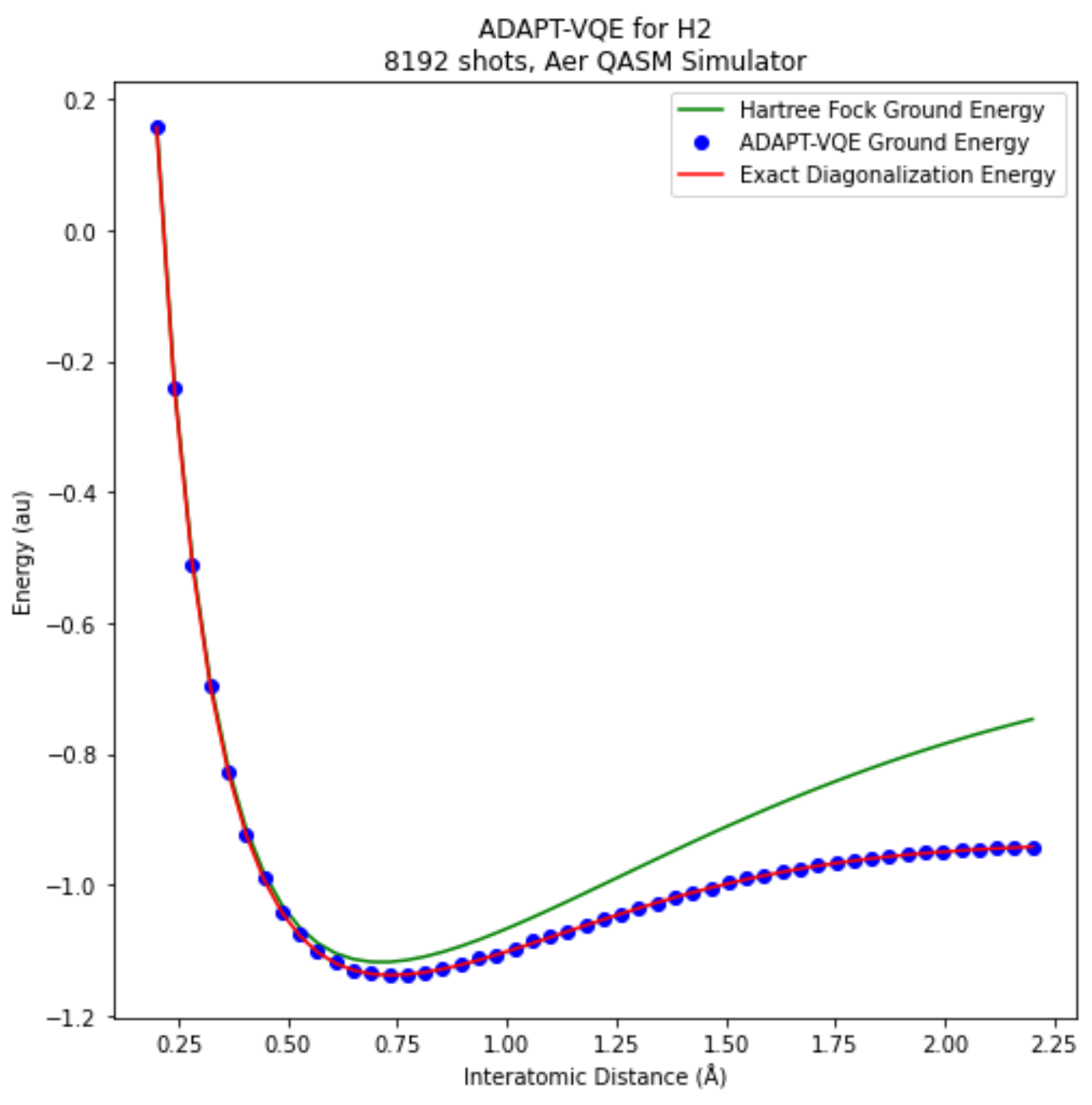}
         \caption{QASM simulator, 8192 shots}
         \label{fig:h2_fullbc_adapt_qasm_8192}
     \end{subfigure}
     \hfill
     \begin{subfigure}[b]{0.45\textwidth}
         \centering
         \includegraphics[width=\textwidth]{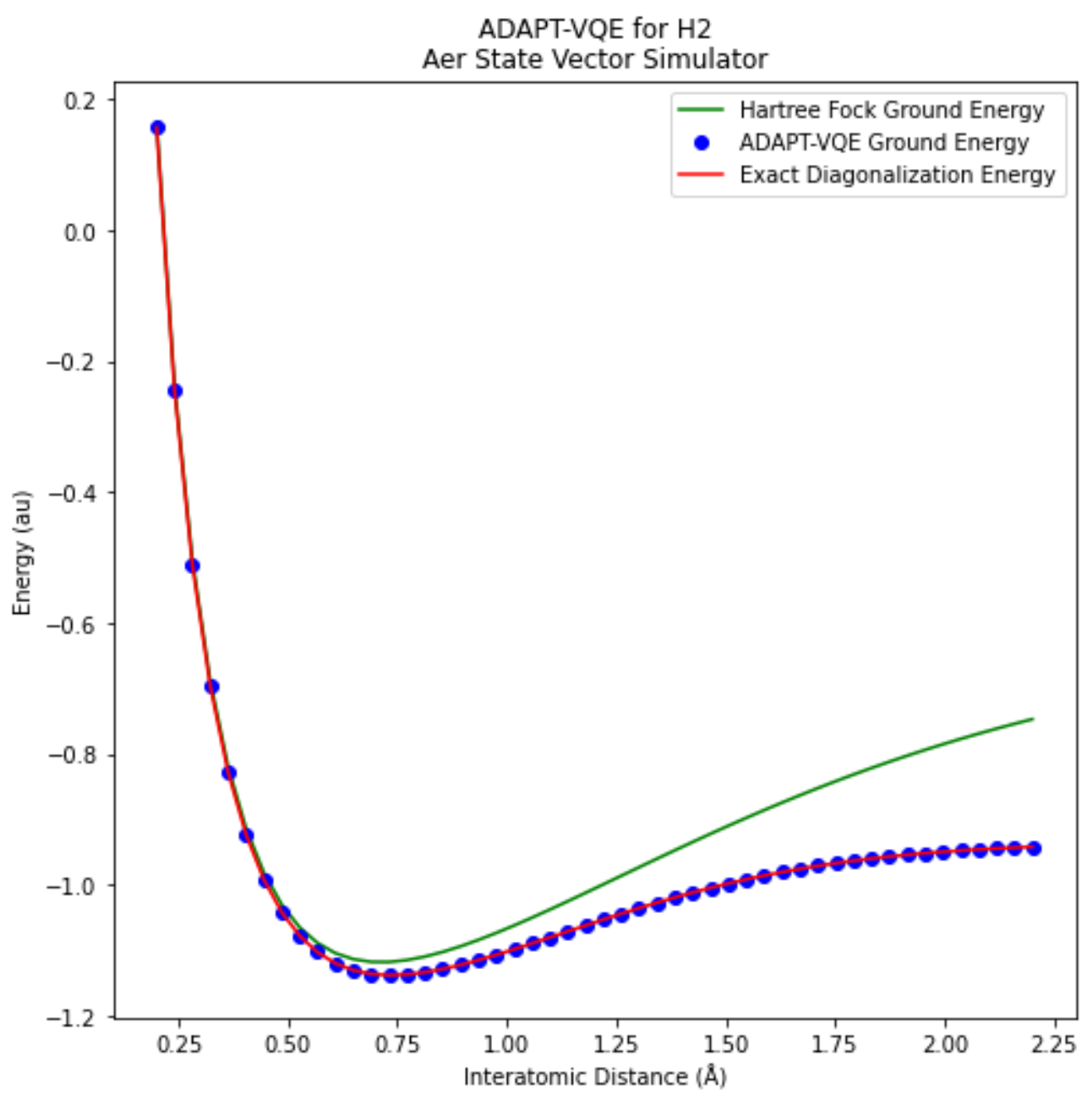}
         \caption{Statevector simulator}
         \label{fig:h2_fullbc_adapt_statevector_simulator}
     \end{subfigure}
    \caption{Single-run, single-iteration \gls{ADAPT2}-\gls{VQE} ground energy along the bond dissociation curve of the hydrogen molecule, obtained using IBMQ \cite{IBMQ} simulators. The results plotted in figures \ref{fig:h2_fullbc_adapt_qasm_256}-\ref{fig:h2_fullbc_adapt_qasm_8192}, ordered by increasing shot count, were obtained using the QASM simulator. The results plotted in figure \ref{fig:h2_fullbc_adapt_statevector_simulator} were obtained using the state vector simulator (equivalent to infinite shots). The red and green curves represent the \gls{FCI} energy and the Hartree-Fock energy, respectively.}
    \label{fig:h2_adapt_fullbc}
\end{figure}

Following the fully noise-free simulation, the bond dissociation graphs of the molecule with sampling noise will now be analysed. The plots in \ref{fig:h2_adapt_fullbc} show the \gls{ADAPT2}-\gls{VQE} energies along the curve, for different shot counts; they should be compared with figure \ref{fig:h2_uccsd_fullbc}, that presents the same plots for \gls{UCCSD}-\gls{VQE}. 

Like before, there is a visible improvement on the performance of \gls{ADAPT2}-\gls{VQE} with the increase of the number of shots from figure \ref{fig:h2_fullbc_adapt_qasm_256} to figure \ref{fig:h2_fullbc_adapt_qasm_8192}. In the limit of infinite shots (figure \ref{fig:h2_fullbc_adapt_statevector_simulator}), the error is of the order of $10^{-8}$ Hartree along the whole curve. 

As did \gls{UCCSD}-\gls{VQE}, \gls{ADAPT2}-\gls{VQE} successfully brings the energy into the \gls{FCI} solution from the Hartree-Fock approximation (which is the reference state). The extra variational freedom (in this case, a single variational parameter) seems to be enough to account for electronic correlation.

What is interesting is that \gls{ADAPT2}-\gls{VQE} clearly suffers less from sampling noise. With only 256 shots (figure \ref{fig:h2_fullbc_adapt_qasm_256}), \gls{ADAPT2}-\gls{VQE} is visibly closer to the \gls{FCI} curve than \gls{UCCSD}-\gls{VQE} with 1024 shots (figure \ref{fig:h2_fullbc_uccsd_qasm_1024}). This is not a circuit depth dependent noise source; what causes the \gls{UCCSD}-\gls{VQE} performance to be worse is the extra variational freedom, that results in a more arduous optimization.

\begin{figure}[htbp]
    \centering
     \begin{subfigure}[b]{0.45\textwidth}
         \centering
         \includegraphics[width=\textwidth]{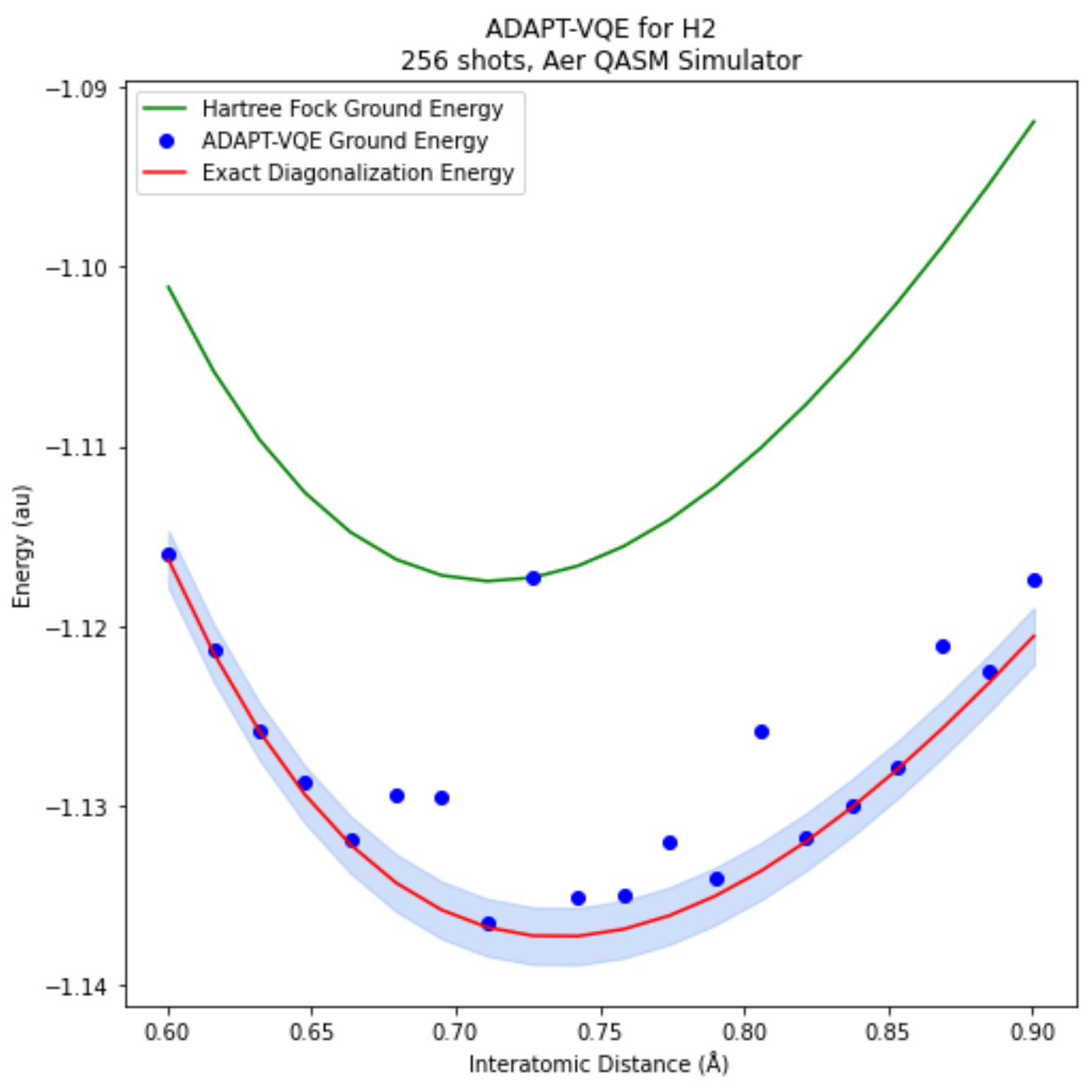}
         \caption{QASM simulator, 256 shots}
         \label{fig:h2_zoombc_adapt_qasm_256}
     \end{subfigure}
     \hfill
     \begin{subfigure}[b]{0.45\textwidth}
         \centering
         \includegraphics[width=\textwidth]{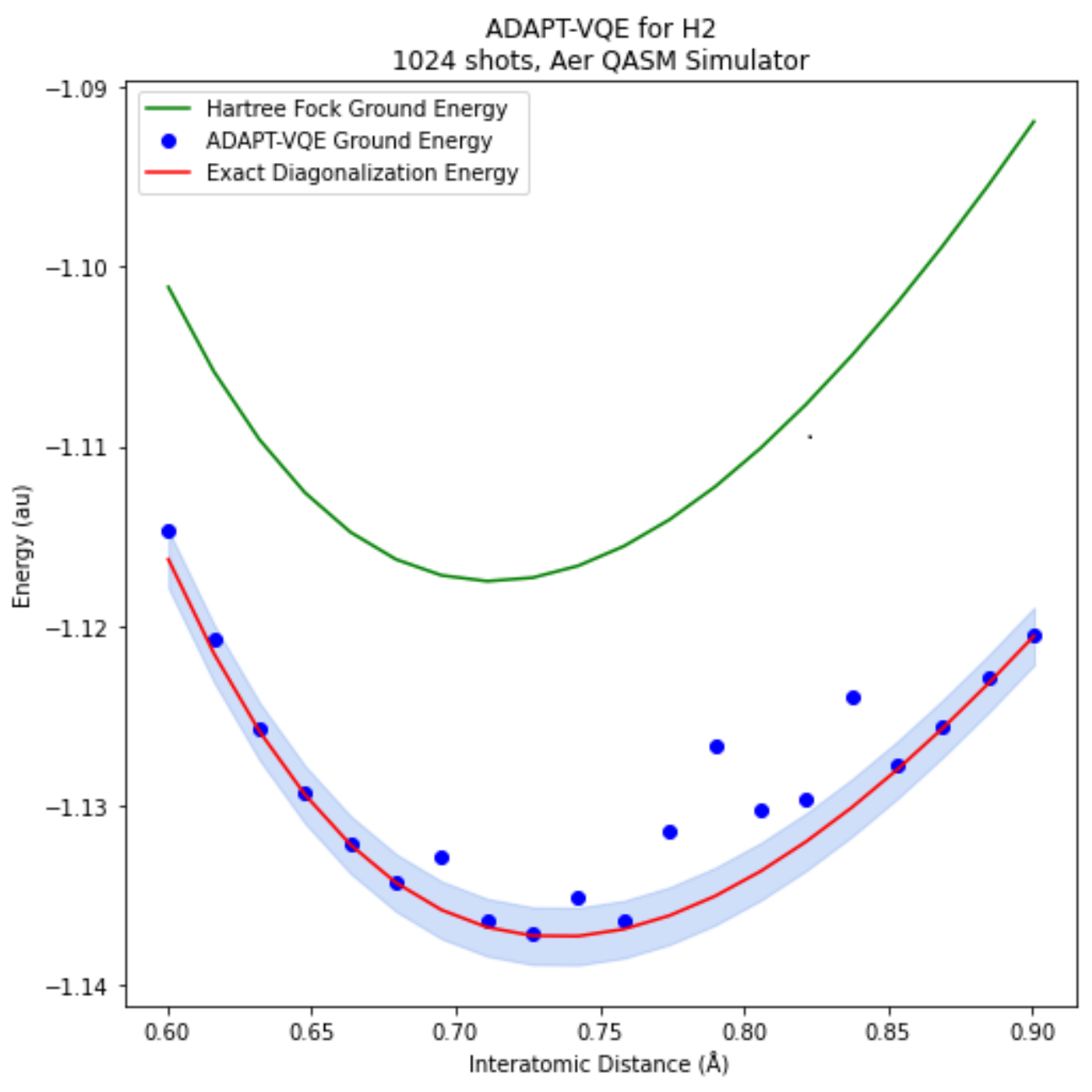}
         \caption{QASM simulator, 1024 shots}
         \label{fig:h2_zoombc_adapt_qasm_1024}
     \end{subfigure}
     \\
     \begin{subfigure}[b]{0.45\textwidth}
         \centering
         \includegraphics[width=\textwidth]{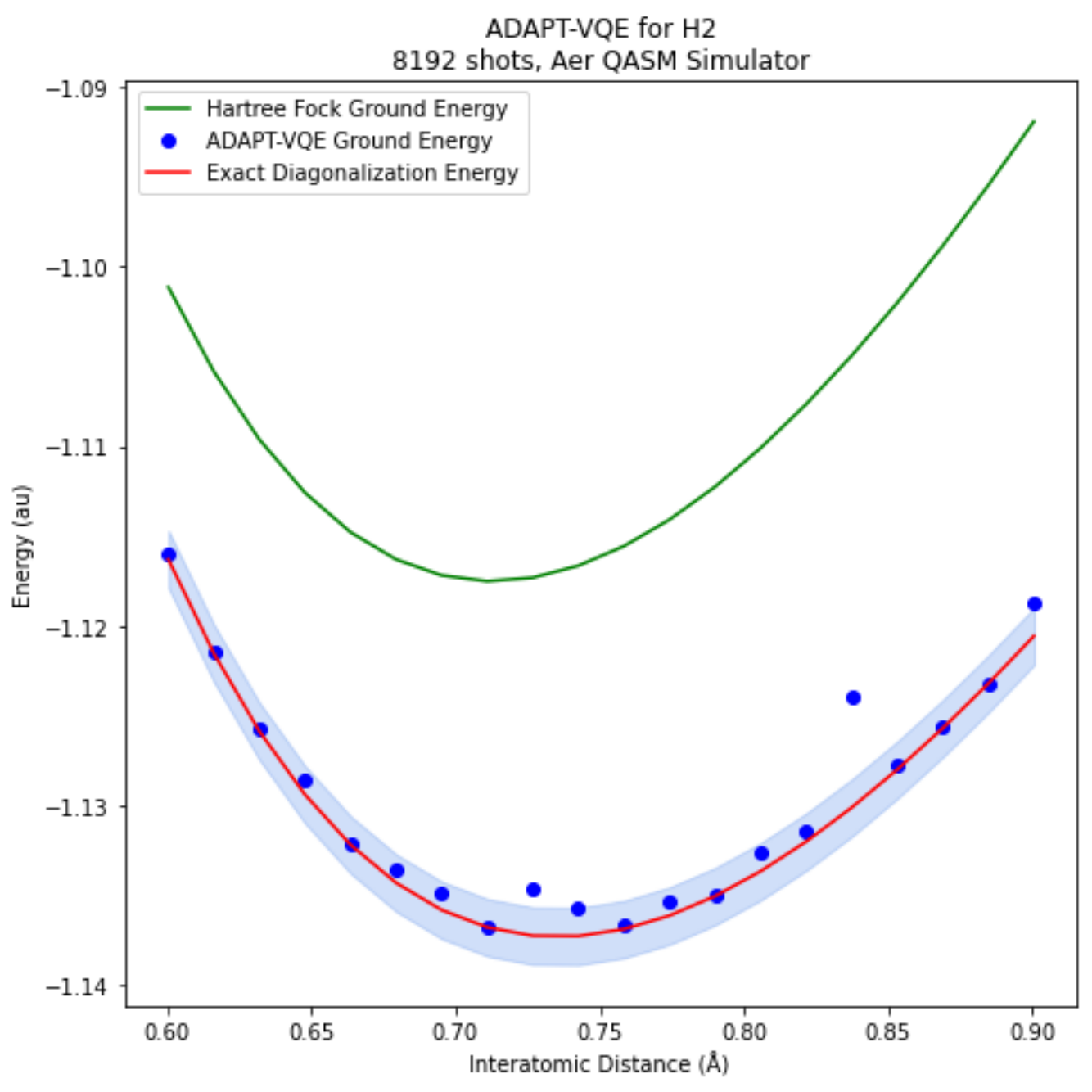}
         \caption{QASM simulator, 8192 shots}
         \label{fig:h2_zoombc_adapt_qasm_8192}
     \end{subfigure}
     \hfill
     \begin{subfigure}[b]{0.45\textwidth}
         \centering
         \includegraphics[width=\textwidth]{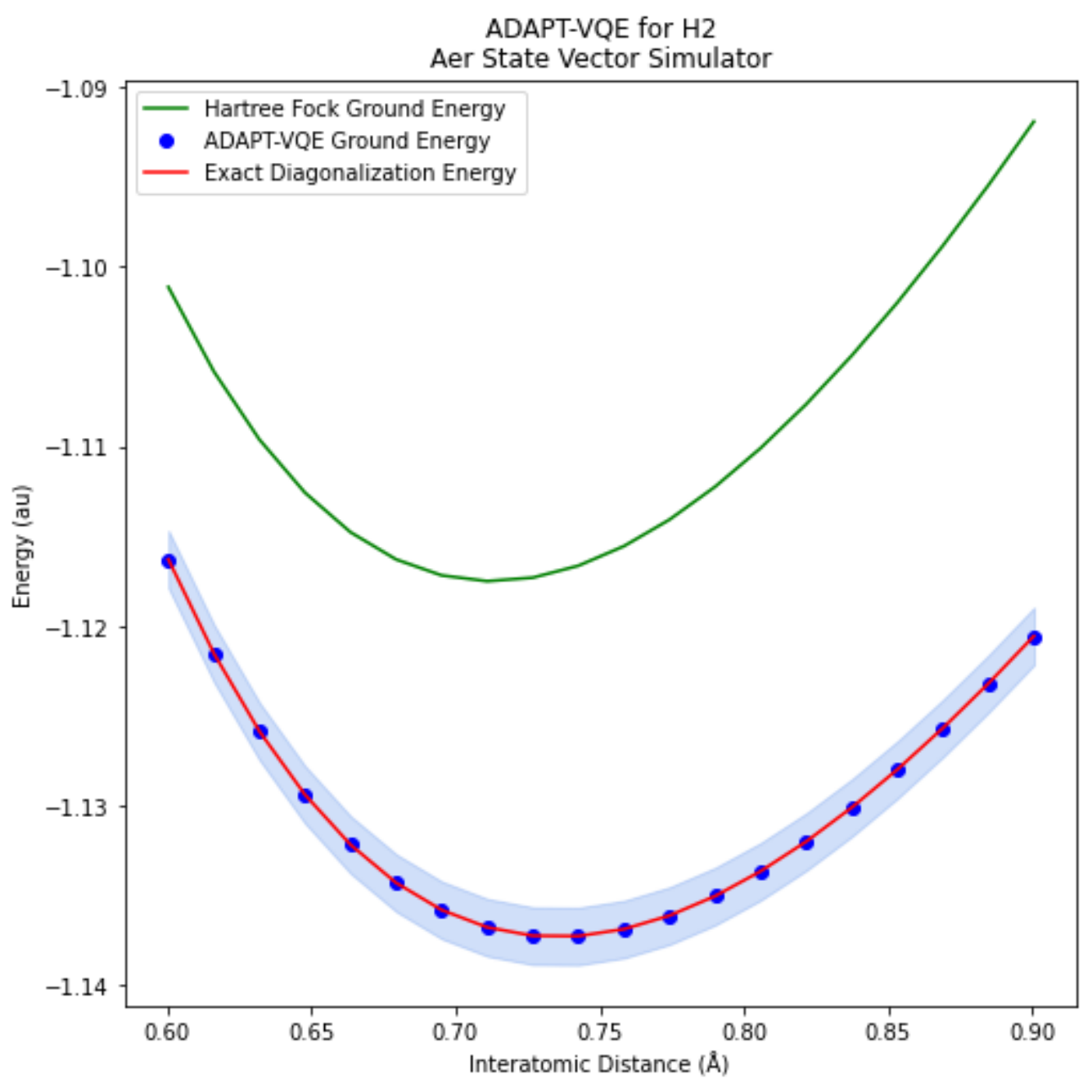}
         \caption{Statevector simulator}
         \label{fig:h2_zoombc_adapt_statevector_simulator}
     \end{subfigure}
    \caption{Bond dissociation curves of figure \ref{fig:h2_adapt_fullbc}, zoomed in on the region nearing the equilibrium bond length. The blue area marks the region of chemical accuracy (error of less than 1kcal/mol).}
    \label{fig:h2_adapt_zoombc}
\end{figure}

Figure \ref{fig:h2_adapt_zoombc} shows the same plots of figure \ref{fig:h2_adapt_fullbc}, magnified around the equilibrium bond length. This can be compared with figure \ref{fig:h2_uccsd_zoombc}, that shows these plots for \gls{UCCSD}-\gls{VQE}. 

The only scenario in which \gls{UCCSD}-\gls{VQE} beats 1-iteration qubit-\gls{ADAPT2}-\gls{VQE} is the fully noise-free one: even though it is not visible in the graph, simulations showed that under those circumstances \gls{UCCSD}-\gls{VQE} could achieve a precision of the order of $10^{-9}$ Hartree against $10^{-8}$ from \gls{ADAPT2}-\gls{VQE}. However, this precision is exceedingly high: not only it is not necessary for most applications, it also can hardly be ever met, considering not only sampling but also other sources of noise prevalent in \gls{NISQ} devices.

With sampling, regardless of the shot count, single-iteration \gls{ADAPT2}-\gls{VQE} manages to outperform \gls{UCCSD}-\gls{VQE}. With 256 shots (figures \ref{fig:h2_zoombc_adapt_qasm_256} and \ref{fig:h2_zoombc_uccsd_qasm_256}) and 1024 shots (figures \ref{fig:h2_zoombc_adapt_qasm_1024} and \ref{fig:h2_zoombc_uccsd_qasm_1024}), the \gls{UCCSD}-\gls{VQE} result does not fall within chemical accuracy in any of the points, while \gls{ADAPT2}-\gls{VQE} falls within chemical accuracy in 11 and 13 points respectively. With 8192 shots (figures \ref{fig:h2_zoombc_adapt_qasm_8192} and \ref{fig:h2_zoombc_uccsd_qasm_8192}), \gls{ADAPT2}-\gls{VQE} only falls outside of chemical accuracy in 3 out of 20 points, against 12 out of 20 fails of \gls{UCCSD}-\gls{VQE}.

\subsection{Running the Algorithm on Cloud Quantum Computers}

\begin{figure}[htbp]
     \centering
     \begin{subfigure}[b]{0.45\textwidth}
         \centering
         \includegraphics[width=\textwidth]{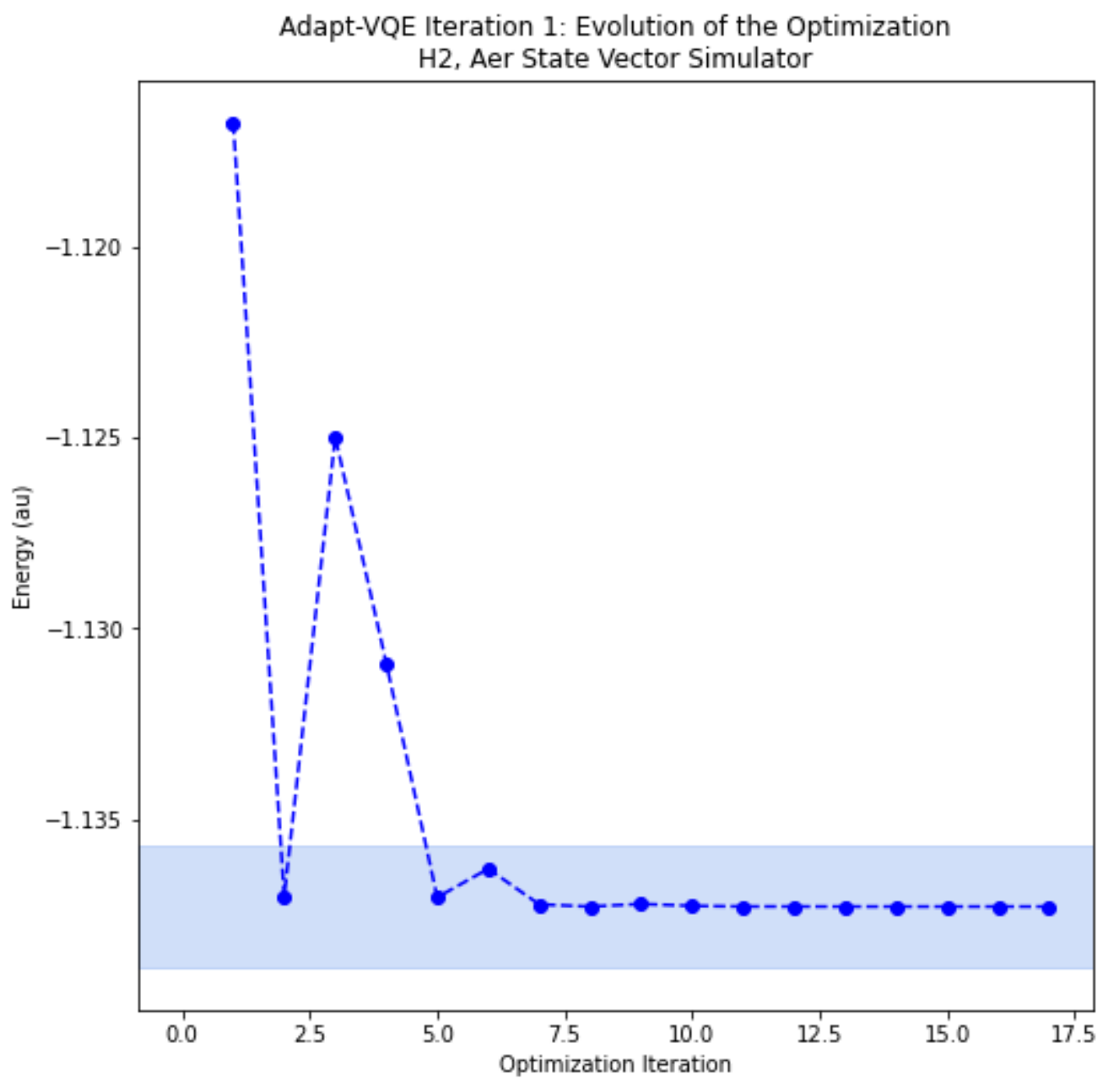}
         \caption{Statevector simulator}
         \label{fig:h2_adapt_StateVectorSimulator}
     \end{subfigure}
     \hfill
     \begin{subfigure}[b]{0.45\textwidth}
         \centering
         \includegraphics[width=\textwidth]{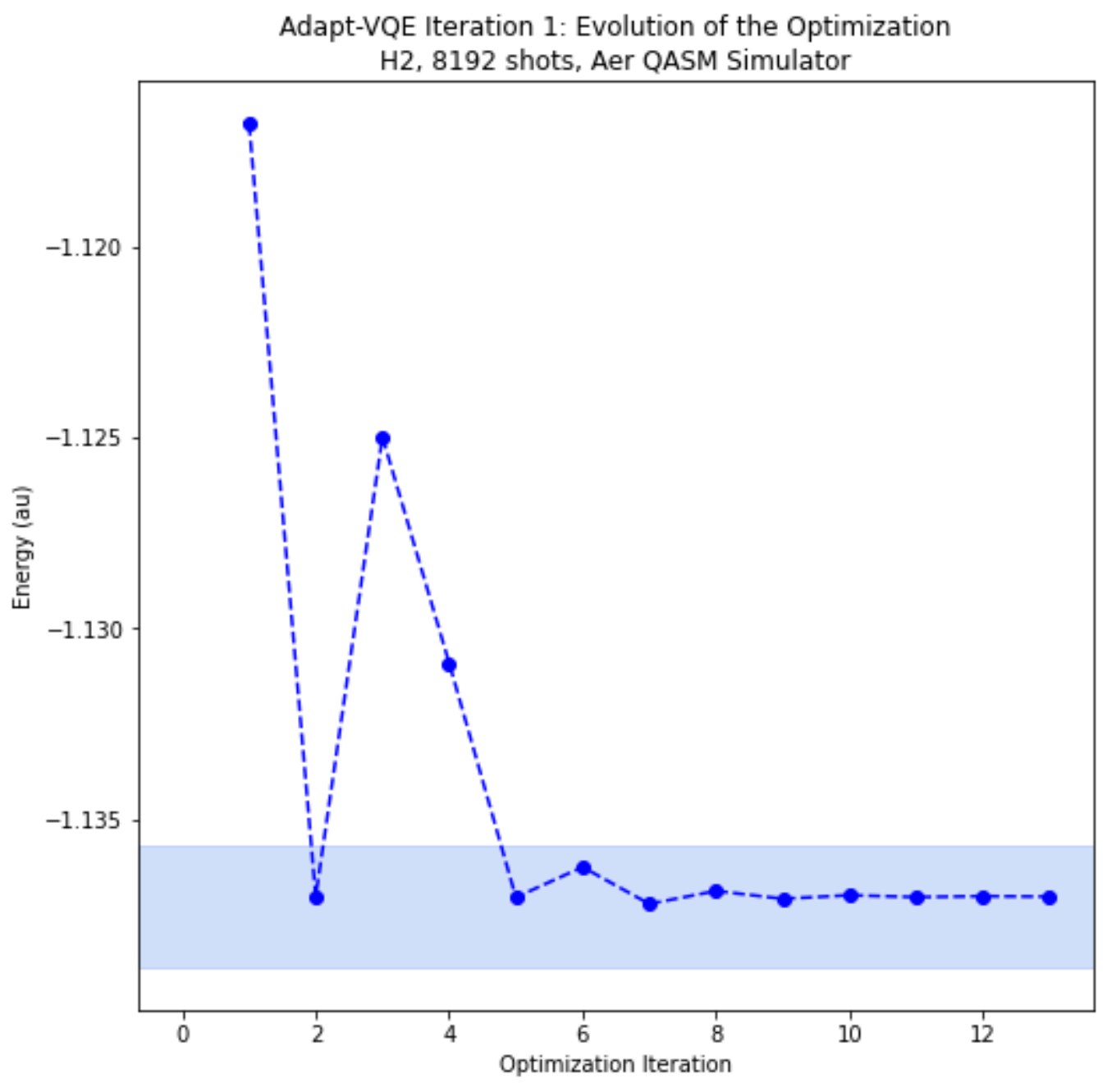}
         \caption{QASM Simulator}
         \label{fig:h2_adapt_qasm}
     \end{subfigure}
     \\
     \begin{subfigure}[b]{0.45\textwidth}
         \centering
         \includegraphics[width=\textwidth]{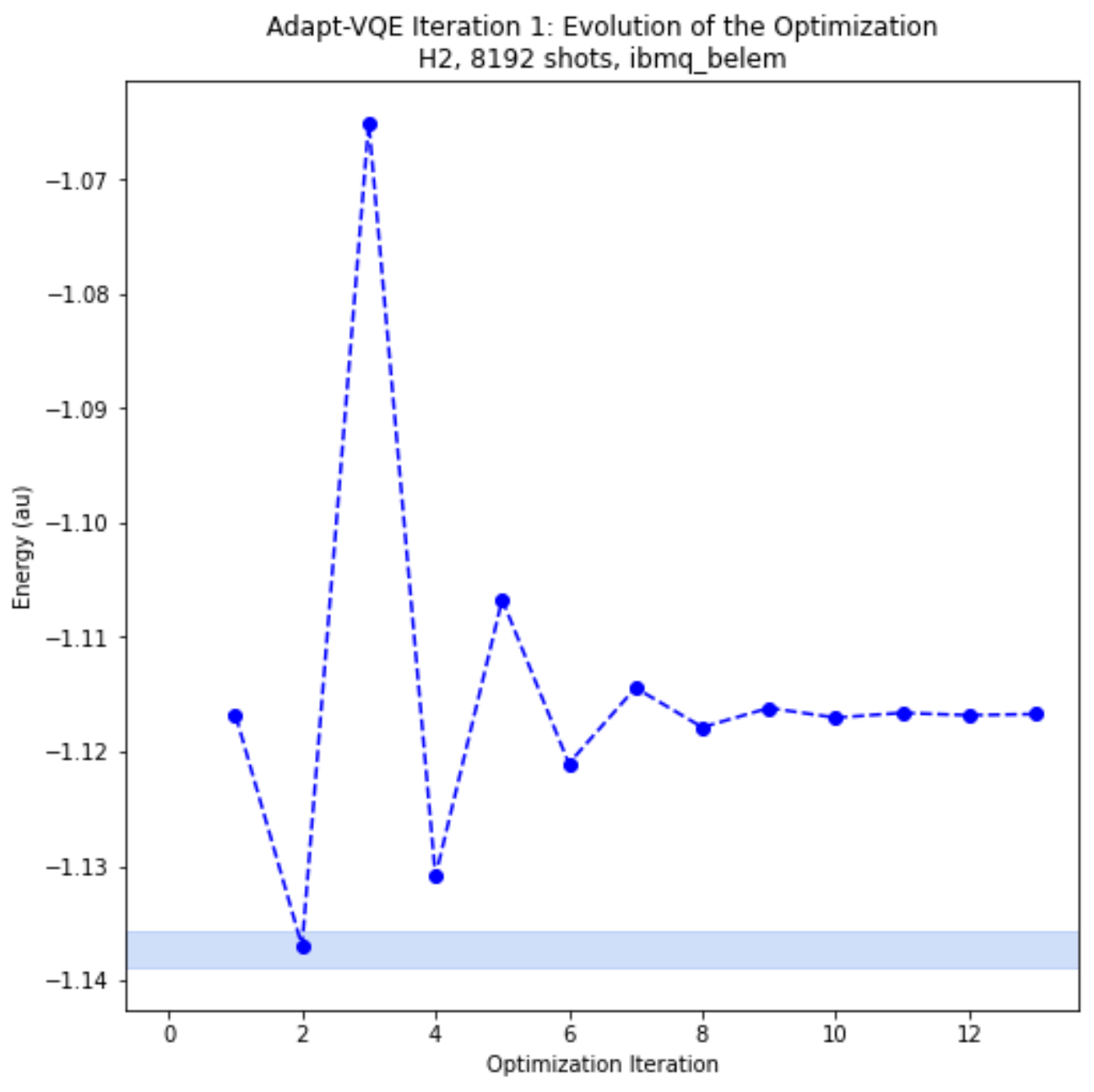}
         \caption{IBMQ Belem}
         \label{fig:h2_adapt_belem}
     \end{subfigure}
     \caption{Evolution of first the \gls{ADAPT2}-\gls{VQE} optimization for $H_2$ at an interatomic distance of 0.74Å, using three different IBM Quantum \cite{IBMQ} backends: the statevector simulator (with no noise of any kind), the QASM simulator (with sampling noise) and the Belem device (a 5-qubit Falcon  processor). In the last two cases, 8192 shots were used. The energy is plotted in blue; the pale blue region marks the area of chemical accuracy. The same starting point, the Hartree-Fock reference state, was used for all backends. The chosen optimizer was COBYLA.}
     \label{fig:h2_adapt_sv_qasm_belem}
\end{figure}

After considering the impact of sampling noise, we can now observe the impact of other types of noise by analysing the performance of \gls{ADAPT2}-\gls{VQE} on a real quantum computer. Figure \ref{fig:h2_adapt_sv_qasm_belem} shows the evolution of the first iteration of \gls{ADAPT2}-\gls{VQE} in three scenarios: no noise (figure \ref{fig:h2_adapt_StateVectorSimulator}), sampling noise only / ideal quantum processor (figure \ref{fig:h2_adapt_qasm}), real quantum processor (figure \ref{fig:h2_adapt_belem}). These can be compared with those from figure \ref{fig:h2_uccsd_sv_qasm_belem}, that show the evolution of the \gls{UCCSD}-\gls{VQE} optimization for the same three backends.

In the setting with sampling noise only, \gls{ADAPT2}-\gls{VQE} already outperforms \gls{UCCSD}-\gls{VQE}. As it was mentioned previously, this is mainly due to the extra strain on the optimization caused by the larger number of variational parameters and greater expressibility. While \gls{ADAPT2}-\gls{VQE} stabilizes the energy inside the region of chemical accuracy after as few as five iterations, \gls{UCCSD}-\gls{VQE} is still in and out of this region by the time convergence is reached, which happens after more than double the iterations of \gls{ADAPT2}-\gls{VQE}. 

When a real quantum computer is used, and the algorithms are exposed to all types of noise, the difference becomes even sharper. The \gls{ADAPT2}-\gls{VQE} and \gls{UCCSD}-\gls{VQE} optimizations in the Belem processor will be compared at a later subsection both graphically and numerically.

\FloatBarrier
\subsection{Impact of Thermal Relaxation, SPAM errors, and Sampling}
\label{s:noise_types_adapt}

\begin{figure}[htbp]
     \centering
     \begin{subfigure}[b]{0.45\textwidth}
         \centering
         \includegraphics[width=\textwidth]{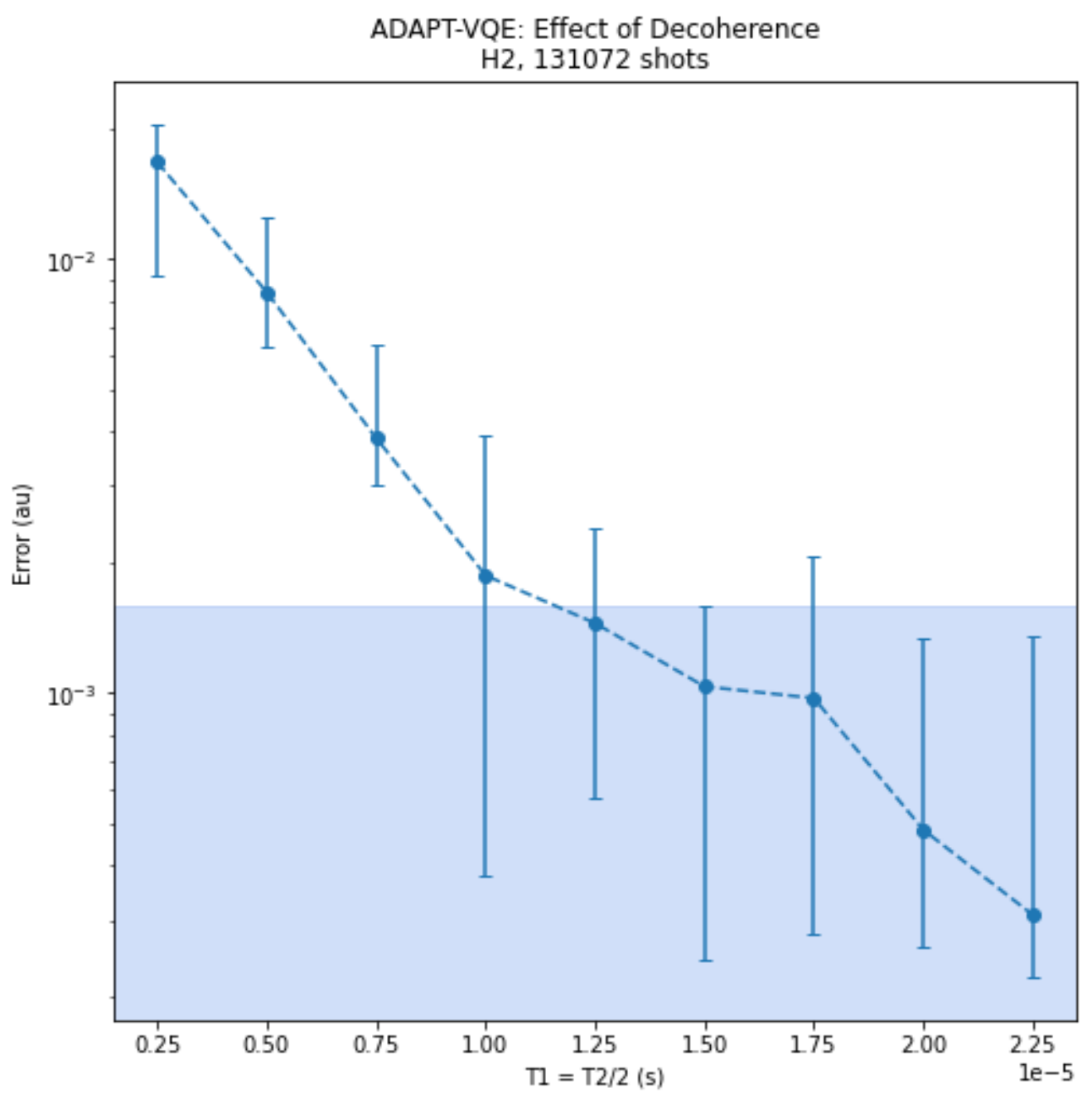}
         \caption{Thermal relaxation}
         \label{fig:h2_adapt_decoherence}
     \end{subfigure}
     \hfill
     \begin{subfigure}[b]{0.45\textwidth}
         \centering
         \includegraphics[width=\textwidth]{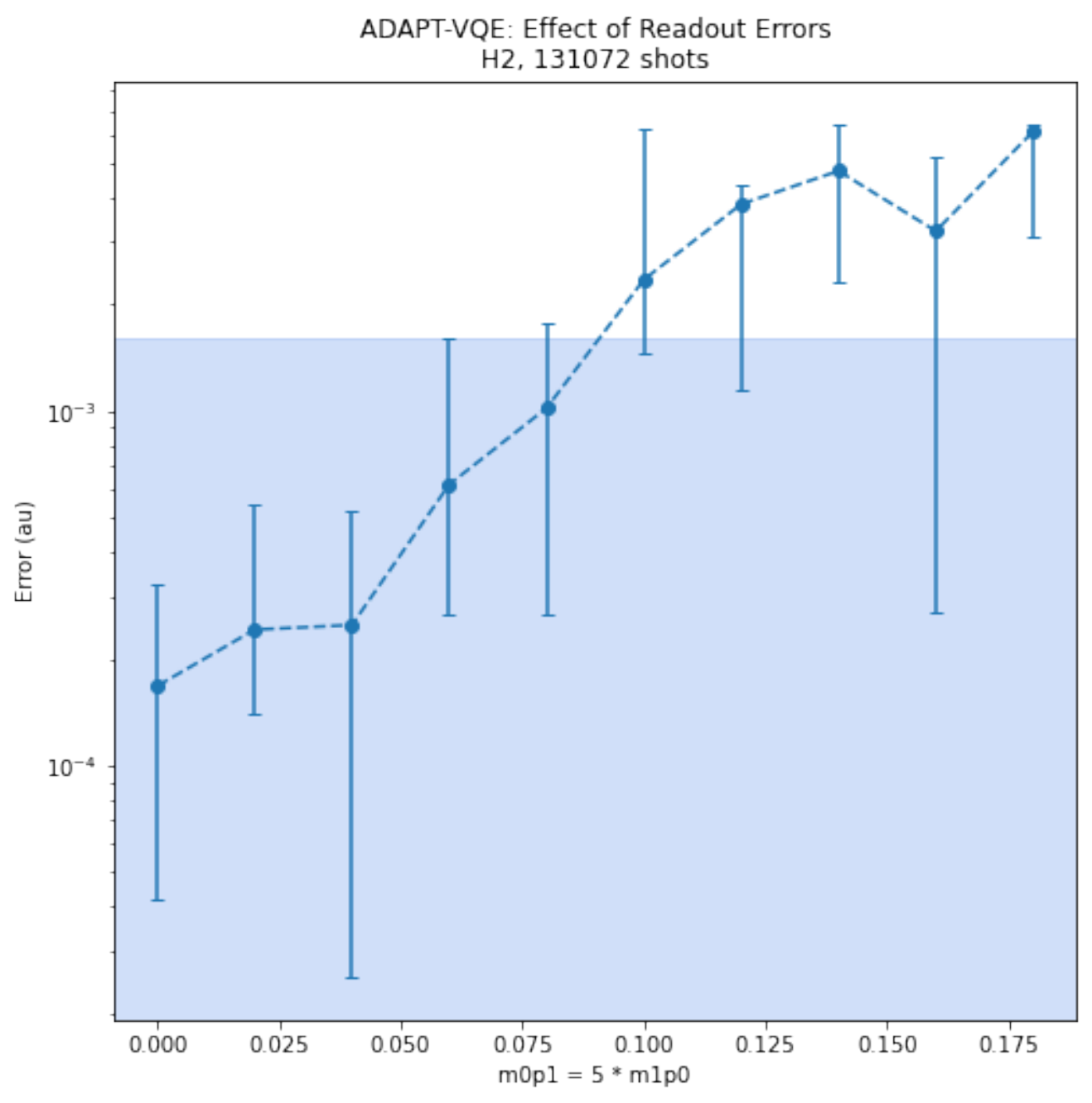}
         \caption{\gls{SPAM} errors}
         \label{fig:h2_adapt_readout}
     \end{subfigure}
     \\
     \begin{subfigure}[b]{0.45\textwidth}
         \centering
         \includegraphics[width=\textwidth]{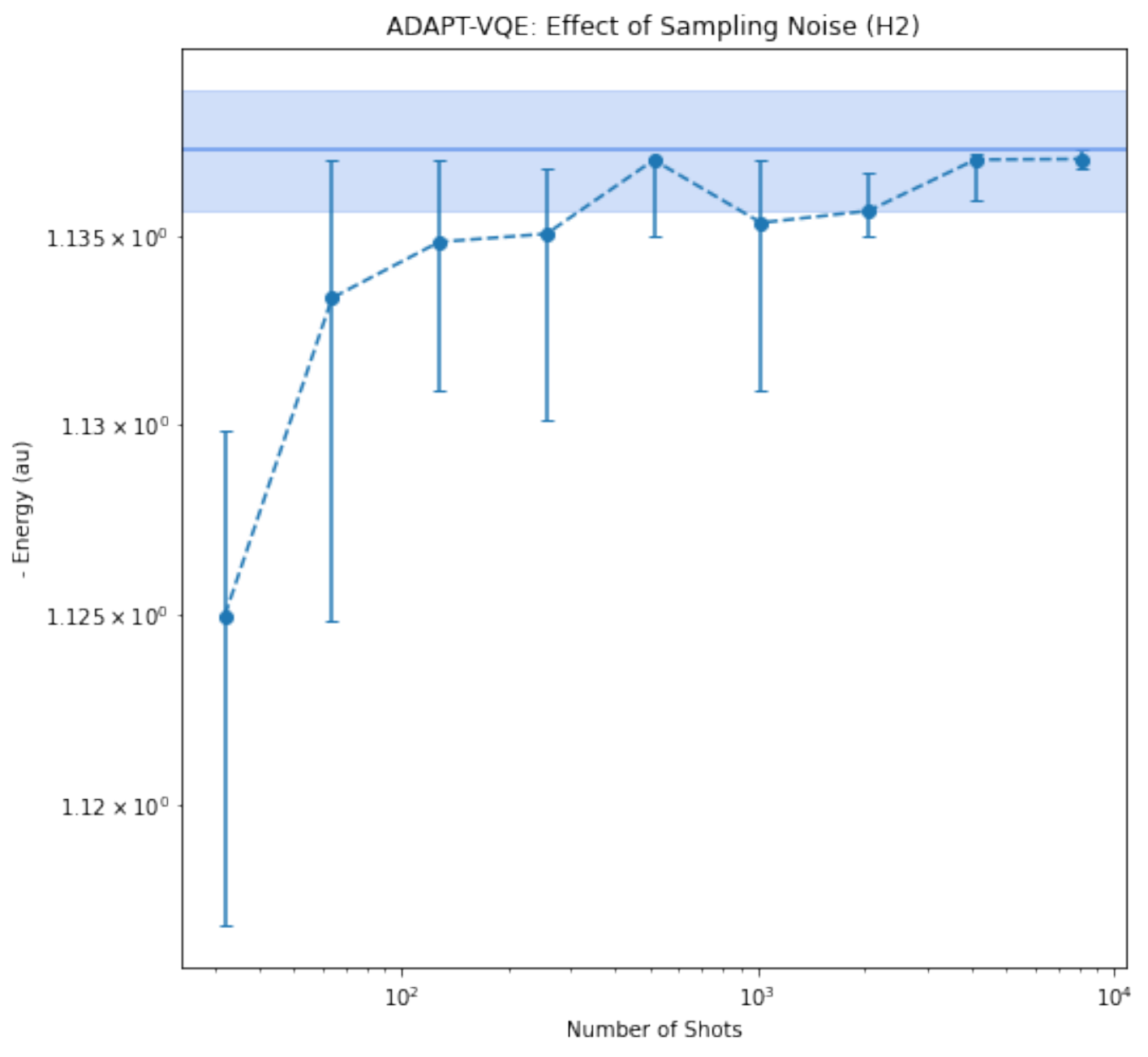}
         \caption{Sampling noise}
         \label{fig:h2_adapt_sampling}
     \end{subfigure}
     \caption{Plots showing the impact of different types of noise on the performance of \gls{ADAPT2}-\gls{VQE} for $H_2$ at an interatomic distance of 0.74Å. For thermal relaxation, a simple model assuming average gate times and a constant ratio between the coherence times T1 and T2 was used. Figure \ref{fig:h2_adapt_decoherence} shows the evolution of the error in the energy as a function of these times. For \gls{SPAM} noise, the ratio between the probability of measuring 0 after preparing 1 and  the probability of measuring 1 after preparing 0 was fixed (and greater than one, for the model to be realistic). Figure \ref{fig:h2_adapt_readout} shows the evolution of the error as a function of these probabilities. Figure \ref{fig:h2_adapt_sampling} illustrates the impact of sampling noise by plotting the error as a function of the number of shots. Twenty runs were used for each point; the median was used as the estimator (as it filters out aberrant runs), and the error bars show the interquartile ranges. All noise models were created using Qiskit \cite{Qiskit}. The chosen optimizer was \gls{COBYLA}.}
     \label{fig:h2_adapt_noisemodels}
\end{figure}

Figure \ref{fig:h2_adapt_noisemodels} shows the impact of thermal relaxation, \gls{SPAM} errors, and sampling noise in \gls{ADAPT2}-\gls{VQE}. The circumstances were chosen to meet exactly those from figure \ref{fig:h2_uccsd_noisemodels}, that show the same plots from \gls{UCCSD}-\gls{VQE}. The same models of thermal relaxation, \gls{SPAM} noise, and sampling noise, created in Qiskit \cite{Qiskit}, were used.

From \ref{fig:h2_adapt_decoherence}, we can see that the majority of the runs fall inside of chemical accuracy when T1 is greater than $2\times10^{-5}$ seconds (with T2 being fixed at $4\times10^{-5}$s). This condition is actually met by IBMQ's devices, that typically offer coherence times of the order of $10^{-4}$ seconds.

The impact of \gls{SPAM} errors can be seen on figure \ref{fig:h2_adapt_readout}. For the runs within the interquartile range to fall inside of chemical accuracy, the probability of measuring $\ket{0}$ after preparing $\ket{1}$ should be 0.06 (with the probability of measuring $\ket{1}$ after preparing $\ket{0}$, made proportional, at 0.012). The Belem backend used to run \gls{ADAPT2}-\gls{VQE} before nearly meets the requirement, with average error probabilities around 0.04 and 0.009, respectively.

Finally, figure \ref{fig:h2_adapt_sampling} shows that \gls{ADAPT2}-\gls{VQE} requires no more than 4000 shots to reach chemical accuracy in the middle 50\% runs in a sampling noise only scenario. This shot count is actually under the maximum allowed in IBMQ's cloud quantum computers.

\FloatBarrier
\subsection{Comparison with UCCSD-VQE}

While the previous \gls{ADAPT2}-\gls{VQE} results were already compared against those from \gls{UCCSD}-\gls{VQE} presented on subsection \ref{ss:noise_types_uccsd}, it will be interesting to compare them side-by-side. 

\begin{figure}[htbp]
     \centering
     \begin{subfigure}[b]{0.45\textwidth}
         \centering
         \includegraphics[width=\textwidth]{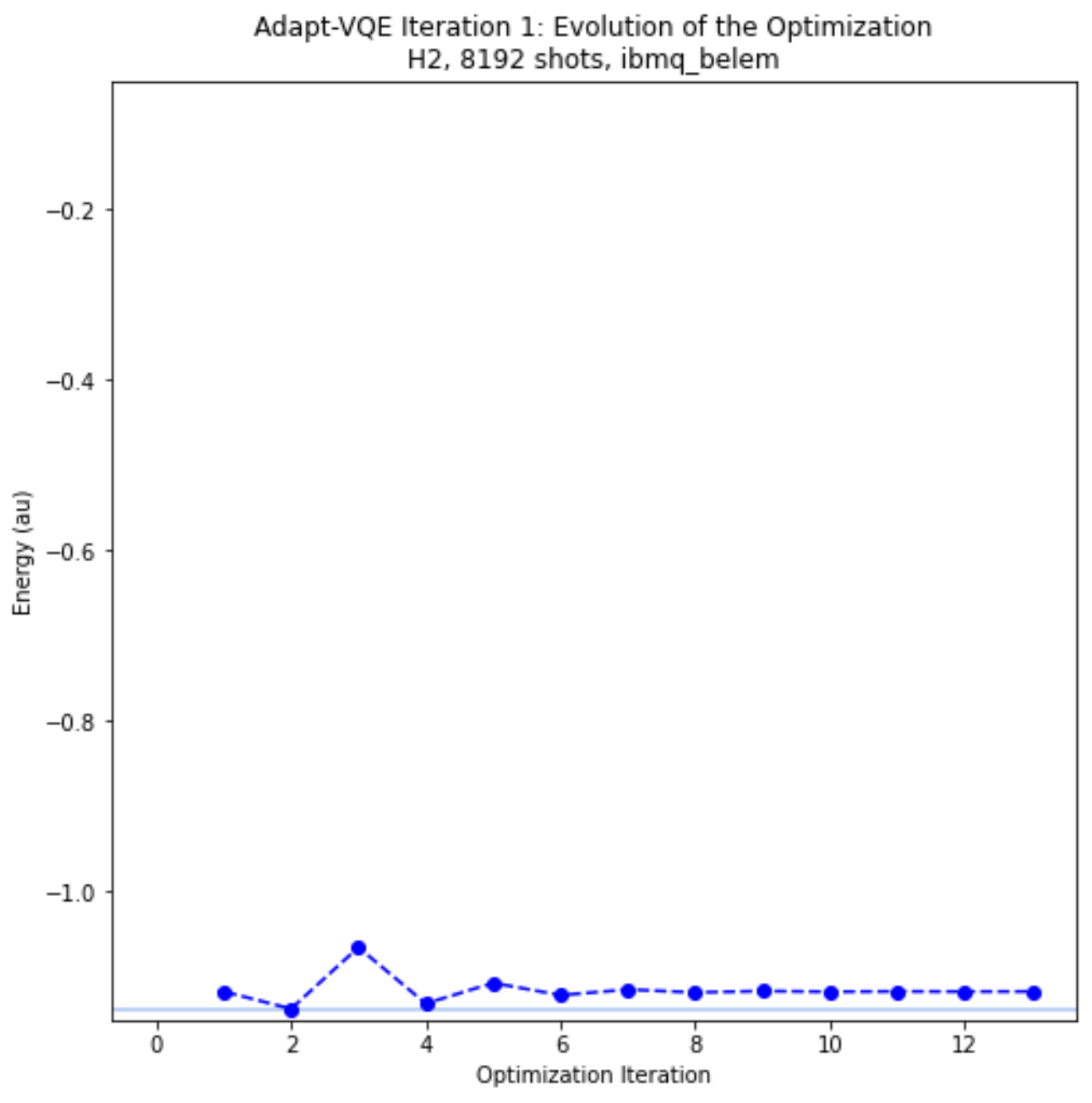}
         \caption{\gls{ADAPT2}-\gls{VQE}}
         \label{fig:h2_uccsdvsadapt_belem8192_adapt}
     \end{subfigure}
     \hfill
     \begin{subfigure}[b]{0.45\textwidth}
         \centering
         \includegraphics[width=\textwidth]{h2_uccsd_belem_8192.pdf}
         \caption{\gls{UCCSD}-\gls{VQE}}
         \label{fig:h2_uccsdvsadapt_belem8192_uccsd}
     \end{subfigure}
     \caption{Evolution of the optimization of the first iteration of \gls{ADAPT2}-\gls{VQE} (figure \ref{fig:h2_uccsdvsadapt_belem8192_adapt}) and of the \gls{UCCSD}-\gls{VQE} algorithm, for $H_2$ at an interatomic distance of 0.74Å. IBMQ's Belem processor was used in both cases.}
     \label{fig:h2_uccsdvsadapt_belem8192}
\end{figure}

Figure \ref{fig:h2_uccsdvsadapt_belem8192} compares the \gls{ADAPT2}-\gls{VQE} (figure \ref{fig:h2_uccsdvsadapt_belem8192_adapt}) and \gls{UCCSD}-\gls{VQE} (figure \ref{fig:h2_uccsdvsadapt_belem8192_uccsd}) optimizations, using the same scale.

Neither of the algorithms manages to reach chemical accuracy with this backend and shot count, at least in this particular iteration. Regardless, \gls{ADAPT2}-\gls{VQE} secures the energy close to the Hartree-Fock energy, with a final error of the order of centesimals of Hartree. In contrast, the \gls{UCCSD}-\gls{VQE} outputs an energy with an error greater than unity. The appearance of the \gls{UCCSD}-\gls{VQE} optimization plot is a symptom of a \textit{noise-induced barren plateau}: due to the excessive circuit depth, noise washes away the quantum information and erases the characteristics of the cost function, causing the optimizer to vainly search through a flat optimization landscape.

When transpiled onto the Belem backend, the single-iteration \gls{ADAPT2}-\gls{VQE} ansatz has 12 \glspl{CNOT}, against 130 from \gls{UCCSD}-\gls{VQE}. This evidently impacts the purity of the state by the end of the circuit.

As was done in the previous chapter, an approximate density matrix representing the state at the end of the noisy \gls{ADAPT2}-\gls{VQE} ansatz was obtained by simulating the algorithm on the QASM simulator with a noise model aiming to mimic the behaviour of the Belem processor.

The fidelity of the final state with the ideal state at the end of the Adapt ansatz was found to be 0.92 (against only 0.081 in \gls{UCCSD}-\gls{VQE}), while the fidelity with the fully mixed state was 0.087 (against 0.40). The final purity of the state was found to be around 0.84 (against 0.16). 

\begin{figure}[htbp]
     \centering
     \begin{subfigure}[b]{0.45\textwidth}
         \centering
         \includegraphics[width=\textwidth]{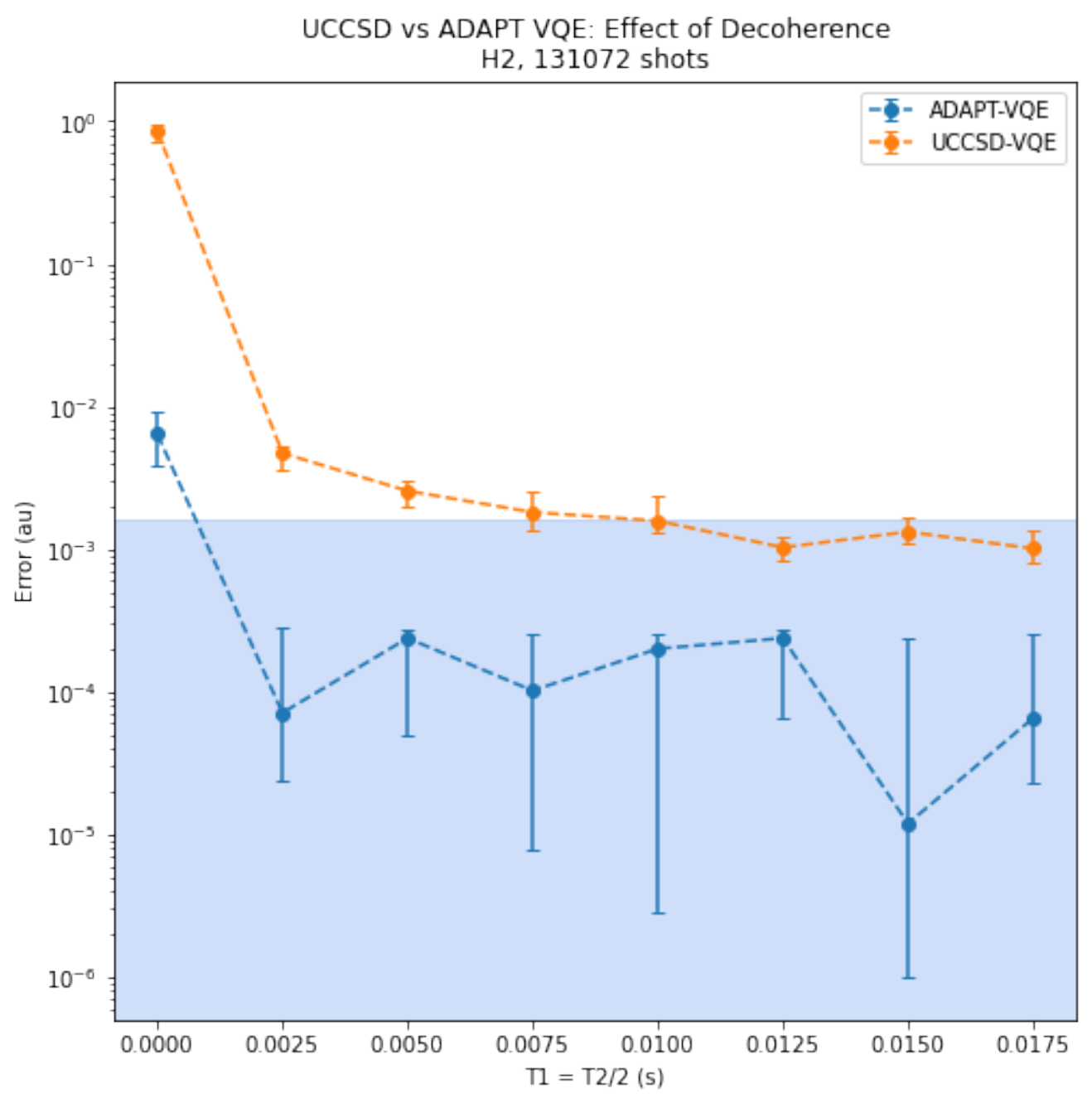}
         \caption{Thermal relaxation}
         \label{fig:h2_uccsdvsadapt_deco}
     \end{subfigure}
     \hfill
     \begin{subfigure}[b]{0.45\textwidth}
         \centering
         \includegraphics[width=\textwidth]{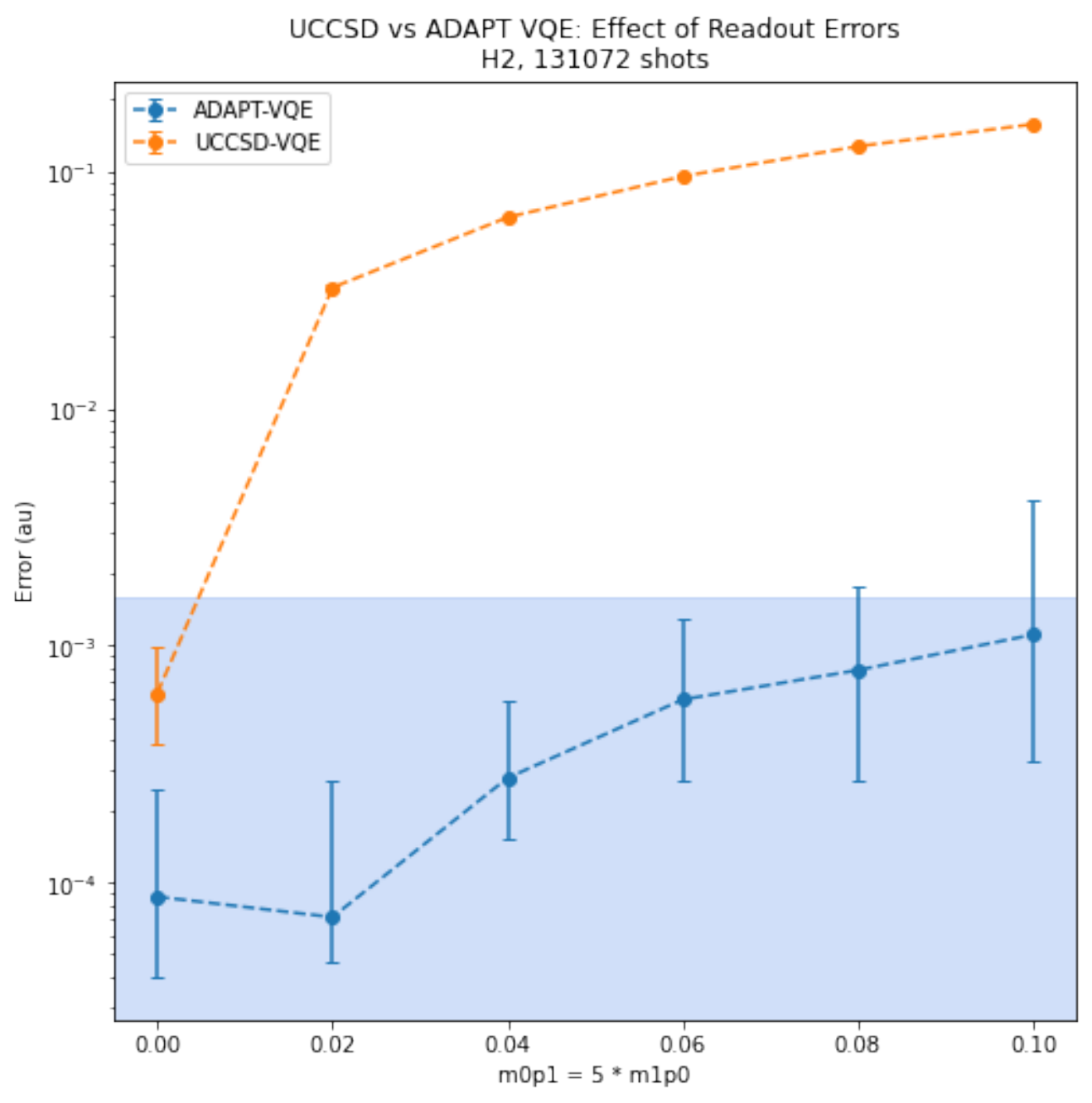}
         \caption{\gls{SPAM} errors}
         \label{fig:h2_uccsdvsadapt_readout}
     \end{subfigure}
     \\
     \begin{subfigure}[b]{0.45\textwidth}
         \centering
         \includegraphics[width=\textwidth]{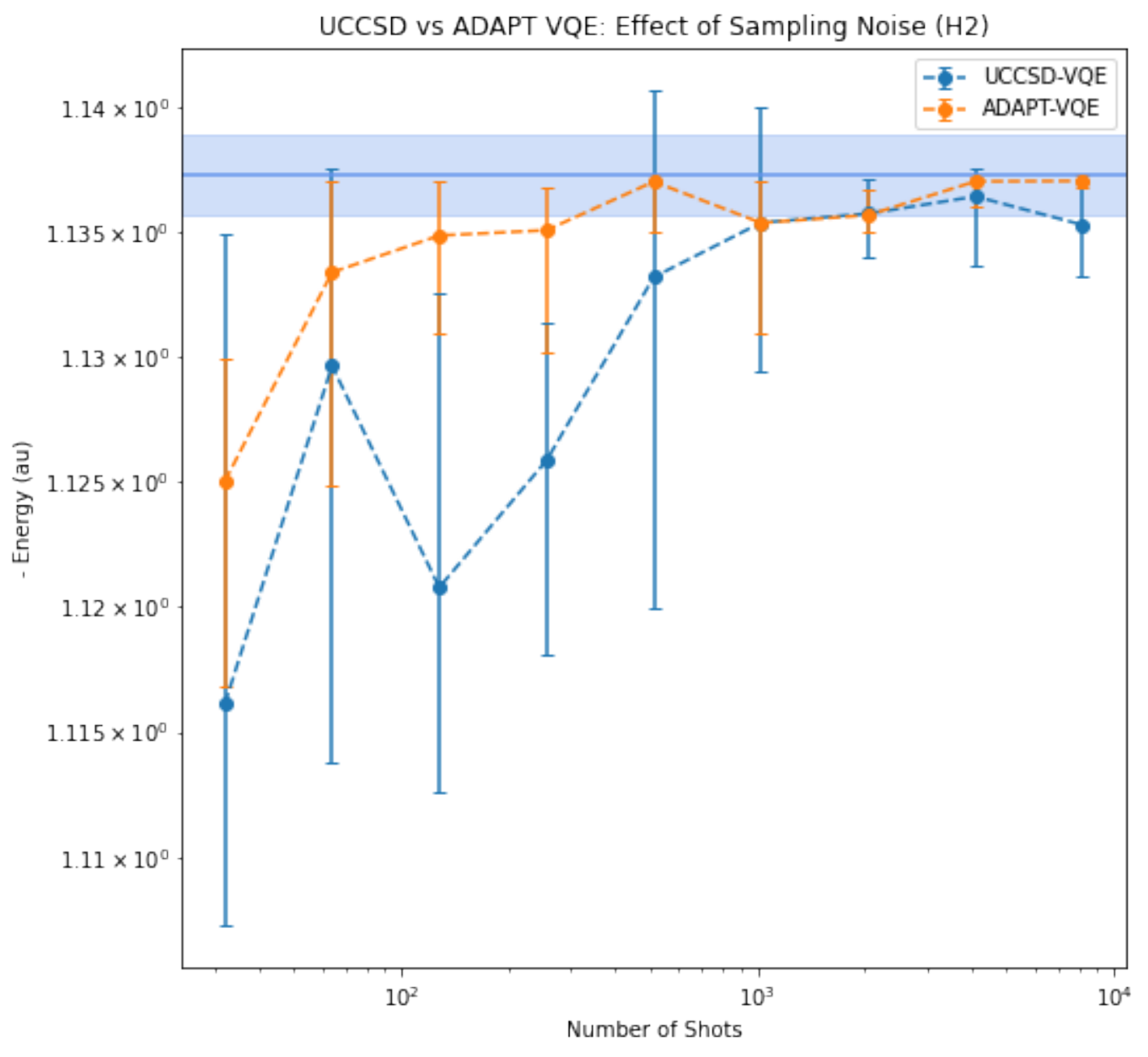}
         \caption{Sampling noise}
         \label{fig:h2_uccsdvsadapt_sampling}
     \end{subfigure}
     \caption{Comparison of the effect of different types of noise on \gls{ADAPT2}-\gls{VQE} and \gls{UCCSD}-\gls{VQE} for $H_2$ at an interatomic distance of 0.74Å. The same noise models described in figures \ref{fig:h2_uccsd_noisemodels} and \ref{fig:h2_adapt_noisemodels} were used.}
     \label{fig:h2_uccsdvsadapt_noisemodels}
\end{figure}

Figure \ref{fig:h2_uccsdvsadapt_noisemodels} compares the noise resilience of \gls{ADAPT2}-\gls{VQE} and \gls{UCCSD}-\gls{VQE} against different sources (thermal relaxation, \gls{SPAM} errors, and sampling). The same noise sources were already explored individually in \ref{fig:h2_uccsd_noisemodels} and \ref{fig:h2_adapt_noisemodels}; however, now the scales were adjusted so that both curves start when \gls{ADAPT2}-\gls{VQE} is still outside of chemical accuracy and finish when \gls{UCCSD}-\gls{VQE} is already inside.

To reach chemical accuracy in more than 50\% of the runs, \gls{UCCSD}-\gls{VQE} required roughly 850 times greater coherence times than \gls{ADAPT2}-\gls{VQE}. The difference in performance is visible in the plot \ref{fig:h2_uccsdvsadapt_deco}, with \gls{ADAPT2}-\gls{VQE} showing around one order of magnitude greater accuracy for any given value of T1 and T2. This is a consequence of the greater circuit depth required by the \gls{UCCSD} ansatz.

Figure \ref{fig:h2_uccsdvsadapt_sampling} shows the impact of sampling noise, and figure \ref{fig:h2_uccsdvsadapt_readout} the impact of \gls{SPAM} noise. In order to reach chemical accuracy in at least half of the runs in the presence of sampling noise exclusively, \gls{ADAPT2}-\gls{VQE} required only 5\% of the number of shots required by \gls{UCCSD}-\gls{VQE}. \gls{ADAPT2}-\gls{VQE} also tolerated roughly 150 times higher error probability in state preparation and measurements. It is interesting to see how both of these noise sources have greater impact in the performance of \gls{UCCSD}-\gls{VQE}, regardless of being independent of circuit depth. Once again it becomes clear that more variational parameters and more variational flexibility also imply an added difficulty in the optimization, especially in the presence of noise. 

The noise resilience of \gls{ADAPT2}-\gls{VQE} against \gls{UCCSD}-\gls{VQE} has become evident, but it must be noted that this is not the only advantage of \gls{ADAPT2}-\gls{VQE} against \gls{UCCSD}-\gls{VQE}: in the original article \cite{Grimsley2019}, it was shown that while the latter often fails to reach chemical accuracy for strongly correlated molecules, \gls{ADAPT2}-\gls{VQE} manages to reach it as long as the convergence criterion is sufficiently ambitious. These simulations were not replicated due to the heavy computational demands involved.

\FloatBarrier

While the results regarding accuracy and noise-resilience favour the latter, comparing the performance of \gls{UCCSD}-\gls{VQE} and \gls{ADAPT2}-\gls{VQE} is a delicate matter that goes beyond the impact of noise or the final error in the energy. There are many different costs at play, and many links between them. \gls{ADAPT2}-\gls{VQE} requires evaluating the expectation value of a significantly larger number of observables, because all gradients must be evaluated each iteration. Additionally, because each iteration requires re-optimizing all the parameters, there will be a significant accumulated number of optimizations by the end of the algorithm, unlike in \gls{UCCSD}-\gls{VQE} (which requires a single optimization). This makes it seem likely that \gls{ADAPT2}-\gls{VQE} will imply a larger number of measurements in total. 

However, attention must be paid to the fact that since \gls{ADAPT2}-\gls{VQE} creates the ansatz from scratch, most of the optimizations will have a significantly lower number of variational parameters. Because they're lower dimensional, these optimizations are likely to be easier, and to require a lower number of function evaluations (i.e. calls to the quantum computer). Further, in the common case that the final \gls{ADAPT2}-\gls{VQE} ansatz includes less variational parameters than the \gls{UCCSD} ansatz, \gls{ADAPT2}-\gls{VQE} will tolerate sampling noise better, and thus require a lower number of shots per term for the same accuracy. Additionally, the measurement overhead from the gradient measurements may be softened by the existence of common factors among the Pauli strings that occur in the $[\hat{H}, \hat{A}_i]$ observables (and in the Hamiltonian itself). This allows reusing measurements.

\begin{figure}[htbp]
     \centering
     \begin{subfigure}[b]{0.45\textwidth}
         \centering
         \includegraphics[width=\textwidth]{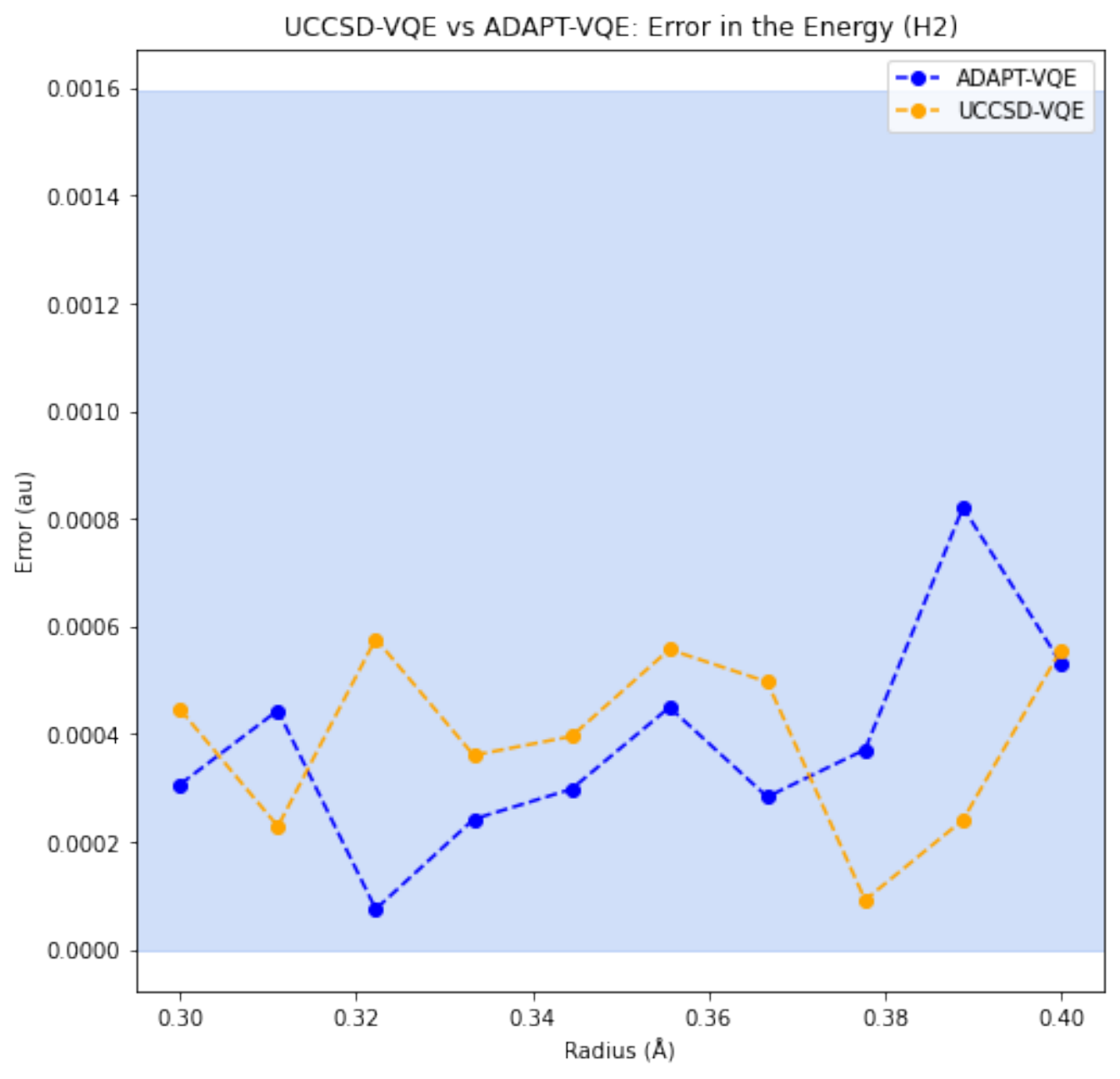}
         \caption{Error in the energy}
         \label{fig:h2_uccsd_vs_adapt_error}
     \end{subfigure}
     \hfill
     \begin{subfigure}[b]{0.45\textwidth}
         \centering
         \includegraphics[width=\textwidth]{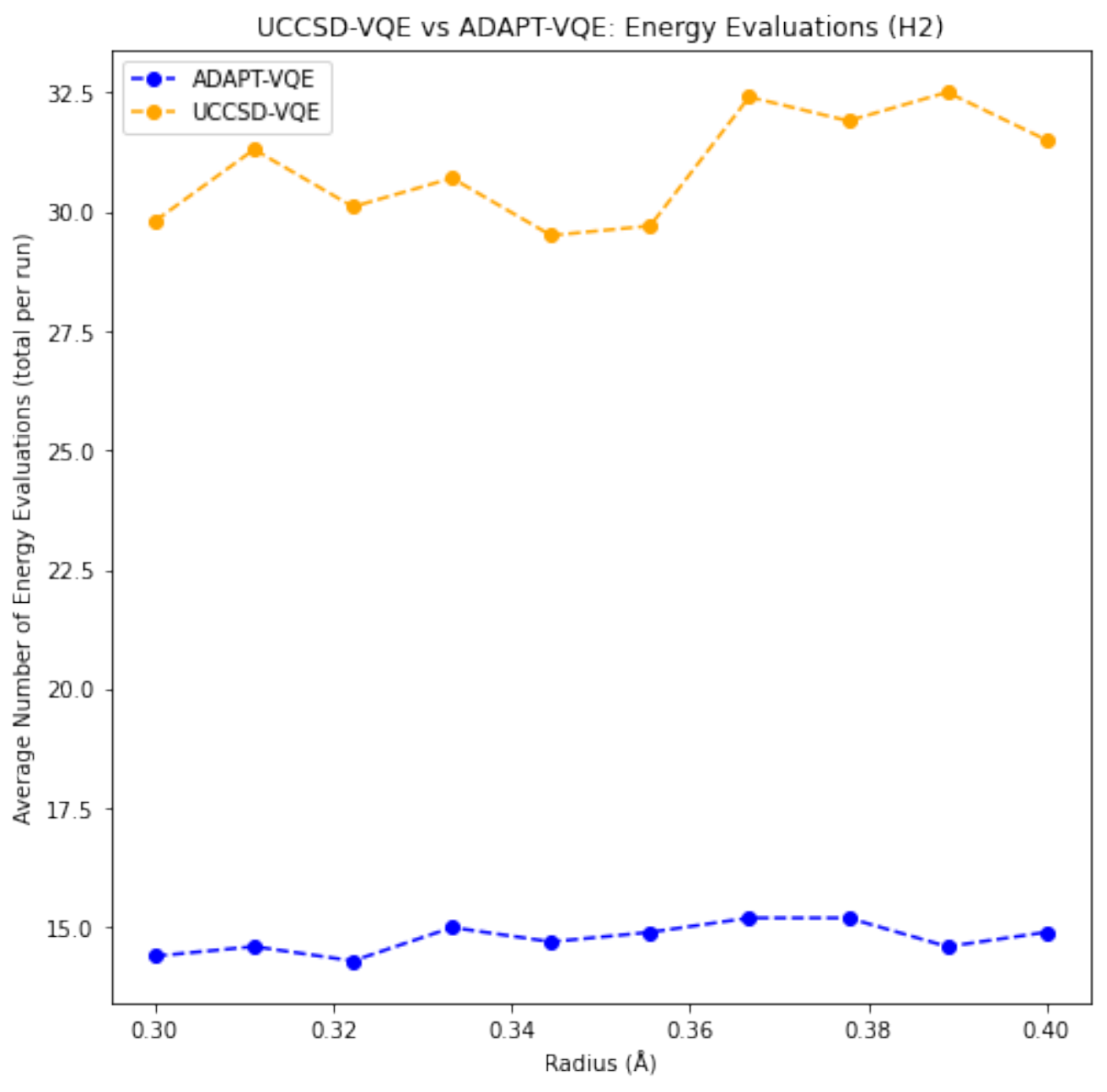}
         \caption{Number of energy evaluations}
         \label{fig:h2_uccsd_vs_adapt_nfev}
     \end{subfigure}
     \\
     \begin{subfigure}[b]{0.45\textwidth}
         \centering
         \includegraphics[width=\textwidth]{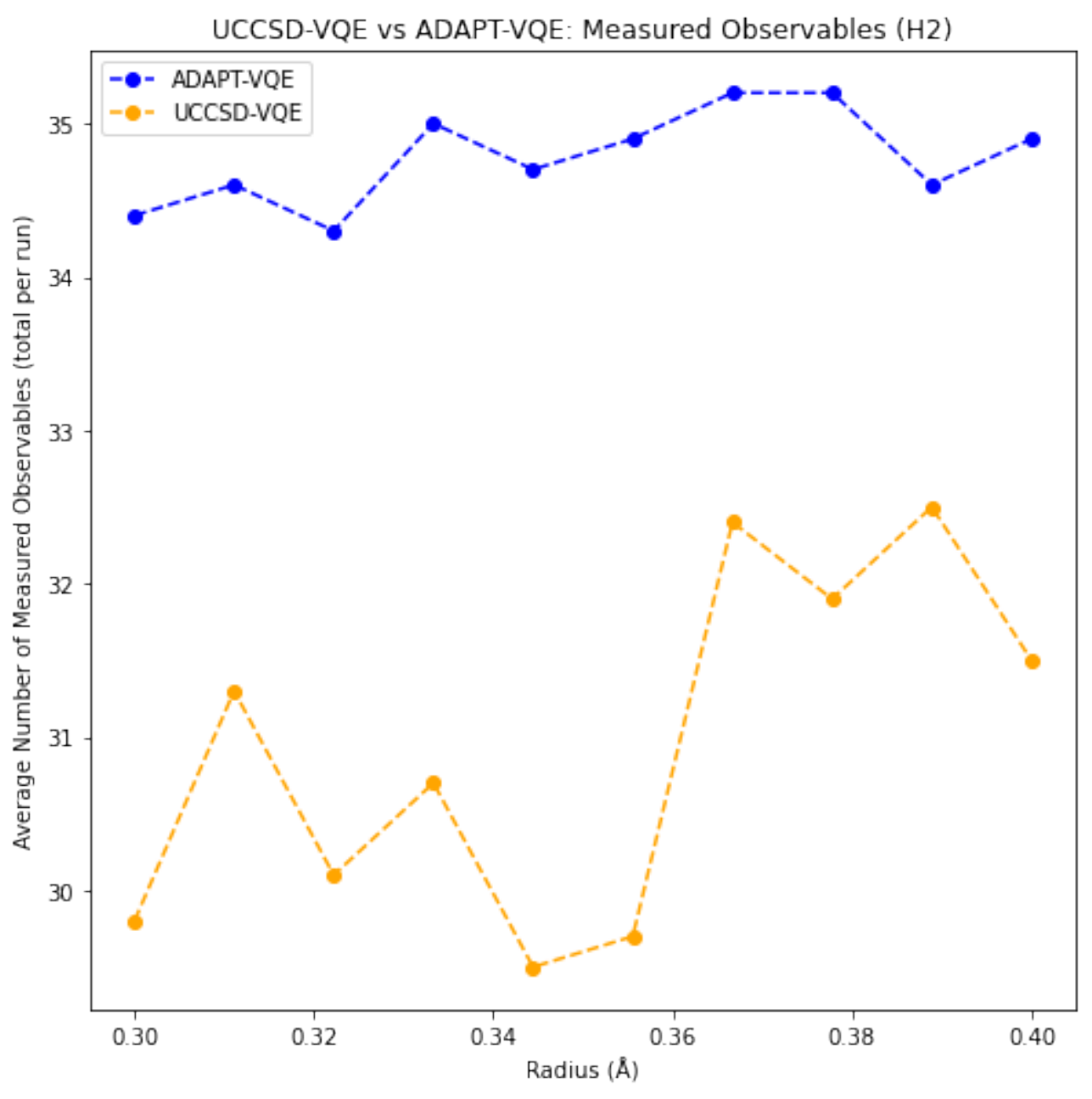}
         \caption{Number of measured observables}
         \label{fig:h2_uccsd_vs_adapt_observables}
     \end{subfigure}
     \hfill
     \begin{subfigure}[b]{0.45\textwidth}
         \centering
         \includegraphics[width=\textwidth]{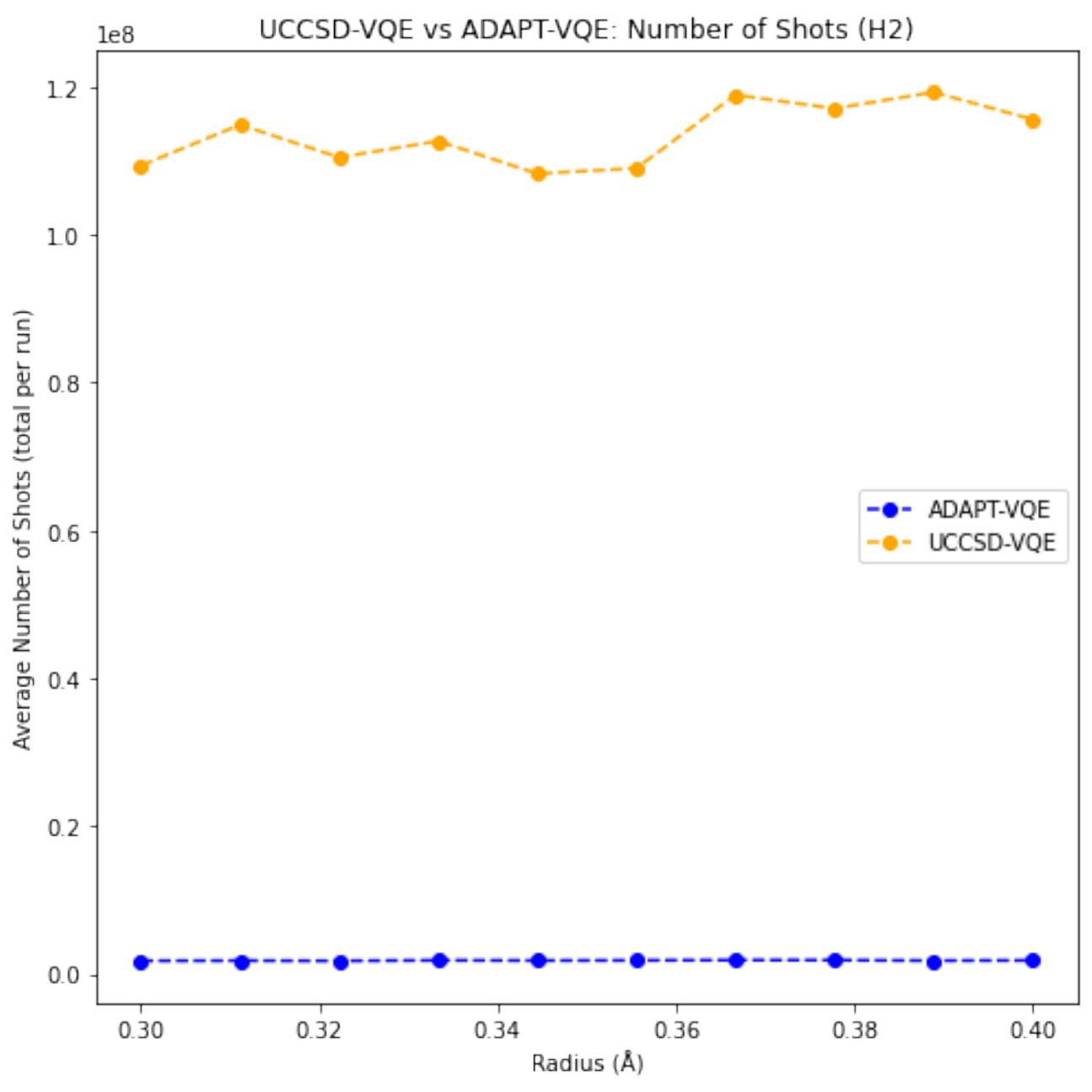}
         \caption{Total number of shots}
         \label{fig:h2_uccsd_vs_adapt_shots}
     \end{subfigure}
     \caption{Plots comparing the costs associated with \gls{UCCSD}-\gls{VQE} and \gls{ADAPT2}-\gls{VQE} for a similar final error, in a setting with no noise other than sampling noise. For each interatomic distance, 10 runs of the algorithms were performed. The median of the errors is plotted in figure \ref{fig:h2_uccsd_vs_adapt_error}. Figures \ref{fig:h2_uccsd_vs_adapt_nfev}, \ref{fig:h2_uccsd_vs_adapt_observables}, and \ref{fig:h2_uccsd_vs_adapt_shots} present respectively the average number of energy evaluations, measured observables, and total shots per run. The number of shots per Pauli string was set to $2^{13}$ and $2^{18}$ for \gls{ADAPT2}-\gls{VQE} and \gls{UCCSD}-\gls{VQE} respectively. These values were chosen so that the error was roughly matched between the two algorithms (the average error was 0.00038 a.u. and 0.00039 a.u. by the same order).}
     \label{fig:h2_uccsd_vs_adapt_costs}
\end{figure}

Figure \ref{fig:h2_uccsd_vs_adapt_costs} aims to carefully analyse all of the different factors that weigh in on the costs. The number of shots for each algorithm was chosen so that the precision was matched; this implied a significantly larger number of shots for \gls{UCCSD}-\gls{VQE} (32 times larger, in this case). As explained before, this is due to the fact that the optimization is higher-dimensional, thus more difficult, and more sensitive to noise.

Figure \ref{fig:h2_uccsd_vs_adapt_nfev} compares the number of energy evaluations throughout the optimizations. This number is larger for \gls{UCCSD}-\gls{VQE}; once more, we can attribute this to the larger number of variational parameters in the optimization. However, this plot does not consider the fact that the number of \textit{observables} measured in \gls{ADAPT2}-\gls{VQE} is larger, because the gradients of the pool operators will also be measured in each iteration.

The additional measurements required for the evaluation of the gradients in \gls{ADAPT2}-\gls{VQE} are factored in in figure \ref{fig:h2_uccsd_vs_adapt_observables}. Here, the number of observables includes \textit{all} observables measured along a complete run of either algorithm. For \gls{ADAPT2}-\gls{VQE}, this includes all energy and gradient measurements; for \gls{UCCSD}-\gls{VQE}, it matches the number of energy evaluations. In both cases, the utilized optimizer was gradient-free, so that all calls to the quantum computer throughout the optimization consisted of energy evaluations. As expected, the total number of measured observables was larger for \gls{ADAPT2}-\gls{VQE}. In this case, the pool had 20 operators, implying an additional 20 observables to be measured each iteration (the $[\hat{H}, \hat{A}_i]$).

Finally, figure \ref{fig:h2_uccsd_vs_adapt_shots} compares a very important metric: the total number of shots in a full run. This number includes all the shots used for the evaluation of \textit{all} Pauli strings in \textit{all} observables in \textit{all} times they were measured. This plot thus reflects a multitude of factors: it weighs in the fact that \gls{UCCSD}-\gls{VQE} requires more shots per string and more energy evaluations per optimization, as well as the fact that \gls{ADAPT2}-\gls{VQE} requires additional measurements per iteration to obtain the gradients. The total number of shots is the total number of circuits, repeated or not, that have to be executed on the quantum computer. Interestingly, \gls{ADAPT2}-\gls{VQE} required a \textit{smaller} number of shots in total than \gls{UCCSD}-\gls{VQE} for the same final error. In addition to implying shallower circuits and a lower number of variational parameters, \gls{ADAPT2}-\gls{VQE} does not in this case imply an overhead in measurements as compared to \gls{UCCSD}-\gls{VQE} - on the contrary. It required only a fraction of the number of shots in total (1.7\%).

In this example, the fact that sampling noise is better tolerated by \gls{ADAPT2}-\gls{VQE} allows for a reduction of the number of shots \textit{per Pauli string} that compensates for the fact that there will be a larger number of observables being measured. Additionally, it must be mentioned that as was conjectured before, there was a significant amount of repeated Pauli strings among these observables. For the molecule in question ($H_2$), the total number of Pauli strings in the $[\hat{H}, \hat{A}_i]$ observables was 152. However, the set of Pauli strings that were \textit{unique} among these and that didn't occur in the Hamiltonian, i.e. the Pauli strings that should actually be measured, was found to consist of only 24 elements. Evidently, acknowledging this fact allows for a reduction of the total number of measured Pauli strings, which in turn reduces the total number of shots. This was already considered in obtaining the results plotted in figure \ref{fig:h2_uccsd_vs_adapt_shots}.

Of course, these results should be interpreted with caution: we're only considering a particular molecule, and a very small one. As the size of the system grows, so will the number of operators in the pool, increasing the measurement overhead of \gls{ADAPT2}-\gls{VQE}. The total number of optimizations required for the same accuracy will grow as well. On the other hand, the difficulty that is brought on by the extra variational parameters in \gls{UCCSD}-\gls{VQE} will also increase with the size of the system. This algorithm will then become even more sensitive to sampling noise, and thus require an even larger precision on the energy evaluations (which will already require a larger number of shots for the \textit{same} precision given that the system is bigger). Additionally, it has been established that \gls{ADAPT2}-\gls{VQE} is better suited for real quantum processors, as the resulting circuits are shallower. The most evident benefit of this is that it improves the viability of the algorithm in near-term devices. But given that additional sources of noise also imply an additional difficulty in the optimizations, their presence is also likely to benefit the relative performance of \gls{ADAPT2}-\gls{VQE} against \gls{UCCSD}-\gls{VQE} in what comes to measurement costs. With more qubits and deeper circuits, errors will accumulate faster, so this is another factor that is affected by the size of the system.

Taking all of this into consideration, it is not certain how the two algorithms will compare in terms of the total number of required shots for larger molecules and in more realistic settings. Such simulations were not attempted on account of the prohibitive computational complexity.

\FloatBarrier
\section{Application to H4}

The $H_4$ molecule is represented by 8 qubits in a minimal basis. While it is already too large a system to allow for meaningful results to be recovered in real quantum computers, it is still is significantly easier to simulate classically than $LiH$. Because it allows for more iterations to be simulated in useful time, $H_4$ was used to analyse the effect of removing the Jordan-Wigner string from the Qubit Pool operators, and to compare fermionic-\gls{ADAPT}-\gls{VQE} against Qubit-\gls{ADAPT2}-\gls{VQE}.

\subsection{Qubit Pool: Effect of the Jordan Wigner String}

\begin{figure}[htbp]
    \centering
    \includegraphics[width=0.8\textwidth]{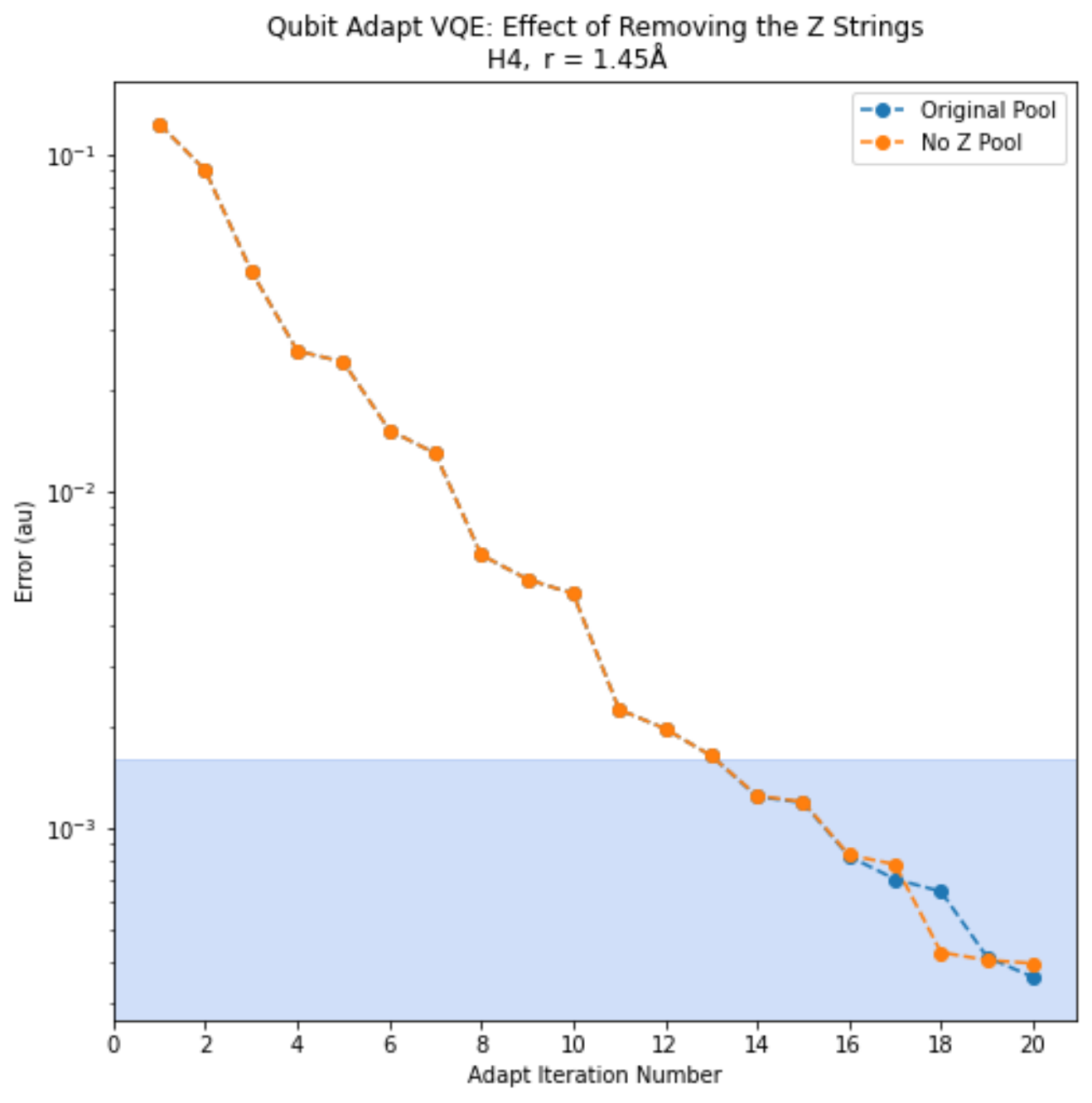}
    \caption{Impact of the Jordan-Wigner string on the convergence of the qubit-\gls{ADAPT2}-\gls{VQE} algorithm. The graphs compare the evolution of the error in the energy along the 20 first iterations of the algorithm, with and without the Z-strings in the pool operators. The molecule in study is the $H_4$ molecule at an interatomic distance of 1.45Å.}
    \label{fig:h4_noZ}
\end{figure}

In one of the Qubit Pools introduced before, the chain of Pauli Z operators responsible for faithful translation of the anticommutation of fermions into the qubit states was eliminated from these Pauli strings. The effect of its removal on the convergence of the algorithm for the $H_4$ molecule is plotted on figure \ref{fig:h4_noZ}. 

The plot shows that there is very little variation between qubit pools with and without the Jordan-Wigner string. This is a remarkable fact. Firstly, it is remarkable because it means that representing the fermionic anticommutation accurately through the ansatz operators brings no benefit. But most of all, it is remarkable because it greatly aids in reducing circuit depth, as was explained before. These Z strings are non-local and act on a number of qubits that scales on average linearly with the size of the system. After relinquishing them, only the qubits corresponding to the one-body wave functions directly involved in the excitations will be acted on by the operators. As a consequence, if the employed excitations are at most doubles, the pool operators will act on at most four qubits regardless of the molecule in study. We saw that the circuit depth required per operator will be $\mathcal{O}(1)$ instead of $\mathcal{O}(N)$.

\subsection{Fermionic vs Qubit Pools: Performance Comparison}
The purpose of the qubit pool is to make the \gls{ADAPT2}-\gls{VQE} algorithm more \gls{NISQ}-Friendly, in hopes of enabling an earlier practical implementation. As it was explained in the previous section, the operators in the qubit pool are in fact implemented by shallower circuits than those in the fermionic pool. However, because they don't preserve fermionic symmetries and consist each of a single Pauli string, they are bound to make convergence slower as a function of the number of iterations. An $N$-operator ansatz from fermionic-\gls{ADAPT}-\gls{VQE} will typically imply a much greater accuracy than an $N$-operator ansatz from qubit-\gls{ADAPT2}-\gls{VQE}. However, it will also imply a much longer circuit. In a fair comparison, circuit depth, the number of variational parameters, and the measurement overhead should all be taken into consideration.
    
\begin{figure*}[htbp]
     \centering
     \begin{subfigure}[b]{0.45\textwidth}
         \centering
         \includegraphics[width=\textwidth]{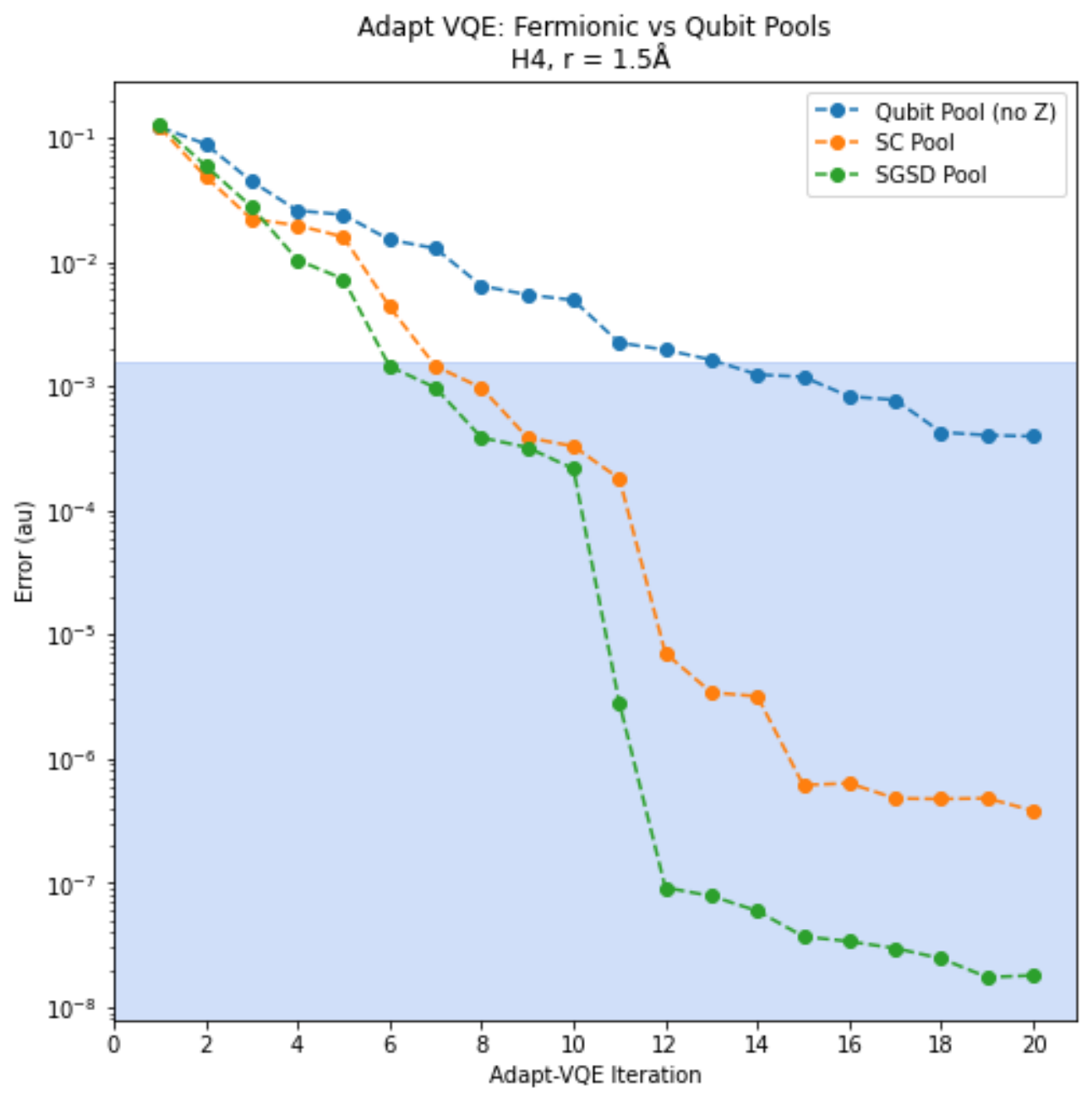}
         \caption{Error as a function of the number of iterations of the algorithm.}
         \label{fig:H4_fvsq_iteration}
     \end{subfigure}
     \hfill
     \begin{subfigure}[b]{0.45\textwidth}
         \centering
         \includegraphics[width=\textwidth]{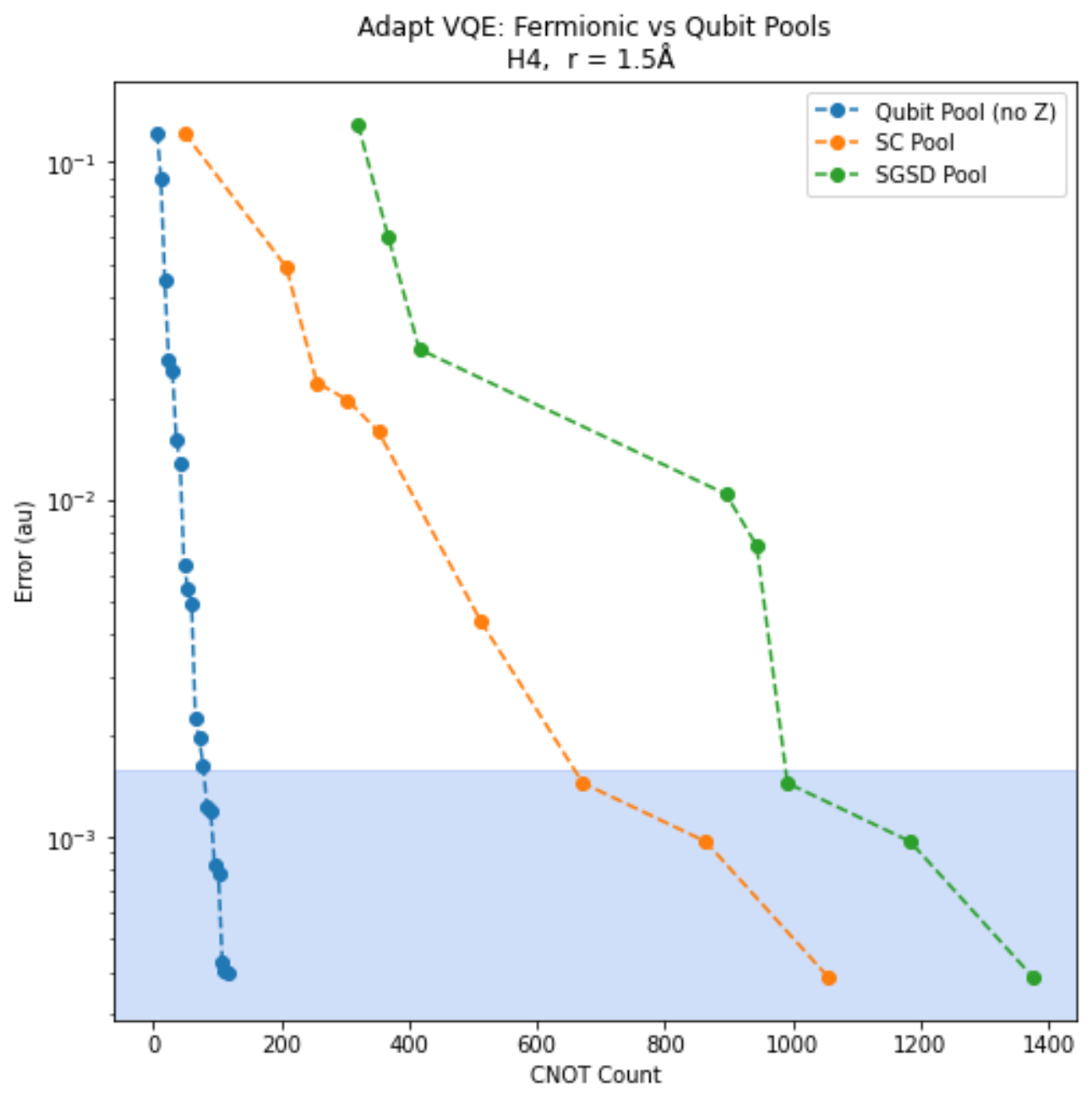}
         \caption{Error as a function of the number of \gls{CNOT} gates in the ansatz circuit.}
         \label{fig:H4_fvsq_cnots}
     \end{subfigure}
     \hfill
     \caption{Evolution of the error in the \gls{ADAPT}-\gls{VQE} energy along the evolution of the algorithm for the \gls{SCGSD} pool, the \gls{SGSD} pool, and the qubit pool. The molecule in study is the $H_4$ molecule.}
     \label{fig:f_vs_q}
\end{figure*}

Figure \ref{fig:f_vs_q} shows the evolution of the error in the \gls{ADAPT}-\gls{VQE} energy for three types of pools, two fermionic ones (\gls{SCGSD} and \gls{SGSD}) and the qubit pool. 

The energy as a function of the iteration number, which corresponds to the number of operators in the ansatz, is plotted in figure \ref{fig:H4_fvsq_iteration}. As expected, we observe that qubit-\gls{ADAPT2}-\gls{VQE} requires more iterations/operators to achieve the same accuracy as fermionic-\gls{ADAPT}-\gls{VQE}. 

In figure \ref{fig:H4_fvsq_cnots}, we can see that the tendency is reversed when we consider the error as a function of the number of \glspl{CNOT} in the circuit, to which the circuit depth will be proportional. Qubit-\gls{ADAPT2}-\gls{VQE} converges much faster than any of the fermionic pools in this case. Further, the molecule in study ($H_4$) is rather small: its representation in a minimal basis set requires only eight spin-orbitals (eight qubits). It was mentioned before that there is a difference in the scaling of the circuit depth necessary to implement qubit pool operators (constant) versus fermionic pool operators (linear). As such, the difference is bound to become even more notable with the increase of the system size (number of molecular orbitals). 

However, the error as a function of the number of \glspl{CNOT} presented in \ref{fig:H4_fvsq_cnots} (and, in general, the circuit depth) does not represent the full picture, because measurement and variational parameter overhead are not considered. Because the qubit pool consists of fermionic operators broken into their constituent Pauli strings (and cleared of the anticommutation operator), it is significantly larger. For example, for $H_4$, we already have 328 operators in the qubit pool, while \gls{SCGSD} and \gls{SGSD} (fermionic pools) have respectively 111 and 66 operators. Since in each iteration the gradients of all the pool operators must be measured, this represents a sizeable measurement overhead. Further, the scaling of the number of variational parameters is as in \ref{fig:H4_fvsq_iteration}, which means that qubit-\gls{ADAPT2}-\gls{VQE} will, for a given accuracy, imply a higher-dimensional (harder) optimization than fermionic-\gls{ADAPT}-\gls{VQE}. This implies more effort of the classical optimizer, which in turn will demand more calls to the quantum computer, resulting in even more measurements being required. 

In essence, Qubit-\gls{ADAPT2}-\gls{VQE} reduces the circuit depth necessary to reach a certain accuracy, at the expense of a greater burden on the classical optimizer and extra measurements on the quantum computer.

%% file: Chapters/chapter5.tex
\pagestyle{plain}
\graphicspath{{./Chapters/Figures/Ch5/}}

\chapter{ADAPT-VQE: Exploring other Pool Options}
\label{ch:adapt_other_pools}

In this chapter, \gls{ADAPT2}-\gls{VQE} will be tested with several pools. The pools will be introduced here and differ both from the ones used in fermionic-\gls{ADAPT}-\gls{VQE} \cite{Grimsley2019} and from the ones used in qubit-\gls{ADAPT2}-\gls{VQE} \cite{Tang2021}.

The purpose is to explore the importance of the similitude between pool operators and fermionic excitations, and investigate the origin of such importance, analysing whether it stems from preserving quantities such as particle number and spin, from respecting antisymmetry, or from other factors.

The constitution of the \gls{ADAPT2}-\gls{VQE} state in terms of Slater determinants, as well as the performance of the algorithm, will be analysed and compared between the different pools.

\section{Motivation}

The operators in the pools from fermionic-\gls{ADAPT}-\gls{VQE} are fermionic excitations that, when transformed into qubit operators, can consist of a superposition of up to 48 Pauli strings, the average length of each scaling linearly with the size of the system. These operators preserve particle number and spin, respect fermionic anticommutation, and properly represent fermionic excitations.

In qubit-\gls{ADAPT2}-\gls{VQE}, the operators are decomposed into the individual Pauli strings, and the Jordan-Wigner string is removed, causing their length to be independent of the system size. These operators do not preserve particle number or spin, do not respect fermionic anticommutation, and do not directly represent any proper fermionic excitation. 

In the previous chapter, we saw that while fermionic-\gls{ADAPT}-\gls{VQE} converges faster as a function of the iteration number, qubit-\gls{ADAPT2}-\gls{VQE} converges faster as a function of circuit depth. The qubit pools have more operators than the fermionic pools, which implies a higher overhead from measuring the gradients in each iteration; they will also typically require a larger number of optimizations and variational parameters for the same accuracy, demanding more effort from the classical optimizer and further increasing the number of calls to the quantum computer. In compensation for this, qubit pools reduce the circuit complexity of each operator, and result in shallower circuits for a given accuracy.

In the previous chapter, no options in-between the fermionic-inspired and qubit-inspired pools were tested. As well as spin-complemented and spin-adapted excitations, simple excitations can be used. These excitations can themselves be cleared of the Jordan-Wigner string while still preserving particle number and spin, and representing proper fermionic excitations up to the antisymmetry principle. And further, it is even possible to create operators that preserve (partially or in full) particle number and spin, but don't correspond directly to a specific fermionic excitation.

In order to assess the relative importance of each of these factors, several other pools are introduced and tested in this chapter. These pools find an intermediary position between those from fermionic-\gls{ADAPT}-\gls{VQE} and those from qubit-\gls{ADAPT2}-\gls{VQE}. Because they have more operators than the former and less than the latter, their overhead due to gradient measurements is in-between these two algorithms; and because the number of Pauli strings in each operator is lower than in the former and higher than in the latter, the depth of their circuit implementation finds also an intermediary position. 

We will begin by analysing the operators in the pool from qubit-\gls{ADAPT2}-\gls{VQE}, as well as the impact that they have on the trial state. This is relevant for the definitions of the pools that ensue. After the pools have been introduced, the evolution of the state when each is used will be analysed. Finally, the performance associated with these pools will be compared.

\section{Analysis of the Qubit Pool Operators}

From the Jordan-Wigner transform in definition \ref{def:jw_transform_paulis} and the fact that excitations are subtracted of their Hermitian conjugate to assure unitarity of the exponential operators, it is evident that any indices q, p will be represented by at most two different Pauli strings in the Qubit Pool (without $Z$ strings):

\begin{align}
i\cdot Y_qX_p,\qquad i\cdot X_qY_p.
\label{def:2_pq_qubitpool}
\end{align}

And double excitations, $q$, $p$, $s$, $r$ will be represented by at most eight different Pauli strings:

\begin{equation}
  \begin{aligned}
    &i\cdot Y_qX_pX_sX_r,\qquad
    &i\cdot X_qY_pX_sX_r,\qquad
    &i\cdot X_qX_pY_sX_r,\qquad
    &i\cdot X_qX_pX_sY_r,\\
    &i\cdot X_qY_pY_sY_r,\qquad
    &i\cdot Y_qX_pY_sY_r,\qquad
    &i\cdot Y_qY_pX_sY_r,\qquad
    &i\cdot Y_qY_pY_sX_r.
  \end{aligned}
\label{def:8_pqrs_qubitpool}
\end{equation}

Apart from the phase factor \textit{i}, the coefficients do not matter because the final coefficients will be the variational parameters. These operators consist exclusively of Pauli X and Y operators, and the number of each is always odd. This is because the other terms are canceled out by the Hermitian conjugate, which is actually what leads to the unitarity of the exponentiated operators: the coefficient of these terms would be real.

It was mentioned that single excitations will lead to \textit{at most} two Pauli strings, and double excitations to \textit{at most} eight Pauli strings in the pool. This is because not all sets of spin-orbitals lead to valid excitations. For example, single excitations will only exist between spin-orbitals of the same type ($\alpha\rightarrow\alpha$ or $\beta\rightarrow\beta$). A similar restriction exists in double excitations; if in a set there is an odd number of $\beta$ orbitals and $\alpha$ orbitals (e.g. $p$, $q$, $r$, $s$ are of types $\alpha$, $\alpha$, $\alpha$, $\beta$), no double excitations involving these four orbitals will exist.

\section{Analysis of the Qubit-ADAPT-VQE State}

If we use the pool without $Z$ strings, the $i$th operator in the Qubit-\gls{ADAPT2}-\gls{VQE} pool consists of an exponentiated Pauli string ($\hat{P}_i=\bigotimes_N\sigma_k $, with $ k\in{\{0,z,x,y\}}$) acting on $N$ qubits ($N$ will be two/four for singles/doubles respectively), multiplied by a real variational parameter $\theta_i$ and the imaginary unit i. We can expand this as

\[
e^{i\theta_i\hat{P}_i} = 1 + \frac{i\hat{P}_i\theta_i}{1} + \frac{(i\hat{P}_i\theta_i)^2}{2!} 
+ \frac{(i\hat{P}_i\theta_i)^3}{3!}  + \frac{(i\hat{P}_i\theta_i)^N}{4!} + \frac{(i\hat{P}_i\theta_i)^5}{5!} + ...  
\]

As do the individual Pauli operators, squared tensor products of them amount to the identity. As such, $(\hat{P}_i)^k$ is the identity if $k$ is even, or simply $\hat{P}_i$ if k is odd. As such, we can rewrite the expression into a simpler form.

\[
e^{i\theta_i\hat{P}_i} = \hat{I}^{\otimes4}\left(1-\frac{\theta_i^2}{2!}+\frac{\theta_i^4}{4!}-...\right) + i\hat{P}_i\left(\frac{\theta_i}{1!}-\frac{\theta_i^3}{3!} + \frac{\theta_i^5}{5!}-...\right)
\]

\[
= \hat{I}^{\otimes N}\cos{\theta_i} + i\hat{P}_i\sin{\theta_i}
\]

Operators in the ansatz of Qubit-\gls{ADAPT2}-\gls{VQE} then act on a computational basis state $\ket{n}$ in a very simple manner.

\begin{equation}
e^{i\theta_i\hat{P}_i}\ket{n} =   \cos{\theta_i}\ket{n} + i\sin{\theta_i}\hat{P}_i\ket{n}
\label{eq:qadapt_operators_on_cbstates}
\end{equation}

In the previous section we saw that each $\hat{P}_i$ has an odd number of $Y$ operators. Since $\ket{n}$ is a computational basis state, acting on it with such an operator will introduce a factor of $\pm i$. Other than that, the effect of $\hat{P}_i$ is to flip the states of the four qubits it acts on (it must be remembered that there are no $Z$ operators in $\hat{P}_i$). So we can rewrite \ref{eq:qadapt_operators_on_cbstates} as

\[
e^{i\theta_i\hat{P}_i}\ket{n} =  \cos{\theta_i}\ket{n} \pm\sin{\theta_i}\ket{m}.
\]

Here, $\ket{m}=iP_i\ket{n}$ is a computational basis state that differs from the original (also computational basis) state $\ket{n}$ in and only in the states of the four qubits $P_i$ acts on. $\ket{m}$, $\ket{n}$ represent Slater determinants with the same particle number if and only if the qubits acted on by $P_i$ in $\ket{n}$ had an equal number of $\ket{0}$ and $\ket{1}$ states.

From this, it's obvious that all eight Pauli strings acting on the same two/four spatial orbitals (for singles/doubles respectively) listed in the previous subsection will have the same effect on a given computational basis state, except for a possible phase factor of -1 associated with $\ket{m}$. Operators acting on different indices may bring different Slater determinants (computational basis states) into the superposition among themselves; operators acting on the same indices will only differ in their impact on the wave function through interference.

For example, in the simple case that we have four spin-orbitals and two electrons, the Hartree-Fock state (using alternate orbital ordering) can be written $\ket{0011}$. The impact of each of the eight operators in the Qubit Pool associated with double excitations is 

\[\ket{0011}\rightarrow\cos{\theta}\ket{0011}\pm\sin{\theta}\ket{1100}.\]

The sign is plus for $XXYX$, $XXXY$, $XYYY$, $YXYY$ and minus for $YXXX$, $XYXX$, $YYXY$, $YYYX$.

\section{Alternative Pools}

In order to test the relative importance of bringing new Slater determinants into the superposition \textit{versus} manipulating interference effects, as well as the impact of particle number and $Z$ spin preservation on convergence, \gls{ADAPT2}-\gls{VQE} was tested with five other pools. The pools were purposefully defined to facilitate the analysis of the impact of these factors, separately, on the performance of the algorithm: they were chosen to have varying degrees of affinity with fermionic excitations. The introduction of each new pool follows. 

\subsection{Generalized Singles and Doubles}

The \gls{GSD} pool includes the simplest possible generalized excitations (that are not spin-adapted nor spin-complemented), up to order two. Of course, they're still forced to be anti-Hermitian. 

Single excitations will consist of a linear combination of two two-body operators,

\begin{equation}
a_p^\dagger a_q - a_q^\dagger a_p,
\label{eq:gsd_single_excitation}
\end{equation} 

and double excitations will consist of a linear combination of two four-body operators,

\begin{equation}
a_p^\dagger a_q^\dagger a_r a_s - a_r^\dagger a_s^\dagger a_p a_q.
\label{eq:gsd_double_excitation}
\end{equation}

\subsection{Eight Pool}

The Eight Pool consists of the same operators as the \gls{GSD} pool, with the Jordan-Wigner string removed from all operators. It was designated this way because each operator will consist of a linear combination of at most eight Pauli strings.

As an example, the generalized double excitation defined in \ref{eq:gsd_double_excitation} gives rise to the Eight Pool operator in \ref{eq:eightpool_operator}.

\begin{align}
\begin{split}
\hat{\tau} = &i(
-X_qX_pX_sY_r
-X_qX_pY_sX_r
+X_qY_pX_sX_r
-X_qY_pY_sY_r\\
&+Y_qX_pX_sX_r
-Y_qX_pY_sY_r
+Y_qY_pX_sY_r
+Y_qY_pY_sX_r
)
\end{split}
\label{eq:eightpool_operator}
\end{align}

We can factorize the operator as in \ref{eq:eightpool_operator_factoredz}.

\begin{align}
\begin{split}
\hat{\tau} = -iX_qX_pX_sY_r
(
&+1
+I_qI_pZ_sZ_r
-I_qZ_pI_sZ_r
-I_qZ_pZ_sI_r\\
-Z_qI_pI_sZ_r
&-Z_qI_pZ_sI_r
+Z_qZ_pI_sI_r
+Z_qZ_pZ_sZ_r
)\\
= -iX_qX_pX_sY_r
(
&+1
+Z_qZ_pZ_sZ_r
)\times\\
(
&+1
+I_qI_pZ_sZ_r
-I_qZ_pI_sZ_r
-Z_qI_pI_sZ_r
)
\end{split}
\label{eq:eightpool_operator_factoredz}
\end{align}

From this, we can easily see the effect of $\hat{\tau}$ on computational basis states. It should be remembered that states $\ket{0}_k$, $\ket{1}_k$ indicate that the spin-orbital that qubit $k$ represents is unoccupied and occupied, respectively (under the Jordan-Wigner mapping as defined in \ref{def:jw_transform_paulis}). It is also worth noting that since the qubits with indices $p$, $q$, $s$, $r$ are the only ones acted on by the operator, we can ignore the state of the remaining qubits. 

The factorized expression with $Z$ operators acts as $8\cdot\hat{I}^{\otimes4}_{qpsr}$ for $\ket{1100}_{qpsr}$ and $\ket{0011}_{qpsr}$, and as $0$ for the remaining computational basis states. The effect of it is then to block transitions that do not correspond to the proper excitation. Thus, $\hat{\tau}$ acts on computational basis states $\ket{1100}_{qpsr}$ and $\ket{0011}_{qpsr}$ to transform them into $+8\ket{0011}_{qpsr}$ and $-8\ket{1100}_{qpsr}$ respectively, and it acts as identity in the remaining computational basis states.

The original generalized double excitation in \ref{eq:gsd_double_excitation} acts to excite (or de-excite) electrons from spin-orbitals $r$, $s$ to spin-orbitals $p$, $q$. The Jordan-Wigner transformed operator then has to be faithful to this action; it can't e.g. transform $\ket{1010}_{qpsr}$ into $\ket{0101}_{qpsr}$, as that would correspond to a different excitation ($q,s\rightarrow p,r$). Such an excitation might even not preserve $S_Z$. Other transitions might not preserve particle number: that is the case if $\ket{0000}_{qpsr}$ is transformed into $\ket{1111}_{qpsr}$, for example.

In a similar way to how we saw the (exponential) ansatz operators act on computational basis states when the pool consists of single Pauli strings with imaginary coefficients (equation \ref{eq:qadapt_operators_on_cbstates}), we can see what the effect of the ansatz operator corresponding to pool operator $\hat{\tau}$ (associated with the excitation ($r,s\rightarrow p,q$) and coefficient $\theta$ is. On the computational basis states $\ket{0011}_{qpsr}$ and $\ket{1100}_{qpsr}$, it acts as


\[
e^{\theta\hat{\tau}}\ket{0011}_{qpsr} = -\sin(8\theta)\ket{1100}_{qpsr} + \cos(8\theta)\ket{0011}_{qpsr},
\]

\[
e^{\theta\hat{\tau}}\ket{1100}_{qpsr} = +\sin(8\theta)\ket{0011}_{qpsr} + \cos(8\theta)\ket{1100}_{qpsr},
\]

and it has trivial action on the remaining computational basis states.

The Eight Pool operators conserve particle number and $S_z$, and properly represent fermionic excitations, up to the anticommutation requirement. This was the pool used in the \gls{QEB}-ADAPT-\gls{VQE} introduced in \cite{yordanov2020}, where they called such operators \textit{qubit excitations} for obvious reasons (they are essentially excitations that don't respect fermionic antisymmetry). 

\subsection{One Pools}

Pools of this type are a subset of the Qubit Pool (the one without the Jordan-Wigner strings). From any fermionic excitation, One Pools keep only one Pauli string (from the two in \ref{def:2_pq_qubitpool} or the eight in \ref{def:8_pqrs_qubitpool}, in the case of single and double excitations respectively). 

The One Pools will hereby be denoted after the strings that are kept from double excitations. For example, if from any four spin-orbital indices $q$, $p$, $s$, $r$ only the string $i\cdot X_qX_pY_sX_r$ is kept, the pool will be called \textbf{XXYX} pool. From single excitations, the string $Y_qX_p$ was the one kept always.

This type of pool does not conserve $S_z$ or particle number, much like the original Qubit Pool. However, keeping only strings of a specific form greatly limits variational flexibility by narrowing the possibilities of control over quantum interference effects. As such, One Pools are likely to perform worse.

\subsection{Two Pools}

Two Pools are obtained from the previous One Pools by multiplying each operator arising from a single excitation by ($1-Z_qZ_p$), and each  arising from a double excitation by ($1+Z_qZ_pZ_sZ_r$). The effect of this is to block some improper transitions between Slater determinants. When only two Pauli strings are used per operator, it is not possible to block them all. 

Regarding double excitations, the effect is to avoid those transitions that attempt to act on a set of four spin-orbitals with an odd number of electrons: the action of Two Pool operators on the eight 4-qubit computational basis states with an odd number of 1 states is trivial. This is a natural option because it can be done with a simple parity check. 

We can see that ($I^{\otimes4}_{qprs}+Z_qZ_pZ_sZ_r$) is one of the factors appearing in \ref{eq:eightpool_operator_factoredz} that prevents improper excitations. Without the other factor, not all of them can be prevented. This one only prevents eight improper transitions between Slater determinants, and allows six improper ones in addition to the correct two. 

In the case of single excitations, the effect is to block those transitions that attempt to act on a set of two spin-orbitals with an even number of electrons; this is enough to conserve particle number and $S_z$. In fact, the single excitation operators in Two Pools have exactly the same effect on the state as those in the Eight Pool.

However, aside from single excitations, these quantities are not in general conserved by the operators in this pool.

\subsection{Four Pools}
Four Pools are obtained from the Two Pools. The operators arising from single excitations are kept as before, and the ones arising from double excitations are multiplied by an operator of the form ($1-Z_aZ_b$). The indices $a$, $b$ depend on the type of spin-orbitals (up or down) the operator acts on. The following specifies what was the utilized factor for each type of spin-orbitals. Case $\alpha\beta\alpha\beta$, for example, is used to mean that the indices $q$, $p$, $s$, $r$ correspond in this order to up, down, up, down spin.

\textbf{Case A, $\alpha\beta\alpha\beta$ or $\beta\alpha\beta\alpha$:}

\[I^{\otimes4}_{qprs}-I_qZ_pI_sZ_r\]

\textbf{Case B, $\alpha\alpha\beta\beta$ or $\beta\beta\alpha\alpha$:}

\[I^{\otimes4}_{qprs}-I_qI_pZ_sZ_r\]

\textbf{Case C, $\alpha\beta\beta\alpha$ or $\beta\alpha\alpha\beta$:}

\[I^{\otimes4}_{qprs}-Z_qI_pI_sZ_r\]

\textbf{Case D, $\alpha\alpha\alpha\alpha$ or $\beta\beta\beta\beta$}:

\begin{align*}
I^{\otimes4}_{qprs}-I_qI_pZ_sZ_r\\
I^{\otimes4}_{qprs}-I_qZ_pI_sZ_r\\
I^{\otimes4}_{qprs}-Z_qI_pI_sZ_r\\
\end{align*}

This pool preserves $S_z$ and particle number, without being completely faithful to a true fermionic excitation. To see this, we can analyse a specific case of spin-orbitals $q$, $p$, $s$, $r$ - e.g. case \textbf{A}. If we had picked the XXYX One Pool, the corresponding operator in the Four Pool will be

\[i\cdot X_qX_pY_sX_r(I^{\otimes4}_{qprs}+Z_qZ_pZ_sZ_r)(I^{\otimes4}_{qprs}-I_qZ_pI_sZ_r).\]

It is easy to see that $(I^{\otimes4}_{qprs}+Z_qZ_pZ_sZ_r)(I^{\otimes4}_{qprs}-I_qZ_pI_sZ_r)$ is $0$ when this operator acts on any but four computational basis states of the involved qubits (the state of the remaining qubits is irrelevant):

\[
\ket{0110}_{qpsr},\quad
\ket{1001}_{qpsr},\quad
\ket{1100}_{qpsr},\quad
\ket{0011}_{qpsr}.
\]

When we add the variational parameter and exponentiate the operator, it will be a conditional rotation $e^{i\theta X_qX_pY_sX_r}$ applied only to these four states.

When acting on $\ket{n}$, $\ket{n}$ being one of these four computational basis states,  this operator rotates it into a superposition $a\ket{n}+b\ket{m}$, where $\ket{m}$ is the same as $\ket{n}$, except qubits $q$, $p$, $s$, $r$ have their states flipped. The new state appearing in the superposition has the same particle number and $S_z$ as the original state, e.g.

\[\ket{0110}_{qpsr}\rightarrow a\ket{0110}_{qpsr}+b\ket{1001}_{qpsr}.\]

The action of the operator is trivial in the cases were this would not happen - e.g., the action of $e^{i\theta X_qX_pY_sX_r}$ in $\ket{1000}_{qpsr}$ would be

\[\ket{1000}_{qpsr}\rightarrow a\ket{1000}_{qpsr}+b\ket{0111}_{qpsr}.\]

The particle number is altered by 2 and $S_z$ is altered by 1 in the computational basis state $\ket{0111}_{qpsr}$. The two factors $(I^{\otimes4}_{qprs}+Z_qZ_pZ_sZ_r)$, $(I^{\otimes4}_{qprs}-I_qZ_pI_sZ_r)$ prevent all these improper rotations. However, they do not prevent transitions that do not correspond to the original excitation. This is the difference between the Four Pool and the Eight Pool. 

For each set of four spin-orbitals $q$, $p$, $s$, $r$ of type \textbf{A}, the Eight Pool will have two operators, each corresponding to specific excitations / de-excitations. The excitation $a_p^\dagger a_q^\dagger a_r a_s - a_r^\dagger a_s^\dagger a_p a_q$ (\ref{eq:gsd_double_excitation}) would give rise to an Eight Pool operator that acts non-trivially on two computational basis states:

\begin{align*}
\ket{0011}_{qpsr}\rightarrow a \ket{1100}_{qpsr},\\
\ket{1100}_{qpsr}\rightarrow b \ket{0011}_{qpsr}.
\end{align*}

Corresponding to the excitation $r, s \rightarrow p, q$ and the respective de-excitation $p, q \rightarrow r, s$. The coefficients a, b are irrelevant because they will be multiplied by the variational parameter. Another excitation $a_p^\dagger a_s^\dagger a_r a_q - a_r^\dagger a_q^\dagger a_p a_s$ would correspond to an Eight Pool operator that acts non-trivially on two other computational basis states:

\begin{align*}
\ket{1001}_{qpsr}\rightarrow c \ket{0110}_{qpsr},\\
\ket{0110}_{qpsr}\rightarrow d \ket{1001}_{qpsr}.
\end{align*}

Corresponding to the excitation $r,q \rightarrow p, s$ and the respective de-excitation $p, s \rightarrow r, q$. In contrast, there will be a single Four Pool operator acting on these four spin-orbitals. This is the operator we saw before, with non-trivial action on four spin-orbitals:

\begin{align*}
\ket{0011}_{qpsr}\rightarrow e \ket{1100}_{qpsr},\\
\ket{1100}_{qpsr}\rightarrow f \ket{0011}_{qpsr},\\
\ket{1001}_{qpsr}\rightarrow g \ket{0110}_{qpsr},\\
\ket{0110}_{qpsr}\rightarrow h \ket{1001}_{qpsr}.
\end{align*}

This operator corresponds simultaneously to all excitations / de-excitations acting on this set of spin-orbitals that are possible (i.e. that preserve particle number and $S_Z$). 

In case \textbf{D}, the One Pool operator yields three operators in this pool. This is because, in the special case that all orbitals are of the same type, enforcing particle number preservation is enough to ensure $S_Z$ preservation. As a consequence, there are less restrictions in the excitations that are allowed to occur within these spin-orbitals.

\subsubsection{Considerations}
Because of the relations between Pauli matrices, the constituents of all of these pools are linear combinations of operators belonging to the original Qubit Pool. As compared to the latter, the number of operators in any of the One, Two, and Four Pools is decreased almost 8-fold, and almost 4-fold in the Eight Pool. The operators in these new pools consist of linear combinations of at most one, two, four and eight Pauli strings, in the same order; the increase in number implies an increase in affinity with proper fermionic excitations.

\begin{figure}[htbp]
    \centering
    \includegraphics[width=0.9\columnwidth]{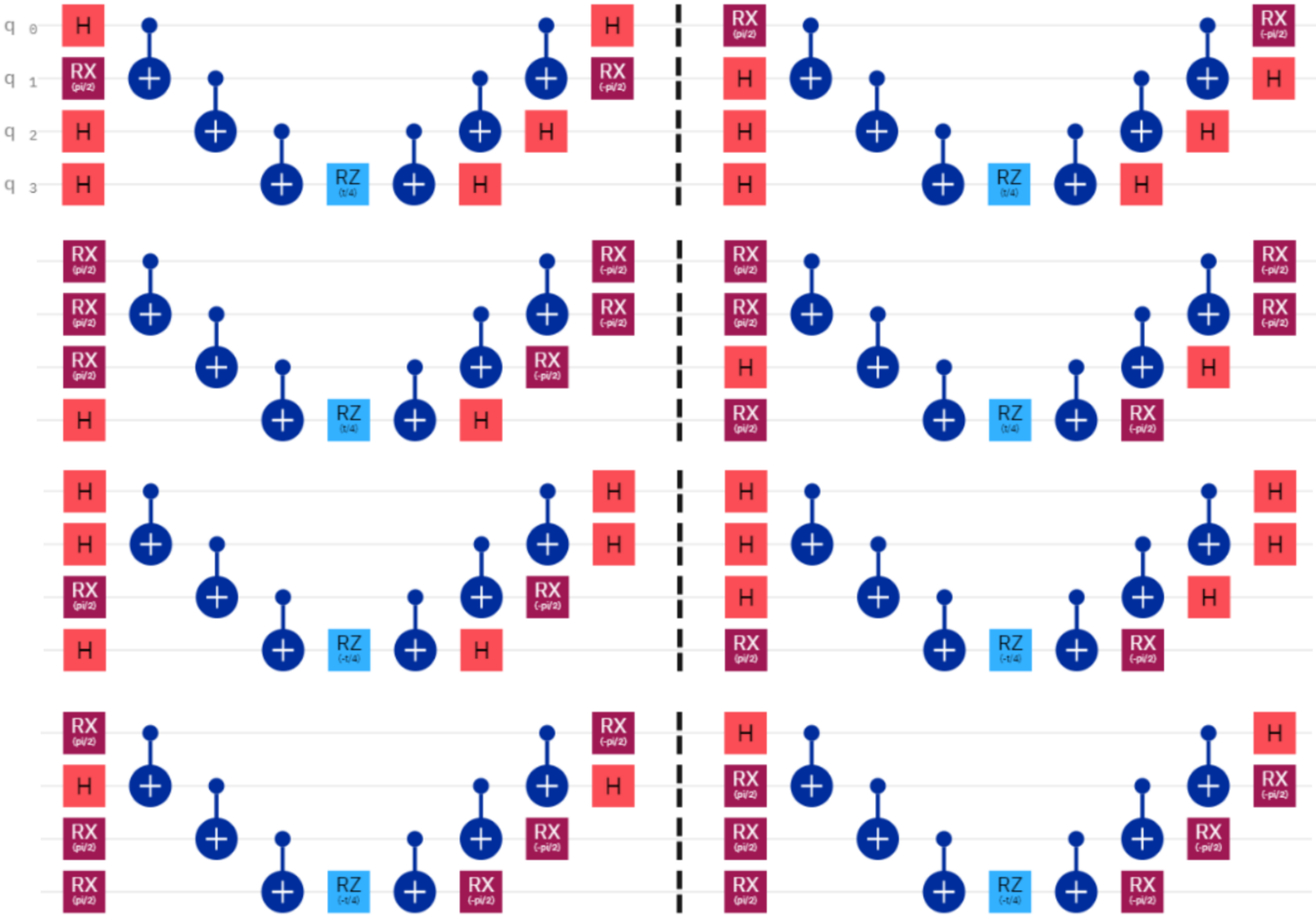}
    \caption{Circuit for implementing the operator in \ref{eq:qubit_excitation_example} using the ladder-of-\glspl{CNOT} method explained in subsection \ref{ss:simulation_trotterization}. The circuit was drawn in the IBM Quantum Composer \cite{IBMcomposer}.}
    \label{fig:QE_ladder_of_cnots}
\end{figure}

\begin{figure}[htbp]
    \centering
    \includegraphics[width=0.9\columnwidth]{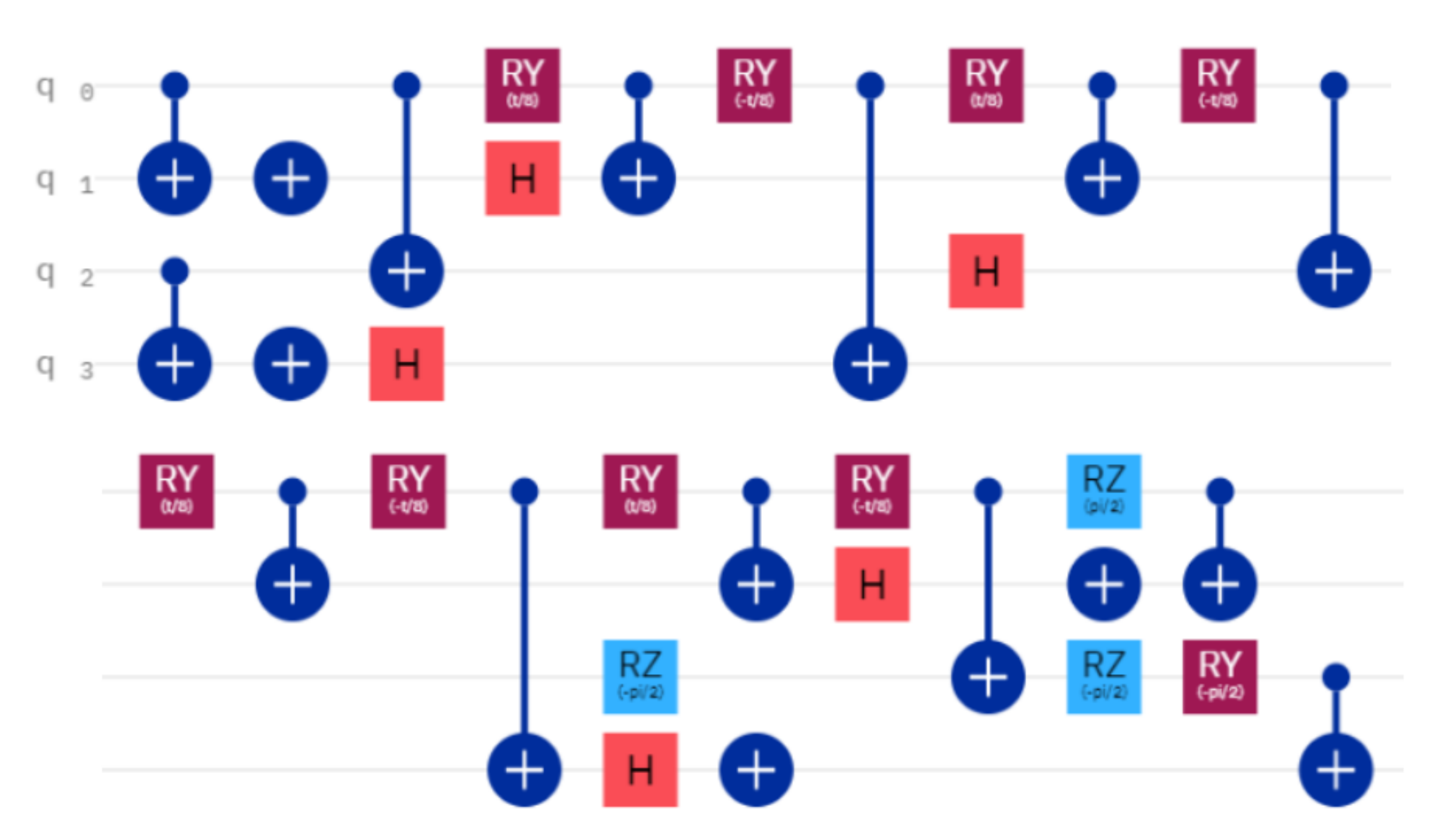}
    \caption{Circuit for implementing the operator in \ref{eq:qubit_excitation_example} using the more efficient method introduced in \cite{yordanov2020circuits}. The circuit was drawn in the IBM Quantum Composer \cite{IBMcomposer}.}
    \label{fig:QE_yordanov}
\end{figure}

It's important to note that implementing the exponential of a linear combination of $N$ Pauli strings does not necessarily correspond to an $N$-fold increase in circuit depth, as compared to the single Pauli strings in the Qubit Pool. Because they have a very specific form resembling controlled rotations, the exponentials of these newly introduced pools can be implemented more efficiently than those consisting of linear combinations of arbitrary Pauli strings. For example, reference \cite{yordanov2020circuits} presented circuits for implementing the exponentiated Eight Pool operators (\textit{qubit excitations}) with a \gls{CNOT} count of 13 and depth of 11, against the 48 \gls{CNOT} count and depth of the ladder-of-\glspl{CNOT} \cite{NielsenChuang} method. Figures \ref{fig:QE_ladder_of_cnots} and \ref{fig:QE_yordanov} exemplify the two different circuit implementations for a specific operator (the operator in \ref{eq:qubit_excitation_example}). The former illustrates the vanilla approach, while the latter shows the more efficient option.

\begin{align}
\begin{split}
\hat{\tau} = &e^{-\frac{it}{8}(
+X_0Y_1X_2X_3
+Y_0X_1X_2X_3
+Y_0Y_1Y_2X_3
+Y_0Y_1X_2Y_3
-X_0X_1Y_2X_3
-X_0X_1X_2Y_3
-Y_0X_1Y_2Y_3
-X_0Y_1Y_2Y_3
)}
\end{split}
\label{eq:qubit_excitation_example}
\end{align}

\FloatBarrier

\section{Graphical Analysis of the State Evolution}

\begin{figure}[htbp]
    \centering
    \includegraphics[width=0.9\columnwidth]{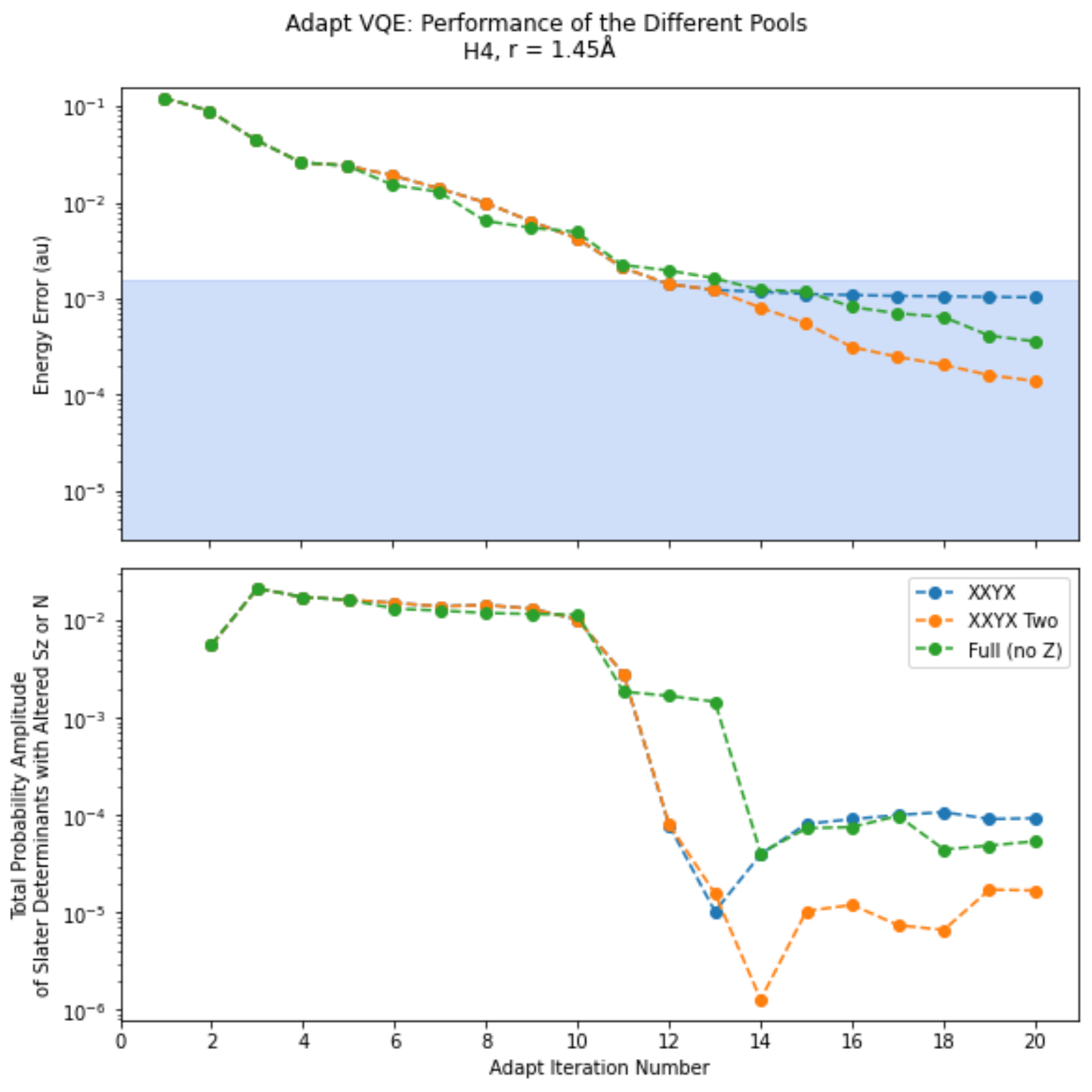}
    \caption{Evolution of the 20 first iterations of the \gls{ADAPT2}-\gls{VQE} algorithm for three different pools: the XXYX One Pool, the XXYX Two Pool, and the original Qubit Pool (without the Jordan-Wigner string). The upper plot shows the evolution of the error in the energy. The lower plot shows the evolution of the total probability amplitude of computational basis states representing Slater determinants with an altered particle number or $Z$ spin projection. The molecule in consideration is the $H_4$ molecule at an interatomic distance of 1.45Å.}
    \label{fig:pools_SDs}
\end{figure}

Before comparing all of these pools, an illustration of the impact of `partial' conservation of $S_z$ and particle number on convergence is presented. Figure \ref{fig:pools_SDs} shows a comparison of the performance of all of the pools that do not preserve in full any of these quantities. Other than the evolution of the error, the plot shows the significance of Slater determinants with altered $Z$ spin projection and particle number in the wave function, for all three pools.

Even though the XXYX One Pool contains operators capable of bringing any Slater determinant into the wave function, it doesn't seem to be capable of convergence, as the error stabilizes at around $10^{-3}$ Hartree. The same does not happen for the original Qubit Pool: it becomes clear that the different Pauli string formats are important for indispensable interference effects. However, the efforts to partially preserve $S_z$ and particle number in the XXYX Two Pool are clearly rewarded: this pool does not only match the performance of the Qubit Pool, it surpasses it. The upper and lower plots in figure \ref{fig:pools_SDs} do indeed suggest a link between convergence and preservation of these quantities. It seems like wave functions that are more contaminated by Slater determinants with altered $S_z$ and particle number imply a slower convergence.

It is interesting to observe that the curves in the lower plot of figure \ref{fig:pools_SDs} are non-monotonic. All curves have a minimum at iteration 13 or 14. It seems as though as while in the beginning it is possible to get closer to the ground state while decreasing the probability amplitude of Slater determinants with altered $S_z$ or N, after a point the energy can't be lowered further without increasing it. Regardless, this increase quickly comes to a halt, and does not seem to impede convergence.

\begin{figure*}[htbp]
     \centering
     \begin{subfigure}[b]{0.37\textwidth}
         \centering
         \includegraphics[width=\textwidth]{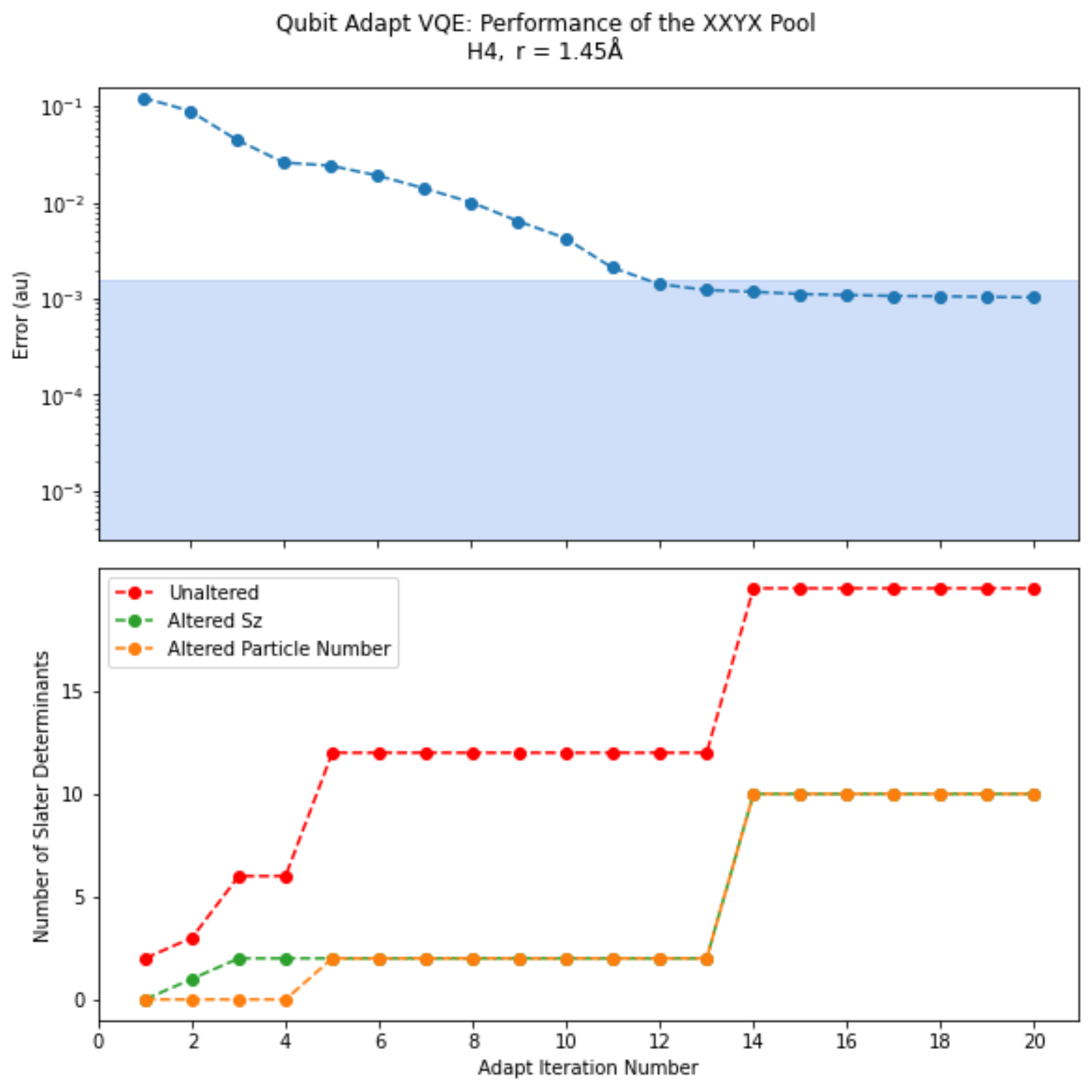}
         \caption{XXYX One Pool}
         \label{fig:XXYX_SDs}
     \end{subfigure}
     \hfill
     \begin{subfigure}[b]{0.37\textwidth}
         \centering
         \includegraphics[width=\textwidth]{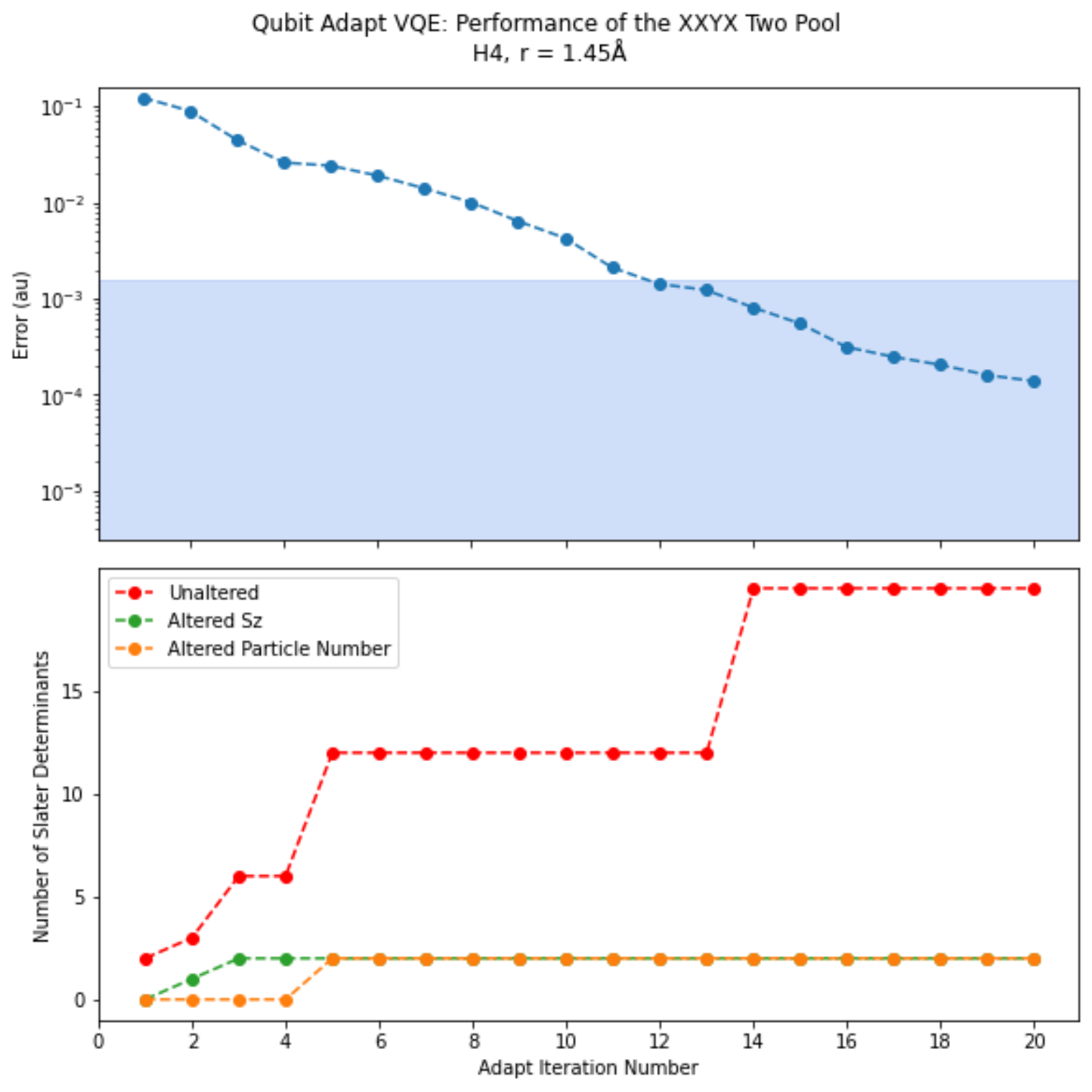}
         \caption{XXYX Two Pool}
         \label{fig:XXYX_Two_SDs}
     \end{subfigure}
     \\
     \begin{subfigure}[b]{0.37\textwidth}
         \centering
         \includegraphics[width=\textwidth]{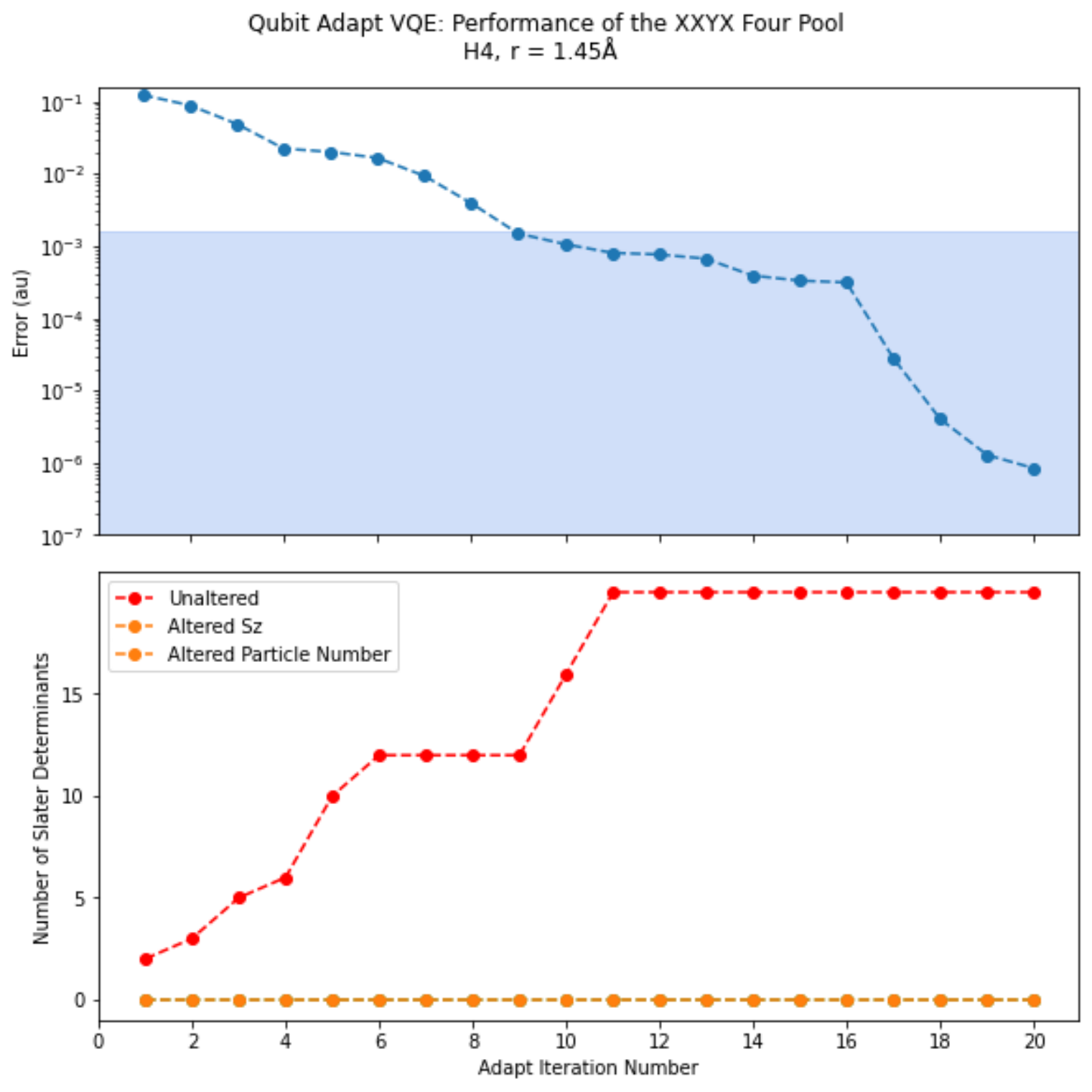}
         \caption{XXYX Four Pool}
         \label{fig:XXYX_Four_SDs}
     \end{subfigure}
     \hfill
     \begin{subfigure}[b]{0.37\textwidth}
         \centering
         \includegraphics[width=\textwidth]{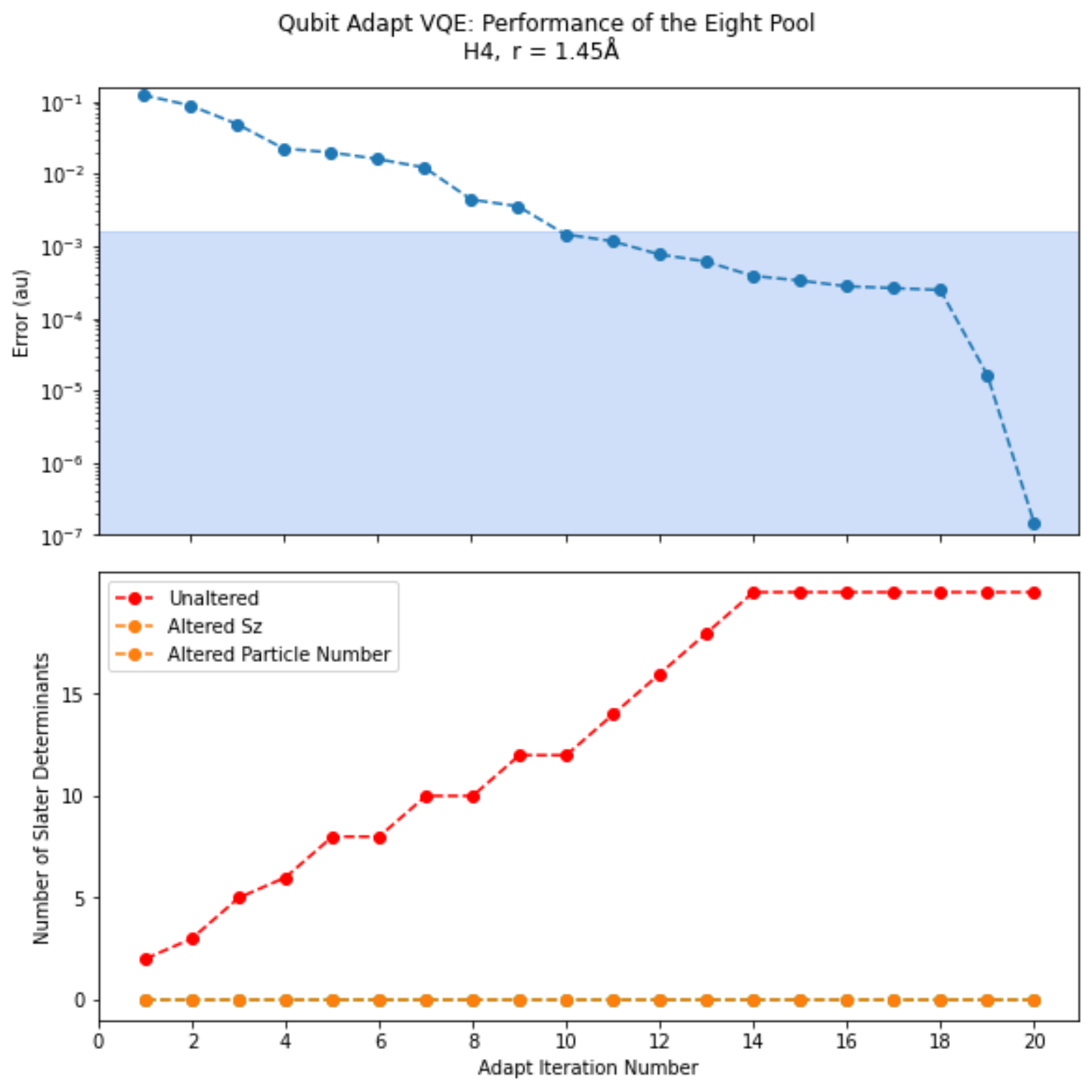}
         \caption{Eight Pool}
         \label{fig:Eight_SDs}
     \end{subfigure}
     \\
     \begin{subfigure}[b]{0.37\textwidth}
         \centering
         \includegraphics[width=\textwidth]{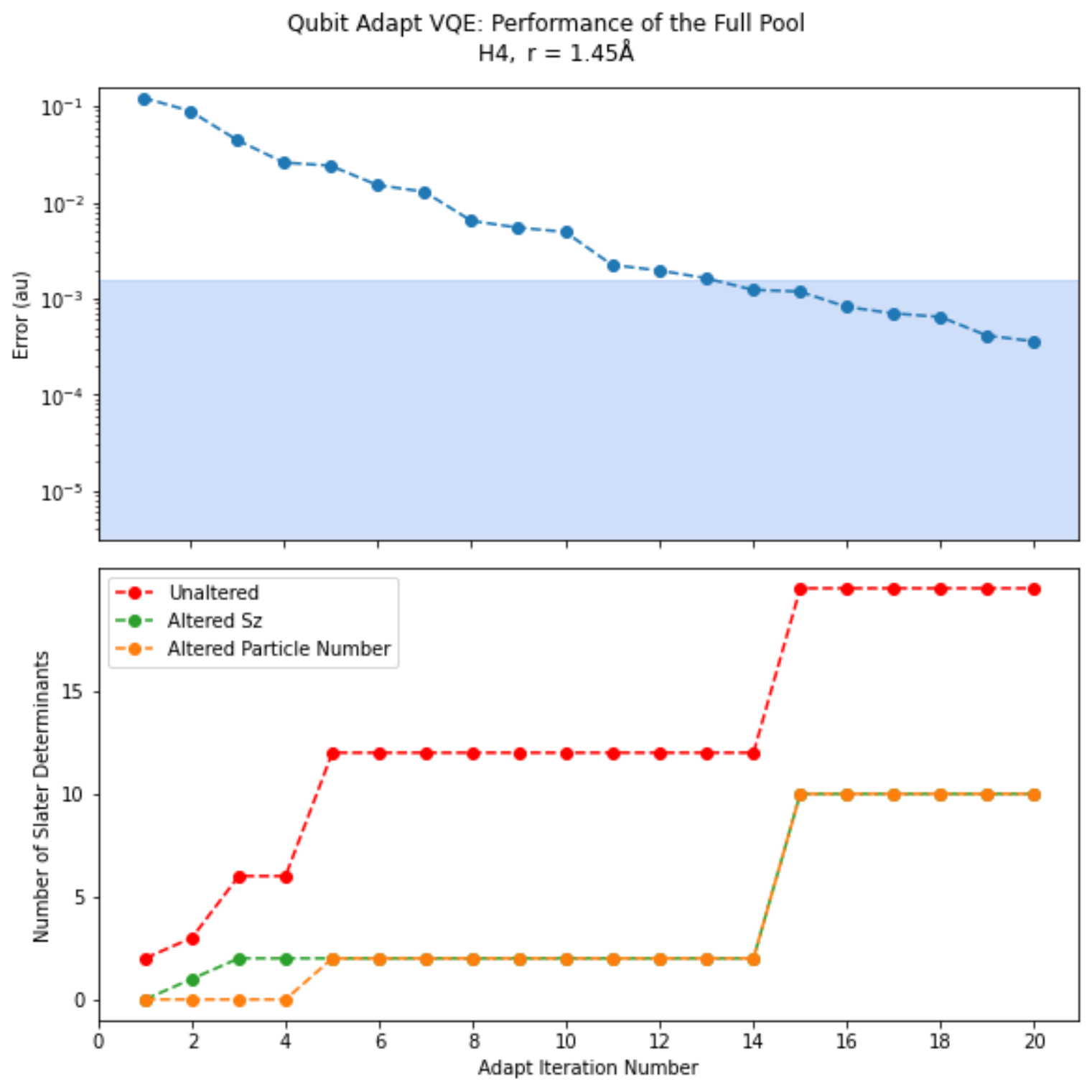}
         \caption{No $Z$ Qubit Pool}
         \label{fig:Full_noZ_SDs}
     \end{subfigure}
     \hfill
     \caption{Evolution of the 20 first iterations of the \gls{ADAPT2}-\gls{VQE} algorithm for different pools. The upper plot shows the evolution of the error in the energy. The lower plot shows the number of Slater determinants in the superposition state with altered particle number, $S_z$, or neither. In \ref{fig:XXYX_Four_SDs} and \ref{fig:Eight_SDs}, altered $S_Z$ and particle number correspond to the same colour because the curves coincide. When the green curve is not visible, it lies under the orange one. The molecule in consideration is $H_4$ at an interatomic distance of 1.45Å.}
     \label{fig:pools_number_SDs}
\end{figure*}

Figure \ref{fig:pools_number_SDs} shows the evolution of the \gls{ADAPT2}-\gls{VQE} algorithm for four different pools defined in the previous section: the XXYX One Pool, the XXYX Two Pool, the XXYX Four Pool, and the Eight Pool. The evolution of the algorithm with the original Qubit Pool (without the Jordan-Wigner string) is also plotted for reference. Other than the evolution of error, the plots show the evolution of the number of Slater determinants (computational basis states) in the superposition state, split through three groups: those with altered $S_Z$, those with altered particle number, and those with unaltered $S_Z$ and particle number. The purpose is to contrast the convergence speed against the composition of the state in terms of Slater determinants.

An interesting aspect all plots have in common is that by the end of these iterations, there are twenty fully correct Slater determinants in the state. In the case of the Eight Pool (figure \ref{fig:Eight_SDs}), this is enough to reach a very high accuracy (almost $10^{-7}$ Hartree, well inside chemical accuracy, of the order of $10^{-3}$ Hartree). Further, this number stabilizes after a certain number of iterations that ranges from 11 (in the Four Pool, figure \ref{fig:XXYX_Four_SDs}) to 15 (in the Qubit Pool, figure \ref{fig:Full_noZ_SDs}); this stabilization does not seem to imply a stabilization of the error in general. This seems to confirm that the relevant limitation after this point is \textit{not} the lack of computational basis states in the superposition that correspond to Slater determinants with relevance for the ground state, rather the lack of variational freedom with respect to their coefficients.

The error only does seem to stabilize after the 20 correct Slater determinants are reached when the XXYX One Pool is used (figure \ref{fig:XXYX_SDs}). The reason was explained before: while this pool allows for enough degrees of freedom to bring all necessary Slater determinants into the superposition, it does not allow for enough degrees of freedom that play with interference effects. Restricting the format of Pauli strings to XXYX forbids the use of different operators in the same four spin-orbitals that would lead to different local phases (in this case, the only options are $\pm1$, because the state is real). This should be contrasted with the results obtained using the original Qubit Pool (figure \ref{fig:Full_noZ_SDs}). The composition of state in terms of correct and incorrect Slater determinants is remarkably similar between these two pools. However, the Qubit Pool continues to decrease the error after the number of Slater determinants has stabilized at 20. This is because the existence of different formats of Pauli strings for the same spin-orbitals (the eight in \ref{def:8_pqrs_qubitpool}) empowers \gls{ADAPT2}-\gls{VQE} with the ability to better control interference effects, and use them to converge: namely, it can leverage them to decrease the probability amplitude of wrong Slater determinants.

When we compare the XXYX One Pool against the corresponding Two Pool (figure \ref{fig:XXYX_Two_SDs}), we can find a reason for the better performance of the latter: while it does not fully preserve particle number and $S_Z$, it decreases the amount of wrong Slater determinants in the state. While the One Pool finishes with ten Slater determinants with wrong $S_z$ and particle number in the state, the Two Pool finishes with only two of each. The error decreases steadily up until the last iteration. Unlike in the Qubit Pool, the improved convergence is not due to different operators in the pool implying different local phases; the Two Pool operators associated with double excitations consist of a linear combination of Qubit Pool operators, but there is still only one operator for each set of four spin-orbitals (the size of the Two Pool is almost an eight of the size of the Qubit Pool). The effect of multiplying the individual Pauli strings from the One Pool by ($1+Z_qZ_pZ_sZ_r$) is to preserve particle number and $S_z$ in most computational basis states. While this does not provide \gls{ADAPT2}-\gls{VQE} with more control over interference effects, it reduces the presence of incorrect Slater determinants in the state. Presumably, this decreases the need for the extra options of variational freedom, leading to improved convergence. In fact, the Two Pool even outperforms the Qubit Pool.

Finally, we can contrast the composition of the state associated with the Four Pool (figure \ref{fig:XXYX_Four_SDs}) against that of the Eight Pool (figure \ref{fig:Eight_SDs}). Their ease of convergence will be compared in the next section. 

Both the Four Pool and the Eight Pool preserve $S_Z$ and particle number; the difference is that the operators in the latter are in direct correspondence with proper fermionic excitations. The former is actually faster to reach the 20 correct Slater determinants in the state, even though it has less and smaller operators in the pool (with an almost 2-fold reduction in both the size of pool and the number of Pauli strings per operator). The reason why this happens is clear from the previous section: the operators in the Four Pool and in the Eight Pool have non-trivial action on four and two Slater determinants, respectively. This is also the reason why the number of Slater determinants in the Eight Pool increases at most by two upon the addition of a single operator.

\section{General Comparison of Pool Performance}

Up to now, we have been analysing the relation between convergence and composition of the state in terms of Slater determinants. This section will abstract from such details and directly compare the performance of different pools, in light of the acquired knowledge.

\begin{figure*}[htbp]
     \centering
     \begin{subfigure}[b]{0.45\textwidth}
         \centering
         \includegraphics[width=\textwidth]{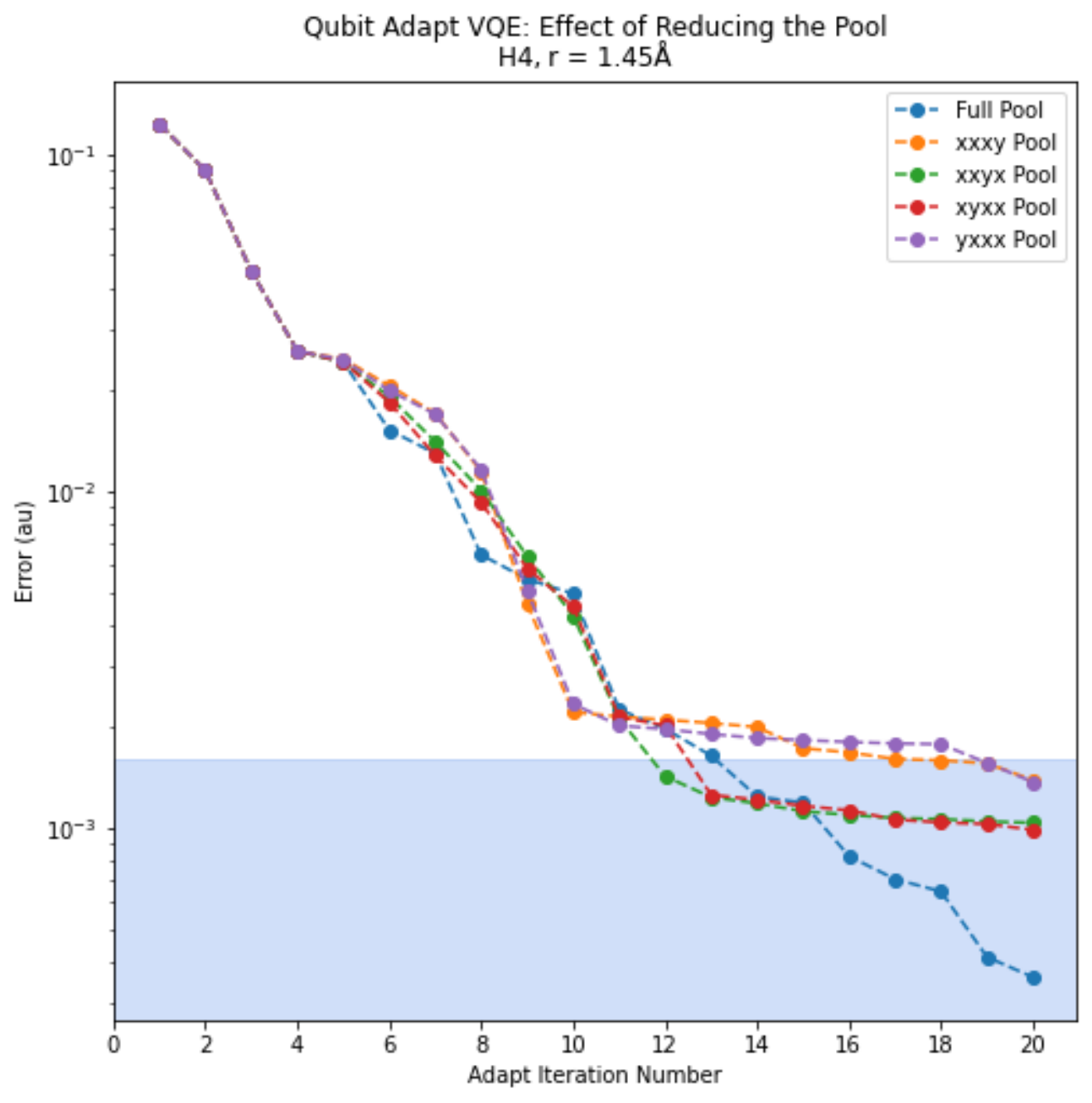}
         \caption{One Pools}
         \label{fig:OnePools_vs_NoZ}
     \end{subfigure}
     \hfill
     \begin{subfigure}[b]{0.45\textwidth}
         \centering
         \includegraphics[width=\textwidth]{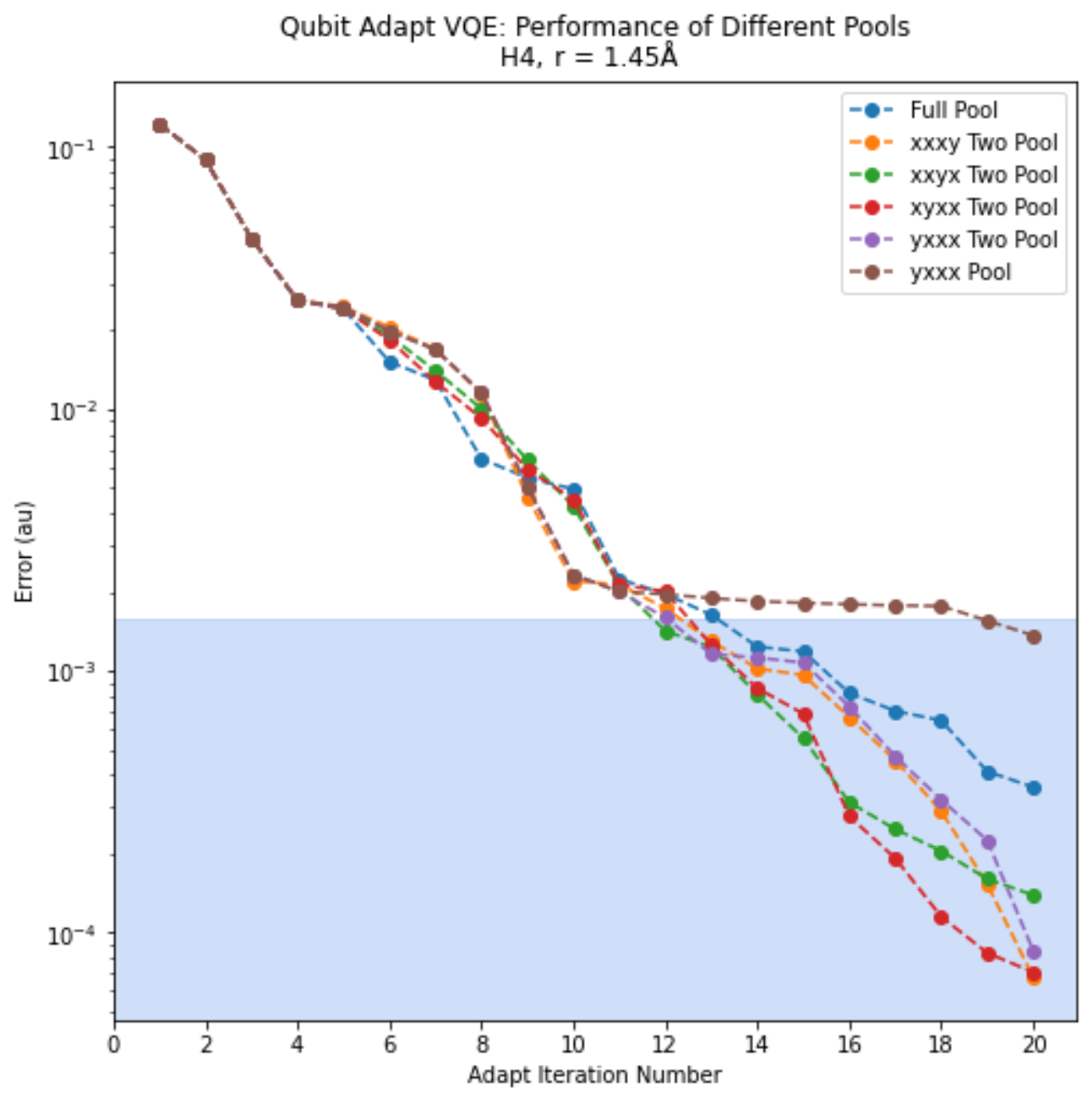}
         \caption{Two Pools}
         \label{fig:TwoPools_vs_NoZ}
     \end{subfigure}
     \hfill
     \caption{Comparison of the convergence of the \gls{ADAPT2}-\gls{VQE} algorithm using several different One Pools (figure \ref{fig:OnePools_vs_NoZ}) and Two Pools (figure \ref{fig:TwoPools_vs_NoZ}) against the full Qubit Pool (without Jordan-Wigner strings). In figure \ref{fig:TwoPools_vs_NoZ}, a curve corresponding to a One Pool was also plotted for reference. The error in the energy is plotted as a function of the iteration number. The molecule in consideration is $H_4$ at an interatomic distance of 1.45Å.}
     \label{fig:one_two_vs_noZ}
\end{figure*}

Firstly, figure \ref{fig:one_two_vs_noZ} compares multiple One Pools and multiple Two Pools. The purpose is to verify that the evolution of the error is not overly dependent on the chosen Pauli string format, and so the results presented previously for the One, Two, and Four pools constructed from XXYX strings hold in general. In fact, while there is evidently some variation between pools (because the operators do imply different local phases), the tendency seems identical between all plotted One Pools (YXXX, XYXX, XXYX, XXXY) and between all corresponding Two Pools.

In figure \ref{fig:OnePools_vs_NoZ}, we can see that all these four One Pools perform worse than the original, full Qubit Pool. In fact, all of them seem to nearly stop decreasing the error after around iteration 10, while the Qubit Pool is steadily improving the approximation to the ground state. This seems to confirm that the different formats of Pauli string are necessary for interference effects to allow convergence, when a single string is used per operator.

In contrast with the performance of the One Pools, all the corresponding Two Pools (figure \ref{fig:TwoPools_vs_NoZ}) converge faster than the original Qubit Pool. This does indeed suggest that the preservation of $S_Z$ and particle number in most computational basis states accelerates convergence in general, regardless of the string format used to construct the pool.

\begin{figure}[htbp]
    \centering
    \includegraphics[width=0.7\columnwidth]{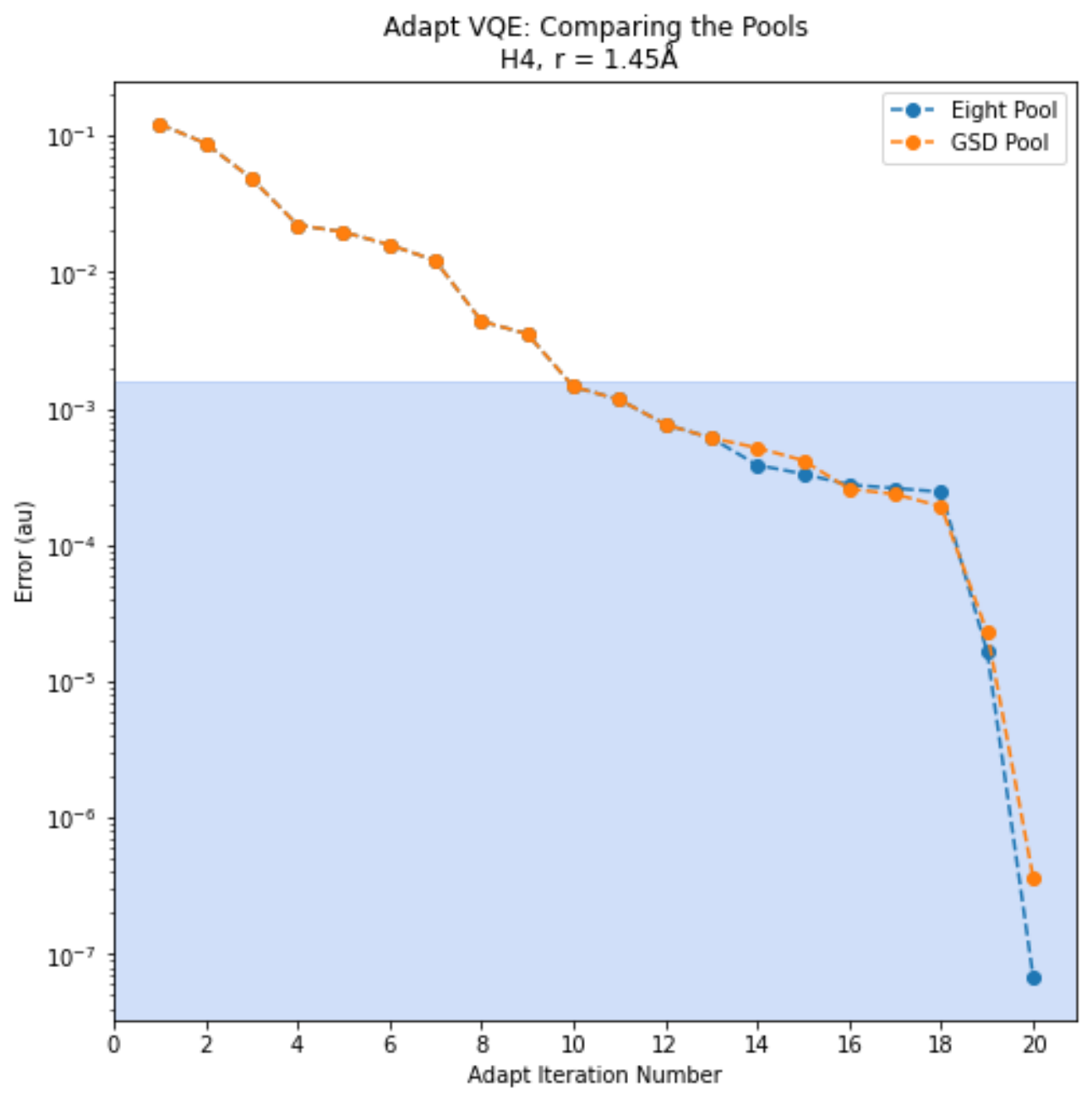}
    \caption{Comparison of the convergence of the \gls{ADAPT}-\gls{VQE} algorithm using the \gls{GSD} Pool and the Eight Pool. The error in the energy is plotted as a function of the iteration number. The molecule in consideration is $H_4$ at an interatomic distance of 1.45Å.}
    \label{fig:gsd_eight}
\end{figure}

In figure \ref{fig:gsd_eight}, we can compare the performance of the \gls{ADAPT}-\gls{VQE} algorithm with the \gls{GSD} Pool and with the Eight Pool. It should be remembered that the difference between the respective operators is that in the latter, all Jordan-Wigner strings responsible for the anticommutation of fermions are removed. As such, this figure is comparable with figure \ref{fig:h4_noZ}, that compared the performance of \gls{ADAPT2}-\gls{VQE} using the full Qubit Pool with the Jordan-Wigner strings and without them. Once again, removing such strings barely seems to have an effect on the algorithm.

As happened with the individual Pauli strings from the Qubit Pool, removing the Jordan-Wigner strings here represents a relevant saving in the circuit depth. Both using the ladder-of-\glspl{CNOT} method \cite{NielsenChuang} and using the more efficient method of \cite{yordanov2020circuits}, the scaling of the depth necessary to implement the Eight Pool and the \gls{GSD} Pool operators is respectively $\mathcal{O}(1)$ and $\mathcal{O}(N)$, $N$ the number of spin-orbitals.

\begin{figure*}[htbp]
     \centering
     \begin{subfigure}[b]{0.45\textwidth}
         \centering
         \includegraphics[width=\textwidth]{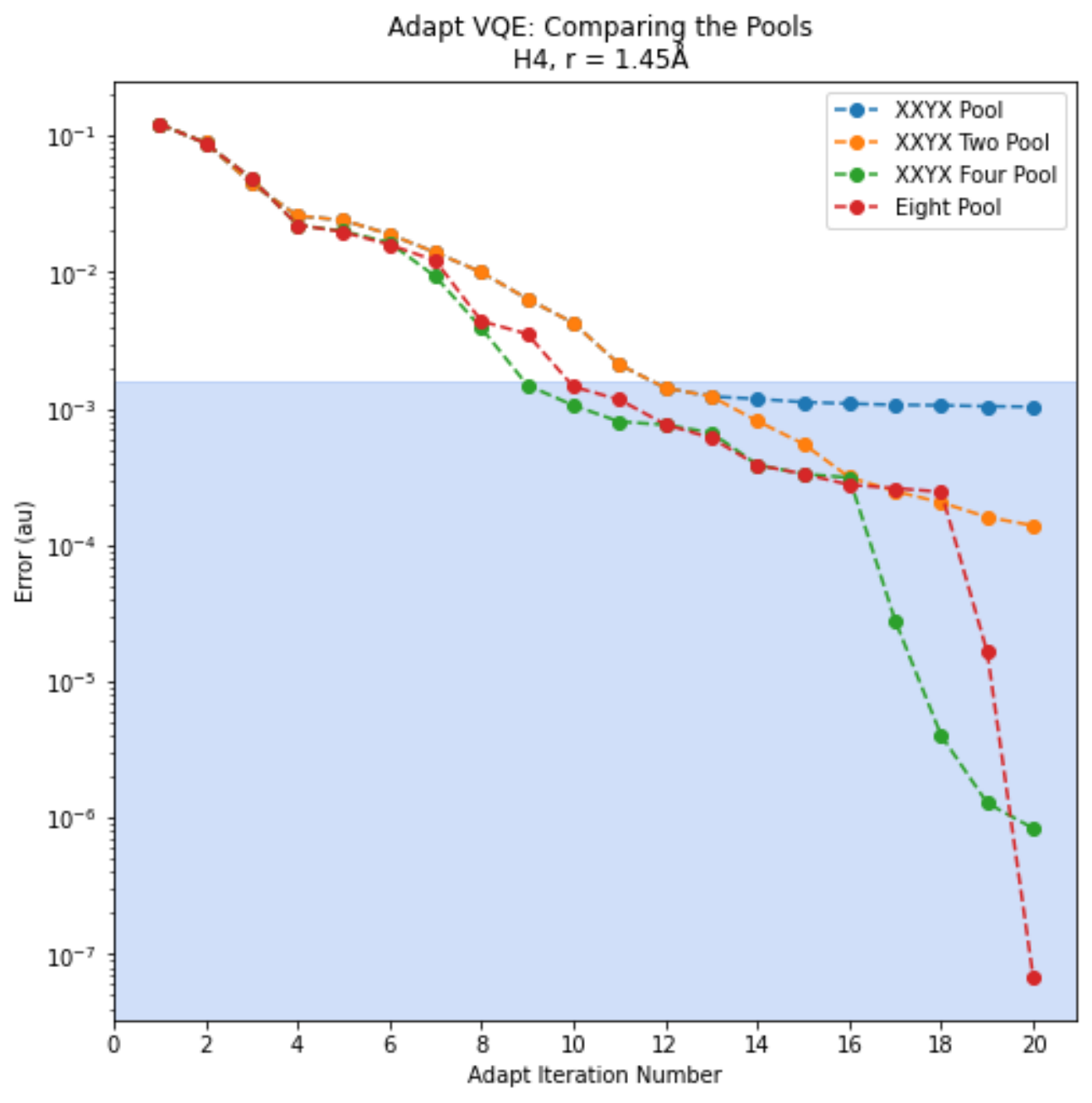}
         \caption{$H_4$ molecule at an interatomic distance of 1.45Å.}
         \label{fig:H4_1_2_4_8_pools}
     \end{subfigure}
     \hfill
     \begin{subfigure}[b]{0.45\textwidth}
         \centering
         \includegraphics[width=\textwidth]{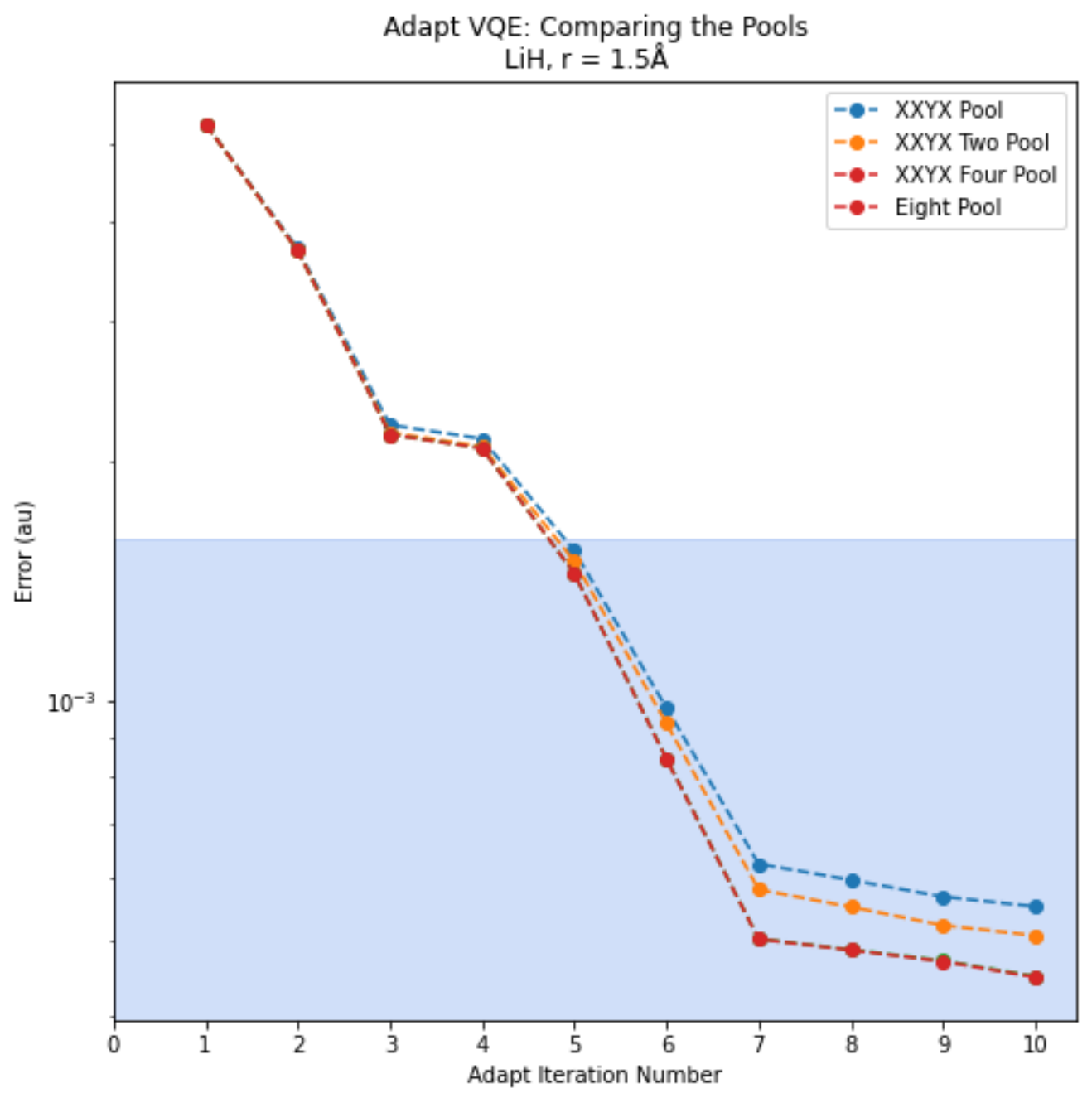}
         \caption{$LiH$ molecule at an interatomic distance of 1.5Å.}
         \label{fig:LiH_1_2_4_8_pools}
     \end{subfigure}
     \hfill
     \caption{Impact of the choice of pool (between One, Two, Four, and Eight Pools) in the \gls{ADAPT2}-\gls{VQE} algorithm, for two different molecules. The error in the energy is plotted against the iteration number. In figure \ref{fig:LiH_1_2_4_8_pools}, the XXYX Four Pool and the Eight Pool correspond to the same color because the curves coincide. These curves also overlap in some iterations in figure \ref{fig:H4_1_2_4_8_pools}.}
     \label{fig:1_2_4_8_pools}
\end{figure*}

Finally, figure \ref{fig:1_2_4_8_pools} compares the performance of all four newly introduced pools. The twenty first iterations of \gls{ADAPT2}-\gls{VQE} for $H_4$ are plotted in \ref{fig:H4_1_2_4_8_pools} and the ten first for $LiH$ are plotted in \ref{fig:LiH_1_2_4_8_pools}. As expected, the observed tendency is an improvement of convergence with increased similarity with fermionic excitations. However, it is interesting to see how the XXYX Four Pool performs similarly to the Eight Pool, even though the operators consist of half the Pauli strings and the pool has nearly half the operators. 

In the case of $LiH$ (figure \ref{fig:LiH_1_2_4_8_pools}), the curve corresponding to the XXYX Four Pool is not visible because it lies under the one corresponding to the Eight Pool. In the case of $H_4$ (figure \ref{fig:H4_1_2_4_8_pools}), the curves only overlap in some iterations. In most iterations in which this doesn't happen, the XXYX Four Pool has a visibly better performance (iterations 7-11 and 17-19). Only in iteration 20 does the Eight Pool ansatz prepare a better approximation to the ground state. 

As there is no clear trend, tests on more molecules would be required to draw conclusions. It seems possible that as the energy plunges well into chemical accuracy, the Eight Pool allows fine-tuning of the wave function by allowing selection of specific double excitations, unlike the XXYX Four Pool, which only guarantees preservation of $S_Z$ and particle number. However, since it allows action on more computational basis states at once, there is also reason to believe that the XXYX Four Pool may hasten convergence until points of extreme accuracy. For $H_4$, this pool outperforms the larger Eight Pool until the error is lowered to under $10^{-6}$, which is roughly 0.07\% of what is actually necessary to be inside chemical accuracy.

%% file: Chapters/chapter6.tex
\pagestyle{plain}
\graphicspath{{./Chapters/Figures/Ch6/}}

\chapter{ADAPT-VQE: Further Ansatz Manipulation}
\label{ch:ansatz_manipulation}

This chapter aims to explore the possibility of introducing extra flexibility and deliberation into the evolution of the \gls{ADAPT}-\gls{VQE} ansatz, instead of having it defined by a single criterion of selection. 

Two options will be outlined, having in common the intent of keeping the preparation circuit shallower than \gls{ADAPT}-\gls{VQE} (while attempting to avoid a prohibitive number of additional optimizations). The first one focuses on removing operators from the ansatz along the evolution of the algorithm, with the purpose of removing the ones that performed worse than expected. The second one attempts to be more conservative in the addition of new operators, in order to avoid getting bad performing operators in the ansatz to begin with. The approaches will be compared against each other, as well as against the original \gls{ADAPT}-\gls{VQE} algorithm, for multiple molecules and multiple pools. This comparison will be done by using the different approaches to obtain ansätze with the same number of operators, and analysing the error resulting from each in light of the number of optimizations accumulated along the execution (which is strongly  related to the measurement cost).

\section{Removing Operators}
\label{s:removing_ops}

\subsection{Motivation}

The \gls{ADAPT}-\gls{VQE} algorithm selects the operators based on a single quantity: the magnitude of the derivative of the energy with respect to the operator coefficients $\theta_i$, at point zero. This is a convenient selection method, since the gradient can be measured in a quantum computer. The circuits used for these measurements at a given iteration are barely deeper than the ansatz itself at that same iteration, thus guaranteeing that the selection method itself isn't the bottleneck of the algorithm. What is more, there being no further information, how much an operator is impacting the energy locally, at $\theta_i = 0$, is the best possible indicator of how much it will have lowered it once it has been added to the ansatz and had its coefficient optimized. 

However, this is not a faultless selection method. While it seems likely that the best performing operator will be among those with the largest gradients, it is certainly possible that its gradient is not the largest one of all. Conversely, the operator with the largest gradient is not guaranteed to have a great impact on the energy. It might even be that the derivative of the selected operator was large at $\theta_i = 0$, but upon optimizing the coefficient and exploring the optimization landscape further, the optimizer will find that the change in energy quickly comes to a halt, as a local minimum is reached, and return a state with little energy difference as compared to the previous one. Of course, there is no way to know in advance, and this criterion allows for a much more informed choice than random selection. 
 
But in spite of their actual impact on the energy, once they are chosen by their gradient, \gls{ADAPT}-\gls{VQE} simply accepts the operators. The selected operators remain in the ansatz until the algorithm terminates, regardless of how much they actually lowered the energy. As a consequence, it is possible that some operators in the ansatz will bring little advantage in converging to the ground state, but nevertheless contribute to deepen the circuit (placing more demands on the quantum computer) and increase the number of variational parameters (placing more demands on the classical optimizer, which in turn will require more function evaluations until convergence, leading to more calls to the quantum computer). 

The problem of finding the optimal ansatz is combinatorial in nature. The best solution could be found by exhaustive search: if we tested \textit{all} n-operator ansatze that could be formed from the operators in our pool, we could find the optimal one. Of course, this approach is not viable due to its computational complexity - hence, \gls{ADAPT}-\gls{VQE} settles for a procedure of growing the ansatz that, while not perfect, is at least tractable. However, there is a lot of space between \gls{ADAPT}-\gls{VQE} and a brute-force solution. It is possible to keep the algorithm tractable, but nevertheless be stricter in allowing the operators to actually remain in the ansatz, instead of simply keeping them once they are selected. 

This subsection builds on the idea of removing bad performing operators from the ansatz along the execution of the algorithm. An attempted solution to the problem of finding a removal procedure that reaches a lower error while keeping the optimization overhead to a minimum will be developed.

\subsection{Heuristics}

The heuristics behind removing terms has to address two major points: when to remove operators, and when to allow them to be added back. If not done carefully, the process can actually harm the evolution of the algorithm by requiring unnecessarily many additional iterations, and consequently also optimizations and measurements. 

The purpose of the following exposition is to develop a set of rules to remove operators which satisfies two conditions: resorting exclusively to data that is available during the process of running \gls{ADAPT}-\gls{VQE}, and resulting in a modest overhead in optimization/measurement costs.

\subsubsection{Removal Criterion}

The first question that should be asked is:

\textbf{When should an operator be susceptible to being removed?}

A sensible first answer is \textit{when it performs badly}, in the sense that it doesn't lower the energy as much as it could. However, defining \textit{bad performance} is in itself cumbersome.

In a realistic run of the algorithm, we never know how far the current energy lies from the \gls{FCI} ground energy. As such, we are ignorant of what consists a reasonable change in energy. What is more, \textit{reasonable} might depend on the particular choice of pool: naturally, large operators as one encounters in the \gls{SGSD} pool will have a greater capacity of changing the energy as compared to the length four Pauli strings in the qubit pool.

A natural option is to characterize performance \textit{relative} to the other operators. Since the operators used in a particular run all belong to the same pool, the latter problem is solved. However, attention must be paid to which operators we are using as standard for the comparison. 

As the state grows closer to the true ground state, and the energy grows closer to the true ground energy, there will be a decreasing trend in the impact of the newly added operators on the energy. It becomes clear that it is not fair to compare an operator to its predecessors in the ansatz: having acted on a state farther away from the ground state, they were in a better position to lower the energy.

\begin{figure}[htbp]
    \centering
    \includegraphics[width=.8\textwidth]{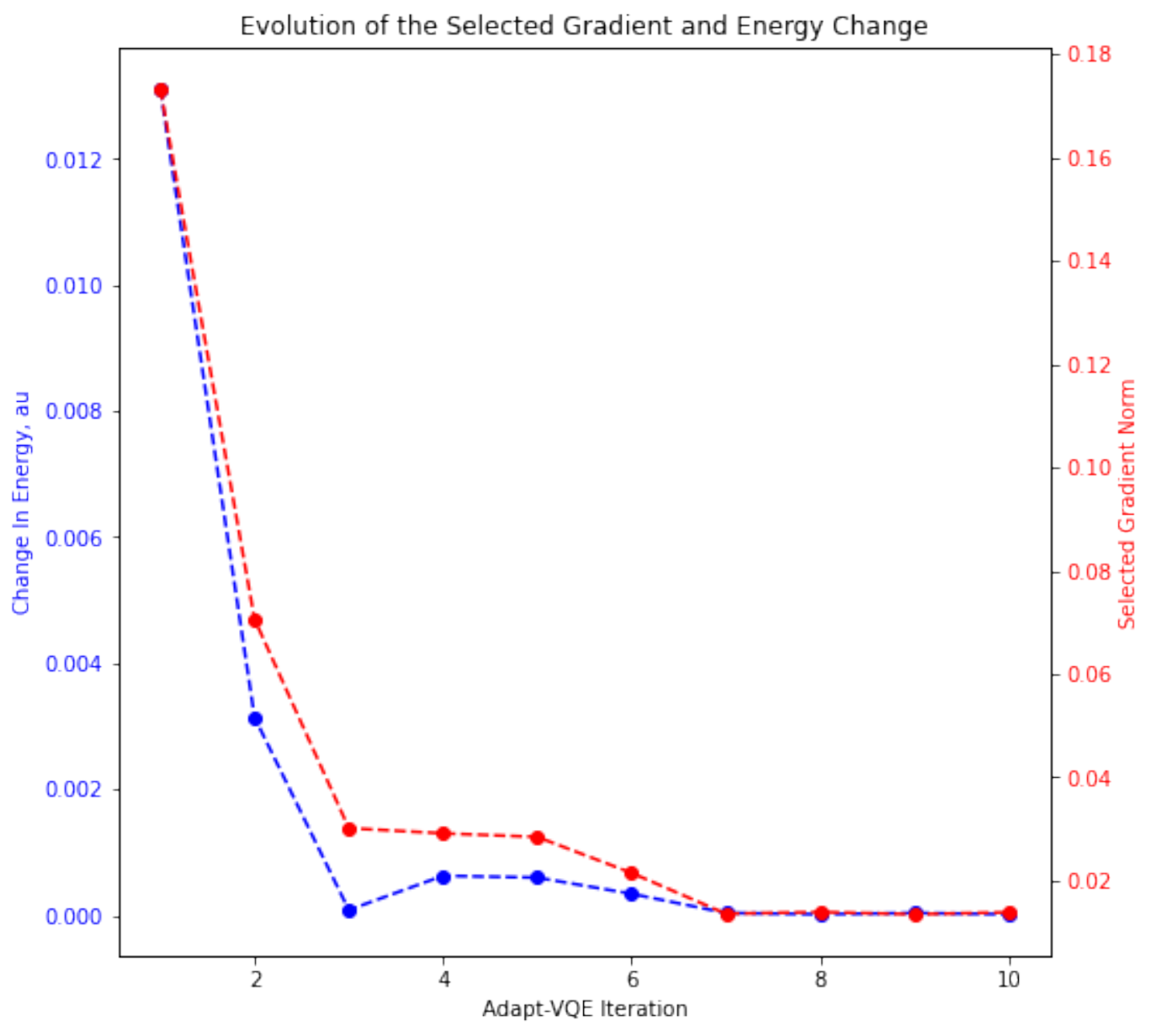}
    \caption{Norm of the selected operator and change in energy in the first ten iterations of \gls{ADAPT}-\gls{VQE}. The molecule in consideration is the $LiH$ molecule at an interatomic distance of 1.45Å. The \gls{SGSD} pool was used.}
    \label{fig:gradient_energy_change}
\end{figure}

This is better illustrated by the plot in figure \ref{fig:gradient_energy_change}, that shows the change in energy in the first ten iterations of the \gls{ADAPT}-\gls{VQE} algorithm for $LiH$. The change in energy is taken as compared to the previous iteration, with the energy at iteration 0 taken to be that of the reference state (the Hartree Fock state).

As expected, the tendency is for the impact that the operators have on the energy to decrease significantly along the evolution of the algorithm. As such, not lowering the energy as much as its predecessor seems like a highly inadequate criterion for removing an operator: taking this case, for example, it would imply removing almost all the operators in the ansatz.

Another, more viable possibility would be comparing two operators by their energy change \textit{relative to their gradient}. After all, by setting it as the selection criterion, we are using the gradient as the sole indicator of how much we expect a given operator to change the energy. The absolute value of the ratio between the change in energy caused by an operator and its the gradient seems like an interesting metric for characterizing the performance of an operator against expected. In addition, the selected gradient norm has, as does the energy change, a tendency to decrease as the algorithm evolves. This can be confirmed from figure \ref{fig:gradient_energy_change}, and understood intuitively: as the possible impact on the energy by the operators decreases, so does the absolute value of the maximum gradient, as a measure of how much the operator is affecting the energy at a certain point. Even though, as it was mentioned, the norm of the gradient of an operator isn't a rigorous indicator of how much it will impact the energy, this statement certainly holds in a broad sense, and the tendency is verified in practice. 

The norm of the ratio between the energy change produced by an operator and its gradient at the time it was selected will hereby be denoted \textit{performance ratio} for simplicity (formula \ref{eqn:performance_ratio}). 

\begin{equation}
\label{eqn:performance_ratio}
\text{performance ratio} = \left|\frac{\text{energy change}}{\text{gradient}}\right|
\end{equation}

\begin{figure}[htbp]
    \centering
    \includegraphics[width=.8\textwidth]{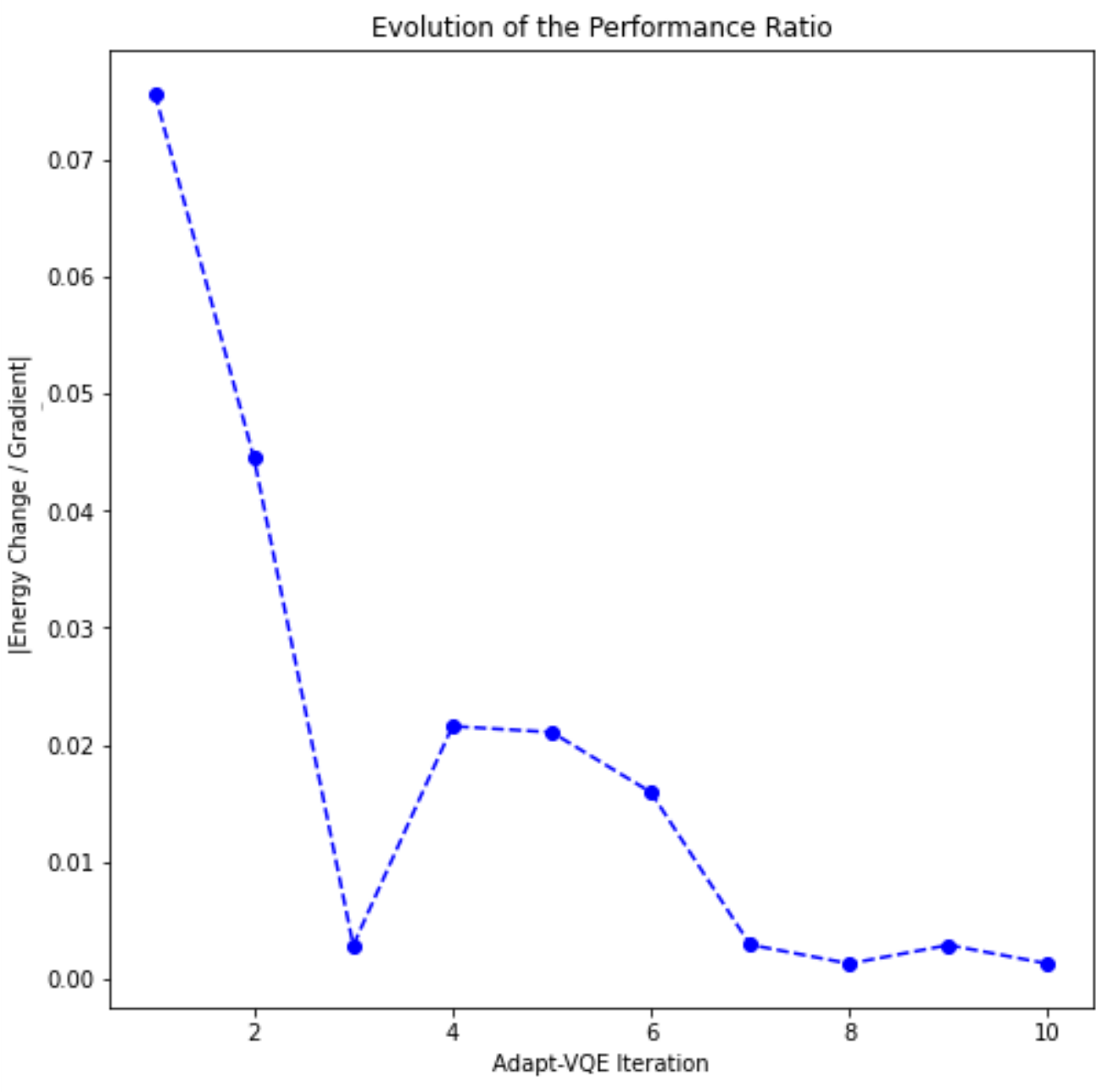}
    \caption{Absolute value of the performance ratio in the first ten iterations of \gls{ADAPT}-\gls{VQE}, for the same pool and molecule as figure \ref{fig:gradient_energy_change}. The performance ratio at a given iteration corresponds to the operator added then.}
    \label{fig:performance_ratio}
\end{figure}

For all that was exposed, one could imagine that the performance ratio would facilitate comparing an operator to its predecessor. However, numerical simulations show that this ratio also has a tendency to decrease along the evolution of the algorithm. Figure \ref{fig:performance_ratio} shows its evolution along an \gls{ADAPT}-\gls{VQE} run. Once again, if one took the decrease in this ratio as compared to the previous iteration as the criterion for removing the operator that was added last, a very large amount of operators would be removed. This would warrant, for example, removing the operator added at iteration 2. However, this is the second best performing operator in all ten iterations. The operator added at iteration 7 would also be removed, as there is a sharp drop in the performance ratio at that iteration; however, the ratio does not go back to previous values in iterations 8, 9 or 10. Looking at the big picture, it does not seem like operator 7 performed particularly bad for its position on the ansatz.

It seems therefore difficult to remove operators by comparing them with their predecessors. However, when we add a new operator to the ansatz, the only information we have available is precisely that regarding the operators previously added: as for the remainder of the operators in the pool, they were not added to the ansatz, which means that there is no way to know how much they're capable of lowering the energy.

The possibility that is left is to compare an operator against its \textit{successors}. If an operator added to the ansatz later on has a larger impact on the energy, there is a reason to believe that, up until that point, it was reasonable to expect changes in energy of at least that order of magnitude. 

Of course, this is a simplified view of things: it is also possible that the operators added previously were necessary for the energy change caused by the new operator to be as large as it was. It is important to remember that the operators are selected based on their energy gradient taken on the state \textit{at the point at which they are added}. So it is even possible that, had the previous operator not been there, the choice of operator would have been different altogether.

As a specific example, an operator might add to the state Slater determinants that did not appear in the superposition before. The term added afterwards can have relied on its impact on those determinants to change the energy as much as it did. In that case, even if the previous operator performed badly, removing it from the ansatz would prove to be a bad decision, as the energy would increase significantly due to the change that this operator
produced not on the energy, but on the state. 

However, there is an easy way to discover whether removing an older operator is a bad decision: removing it, re-optimizing the energy, and comparing the increase in energy from removing the operator with the decrease that had been caused by adding it. If the increase is significantly larger, then the operator is more important to the ansatz than one could have imagined from its impact on the energy.

It is then interesting to contemplate the possibility of removing an operator at a later iteration, once a \textit{successor} has a greater impact on the energy, rather than immediately in the iteration at which it was added. Looking once again at figure \ref{fig:gradient_energy_change}, it can be seen that the operator added at iteration 3 has quite lower an impact on the energy than its immediate successor, added in iteration 4. As much as we don't have a view on how the future operators will impact the energy as the algorithm proceeds, at iteration 4 we can certainly look back at the previous iterations and notice that the third one produced a remarkably low difference in energy as compared to the fourth one. Looking at the numerical data, one finds that the energy change of iteration 4 is 739\% that of iteration 3.

Considering that removing terms also slows down the evolution of the algorithm, and implies an added cost in measurements and optimizations, it is important that the criterion for removal is chosen to be strict enough that reasonably performing operators will not be removed. If operator A is followed by operator B that produces 101\% its energy change, it doesn't seem fruitful to remove operator A: that would demand an extra optimization step after removing it, possibly only to add it back to the ansatz soon (how and when this should happen will be discussed shortly) and re-optimize all the coefficients once more. This would imply two extra optimizations for little to no gain. 

With all of this in consideration, the following rules were chosen as the procedure for removing operators:

\begin{itemize}
  \item Upon adding an operator to the ansatz at iteration $j$, compare the resulting energy change $\Delta E_j$ with that of all previous iterations $i<j$. 
  \item If an operator added at iteration $i$ caused an energy change $\Delta E_i$ such that $\Delta E_i > r\Delta E_j$\footnote{It should be noted that the $\Delta E_{i,j}$ are expected to be negative.}, attempt to remove this operator and re-optimize the coefficients of the ansatz without it. Calculate the energy change produced by this action, $\Delta E_i'$.
  \item Compare the increase in energy caused by removing the operator with the decrease that had been caused by adding it\footnote{Being caused by the removal of an operator, $\Delta E_i'$ will usually be positive.}. If $-\Delta E_i'>t\Delta E_i$, remove the operator.
\end{itemize}

The parameter r, $0<r<1$, determines what energy change is considered acceptable for previous iterations when a new operator is added and a new energy change is known of. As mentioned previously, a value of $r$ too close to 1 would often cause the removal process not to compensate its measurement/optimization overhead. In the following results, $r=0.5$ was used. This means that when an operator is added to the ansatz and produces a certain change in energy, all its predecessors that lowered the energy by under 50\% of this are liable to be removed.

Once an operator is considered suitable for removal, the parameter $t>1$ determines the tolerance for the difference in impact that there might be as a consequence of this operator not being the last one in the ansatz, and as such affecting the energy also indirectly through its successors. In the simulations, $t=1.5$ was used. This means that, once the removal has been attempted, an operator is effectively removed if the increase in energy resulting from removing it is no greater than 150\% the decrease in energy that had been caused by adding it. As explained before, not using this limit would make the process of removing terms blind to the possibility of operators with little impact in the energy actually being important for the wave function that was built after them. One could also imagine different alternatives: for example, comparing the change in energy from removing the operator \textit{i} with the change in energy from adding operator \textit{j} instead (removing operator \textit{i} if $-\Delta E_i'>t\Delta E_j$), or the sum of both ($-\Delta E_i'>t(\Delta E_j+\Delta E_i)$), adjusting the parameter $t$ accordingly. In essence, what we want to do is guarantee that the operator is only removed if there is something to be gained from it; that can be assured in several different ways.

\subsubsection{Blocking Operators}

Once the question of when to add operators has been dealt with, it is important to decide when and where they will be allowed to be added back to the ansatz.

At the point we've removed an operator, we know that even though its gradient was high, it didn't impact the energy as much as we could have hoped. Thus, it seems unwise not to make sure it doesn't get picked again too soon. This is likely to happen if we use the gradient as the sole selection method. 

It is possible that the removed operator doesn't have the highest gradient anymore, since another operator has been added to the ansatz and so the state in which we are calculating the gradient is different; however, in practice, that is seldom the case. Usually, the change that a single operator produces on the wave function is not enough to cause significant changes to the gradients of the other operators in the pool. If nothing is done, a removed operator will almost always be added back to the ansatz in the next iteration, and once again have little impact on the energy. This behaviour is not desirable, as it prevents the removal of the term from being advantageous at all: once the term is removed, we should be giving the opportunity of being in the ansatz to other operators, that despite having lower gradients might perform better.

One solution could be blocking removed operators, taking them off the pool altogether and not allowing them to be added to the ansatz ever again. But this is another extreme that might jeopardize the performance of the algorithm: as was discussed, as the ansatz evolves, the change in energy produced by each operator will have a tendency to decrease. What consists a bad performance early on might be a great one later.

Taking figure \ref{fig:gradient_energy_change} once again, we see that while the energy change at iteration 3 is the smallest in the first 6 iterations, it seems comparable to the change observed in iterations 7 to 10. In fact, the numerical data shows that all these iterations have associated energy changes of the order of $10^{-5}$, with the change in 3 being the greatest one of all. What is more, it was verified that if the operator added in iteration 3 is removed and added back after a few iterations, the energy change it produces is very similar to the one it had produced back in position 3 - meaning that even after being removed, it seems like the best possible addition to the wave function at any of iterations 7 to 10. As an example, removing this operator and adding it back at iteration 8 would produce an energy change corresponding to 97\% the change it had produced back at iteration 3. Importantly, at this point the removed operator is capable of producing more than double the change in energy of the operator added to the ansatz in case this one is not allowed to reenter it at all. This is only due to worsen as the ansatz evolves: the operator that was removed might be crucial to convergence after a certain point.

In dealing with terms that have been removed, there are thus two main requirements: they can't be allowed to be added back too soon, and they can't be forbidden from being added back forever.

An approach that seems reasonable is blocking the operators that are removed until an operator is added that performs similarly. Since operators are removed when they have little impact on the energy as compared to operators added after them, it seems natural that they would be allowed back once a newcomer has an impact comparable to the one they had.

There are, however, a few problems with this method, that are better illustrated by an example. The following is a depiction of the evolution of an actual simulation of \gls{ADAPT}-\gls{VQE} with term removal (again for the LiH molecule at an interatomic distance of 1.45Å, using the \gls{SGSD} pool). Aside from showing the drawbacks of blocking operators, this will serve to demonstrate how the term removal procedure flows. The criterion for removing operators is as defined before, and a removed operator $i$ is allowed to be selected again once an operator $k$ added to the ansatz later produces an energy change that is no more than the double the one operator $i$ had produced ($\Delta E_i<0.5\Delta E_k$).

\begin{figure}[htbp]
    \centering
    \includegraphics[width=\textwidth]{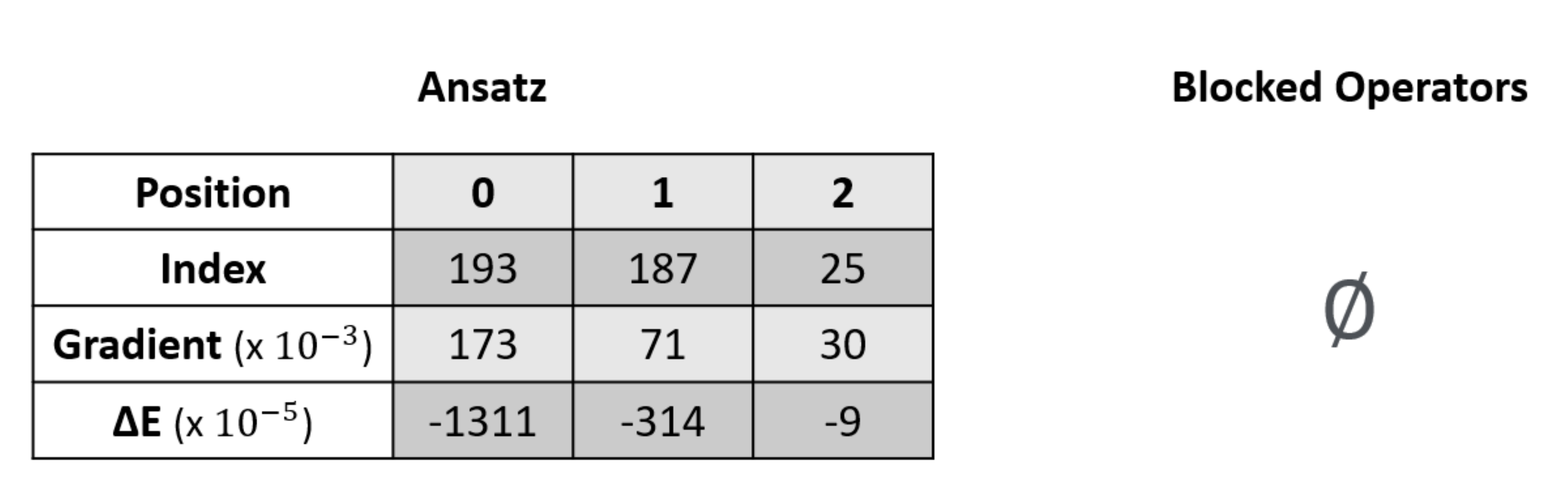}
    \caption{Relevant data structures at iteration 3 of the \gls{ADAPT}-\gls{VQE} algorithm, with an additional term removal feature based on blocking and unblocking operators. The molecule in consideration is $LiH$, and the pool the \gls{SGSD} pool. The units of energy, omitted to avoid overcrowding of the figures, are Hartree throughout the whole chapter. Because the variational parameters are dimensionless, the gradient shares the energy units.}
    \label{fig:blocking_it3}
\end{figure}

Figure \ref{fig:blocking_it3} represents the relevant data structures at iteration three of the algorithm. The most important one is the ansatz, that specifies the current state preparation circuit. 

As the name suggests, \textit{position} represents the position of each operator in the ansatz: the column labeled with 0 concerns the first operator in the ansatz, the one labeled with 1 concerns the second one, and so on. The first operator is both the one that was first added and the one that comes first in the circuit. 

The index identifies which operator in the pool was selected; what fermionic operator this corresponds to is irrelevant, as the purpose is just attributing an identifier to each pool operator. For a matter of simplicity, this identifier is a number rather than a formula.

The first three iterations are uneventful: the criterion for attempting to remove a term is never met. As can be seen in figure \ref{fig:blocking_it3}, the energy change always decreases from an iteration to the following one. As such, no operator is suitable to be removed, and the algorithm proceeds just as the original \gls{ADAPT}-\gls{VQE}. The set of blocked operators is empty.

\begin{figure}[htbp]
    \centering
    \includegraphics[width=\textwidth]{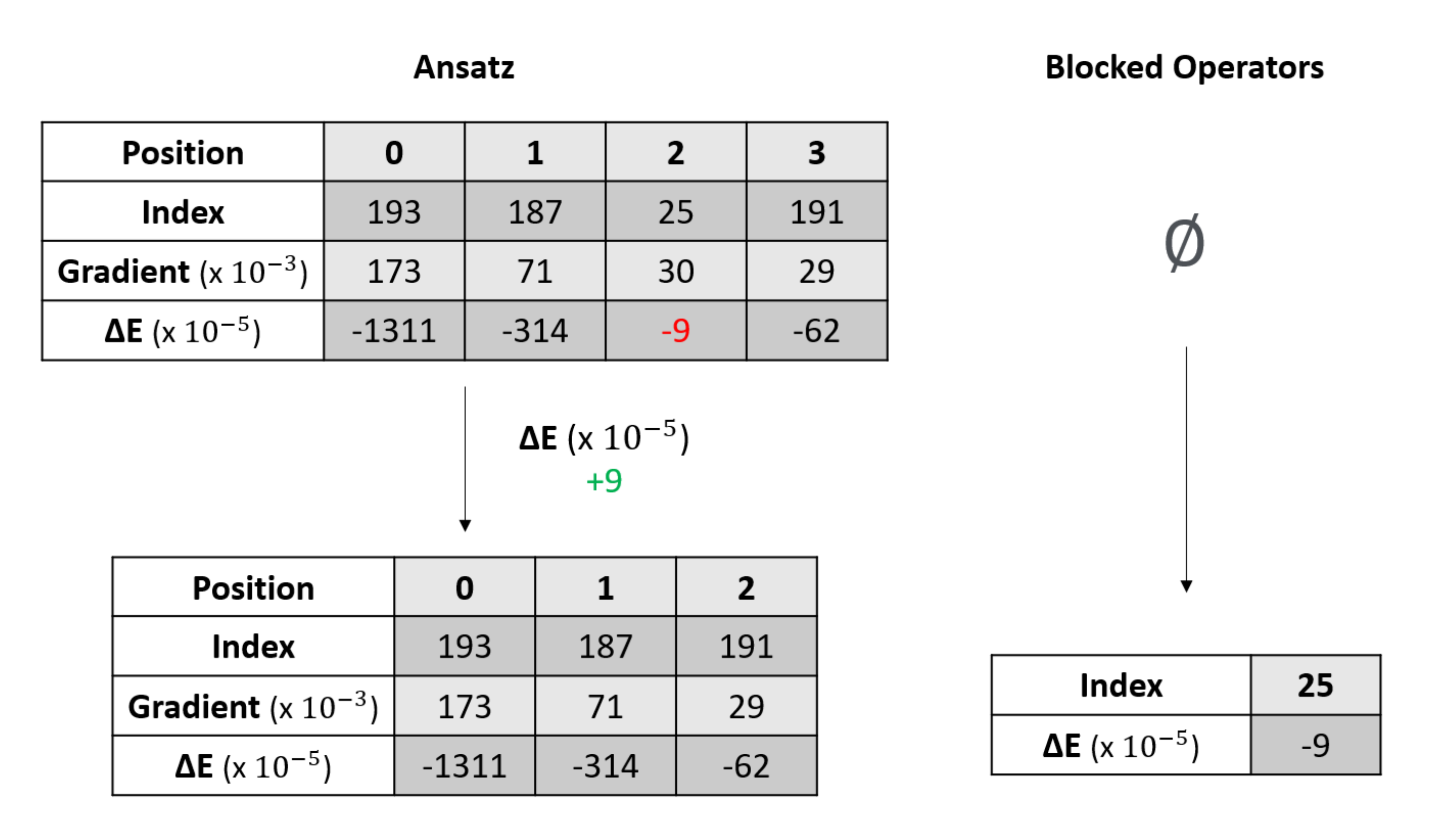}
    \caption{Relevant data structures at iteration 4.}
    \label{fig:blocking_it4}
\end{figure}

Finally, at iteration 4 (figure \ref{fig:blocking_it4}), the condition necessary to attempt to remove an operator is met. The condition reads 

\[\Delta E_i > r\Delta E_j \Leftrightarrow -9 > 0.5 \times (-62),\]

which does indeed hold: the operator 191 (\textit{j}), added in this iteration, produces a change in energy significantly higher than that of operator 25 (\textit{i}), added to the ansatz in the previous iteration.

As such, operator 25 is removed from the ansatz, and the coefficients of the remaining operators are re-optimized. From this, it is possible to calculate the increase in energy caused by removing this operator, which turns out to be $9\times10^{-5}$ a.u.. The condition for effectively removing the operator reads 

\[-\Delta E_i'>t\Delta E_i \Leftrightarrow -9>1.5\times(-9),\]

which once again holds. It can be concluded that the impact of this operator in the energy isn't much changed by the posterior addition of operator 191 into the ansatz. This will not necessarily always be the case: sometimes, one might find that removing an operator from the middle of the ansatz will cause a drastic increase in energy, even if adding that operator in the first place had barely lowered it. This is a consequence of the presence of the operator in the ansatz being crucial for the impact on the energy of the next ones, and possibly decisive for their selection.

By the end of iteration 4 operator 25 is blocked, so as to seek operators that might have a greater impact on the energy, even if their gradient is lower.

\begin{figure}[htbp]
    \centering
    \includegraphics[width=\textwidth]{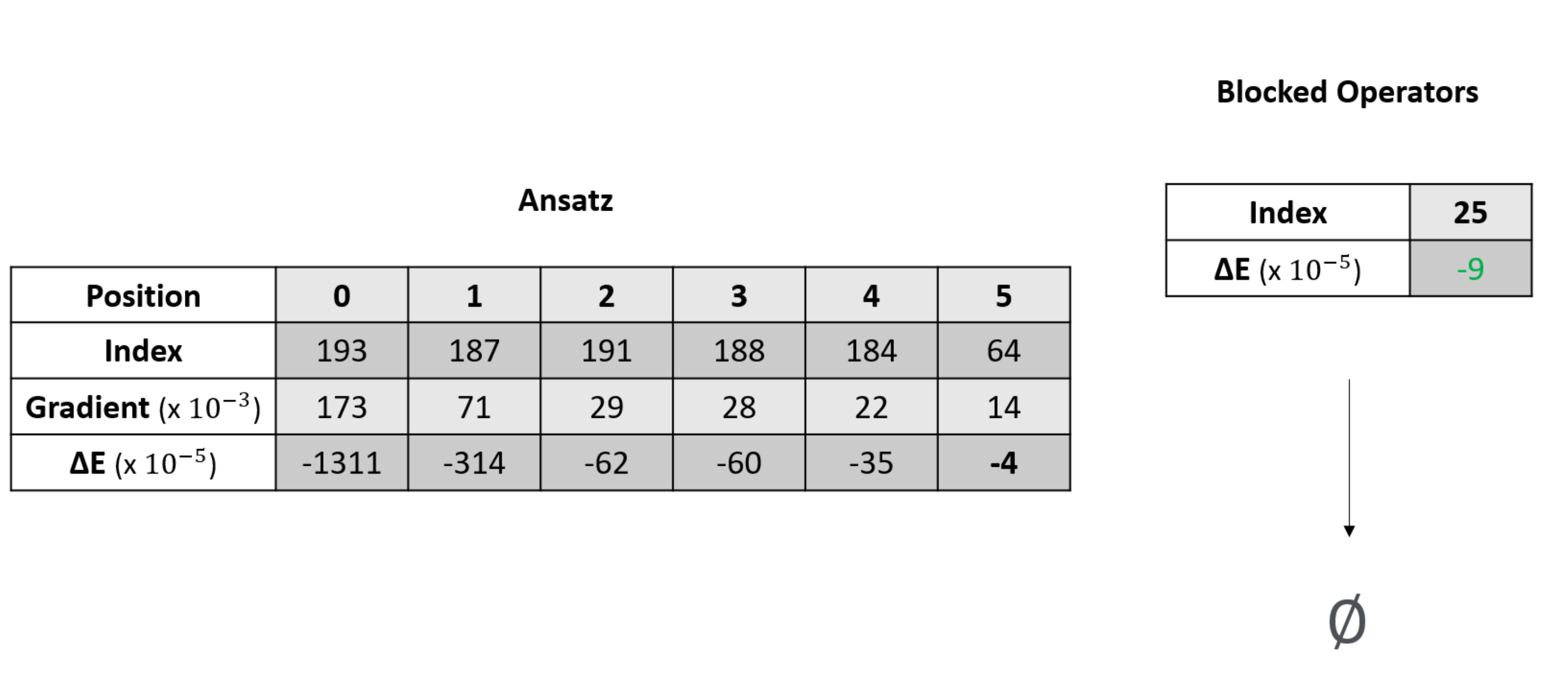}
    \caption{Relevant data structures at iteration 7.}
    \label{fig:blocking_it7}
\end{figure}

The algorithm proceeds without noteworthy developments until iteration 7. At iterations 5 and 6, operators 188 and 184 are added. They lower the energy roughly 7 and 4 times more than 25 (the blocked operator) respectively. This is a sign that the removal was worthwhile. Finally, at iteration 7, an operator is added that causes a significantly smaller change in energy. The condition for unblocking operator 25 reads

\[\Delta E_i<0.5\Delta E_k\Leftrightarrow-9<0.5\times(-4),\]

so that operator 25 is released and allowed to be selected again in future iterations. The set of blocked operators is once again found empty.

\begin{figure}[htbp]
    \centering
    \includegraphics[width=\textwidth]{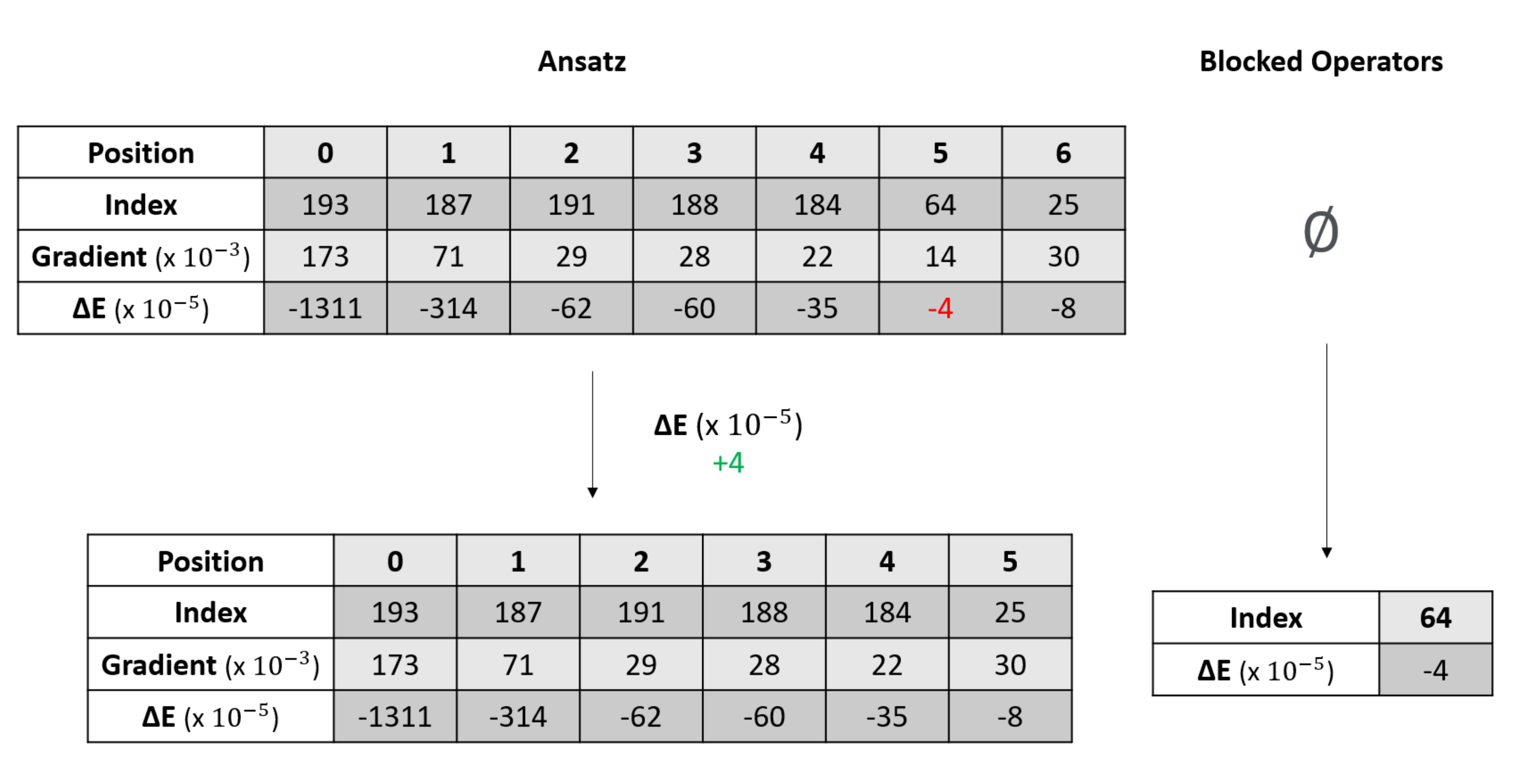}
    \caption{Relevant data structures at iteration 8.}
    \label{fig:blocking_it8}
\end{figure}

Unsurprisingly, 25 is the operator selected in the next iteration (iteration 8). It is interesting to note how the gradient and energy change of this operator are now, at iteration 8 (figure \ref{fig:blocking_it8}), very similar to those back at iteration 3 (figure \ref{fig:blocking_it3}), when the operator was first added. Despite the position in the ansatz being changed (from 2 before, to 6 now), the effect of the operator on the ansatz seems to be about the same.

Since the previous operator (64) had produced an energy change of $-4\times10^{-5}$ a.u., roughly half of that seen in iteration 8, that operator is removed from the ansatz. After re-optimizing the coefficients, the energy has only increased by $4\times10^{-5}$ a.u., so the algorithm goes forward with the removal. At the end of iteration 8, operator 64 is blocked.

By this iteration, one can already notice undesirable behaviour. Operator 25 had been removed and blocked, but hadn't it been for that, it would precede operator 64. The removal criterion is based on subsequently added operators, but if they are only added later because they had been removed from an earlier position in the ansatz, this seems rather inadequate. 

An easy solution for this would be not allowing operators to be removed upon the addition of a previously blocked operator. However, this is not the only problem with the approach, as will become clear soon.

\begin{figure}[htbp]
    \centering
    \includegraphics[width=\textwidth]{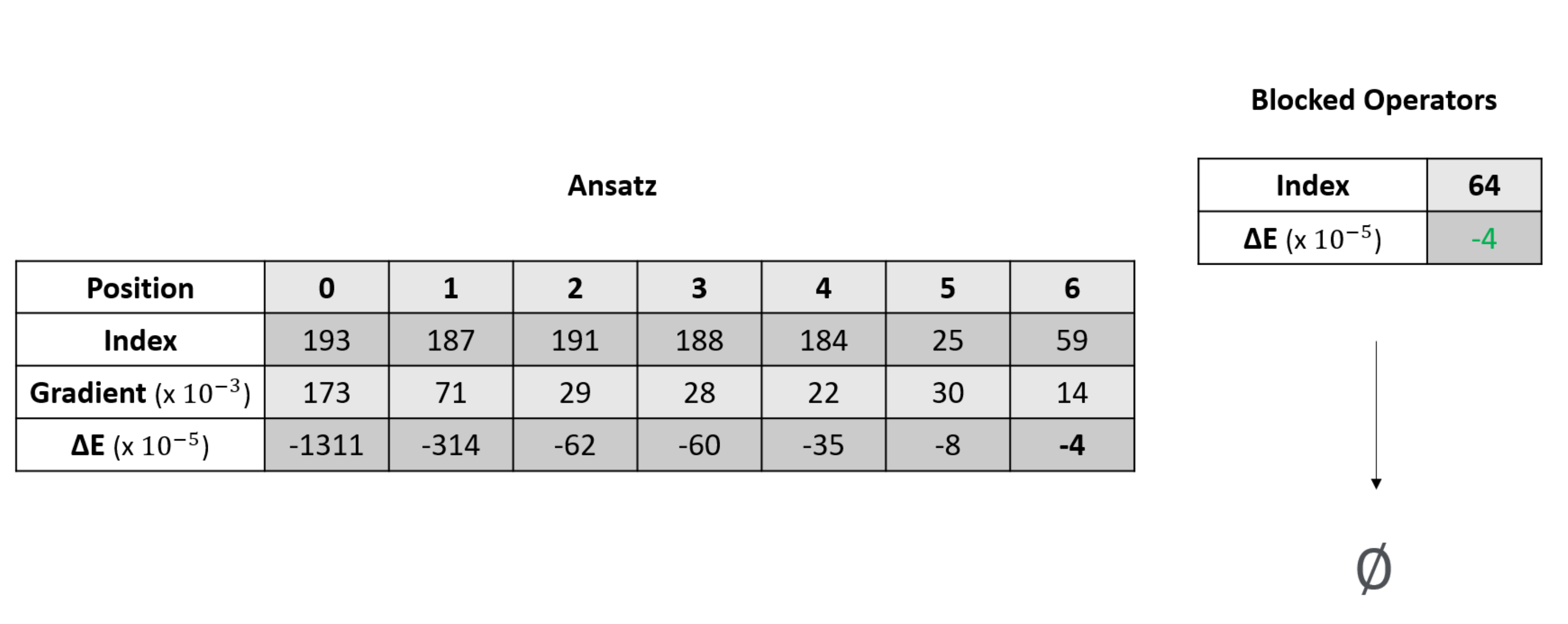}
    \caption{Relevant data structures at iteration 9.}
    \label{fig:blocking_it9}
\end{figure}

By iteration 9 (figure \ref{fig:blocking_it9}), operator 59 is added. This operator produces an energy change similar to the one operator 64, added in iteration 7 (figure \ref{fig:blocking_it7}) and blocked in the ensuing iteration (figure \ref{fig:blocking_it8}). As such, the operator is unblocked.

\begin{figure}[htbp]
    \centering
    \includegraphics[width=\textwidth]{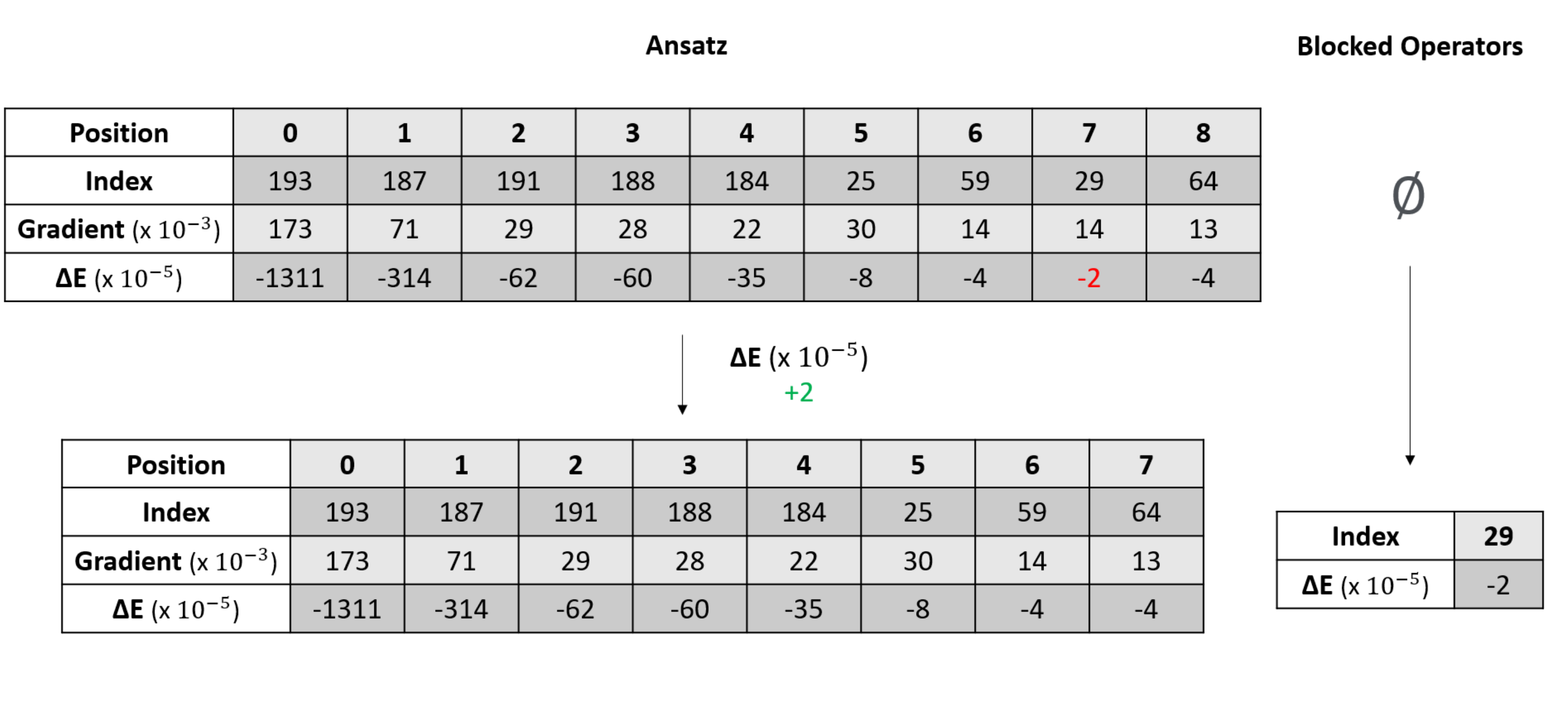}
    \caption{Relevant data structures at iteration 11.}
    \label{fig:blocking_it11}
\end{figure}

Nothing remarkable happens at iteration 10: operator 29 is added, and no operator is suitable to be removed upon this addition. That changes by iteration 11 (figure \ref{fig:blocking_it11}), when the previously removed operator 64 is added back to the ansatz. At this point, operator 29 is removed and blocked.

Once again, it should be pointed out that this is really inefficient. Operator 64 had been removed from being outperformed by an operator that belonged to an earlier point of the ansatz, and this proved to be a bad choice: it was added shortly after, with no benefits and a cost of two extra optimization steps. Now, this very operator causes the same problem: it removes operator 29 from the ansatz, even though it would have come before it if we hadn't removed it.

\begin{figure}[htbp]
    \centering
    \includegraphics[width=\textwidth]{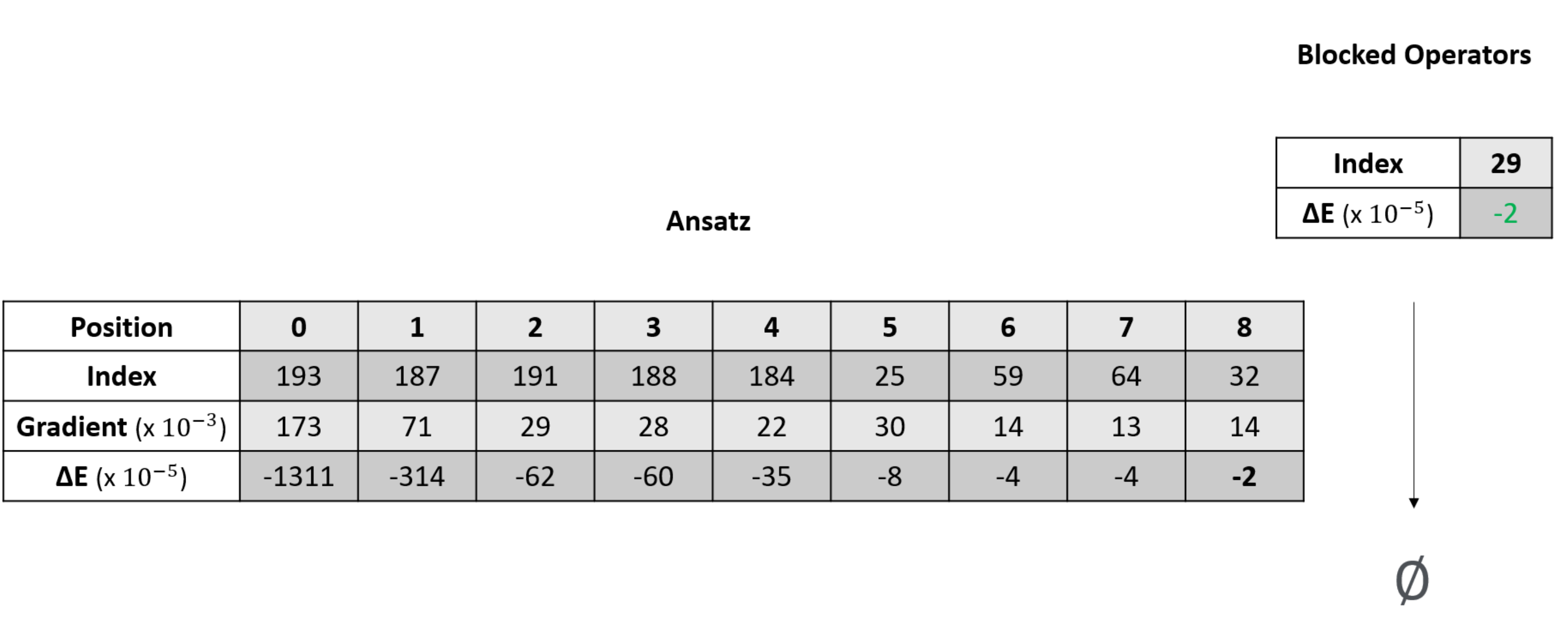}
    \caption{Relevant data structures at iteration 12.}
    \label{fig:blocking_it12}
\end{figure}

In fact, this operator (29) is unblocked right in the following iteration (iteration 12, figure \ref{fig:blocking_it12}), once another operator is added (32) that causes about the same impact in the energy. And it will be added again immediately, by iteration 13. Once again, removing the operator costed us two extra optimization steps, and had no effect other than moving it a couple of positions forward in the ansatz (with no benefit in the energy).

\begin{figure}[htbp]
    \centering
    \includegraphics[width=\textwidth]{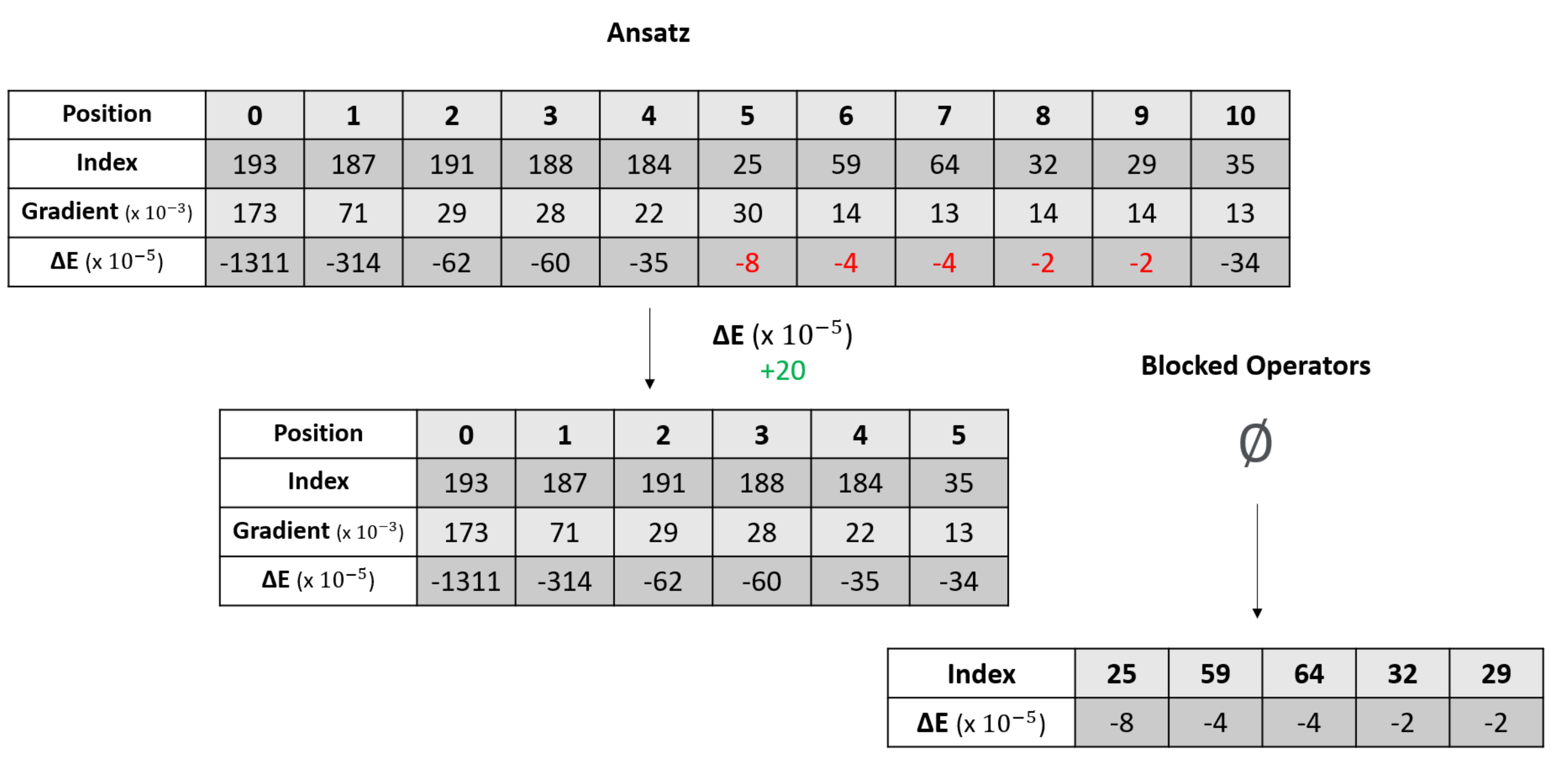}
    \caption{Relevant data structures at iteration 14.}
    \label{fig:blocking_it14}
\end{figure}

At iteration 14 (figure \ref{fig:blocking_it14}), something very interesting happens: the selected operator (35) causes an energy change of $-34\times10^{-5}$ a.u., a change so large that it is only surpassed by that of iteration 6. Despite having a lower gradient than any other operator so far, it largely outperforms any of the five last operators, with an energy change of up to almost 20 times that of its predecessors.

Thus, unprecedentedly, there are five operators that are liable to be removed. They are removed one by one, the coefficients being re-optimized each time; and each time, it is verified that the operator can be effectively removed. This is remarkable: removing these five operators caused an increase in energy roughly equal to the sum of the decreases that they had individually caused. This means that operator 35, added only in this iteration, was by far a better candidate to hold position 5 of the ansatz that any of those in positions 5-9 up to now. However, it was not picked until as late as iteration 14, because its gradient was not large enough.

Looking at figure \ref{fig:blocking_it14}, and focusing especially on the data regarding the ansatz at the begging of the iteration, one can notice that operators in positions 6-10 all had very similar gradients upon being selected. However, there are relevant differences in the energy change that they produced: and in fact, it was the operator with the lowest gradient among these (35) that caused the greatest change in energy. Moreover, selected with a gradient of more than double and holding position 5 in the ansatz, operator 25 produced only a fraction of this energy change. This further highlights the downsides of using the gradient as the selection method, and proceeding in disregard of the energy change that the selected operator actually produced.

At this point, we have an ansatz with only 6 operators that prepares a state with lower energy than any before. It is a significantly closer approximation to the ground state than the ansatz with 10 operators we had back at iteration 13, despite having a 40\% decrease in the number of variational parameters and (approximately) circuit depth.

But in spite of the the ansatz looking favorable at this point, the way it was reached was certainly not the best. By iteration 14, three operators (25, 29 and 64) have been removed from the ansatz twice. Removing an operator and adding it back represents an overhead in optimization costs (coefficients must be re-optimized upon removal and upon re-addition), so it is certainly beyond optimal to have them removed twice. 

There could be no greater sign that removed operators are being allowed back into the ansatz too soon than seeing them removed a second time. In conclusion, it seems that the strategy of blocking them until a new operator produces a change in energy similar to the one they had produced is not the most appropriate. After examining this example run, this became exceedingly evident.

This is not quite so unpredictable. After we remove an operator, we know that it didn't perform as well as could be expected given its gradient. In a first analysis, blocking it until there is a similar energy change seems adequate. However, unblocking operators based on the energy change produced by a single new operator is dangerous: there is the possibility that the new operator also impacted the energy significantly less than another one with the same gradient could. What is more, once unblocked, the operator will go back to competing with the remaining operators in the pool on the same standing. The selection method is still the gradient, so it is likely that it will be picked very soon after it's unblocked (as did happen multiple times in the example). The information that an operator had little impact in energy is being underused: we are only taking it into account to remove and block the operator. Once it is unblocked, we allow it to be selected again by its gradient, even though we now know that it is an inadequate indicator for that operator especially.

In the previous exposition, we can see that operator 25, the first being removed (iteration 4, figure \ref{fig:blocking_it4}), is allowed back into the ansatz by the addition of operator 64 (iteration 7, figure \ref{fig:blocking_it7}). However, operator 64 is later removed, signaling that it is not a good reference for unblocking others. What is more, this last operator is itself unblocked by the addition of 59 (iteration 9, figure \ref{fig:blocking_it9}), later removed (iteration 14, figure \ref{fig:blocking_it14}). By iteration 14 all the aforementioned operators are removed once again, because an operator with a smaller gradient turns out to be a far better choice than any of them. However, this operator did not have a chance to be added to the ansatz before all the removed operators were added again, because they did still have higher gradients. 

This is particularly bad considering how close the gradient of this operator (35) was to that of the bad performing operators. For example, at iteration 11 (figure \ref{fig:blocking_it11}), previously removed operator 64 is added back to the ansatz. In this iteration, the gradient of operator 64 is only 107\% that of operator 35, the above average performing one. And even though we have no way of knowing that this operator will perform so well, we do know that operator 64 did not perform as well as it could be expected from its gradient. As such, it would definitely be interesting to give an opportunity to others with close, but slightly smaller gradients.

The goal of removing terms was precisely to explore operators with lower gradients, and this goal is not being met: blocking removed operators was proved to be an inadequate option. As soon as a bad performing operator is added to the ansatz, all of the previously blocked ones are unblocked and likely to be selected. 

After this discussion, the solution that seems the most adequate is to somehow change the selection criterion so as to incorporate information regarding previously removed operators: ideally, an operator that has been removed would be less likely to be selected than others. Something as binary as blocking and unblocking operators doesn't seem to take advantage of the information that we have to the full extent.

\FloatBarrier

\subsubsection{Performance Penalty}

The best way to incorporate extra factors into the selection criterion without drastically changing it would be to impose a penalty on the gradient of the removed operators. The selected operator would be the one having the highest `effective' gradient, with this quantity reflecting the below average performance of those operators that had been previously removed. Instead of just having the operator blocked and later unblocked, we'd have its effective gradient lower than the actual gradient, in accordance with how poorly it performed. Evidently, it would be interesting to have some quantification of just how bad the energy change was, so as to apply a proportional penalty.

The purpose would be signaling to the algorithm that, even if a removed operator has a high gradient, we know that it doesn't change the energy as much as one could expect. As such, we want to attempt to add other operators to the ansatz, even if they have lower gradients, in hope of finding some that have a greater impact on the energy. The goal is using the knowledge that we now have (regarding the removed operators) to make a more informed selection: this penalty would be a way of correcting, in part, the flaws of the gradient as an indicator of the potential energy change. In those cases in which the gradient was a particularly poor indicator, to the point we removed the operator, the effective gradient would be lowered accordingly.

The aforementioned performance ratio seems interesting for this purpose. After we select a given operator and re-optimize the ansatz, we can calculate the associated performance ratio from its gradient and the resulting energy change (as compared to the previous iteration) from formula \ref{eqn:performance_ratio}.

The energy change is what we actually want from an operator: that it brings us closer to the ground energy. As the selection method, the gradient quantifies in a way how much we expect it to change the energy: we select the operator with the highest gradient hoping that it is also the one with the greatest impact on the energy. So intuitively, this ratio can be interpreted as something like

\[
\text{performance ratio} = \frac{\text{actual performance}}{\text{expected performance}}.
\]

While the performance ratio seems useful and has the nice property of being dimensionless, it doesn't allow us in itself to establish a penalty: we don't know what is a good or bad performance ratio.

Once again, the easiest solution seems to be assessing performance relative to that of the remaining operators. If we take the average over all known performance ratios (meaning, over the ratios of all of the operators ever added to the ansatz), we get a very interesting quantity: the average (absolute) energy change per unit gradient. At last, we know what energy change it is reasonable to expect from an operator with a given gradient.

One can then define a dimensionless penalty for removed operators as 

\begin{equation}
\label{eqn:penalty}
\text{penalty} = \frac{\text{performance ratio}}{\text{standard performance ratio}},
\end{equation}

where the standard performance ratio is our reference - it can just be the average taken over all known performance ratios. This metric tells us just how below average the performance of the operator was. If the energy change per unit gradient of the removed operator was half of the average, the effective gradient will be half of the actual gradient. This `sanction' will prevent it from being selected back into the ansatz as soon.

As long as there are other operators in the pool that, considering their gradient, will have a greater impact on the energy \textit{assuming that their energy change per unit gradient is average}, the removed operator will not be selected again. Only when, from the gradients of the other operators in the pool and under the same assumption, we don't expect that any will impact the energy more, will the removed one be selected. 

As it was exposed before in figure \ref{fig:performance_ratio}, the performance ratio has a tendency to decrease along the iterations; that is why it wasn't a suitable metric to decide which operators to remove. Taking the average will already soften this effect: the value will be constantly updated as more performance ratios are known, and as such will follow their tendency to decrease. To further avoid over-penalizing the removed operators, it will be interesting to use the moving average, rather than the average, as the standard performance ratio. This will prevent data from too old iterations to affect the calculation of the penalty.

The window over which we calculate the average shouldn't fall in the other extreme and be too small. If we only consider the ratio of two or three of the last added operators, and one of them performs particularly bad, it will have a significant impact on the average - thus causing the penalty to be too soft, and the removed operator to be added back too soon. It is better to be cautious, because the resulting optimization overhead can significantly worsen how fast the algorithm is to converge, by wasting iterations adding and removing the same operator multiple times. In the following examples, a window of 10 was used.

The final version of the procedure of operator removal can now be stated. The basic algorithm is the same as the original \gls{ADAPT}-\gls{VQE}, except at every iteration $j$, when a new operator is added, the following must be done:

\begin{itemize}
    \item Calculate the performance ratio of the new operator (equation \ref{eqn:performance_ratio}). Store this value (associated with the operator) and update the standard performance ratio.
    \item Go through the previous iterations and compare the energy changes produced in them with that produced in this one. If an operator added at iteration $i$ caused an energy change $\Delta E_i$ such that $\Delta E_i > r\Delta E_j$, attempt to remove this operator and re-optimize the coefficients of the ansatz without it. Calculate the energy change produced by this action, $\Delta E_i'$.
    \item Compare the increase in energy caused by removing the operator with the decrease that had been caused by adding it. If $-\Delta E_i'>t\Delta E_i$, remove the operator. In the following iterations, multiply its gradient by a penalty calculated as in equation \ref{eqn:penalty}. The penalty should be recalculated each iteration, to use the most recent value of the standard performance ratio.
\end{itemize}

As it was mentioned before, one should choose $0<r<1$ and $t>1$. The examples to follow used $r=0.5$ and $t=1.5$. The 10-iteration moving average of the performance ratio was used as the standard.

In order to compare this approach with the previous one (blocking / unblocking operators), the execution of the algorithm will be exemplified for the same test case as before, albeit more succinctly.

\begin{figure}[htbp]
    \centering
    \includegraphics[width=\textwidth]{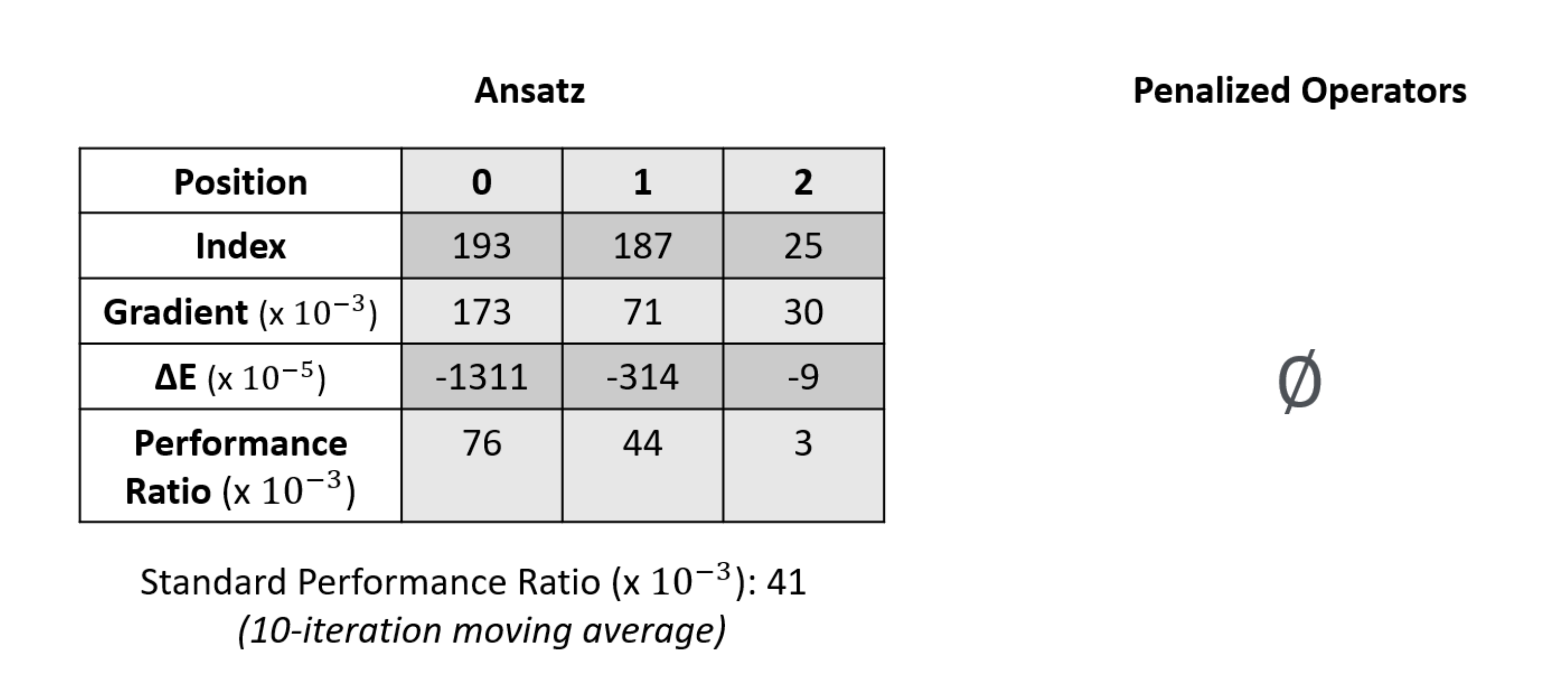}
    \caption{Relevant data structures at iteration 3 of the \gls{ADAPT}-\gls{VQE} algorithm, with an additional term removal feature based on a performance penalty. The molecule in consideration is $LiH$, and the pool the \gls{SGSD} pool.}
    \label{fig:penalty_it3}
\end{figure}

At iteration 3 (figure \ref{fig:penalty_it3}), the ansatz is exactly the same as previously (figure \ref{fig:blocking_it3}). Until this point, as was pointed out before, no operators have met the removal criterion. 

The only difference is that now we're keeping track of the performance ratio of the operators, easily calculated from their gradient upon being selected and the energy change that they caused (equation \ref{eqn:performance_ratio}). This ratio has decreased from iteration to iteration up to now. It was already known that the produced energy change and the gradient of the selected operator were strictly decreasing until iteration 3, but it is also interesting to note that the absolute value of the change in energy per unit gradient is also decreasing.

At this point, the standard performance ratio sits at $41\times10^{-3}$. Evidently, it has also been on a monotonous decrease.

\begin{figure}[htbp]
    \centering
    \includegraphics[width=\textwidth]{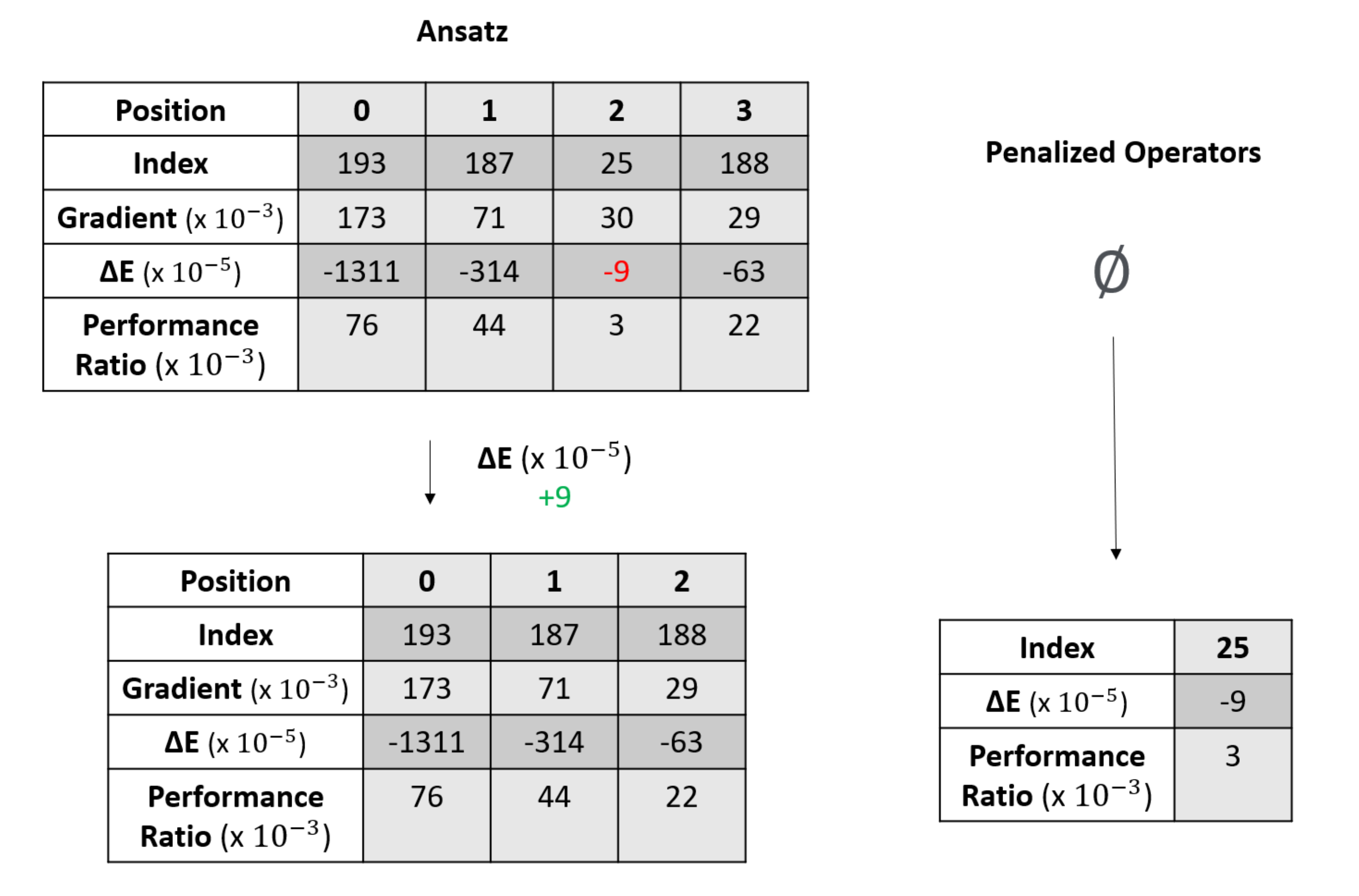}
    \caption{Relevant data structures at iteration 4.}
    \label{fig:penalty_it4}
\end{figure}

As before, the monotony is broken at iteration 4 by an operator that impacts the energy significantly more than its predecessor. We attempt to remove it and do so successfully, as the impact in the energy is still low after the addition of the last operator.

But now we don't `block' operator 25: we just mark its bad performance, and keep a note that it should be penalized in the following iterations.

\begin{figure}[htbp]
    \centering
    \includegraphics[width=\textwidth]{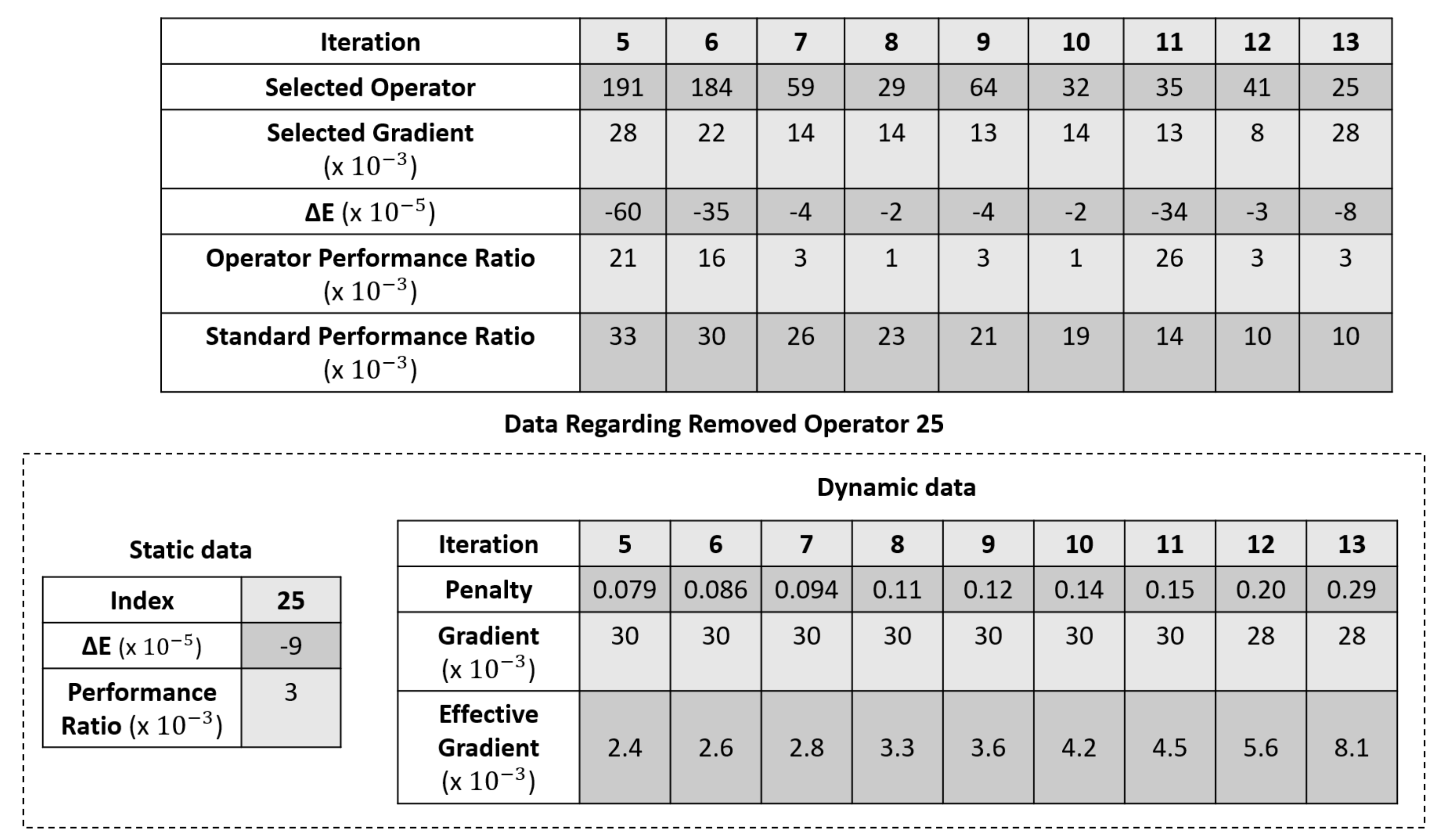}
    \caption{Evolution of the relevant data, iteration 5 through 13. On top, we have data on the evolution of the algorithm and on the selected operators. On the bottom, we have data specifically regarding operator 25, that was removed at iteration 4.}
    \label{fig:penalty_it5-13}
\end{figure}

The developments of iterations 5 to 13, with a focus on the penalized operator, are represented in figure \ref{fig:penalty_it5-13}.

The relevant static data regarding this operator are simply the energy change it produced and its performance ratio. The energy change is only presented for its relevance here, in the context of analysing and comparing the changes produced by different operators; it is not directly used by the algorithm. Only the performance ratio is used in calculating the penalty.

The rest of the data concerning operator 25 changes from iteration to iteration. In each of them, equation \ref{eqn:penalty} is used to calculate the penalty to apply to its gradient. The gradient is then multiplied by this penalty to obtain the \textit{effective gradient}. This is, as should, lower than the original value: we know that this operator produced an energy change per unit gradient under average, and want to use that information to make a more informed selection. As such, we lower its gradient by exactly how under average the energy change by unit gradient was.

As can be seen from the figure, the gradient itself might suffer slight changes, but these changes are not very relevant. What changes significantly is the penalty, as the standard performance ratio decreases with the addition of new operators. In the iterations that immediately follow its removal, this penalty greatly damages the effective gradient of the operator, causing it to be as low as 2.4 (7.9\% of the actual gradient) in iteration 5. This is because all operators added so far impacted the energy much more in proportion to the gradient than operator 25. As such, it is interesting to attempt adding other operators rather than this one into the ansatz, despite them having lower gradients: they might produce a greater energy change per unit gradient. This is, as discussed, the purpose of the penalty. Even though we didn't block operator 25, the lowered effective gradient postponed it being selected to the ansatz a second time. 

At iteration 5, the selected operator is 191, with a gradient of $28\times10^{-3}$, 93\% of the gradient of the removed operator. Adding this operator with a slightly lower gradient proved profitable: it produced an energy change almost seven times bigger than that of the sanctioned operator. Operator 184 follows in iteration 6, with a gradient of 73\% that of the removed operator. Once again, it outperforms it, impacting the energy roughly four times more.

From iteration 7 to 10, that changes: the selected operators not only have lower gradients, but also lower impacts on the energy than the removed operator. Accordingly, their performance ratio is below average. One can see that, as a consequence, the standard performance ratio (as the 10-iteration moving average) is significantly decreasing: it goes from 30 at the beginning of iteration 7 to 19 at the end of iteration 10.

The decrease softens the penalty on operator 25, but it's not enough to have it selected again yet. Even though there are 4 iterations in a row with less beneficial operators, the 10-iteration moving average of the performance ratio is still quite over that of the removed operator (because of the previous operators). That causes the algorithm to still `have hope' that a better operator with a lower gradient will be found.

That is actually the case. By iteration 11, operator 35, with a gradient of 13 (only 43\% that of operator 25), is selected and lowers the energy by $-34\times10^{-5}$ a.u. - an almost fourfold improvement as compared to operator 25. 

Obviously, there was no way to know beforehand which operators were going to be a better addition to the ansatz: all operators added in iteration 7 through 11 had very similar gradients, and before adding them, that is all the information we had about them. However, there was a great disparity in the energy changes they produced, as we had seen before.

We now see that adding poorly performing operators 59, 29, 64 and 32 (iterations 7-10) was a `necessary evil': there was no other way to find operator 35 in iteration 11 than to test them all.

This time, it was in fact avoided to add operator 25 to the ansatz too soon. We managed to find better operators not only in the iterations immediately after (5 and 6) but also in iteration 11, still before adding operator 25 once again. This didn't happen with the blocking / unblocking heuristic, in which the addition of a single operator with a poor impact on the energy was enough to allow adding the removed operator back to the ansatz. With the performance penalty, it was indeed accomplished that more operators were given a chance to before the re-addition of 25.

Finally, at iteration 13, the removed operator 25 is added back to the ansatz. This is allowed by the eased penalty (from the decrease of the performance ratio), and the fact that the gradients of the not-yet-selected operators in the pool are lower now.

It should be noted that, having the purpose of illustrating the \textit{addition} of operators, figure \ref{fig:penalty_it5-13} glosses over the fact that some of the operators added in iterations 5 to 13 are later removed from the ansatz. By the time operator 25 is added, operators 29, 32, 59 and 64 have been removed and have a sanction associated with their gradient. Because their gradient and performance ratio are lower than those of operator 25, their effective gradient is significantly lower. Even if they all belong to the set of penalized operators, they are penalized differently according to how they performed, which is a way of putting to use all the information we have.

\begin{figure}[htbp]
    \centering
    \includegraphics[width=\textwidth]{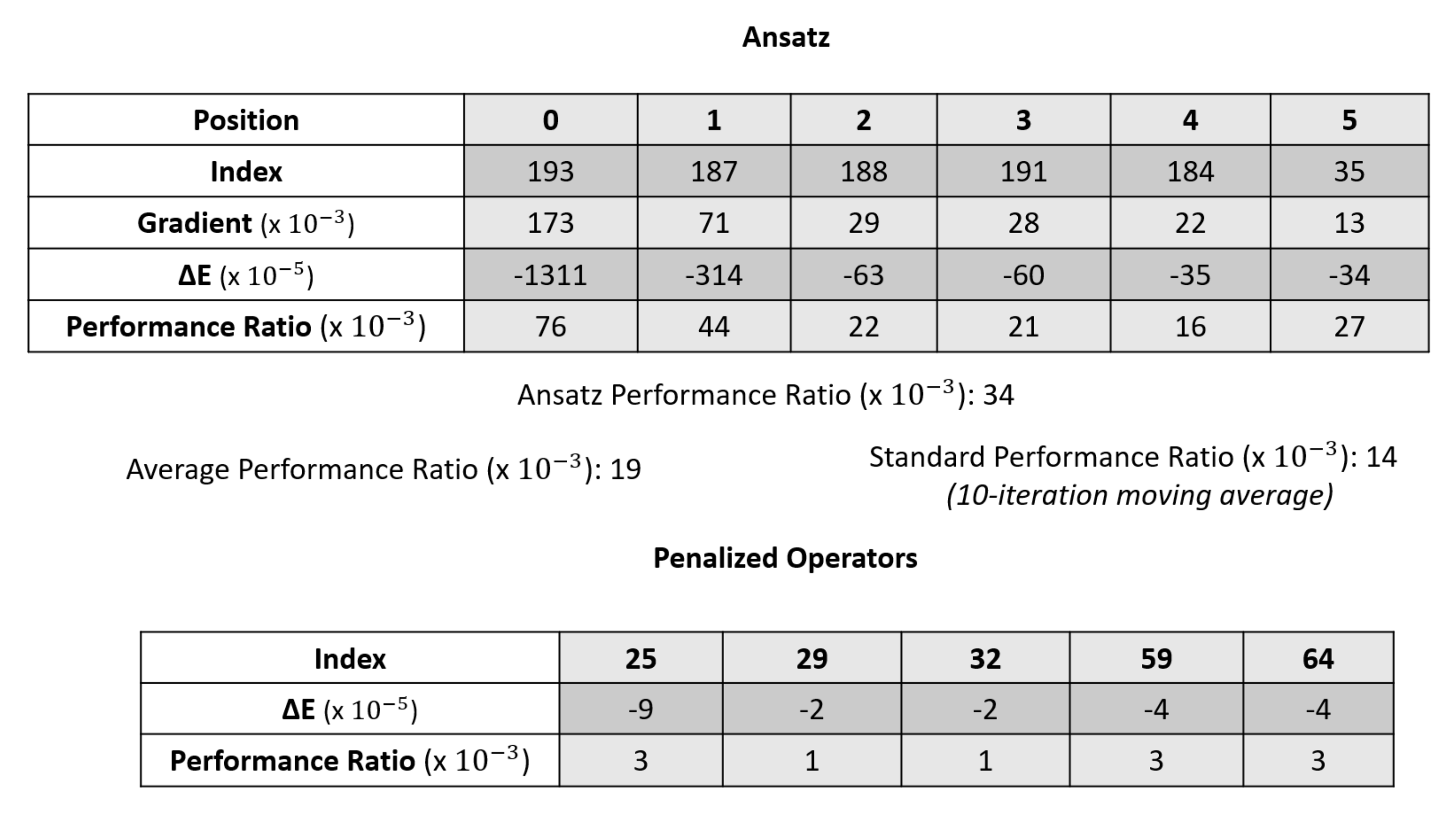}
    \caption{Relevant data structures at the end of iteration 11.}
    \label{fig:penalty_it11}
\end{figure}

Once operator 35 is added and outperforms many of its predecessors (iteration 11), the ansatz and the set of penalized operators is as represented in figure \ref{fig:penalty_it11}.

Interestingly, this is a very similar ansatz to that obtained in the end of iteration 14 using the blocking / unblocking heuristic. However, it took us three less iterations to get to it (11 now against 14 then) to get there, because we didn't waste time removing and adding operators multiple times. Previously, three iterations had been spent adding back operators that had been removed recently and would be removed again soon. What is more, there is an extra optimization per removed operator, meaning that we saved a total of six optimizations by improving how and when the operators are added back to the ansatz.

There is additional interesting information in figure \ref{fig:penalty_it11}. For example, it can be seen that the average performance ratio of the operators in the ansatz is 179\% the average taken over \textit{all} the operators we have information about (in the ansatz or penalized). In particular, it is surprising how under average the performance ratio of the penalized operators is: each of them has a ratio \textit{at least} five times lower than any of the operators in the ansatz. These operators change the energy very little in proportion to their gradient as compared to the other ones. Because their gradient is misleadingly high, the gradient as a selection method recognizes them as good options; but for our purpose (lowering the energy) they are not. Actually, \textit{all} five operators that were removed and penalized produce together an energy change of $-21\times10^{-5}$ a.u.. This is only 62\% of the energy change of the \textit{worst} operator in the ansatz \textit{alone}. Even though they represent about five times the cost in terms of the number of variational parameters and circuit depth, they are together quite less important for lowering the energy than any single operator still in the ansatz by iteration 11.

All of this is a good indicator that we have picked interesting heuristics for removing operators and handling them afterwards. However, the exemplification of a few iterations is not enough to say that the procedure is sound. 

It should be asked whether the advantage of removing terms is limited to certain molecules and/or operator pools. 

The simple Qubit Pool that results from using all tensor products of Pauli operators occurring in the fermionic excitations (with or without the anti-commutation string) is redundant, and therefore not a suitable choice here. We would often find that we removed an operator only to add one with almost exactly the same effect on the state and energy. Regardless, this pool is also excessive and was proven to be significantly larger than necessary for convergence. As for all the other pools, there seems to be no reason to believe that they would not allow for removing terms.

With a reasonable choice of pool, there are still two possibilities for the procedure not being advantageous. 

The most obvious one is the energy change in any given iteration not being significantly higher than that of any of the previous iterations. In case the removal condition is never verified, never will it be attempted to remove an operator. In this case, there is no benefit to the version of \gls{ADAPT}-\gls{VQE} with term removal, but there is also no extra cost in optimizations or measurements. If this happens, the algorithm perform exactly the same with and without term removal. However, it seems highly unlikely: even in the small example that was used before multiple terms were suitable to be removed. In a minimal basis, the representation of LiH has only 12 spin orbitals; bigger molecules will have more spin orbitals, and as a consequence more operators in the pools to choose from. It is also likely that there will be more operators with comparable gradients, in which case the selection using the gradient seems even less fitting. Additionally, only 10 iterations were considered, which is a fairly small number; in larger molecules, more iterations will be necessary to reach a low error, comprising more possibilities for removing terms.

There is, however, another possibility: having the condition for operator removal verified, but upon removal, seeing the energy increase to the point that it doesn't compensate to go through with the removal. It was a part of the method presented in this chapter that, upon removing an operator, it should be verified that the removal did not produce an increase in energy over 150\% the decrease that the operator had caused in the first place - thus avoiding to remove an operator that did not impact the energy much, but was an important part of the variational wave function once other operators were added.

Up to this point, that was never the case. In the example presented before, with lithium hydride, every time the removal condition was met the removal came to fruition. Each time, removing the operator produced a very similar change in energy to that observed when adding it. However, this shouldn't be taken for granted, as there is no reason to believe it should always be the case.

If we attempt to remove an operator and optimize the parameters of the wave function without it, only to conclude that it should not be removed on account of the energy having increased by too much, then we have absolutely no gain to compensate for the extra optimization step. Of course, this is a necessary evil because we can't know in advance what will happen. In this scenario, operator removal brings only extra cost to the algorithm, but this extra cost does not increase maximum requirements in what comes to circuit depth or number of variational parameters.

In order to better assess the performance of \gls{ADAPT}-\gls{VQE} with term removal, the next few subsections aim to present graphical results for four different molecules, and different pools.

\FloatBarrier

\subsection{Application to LiH}

The first example will be the one used in the previous subsection to illustrate the result of the different heuristics: LiH at an interatomic distance of 1.45Å, with the \gls{SGSD} pool.

\begin{figure}[htbp]
    \centering
    \includegraphics[width=\textwidth]{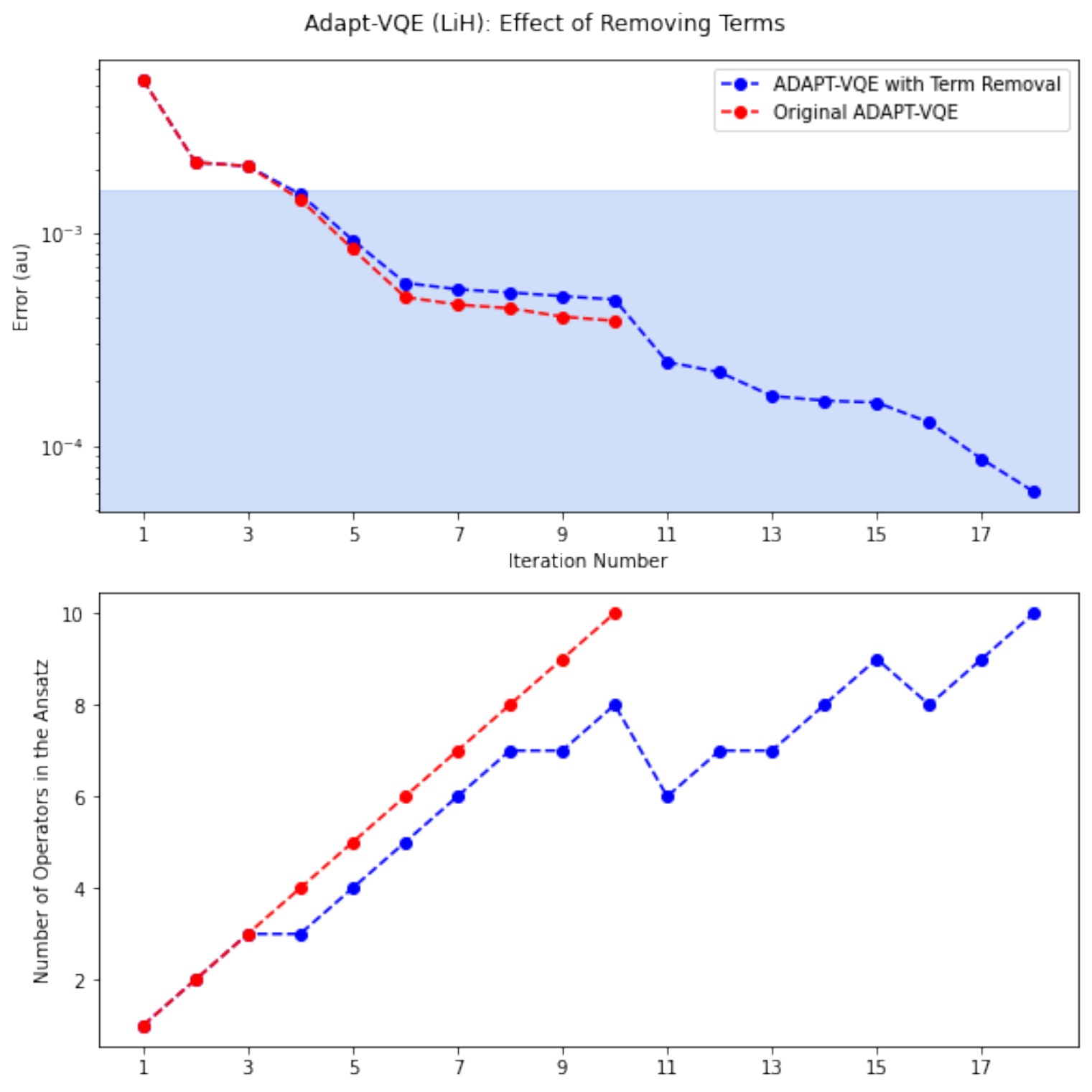}
    \caption{Plot comparing the original \gls{ADAPT}-\gls{VQE} and the version with term removal, for LiH at an interatomic distance of 1.45Å and using the \gls{SGSD} pool. The error in the energy (top) and the number of operators in the ansatz (bottom) are plotted against the iteration number. The shaded blue area on the upper plot marks the error values within chemical accuracy (less than 1kcal/mol).}
    \label{fig:LiH_singlet_error_iteration}
\end{figure}

In figure \ref{fig:LiH_singlet_error_iteration}, we can compare the performance of the original \gls{ADAPT}-\gls{VQE} with the version allowing term removal according to the method outlined before.

It is difficult to compare the two approaches, for in each iteration they may differ both in the number of operators in the ansatz and the number of optimizations performed so far. Since currently the bottleneck of these hybrid algorithms concerns the quantum computers, it was privileged to contrast clearly the maximum number of operators in the ansatz, which is closely related to the maximum circuit depth. The number of operators also corresponds to the number of variational parameters, and thus sets the dimensionality of the optimization.

As such, the plot contemplates 10 iterations of the original \gls{ADAPT}-\gls{VQE} procedure, and as many iterations of the version with term removal as occur \textit{before an iteration finishes with 11 operators}. This means that none of the iterations included for the term removing version ended with an ansatz having more elements than the ansatz of the original \gls{ADAPT}-\gls{VQE} protocol did by iteration 10. This can be confirmed in the bottom plot, that represents the number of operators in the ansatz.

Evidently, the original \gls{ADAPT}-\gls{VQE} outperforms the version with term removal in any given iteration: the latter `wastes' iterations removing terms. However, the iteration count is not in itself an interesting cost metric. 

In what concerns maximum circuit depth, we can expect a similar cost from the 10 iterations of the original protocol as from the 18 of the version with term removal. Of course, there might be small variations arising from the difference between the circuits that implement the different excitations. What is more, the particular architecture of the quantum computer will have an influence. Qubit connectivity will play a big role: operators that represent excitations between spin-orbitals corresponding to qubits that are directly connected will naturally be implemented by shallower circuits. However, there is no reason to believe that those variations would benefit either procedure. As such, assuming an average circuit depth per operator seems reasonable.

The maximum number of variational parameters is also matched between the 10 iterations of the original \gls{ADAPT}-\gls{VQE} and the 18 iterations of the version with term removal. This represents a similar maximum level of effort required of the classical optimizer, which in turn is related to the number of calls to the quantum computer (an optimization with more variational parameters will in principle require more energy evaluations). 

It is interesting to notice how removing terms helps reaching a higher accuracy than that obtained by iteration 10 of the original \gls{ADAPT}-\gls{VQE}, even though the maximum ansatz size and number of variational parameters are matched. Numerically, we find that by the time the ansatz has 10 operators, the original \gls{ADAPT}-\gls{VQE} has reached a state with an energy approximately $3.9\times10^{-4}$ a.u. over the ground state energy. For contrast, the version with term removal has an error of only $6.1\times10^{-5}$ a.u. - only 16\%, all the while keeping the maximum requirements on the quantum computer and the classical optimizer. The extra cost comes in the number of measurements and optimizations, which does not affect how viable the implementation is.

\begin{figure}[htbp]

    \centering
     \begin{subfigure}[b]{0.45\textwidth}
         \centering
         \includegraphics[width=\textwidth]{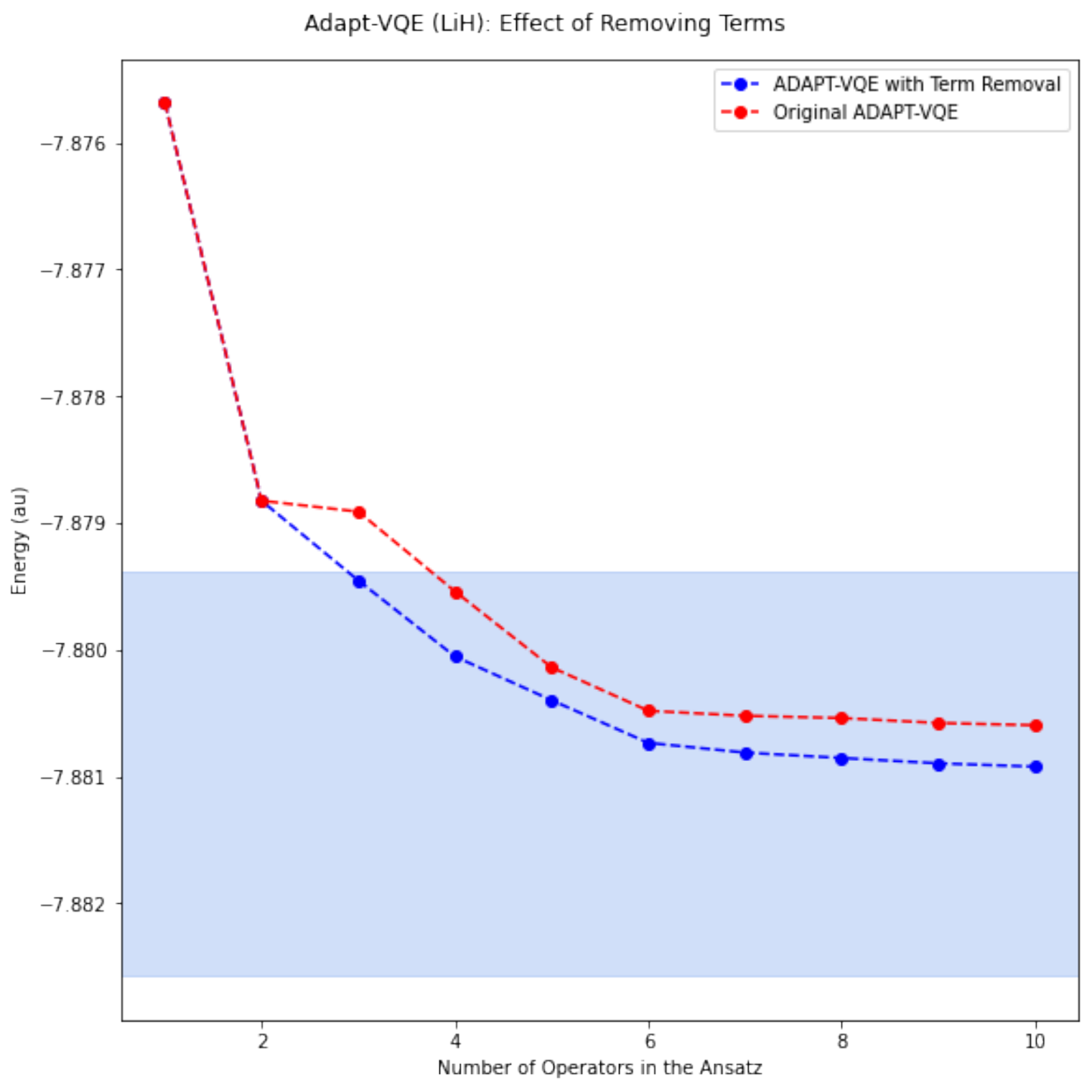}
         \caption{Energy plotted against the number of operators in the ansatz.}
         \label{fig:liH_singlet_energy_size}
     \end{subfigure}
     \hfill
     \begin{subfigure}[b]{0.45\textwidth}
         \centering
         \includegraphics[width=\textwidth]{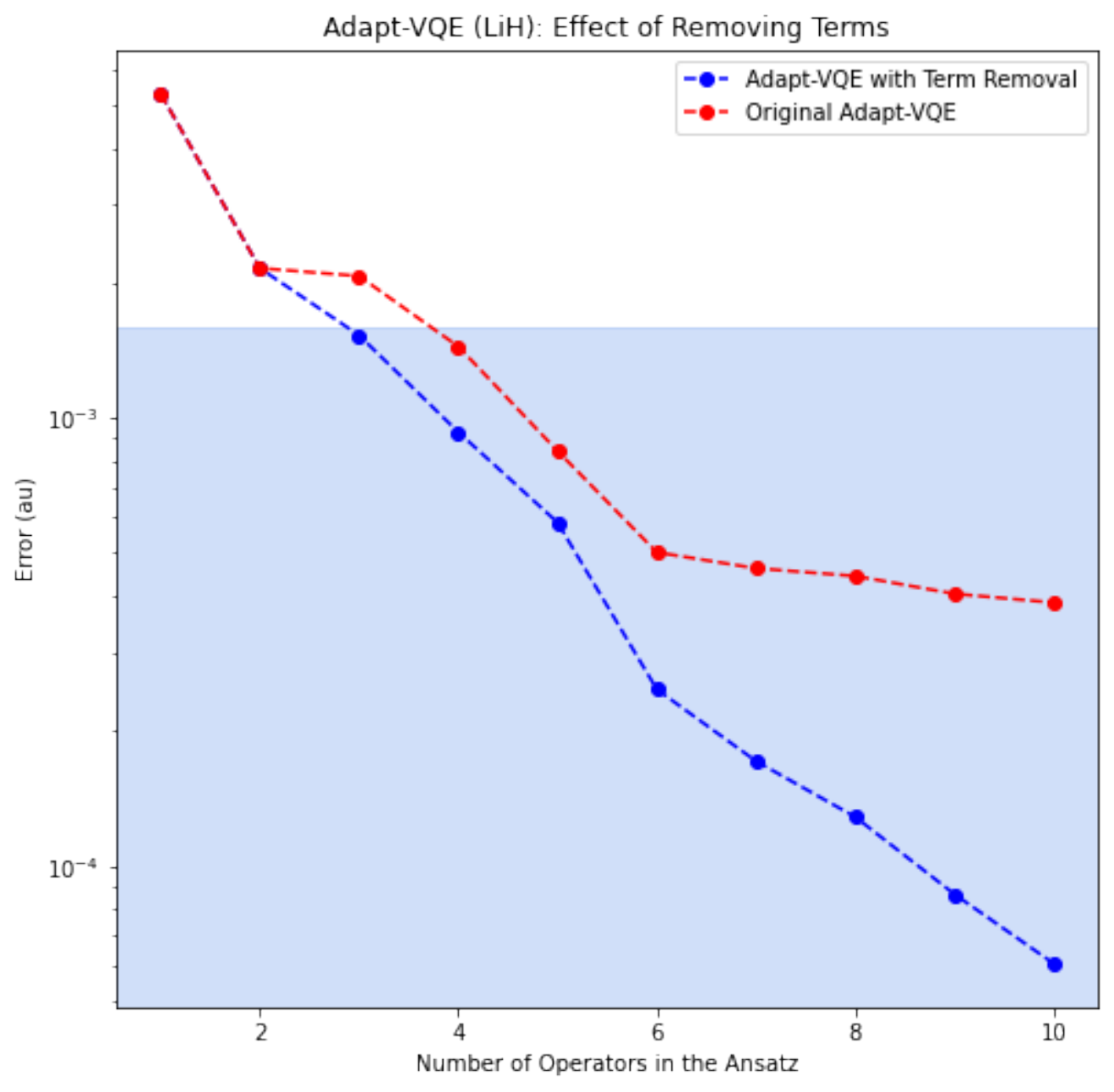}
         \caption{Error plotted against the number of operators in the ansatz.}
         \label{fig:liH_singlet_error_size}
     \end{subfigure}
    \caption{Plots comparing the energy and energy error in the original \gls{ADAPT}-\gls{VQE} and the version with term removal, for LiH at an interatomic distance of 1.45Å and using the \gls{SGSD} pool. The shaded blue areas on the plots mark the values within chemical accuracy both for the energy (less than 1kcal/mol away from the \gls{FCI} energy) and the error (less than 1kcal/mol).}
    \label{fig:LiH_singlet_size}
\end{figure}

In figure \ref{fig:LiH_singlet_size}, the energy and error plotted against the number of operators are shown. In the case of the term removing version, the ansatz in consideration for each operator number is the last occurring ansatz of that size. As before, to allow a fair comparison, iterations were only considered up until (and excluding) the first one ending with 11 operators. 

These graphs do not have present the extra cost in measurements and optimizations brought on by term removal; however, it is interesting to note how it is possible to reach a lower error with more compact ansätze. The version of \gls{ADAPT}-\gls{VQE} with term removal manages to find better ansätze for any number of operators larger than two. What is more, as it was mentioned before, the maximum demands on the quantum computer and the classical optimizer in obtaining either curve are identical.

\subsection{Application to OH-}

The second example is OH$^-$ at an interatomic distance of 1.45Å. At first, the same pool will be used: the \gls{SGSD} pool. In the STO-3G minimal basis set, the pool for this molecule contains exactly the same excitations (in terms of operator representation) as the one for LiH. This is because they have the same number of spin orbitals and we are considering all generalized excitations, so that the particle number is irrelevant in creating the pool. Evidently, what differs is the spatial part of the orbitals each qubit represents; and because the Hamiltonian is different, the cost function is different altogether.

Using the same pool for a different molecule allows obtaining some more insight as to how general the success of the method for removing terms is. After this, a different pool will be tested, to further test generality of the procedure.

\begin{figure}[htbp]
    \centering
    \includegraphics[width=\textwidth]{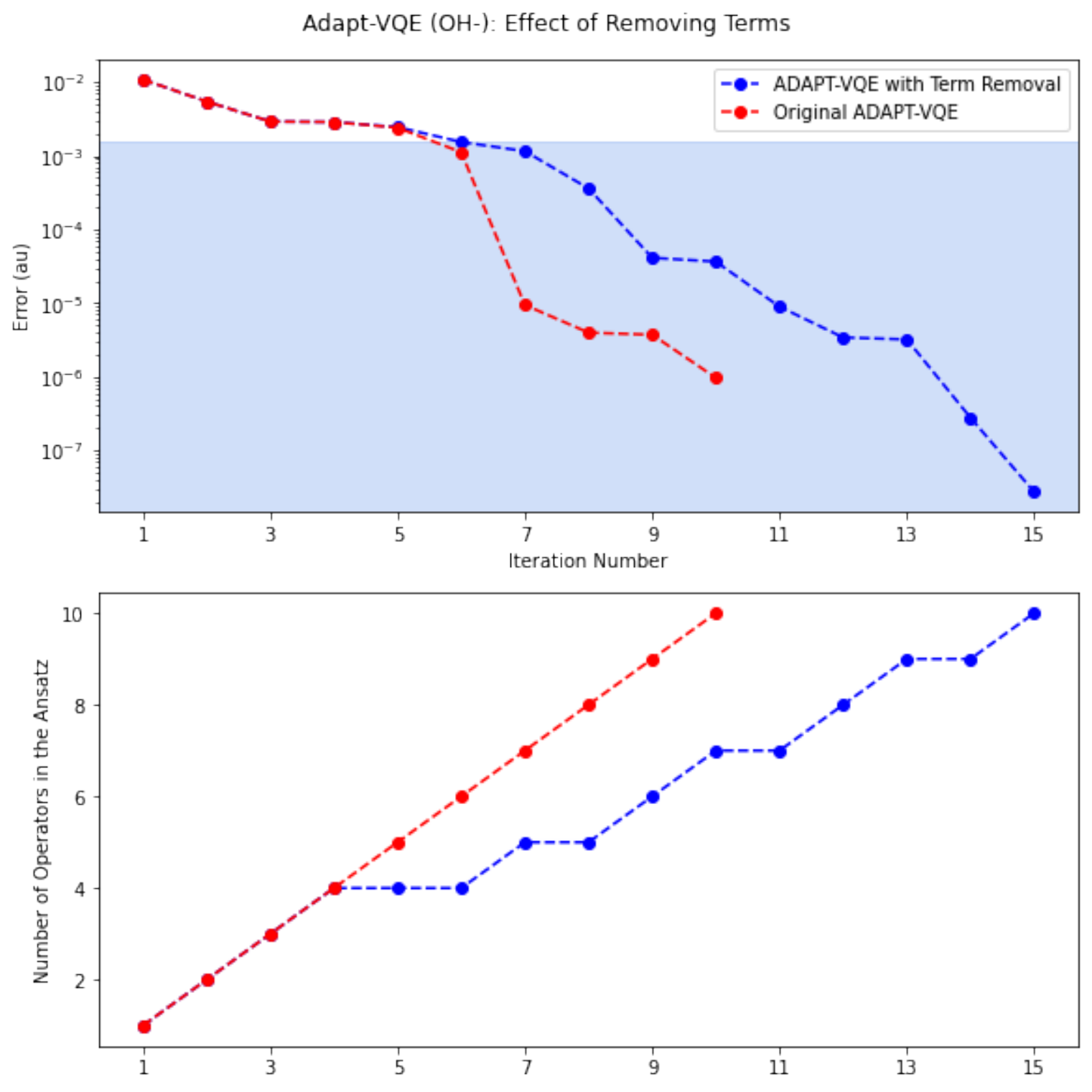}
    \caption{Plot comparing the original \gls{ADAPT}-\gls{VQE} and the version with term removal, for $\text{OH}^{-}$ at an interatomic distance of 1.45Å and using the \gls{SGSD} pool. The error in the energy (top) and the number of operators in the ansatz (bottom) are plotted against the iteration number. The shaded blue area on the upper plot marks the error values within chemical accuracy (less than 1kcal/mol).}
    \label{fig:OH-_singlet_error_iteration}
\end{figure}

As before, the plot in figure \ref{fig:OH-_singlet_error_iteration} illustrates the evolution of the two versions of the algorithm (with and without term removal). Once more, both were allowed to grow the ansatz until an iteration finished with 10 operators in it. 

Again, we see that by the time the size of the ansatz is matched between the two, the version with term removal has reached a lower error. It takes 15 iterations to get there, meaning that there were 5 operators removed in total. At this point, all operators which removal was attempted were in fact removed, because the condition to effectively remove a term was never violated. Removing an operator always resulted in about the same energy change as it had caused back in the iteration in which it was added.

The original \gls{ADAPT}-\gls{VQE} finishes with a 10-operator ansatz that prepares a state with an energy $9.7\times10^{-7}$ a.u. away from the \gls{FCI} ground energy. As for the version with term removal, its final 10-operator ansatz is associated with an error of only $2.7\times10^{-8}$ a.u. - only 2.8\%.

\begin{figure}[htbp]

    \centering
     \begin{subfigure}[b]{0.45\textwidth}
         \centering
         \includegraphics[width=\textwidth]{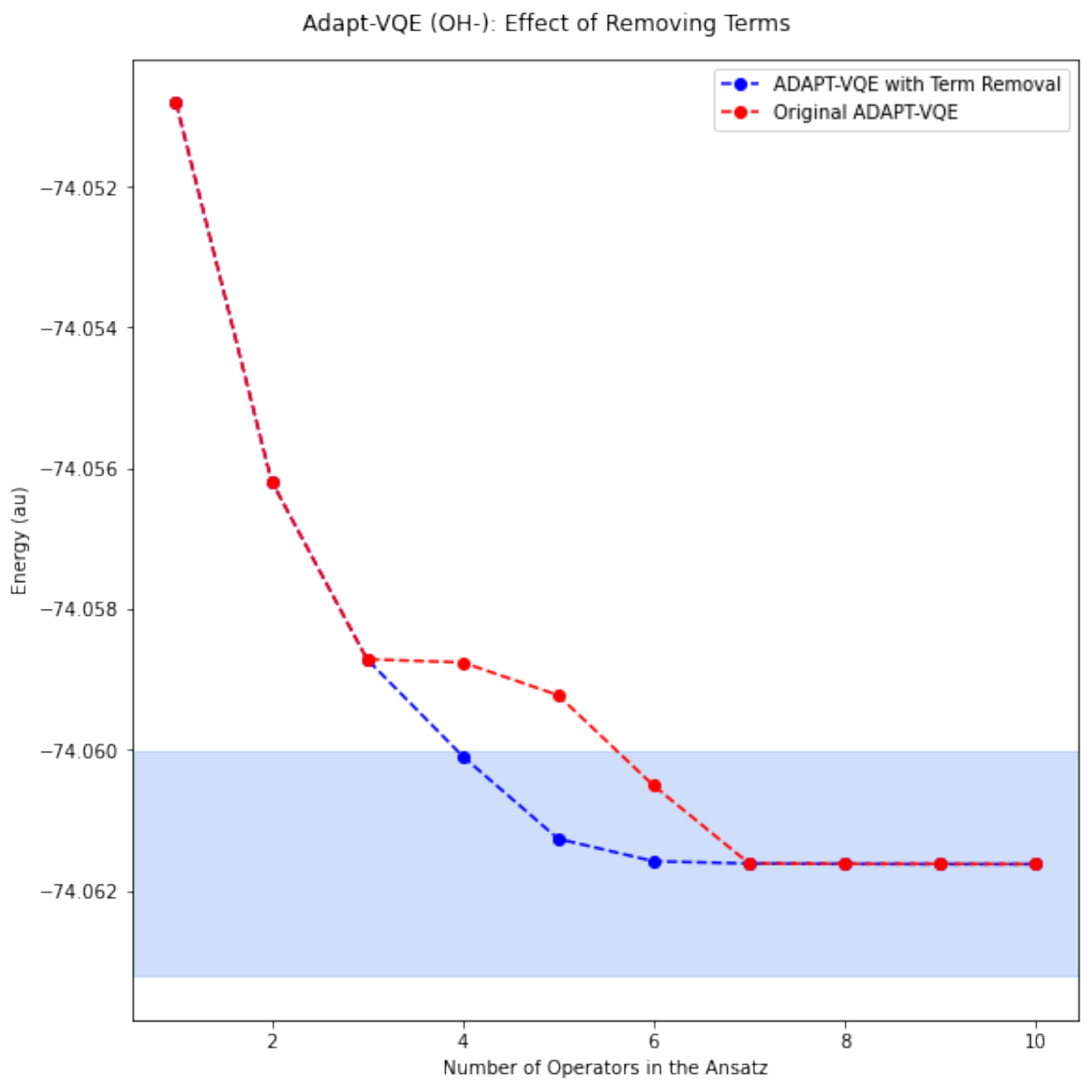}
         \caption{Energy plotted against the number of operators in the ansatz.}
         \label{fig:OH-_singlet_energy_size}
     \end{subfigure}
     \hfill
     \begin{subfigure}[b]{0.45\textwidth}
         \centering
         \includegraphics[width=\textwidth]{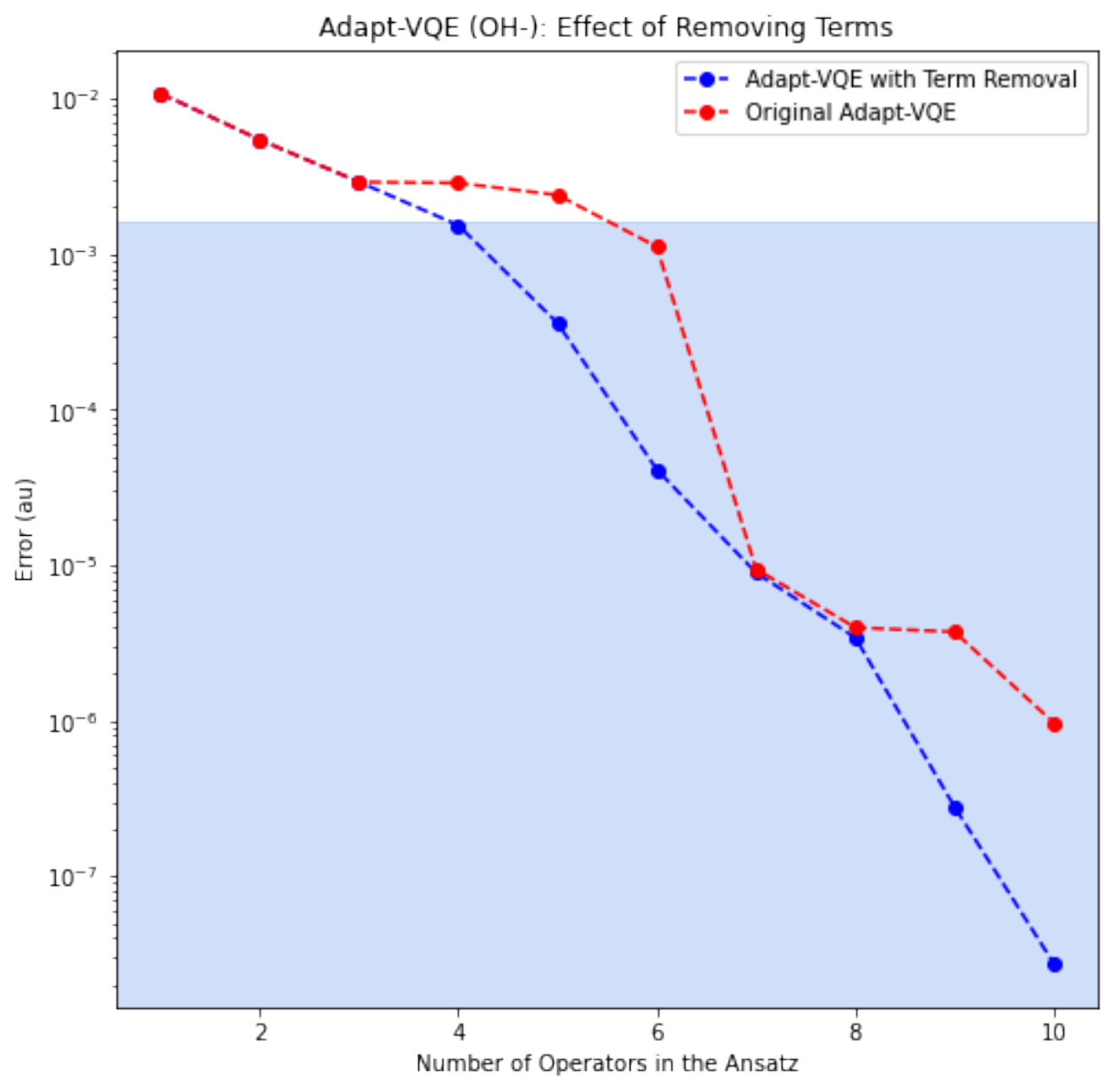}
         \caption{Error plotted against the number of operators in the ansatz.}
         \label{fig:OH-_singlet_error_size}
     \end{subfigure}
    \caption{Plots comparing the energy and energy error in the original \gls{ADAPT}-\gls{VQE} and the version with term removal, for $\text{OH}^{-}$ at an interatomic distance of 1.45Å and using the \gls{SGSD} pool. The shaded blue areas on the plots mark the values within chemical accuracy both for the energy (less than 1kcal/mol away from the \gls{FCI} energy) and the error (less than 1kcal/mol).}
    \label{fig:OH-_singlet_size}
\end{figure}

The energy plotted as a function of the number of operators in the ansatz is in figure \ref{fig:OH-_singlet_energy_size} (the ansatz considered for each size was once again the last ansatz of that size). Interestingly, it seems like those iterations in which the energy does not change much as compared to the surrounding ones (iterations 4 and 5) are entirely avoidable. One could have imagined that the slowdown was inevitable, and that no operator would be capable of changing the energy more than the one chosen. However, this is not the case. Better choices of operators can help convergence be significantly faster as a function of the ansatz size.

In figure \ref{fig:OH-_singlet_error_size}, we can remark on how the version with term removal succeeds in finding a better ansatz for many numbers of operators. Interestingly, in contrast with what happened for LiH, one can see that the 7 and 8-operator ansätze are very comparable between the two versions of the algorithms. This is a sign that, for those numbers of operators, the simple gradient-based selection is nearly optimal.

In the following, results will be presented for the same molecule, but using the Eight Pool introduced in the previous chapter. The purpose is verifying whether the success of removing terms might be dependent on the pool; once the fermionic anti-commutation strings are removed and simpler operators are used, results might differ.

\begin{figure}[htbp]
    \centering
    \includegraphics[width=\textwidth]{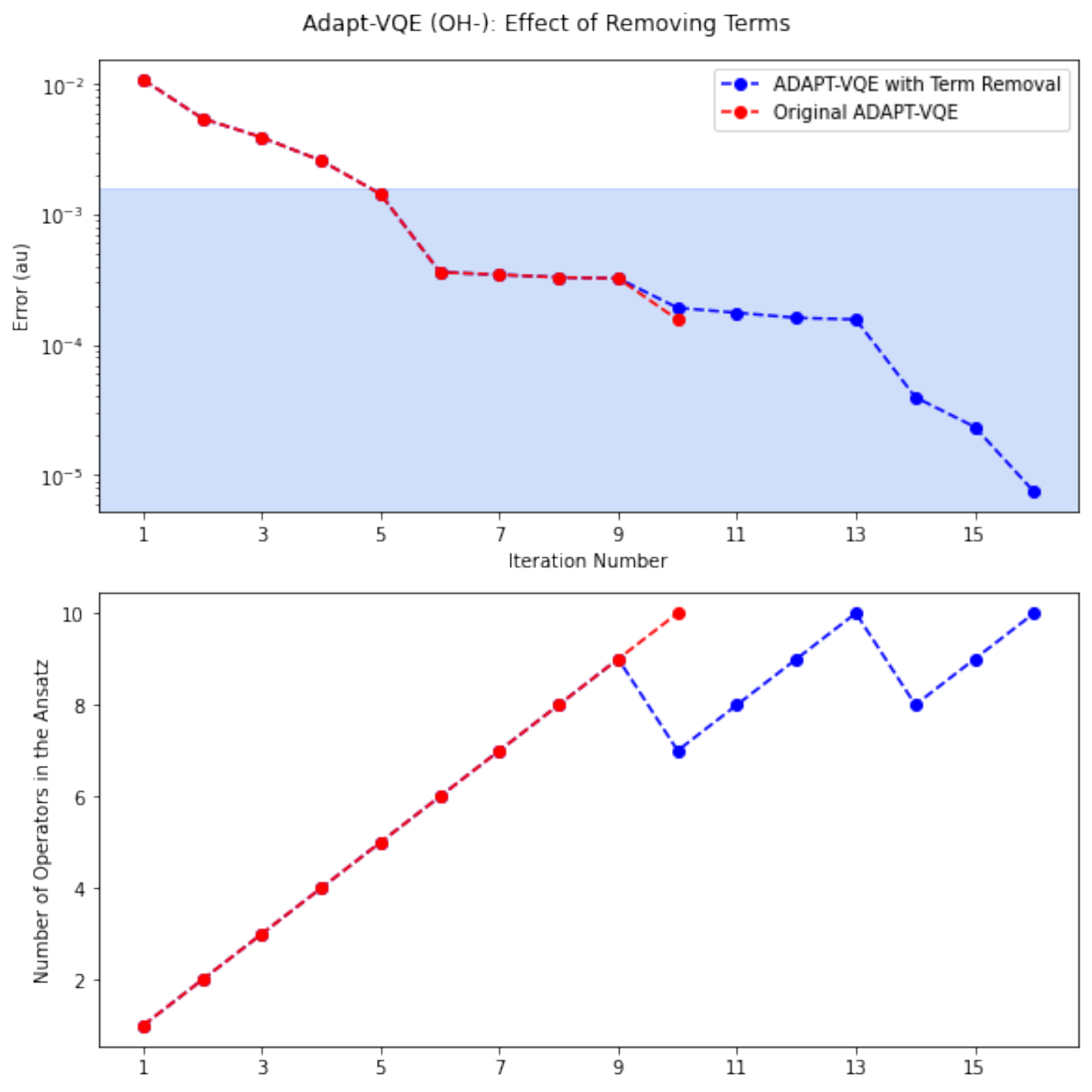}
    \caption{Plot comparing the original \gls{ADAPT}-\gls{VQE} and the version with term removal, for $\text{OH}^{-}$ at an interatomic distance of 1.45Å and using the Eight Pool. The error in the energy (top) and the number of operators in the ansatz (bottom) are plotted against the iteration number. The shaded blue area on the upper plot marks the error values within chemical accuracy (less than 1kcal/mol).}
    \label{fig:OH-_eightPool_error_iteration}
\end{figure}

In figure \ref{fig:OH-_eightPool_error_iteration}, we can see that significant advantage can be achieved with this pool as well. By iteration 16 of \gls{ADAPT}-\gls{VQE} with term removal, we have a 10-operator ansatz that prepares a state with an energy $7.4\times10^{-6}$ a.u. away from the ground energy - only 4.6\% of the error of the 10-operator ansatz that the original \gls{ADAPT}-\gls{VQE} algorithm has reached by iteration 10 ($1.6\times10^{-4}$ a.u.).

\begin{figure}[htbp]
    \centering
     \begin{subfigure}[b]{0.45\textwidth}
         \centering
         \includegraphics[width=\textwidth]{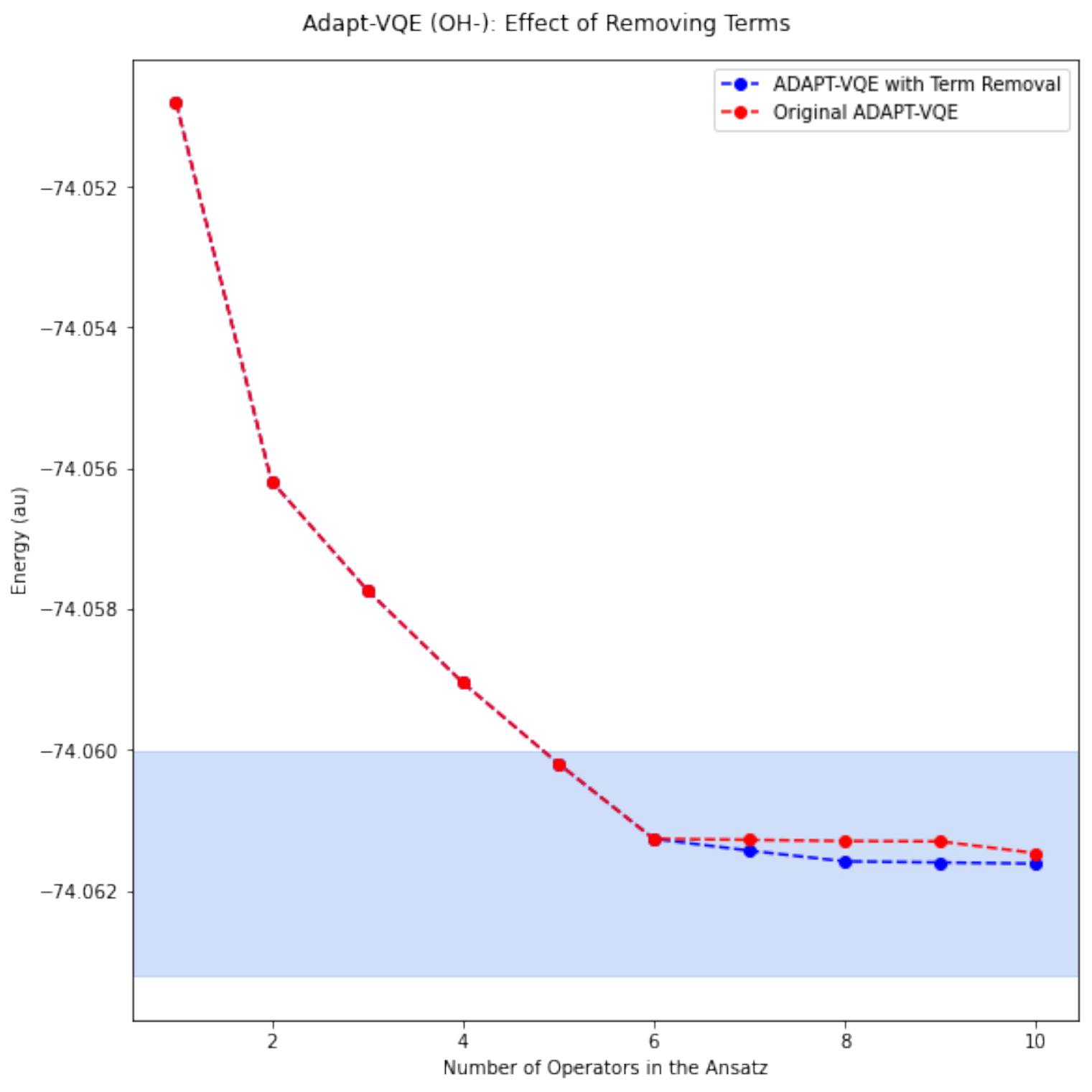}
         \caption{Energy plotted against the number of operators in the ansatz.}
         \label{fig:OH-_eightPool_energy_size}
     \end{subfigure}
     \hfill
     \begin{subfigure}[b]{0.45\textwidth}
         \centering
         \includegraphics[width=\textwidth]{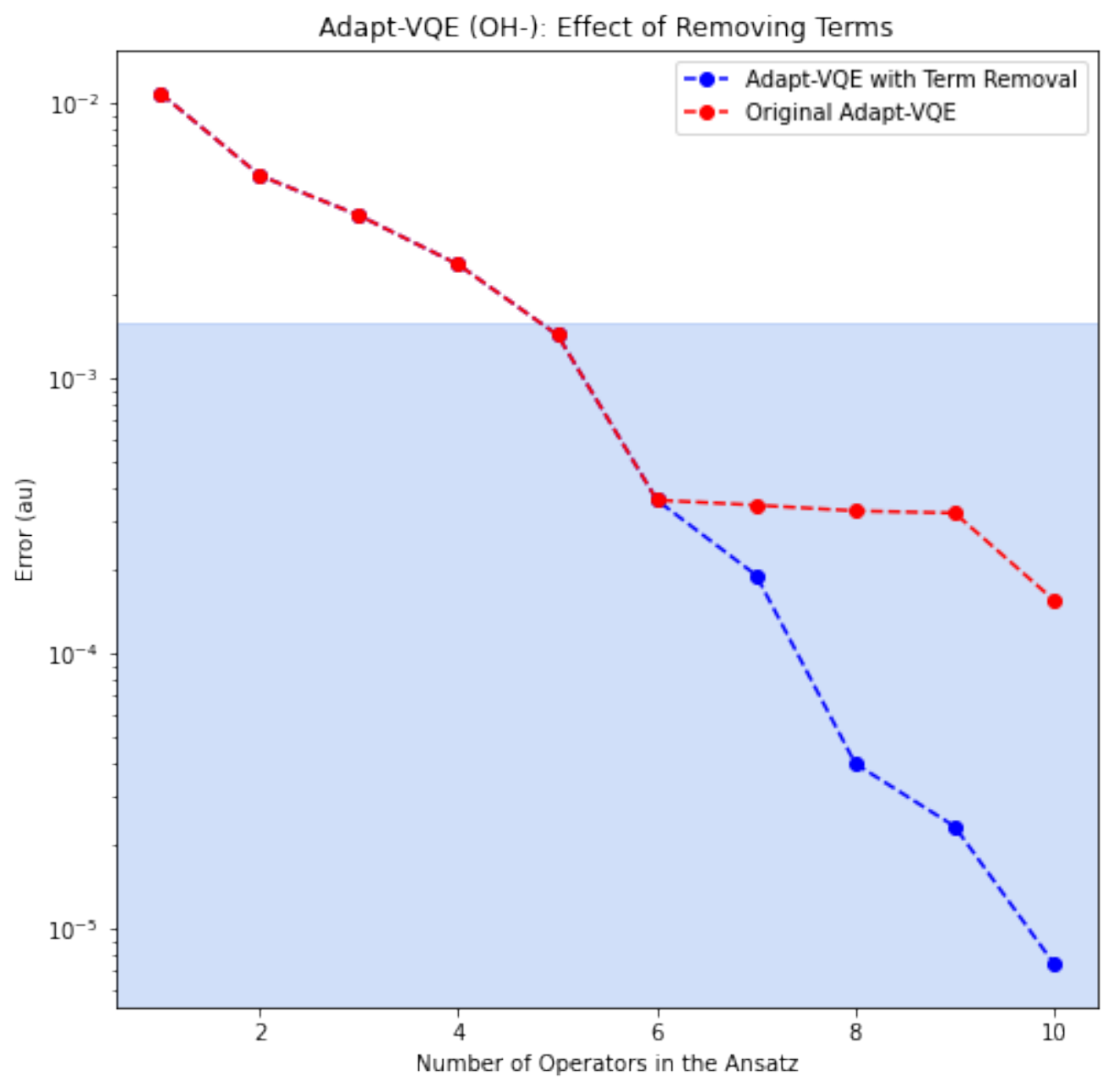}
         \caption{Error plotted against the number of operators in the ansatz.}
         \label{fig:OH-_eightPool_error_size}
     \end{subfigure}
    \caption{Plots comparing the energy and energy error in the original \gls{ADAPT}-\gls{VQE} and the version with term removal, for $\text{OH}^{-}$ at an interatomic distance of 1.45Å and using the Eight Pool. The shaded blue areas on the plots mark the values within chemical accuracy both for the energy (less than 1kcal/mol away from the \gls{FCI} energy) and the error (less than 1kcal/mol).}
    \label{fig:OH-_eightPool_size}
\end{figure}

In figure \ref{fig:OH-_eightPool_size}, we can see that removing terms allows reaching a better ansatz for any number of operators after six. The difference in energy seems small on \ref{fig:OH-_eightPool_energy_size}, which is a consequence of the linear scale being unfit for showing the small but important changes that the energy suffers by the last iterations; once the error is plotted on a logarithmic scale (figure \ref{fig:OH-_eightPool_error_size}), the difference becomes evident.

\FloatBarrier
\section{Conservative Ansatz Growth}
\label{s:conservative_growth}

An alternative approach to removing operators is to attempt to be more cautious in adding them to the ansatz in the first place. Once again, the idea is not to fully trust the gradient norm as an indicator of the possible impact on the energy. However, here the correction is done preemptively, rather than \textit{after} the operators have been added to the ansatz.

Instead of blindly accepting the operator with the highest gradient into the ansatz in each iteration, this approach tests several operators before settling for one. At each iteration, the N operators with the highest gradients are independently added to the ansatz, and their coefficients are optimized. In the end, only the one that lowered the energy the most is effectively added. In the simulations to follow, $N=5$ was used. Evidently, this implies a 5-fold increase in the number of optimizations that must be performed.

This method was employed in \gls{QEB}-ADAPT-\gls{VQE} \cite{yordanov2020}.

\subsection{Comparison with Operator Removal}

\subsubsection{Application to LiH}
In order to compare these three approaches (original; term removal; conservative ansatz growth), simulations involving three different molecules were done.

\begin{figure}[htbp]
    \centering
    \includegraphics[width=0.9\columnwidth]{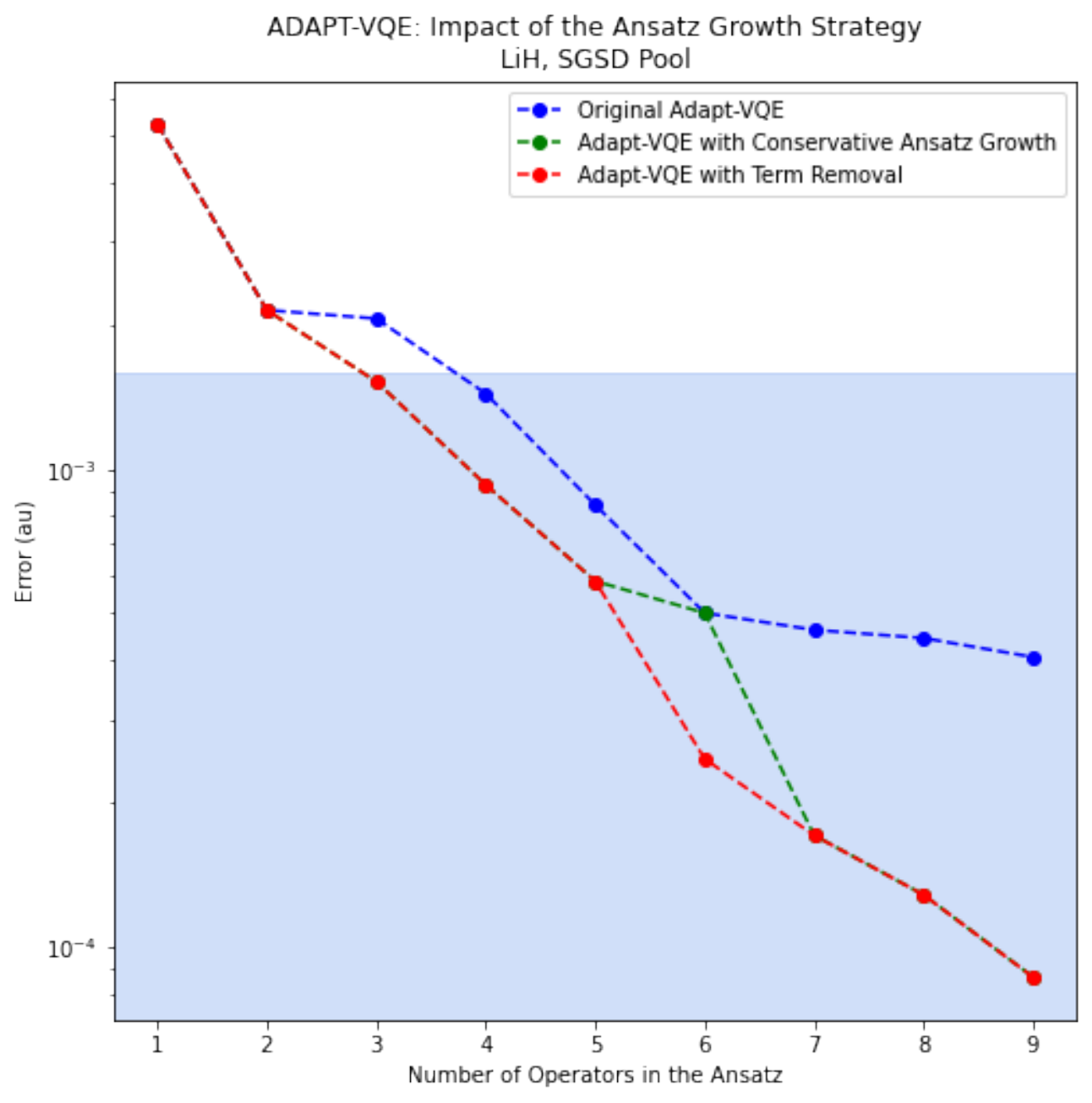}
    \caption{Plot showing the minimum error reached for each ansatz size, for multiple ansatz growth strategies. The molecule in case is LiH at an interatomic distance of 1.45Å. The \gls{SGSD} pool was used. In the points where the green curve, corresponding to conservative ansatz growth, is not visible, it lies directly under the term removal curve.}
    \label{fig:lih_sgsd_growth_strategy}
\end{figure}

Figure \ref{fig:lih_sgsd_growth_strategy} shows the familiar example of $LiH$, now complete with all three ansatz growing approaches. It is remarkable how similarly the term removal and conservative ansatz growth strategies perform. The latter is significantly behind at iteration 6, but recovers immediately after; further, even this difference would be effaced if one allowed a greater N in the strategy. In the other iterations, only one of the curves is visible because they overlap.

The figure does not show the extra overhead in the optimizations. As for the conservative ansatz growth strategy, it is straightforward that it implies a 5-fold increase in the total number of optimizations. The term removal version is not as easy to analyze, because the number of optimizations depends on the number of terms that were removed. In this case, term removal implied 16 extra optimizations to achieve a 10-operator ansatz. So the total number of optimizations for the original \gls{ADAPT}-\gls{VQE}, the version with term removal, and the version with conservative ansatz growth was respectively 10, 26, and 50. The maximum circuit depth, as well as the maximum number of variational parameters, are roughly the same for all. The term removal and conservative ansatz growth strategies achieve a greater accuracy for comparable circuit depth.

\subsubsection{Application to OH-}

\begin{figure}[htbp]
    \centering
    \includegraphics[width=0.9\columnwidth]{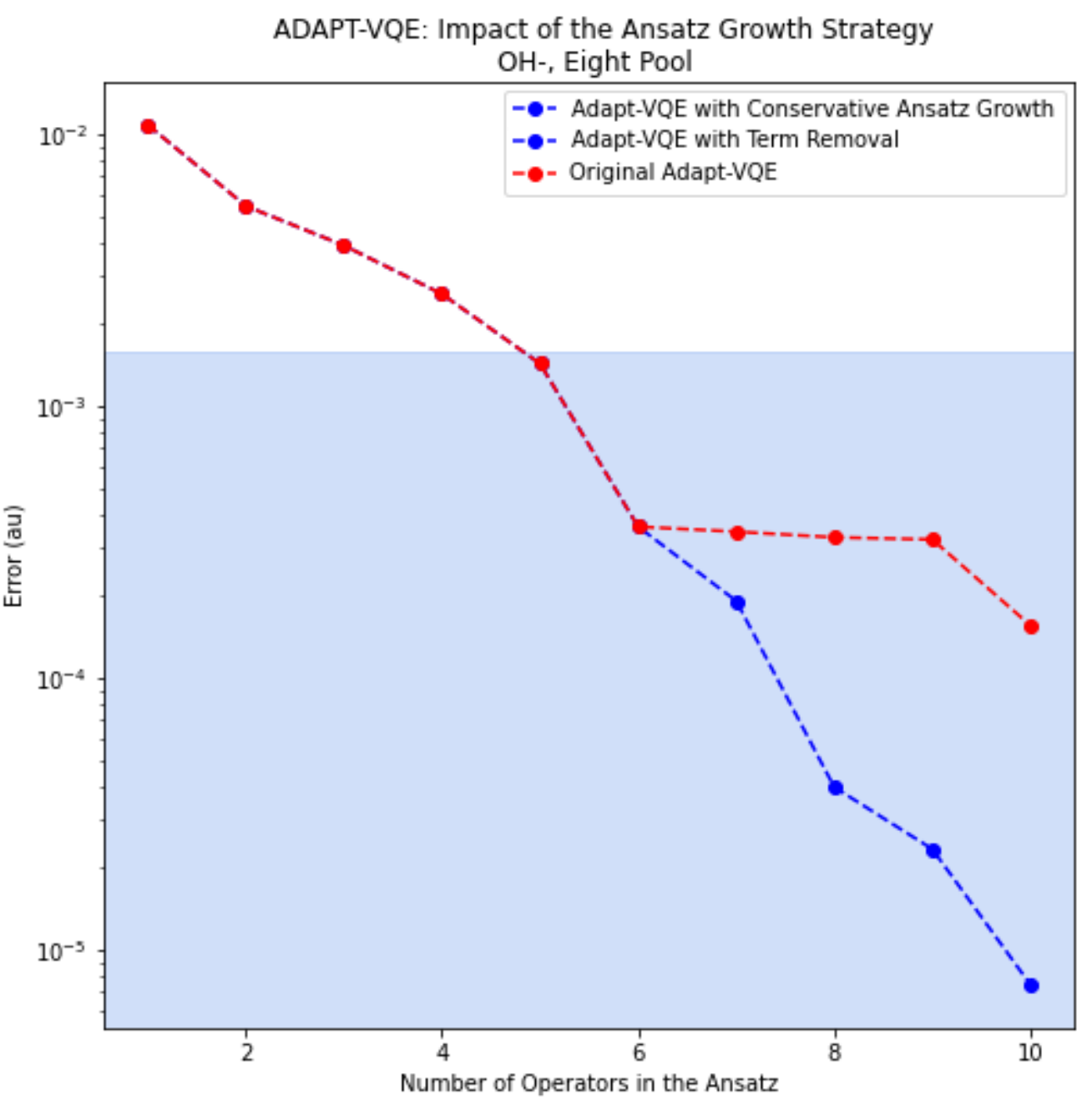}
    \caption{Plot showing the minimum error reached for each ansatz size, for multiple ansatz growth strategies. The molecule in case is $OH^-$ at an interatomic distance of 1.45Å. The Eight Pool was used. The term removal and conservative ansatz growth strategies correspond to the same color because the curves coincide.}
    \label{fig:lOH_eightPool_ansatz_growth_strategies}
\end{figure}

Figure \ref{fig:lOH_eightPool_ansatz_growth_strategies} presents the results for a different molecule ($OH^-$) and a different pool (the Eight Pool). The term removal curve coincides with the conservative ansatz growth curve. Here the original \gls{ADAPT}-\gls{VQE}, the version with term removal, and the version with conservative ansatz growth required respectively 10, 22, and 50 optimizations. The maximum circuit depth, as well as the maximum number of variational parameters, are again roughly the same for all. By the end, for the same size of ansatz, the term removal and conservative ansatz growth strategies reach an error of only 4.6\% that of the original \gls{ADAPT}-\gls{VQE} algorithm.

\subsubsection{Application to H4}
\begin{figure}[htbp]
    \centering
    \includegraphics[width=0.9\columnwidth]{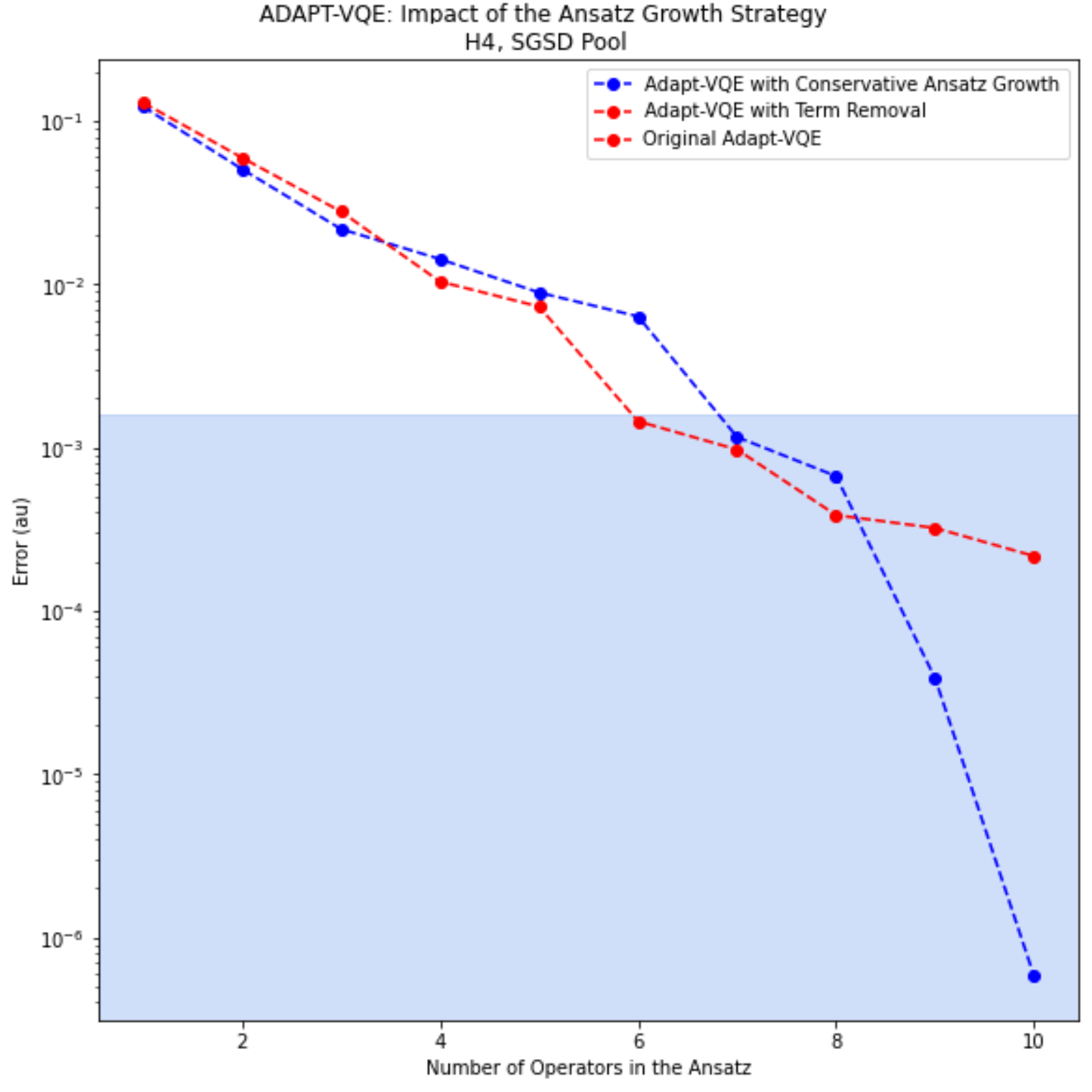}
    \caption{Plot showing the minimum error reached for each ansatz size, for multiple ansatz growth strategies. The molecule in case is $H_4$ at an interatomic distance of 1.45Å. The original \gls{ADAPT}-\gls{VQE} and the version with term removal correspond to the same color because the curves coincide.}
    \label{fig:h4_ansatz_growth_strategies}
\end{figure}

Figure \ref{fig:h4_ansatz_growth_strategies} presents the results for yet another molecule, $H_4$. Interestingly, the results are very different here - the term removal version is not visible, but now because it lies directly under the \textit{original} \gls{ADAPT}-\gls{VQE} curve. There are two terms that fit the criterion for being removed, but on the last step of the removal procedure, it is verified that removing them causes a significantly larger increase in the energy than the prior decrease. Remarkably, by the end, conservative ansatz growth results in an error of only 0.3\% that of the other strategies (original \gls{ADAPT}-\gls{VQE} and term removal).

Here, the optimizations required were 10 for \gls{ADAPT}-\gls{VQE}, 12 for the term removal version, and 50 for the conservative ansatz growth. As much as attempting to remove terms does not improve accuracy, there is barely an optimization overhead. Additionally, the tendency in this graph it is not as clear as before in what comes to determining the success of the conservative ansatz growth strategy: through the first 10 iterations, there are many in which it actually corresponds to a greater error than the original \gls{ADAPT}-\gls{VQE}, despite the sizeable optimization overhead. There is a sharp drop in the error by iterations 9 and 10, seeming to hint that the procedure might become more advantageous as the algorithm evolves; but unfortunately, the simulations become increasingly expensive computationally as more terms are added. This behaviour was also not encountered in simulating any other molecule, which complicates the task of determining whether conservative growth is indeed advantageous when term removal fails. 

Table \ref{tb:growth_strategies} presents a succinct comparison of the three ansatz growth strategies, for the case of the three systems analysed in this subsection. The final error is used as the figure of merit for characterizing performance, and the number of optimizations for characterizing measurement and optimization costs.

\begin{table}[htbp]
\begin{center}
\begin{tabular}{| c | c | c | c | c | c | c |}
 \hline
  \cellcolor[HTML]{EBECF0} & \multicolumn{6}{|c|}{\cellcolor[HTML]{BEBEBE}\textbf{System}}  \\
 \cline{2-7}
  \cellcolor[HTML]{EBECF0} & \multicolumn{2}{|c|}{\cellcolor[HTML]{BEBEBE}$LiH$} & \multicolumn{2}{|c|}{\cellcolor[HTML]{BEBEBE}$OH^{-}$} & \multicolumn{2}{|c|}{$\cellcolor[HTML]{BEBEBE}H_4$}\\ 
 \cline{2-7}
  \cellcolor[HTML]{EBECF0}\textbf{Method} & $|\Delta E|$ & \# Opts & $|\Delta E|$ & \# Opts & $|\Delta E|$ & \# Opts \\  
  \cellcolor[HTML]{EBECF0}  & ($\times 10^{-5}$) & & ($\times 10^{-6}$) & & ($\times 10^{-7}$) &\\
 \cline{1-7}
   \cellcolor[HTML]{EBECF0} Original & 39.0 & 10 & 160.2 & 10 & 2166.5 & 10\\
 \cline{1-7}
   \cellcolor[HTML]{EBECF0} Term Removal & 6.1 & 26 & 7.4 & 22 & 2166.5 & 13 \\
 \cline{1-7}
 \cellcolor[HTML]{EBECF0} Conservative & 6.1 & 50 & 7.4 & 50 & 5.8 & 50\\
 \hline
\end{tabular}
\end{center}
\caption{Summary of the performance of the three methods for manipulating the ansatz (original \gls{ADAPT}-\gls{VQE}, term removal, conservative ansatz growth). The table presents the absolute error in the energy and the accumulated number of optimizations at the point a 10-operator ansatz is reached. The data concerns the three systems explored in this subsection ($LiH$, $OH^{-}$, and $H_4$) with the considered pools (\gls{SGSD}, Eight Pool, and \gls{SGSD}, respectively). The units of energy are Hartree, as always.}
\label{tb:growth_strategies}
\end{table}

\FloatBarrier
\section{Discussion}
\label{s:growth_vs_removal}

Conservative ansatz growth seems like a solid procedure for creating compact ansätze. Unlike term removal, it has guarantees in what comes to maximum circuit size: at any level of accuracy, the maximum size of the ansatz is the current one. By being cautious with growing the ansatz, it is possible to burden the classical optimizer with the compensation of decreasing circuit depth. Further, growing the ansatz conservatively is viable even when the term removal procedure fails at the third step.

On the other hand, term removal has its advantages. By only removing terms \textit{after} they are outperformed by their successors, the procedure avoids an overhead in optimizations at every single iteration. Extra optimizations only occur when there is a good chance that they are in fact benefiting for the ansatz, which results in a more modest overhead in total optimizations. Further, as the algorithm evolves, this saving becomes more and more rewarding; in the results presented before, even when limiting the \textit{maximum} number of operators in the ansatz to 10, the term removal version matched or surpassed the accuracy of the version with conservative growth (with 5 optimizations per step) for $OH^{-}$ and $LiH$, for ansätze with up to only 10 operators. This suggests that the conditions for removing a term are met sufficiently often that the size of the ansatz necessary for a given accuracy can be kept as modest with term removal as with conservative ansatz growth.

Further, it is expected that as the size of the system grows, testing a relevant portion of the operators in the pool per iteration will become harder and harder: larger molecules imply larger pools. In the cases presented here, testing the five operators with largest gradient in each iteration ($N=5$) corresponded to testing up to 7.6\% of the operators in the pool, which is a significant portion. If we increase $N$ to always account for the same portion of the pool, we will have an added cost in optimizations that increases with the system size. In contrast, term removal is size-agnostic, because it will attempt to remove under-performing operators, rather than find out in advance which they are. Removing operators after they're outperformed (instead of testing several \textit{a priori}) can keep the ansatz compact as other operators are explored, while saving on the extra optimizations. 

What is more, because removed operators are penalized, they will not be added too soon after they're removed. In contrast, when the conservative ansatz growth procedure finds an operator that performs very poorly, it will likely still attempt to add it in the next iteration, instead of giving the opportunity to other operators with lower gradients. This approach does not attempt to leverage any information regarding the performance of the operators that was obtained in the previous iterations, which may prevent it from exploring more advantageous candidates and `clog' several of the $N$ attempts per iteration with operators that were already shown to perform badly. 

Of course, removing terms is always more fallible than growing the ansatz conservatively. If, as happened with $H_4$, term removal consistently fails on step 3, there is absolutely no added benefit to the procedure, only extra cost.

%% file: Chapters/chapter7.tex
\pagestyle{plain}

\chapter{Conclusions and Future Work}
\label{ch:conclusions}

The purpose of this dissertation was to explore several proposals of ansätze for the Variational Quantum Eigensolver in chemistry applications, with a focus on the noise-resilience that is paramount in improving short-term viability of the algorithm. 

Ansätze can be split into three broad classes: predetermined and problem agnostic; predetermined and problem tailored; dynamic and problem-tailored. An ansatz from each class was implemented, and each was tested both in simulators and in real quantum computers made accessible on the cloud by IBMQ \cite{IBMQ}. The effect of sampling noise, decoherence, \gls{SPAM} errors and generic noise was analysed \textit{via} simulations on different backends and creation of noise models. With this, we could compare the noise-resilience of different type of ansätze. In particular, the effect of noise on the predetermined \gls{UCCSD} ansatz (considered a `gold standard' in quantum chemistry) and the dynamic ansatz built in the \gls{ADAPT2}-\gls{VQE} algorithm (a more recent proposal) was compared.

The benefits of \gls{ADAPT2}-\gls{VQE} became evident in our simulations. For the $H_2$ molecule, a 1-operator qubit-\gls{ADAPT2}-\gls{VQE} ansatz was only surpassed by \gls{UCCSD}-\gls{VQE} in a fully noise-free scenario, when even sampling noise was removed from the simulations. Such scenario would be no less than impossible to actually create, even if we had an ideal quantum processor available (which is far from reality). Two unavoidable factors, quantum projection noise and a finite number of shots, are enough to deteriorate the performance of \gls{UCCSD}-\gls{VQE} significantly. In the presence of generic noise, the difference between \gls{UCCSD}-\gls{VQE} and \gls{ADAPT2}-\gls{VQE} became even more pronounced.

Problem-tailored (but predetermined) ansätze already help avoid some problems that arise in the context of problem-agnostic ansätze, \textit{barren plateaus} being one example. This notwithstanding, \textit{dynamic} ansätze are one step ahead: in addition to being problem-tailored, they are \textit{system}-tailored, which brings along an array of benefits. Ansätze grown from scratch, like the one from \gls{ADAPT}-\gls{VQE}, can provide high-accuracy results with shallower circuits and less variational parameters than any predetermined alternative. Both of these factors increase the noise-resilience of \gls{VQE}, resulting in it being an even more \gls{NISQ}-Friendly algorithm when dynamic ansätze are used.

The choice of operator pool in \gls{ADAPT}-\gls{VQE} is the part of the algorithm that implies the biggest human involvement. In order to assess its importance, \gls{ADAPT2}-\gls{VQE} was simulated resorting to a variety of pools in addition to those originally proposed. The constitution of the superposition state in terms of Slater determinants was analysed along with ease of convergence, and the relative importance of fermionic antisymmetry, preservation of quantities such as particle number and spin, and similarity with fermionic excitations for the success of the pool was explored.

This analysis allows us to conclude that there is a delicate balance between choosing operators that respect fermionic symmetries, and choosing operators that can be implemented by shallow and convenient circuits. The former helps decrease the number of iterations and variational parameters, while the latter can reduce the circuit depth; but the two are tightly interwoven, and the problem at hand is to find the optimal compromise. Regardless, there seem to be aspects in which being faithful to the proper representation of fermionic operators brings no advantage, and we find that relinquishing them lowers costs free of negative repercussions. This is the case of anticommutation: relinquishing such principle eases, from linear to constant (on the size of the system), the scaling of the average circuit depth required to implement an operator, with no observable negative effect on the success of the algorithm.

Independently of the pool choice, there are also possibilities of improvement through additional manipulation of the ansatz. No selection criterion is unerring; but adjustments and corrections can aid in preventing the misuse of circuit depth on portions of the ansatz that bring little benefit. Two approaches were analysed to address this. Heuristics for removing operators from the ansatz along the evolution of the algorithm were developed. The main goal was keeping the ansatz shallow without hindering the efficiency of the algorithm by incurring in unnecessary measurements and optimizations, a consequence we showed could be brought about by an unpolished strategy. The second attempted strategy was simpler, consisting of testing the parallel addition of distinct operators to the ansatz per iteration, only to choose and add the best performing one. 

The first strategy (term removal) was compared against the second one (conservative ansatz growth), with the original \gls{ADAPT}-\gls{VQE} as a reference. Tests encompassed simulations with several molecules and pools. Both strategies were shown to be able to bring a significant reduction to the circuit depth required to achieve a given accuracy. Term removal seems to have a more modest optimization overhead, since it only entails extra optimizations when it can sense that they may be advantageous, unlike conservative ansatz growth, which performs a constant number of optimizations per iteration irrespective of the evolution of the algorithm. Despite the reduced number of additional optimizations, removing terms proved to be often capable of maintaining the maximum ansatz depth as shallow as conservative ansatz growth. Further, the term removal procedure seems in principle to be unaffected by the size of the pool, unlike conservative ansatz growth. The portion of the pool we can test with a constant number of allowed optimizations per iteration will become smaller and smaller as the size of the system increases, leading to a degradation of the performance of the conservative ansatz growth strategy. However, this method is safer than term removal, as it does not rely on the evolution of the algorithm or on the characteristics of the commutator between operators to keep the ansatz compact. In one of the tested molecules, $H_4$, the term removal procedure was incapable of changing the evolution of the \gls{ADAPT}-\gls{VQE} algorithm for the better. For this molecule, the success of conservative ansatz growth was also less clear; in spite of this, for the maximum simulated ansatz size (twenty operators), it did achieve a lower error.

Variational quantum algorithms are a recent class, created specifically in the context of near-term quantum computing. \gls{VQE} was the first algorithm of this class, proposed in 2014 \cite{Peruzzo2014}; \gls{ADAPT}-\gls{VQE} introduced the possibility of using dynamic ansätze for chemistry applications in 2019 \cite{Grimsley2019}, with further developments in 2020 \cite{Tang2021} and September 2021 \cite{shkolnikov2021} bringing remarkable decreases in both the size of the pool and the circuit depth per operator. 

The dates of these proposals speak already of the novelty and intensity of the research around variational quantum algorithms in general and dynamic ansätze in particular. For the former case, noise-resilient optimization strategies, improved ways of measuring relevant observables, and strategies for avoiding barren plateaus are doubtlessly of great interest. For the latter, and more relevant in the context of this dissertation, methods for developing and manipulating ansätze on the fly can certainly be expanded and improved upon.

The proposals of dynamic ansätze explored in this project were mainly directed at chemistry applications; it would be interesting to generalize the algorithm to other areas, such as nuclear and condensed matter physics, or even mathematics and machine learning. Since in \gls{ADAPT}-\gls{VQE} the specifics of the problem are tied to the content of the operator pool, the algorithm must be formulated differently for other problems. Further, recent results showed that the size of the pool can be significantly decreased by leveraging the characteristics of the specific molecule. We can speculate that, in general, leveraging problem-specific knowledge in the creation of the set of operators that serve as `pieces' in the construction of the ansatz may allow for improving results and decreasing the measurement costs in the creation of adaptive wave functions. 

Even within chemistry problems, \gls{ADAPT}-\gls{VQE} could be further explored. For example, pools could be chosen taking into account the native interactions and the connectivity of the device, with the purpose of obtaining a more hardware-friendly ansatz. Mappings other than Jordan-Wigner could be employed, changing the formulation of the problem, affecting the gradient measurements, the operator pool, and the qubit requirements, among other aspects. 

And finally, as it was explored here, manipulating the ansatz beyond the steady addition of operators of \gls{ADAPT}-\gls{VQE} can improve the approximation to the solution while keeping the ansatz compact, by keeping in it only the most important operators. Removing terms or testing several per iteration are merely examples of what can be done. It would be interesting to test and compare these approaches on more molecules and pools, and for longer runs. 

Explored or unexplored, there are many options and possibilities in designing ansätze for the variational quantum eigensolver - even more when we expand the scope to variational quantum algorithms in general. Finding increasingly more compact and convenient state preparation circuits is a fundamental piece in the pursuit of a chance at quantum advantage with variational quantum algorithms: if they are naturally \gls{NISQ}-friendly, they can become even more so through a refinement of the ansatz. A proper choice can decrease the depth of the required circuits, facilitate the task of the classical optimizer, increase the chances of convergence, and reduce the number of calls to the quantum computer. All efforts done to improve the ansatz pave the way for the success of general variational approaches to quantum algorithms.

%% file: Chapters/appendix1.tex
\pagestyle{plain}
\chapter{Implementation of Unitary Pauli String Exponentials}
\label{ap:exponentiation}

We wish to obtain a circuit to implement $e^{-i tH_k}$ for $H_k$ a Pauli string ($\bigotimes_iP_i$, $P_i\in\{I_i,Z_i,X_i,Y_i\}$). This can be done from the circuit used for $H_k=\bigotimes_iZ_i$. The latter, as it was explained in subsection \ref{ss:simulation_trotterization}, only needs to apply a phase shift conditional on the parity of the computational basis states. In that section, a simple circuit that fit the purpose was presented, consisting of two ladders of CNOT gates (for computing and subsequently uncomputing the parity) and a single-qubit rotation.

The circuit for $H_k=\bigotimes_iP_i$ can be obtained by a simple modification. For any qubit $i$ acted on by $P_i$ in $H_k$, we find a basis rotation $U_i$ that obeys

\[U_i^\dagger Z_iU_i=P_i\]

The unitary $U_i$ will be different from the identity for any qubit that is acted on by a Pauli operator $P_i$ different from $Z_i$. Examples of $U_i$ include the Hadamard gate (for $P_i=X_i$) and a rotation around the X axis by $\pi/2$ (for $P_i=Y_i$).

This allows us to rewrite the expression as

\[e^{-i tH_k}=e^{-i t\bigotimes_iP_i}=e^{-i t\bigotimes_iU_i^\dagger Z_iU_i}\]

Decomposing the exponential series, we obtain
\[=\sum_{n=0}^{\infty}\frac{(-i t\bigotimes_iU_i^\dagger Z_iU_i)^n}{n!}\]

\[=I_1\otimes I_2 \otimes I_3 \otimes I_4-i t\bigotimes_iU_i^\dagger Z_iU_i+\frac{(i t)^2}{2}(\bigotimes_iU_i^\dagger Z_iU_i)^2+...\]

Using the properties of the tensor product ($(a\otimes b)(c\otimes d)=ac\otimes bd$) and the fact that the $U_i$ are unitary ($U_i^\dagger U_i = U_i U_i^\dagger = I$), we can isolate an exponential series with argument free of the $U_i$:

\[\bigotimes_i U_i^\dagger\left(\sum_{n=0}^{\infty}\frac{(-i t\bigotimes_iZ_i)^n}{n!}\right)\bigotimes_i U_i = \bigotimes_i U_i^\dagger\left( e^{-i t\bigotimes_iZ_i}\right) \bigotimes_i U_i \]

So concluding, we have the identity

\[e^{-i t\bigotimes_iP_i}\\
=\bigotimes_i U_i^\dagger\left( e^{-i t\bigotimes_iZ_i}\right) \bigotimes_i U_i\]

And we can apply any operator of the form $e^{-it\bigotimes_iP_i}$ by applying the circuit for $e^{-it\bigotimes_iZ_i}$, preceded and succeeded by single qubit rotations.

%% file: Chapters/appendix2.tex
\pagestyle{plain}
\chapter{Hamiltonian of H2}
\label{ap:h2_hamiltonian}

This appendix contains the Hamiltonian of the $H_2$ molecule at an interatomic distance of 0.37Å, obtained via OpenFermion \cite{openfermion}. The utilized basis set was \textit{sto-3g}. Because it is a minimal basis set and no approximations were used, the molecule is represented by four spin-orbitals/qubits. 

The choice to present the Hamiltonian of the simplest molecule was made due to the quartic growth of the number of terms with the number of orbitals: the Hamiltonian of a molecule such as $LiH$ would already have almost two thousand operators.

The molecular Hamiltonian in the second quantization formalism is the following:
\begin{flalign*}
&0.7151043390810812+\\
&a_0^\dagger a_0 \cdot (-1.2533097866459773)+\\
&a_1^\dagger a_1 \cdot (-1.2533097866459773)+\\
&a_2^\dagger a_2 \cdot (-0.47506884877217576)+\\
&a_3^\dagger a_3 \cdot (-0.47506884877217576)+\\
&a_0^\dagger a_0^\dagger a_0 a_0 \cdot 0.3373779634072241+\\
&a_0^\dagger a_0^\dagger a_2 a_2 \cdot 0.09060523100759854+\\
&a_0^\dagger a_1^\dagger a_1 a_0 \cdot 0.3373779634072241+\\
&a_0^\dagger a_1^\dagger a_3 a_2 \cdot 0.09060523100759854+\\
&a_0^\dagger a_2^\dagger a_0 a_2 \cdot 0.09060523100759854+\\
&a_0^\dagger a_2^\dagger a_2 a_0 \cdot 0.3318557006754069+\\
&a_0^\dagger a_3^\dagger a_1 a_2 \cdot 0.09060523100759854+\\
&a_0^\dagger a_3^\dagger a_3 a_0 \cdot 0.3318557006754069+\\
&a_1^\dagger a_0^\dagger a_0 a_1 \cdot 0.3373779634072241+\\
&a_1^\dagger a_0^\dagger a_2 a_3 \cdot 0.09060523100759854+\\
&a_1^\dagger a_1^\dagger a_1 a_1 \cdot 0.3373779634072241+\\
&a_1^\dagger a_1^\dagger a_3 a_3 \cdot 0.09060523100759854+\\
&a_1^\dagger a_2^\dagger a_0 a_3 \cdot 0.09060523100759854+\\
&a_1^\dagger a_2^\dagger a_2 a_1 \cdot 0.3318557006754069+\\
&a_1^\dagger a_3^\dagger a_1 a_3 \cdot 0.09060523100759854+\\
&a_1^\dagger a_3^\dagger a_3 a_1 \cdot 0.3318557006754069+\\
&a_2^\dagger a_0^\dagger a_0 a_2 \cdot 0.3318557006754069+\\
&a_2^\dagger a_0^\dagger a_2 a_0 \cdot 0.09060523100759854+\\
&a_2^\dagger a_1^\dagger a_1 a_2 \cdot 0.3318557006754069+\\
&a_2^\dagger a_1^\dagger a_3 a_0 \cdot 0.09060523100759854+\\
&a_2^\dagger a_2^\dagger a_0 a_0 \cdot 0.09060523100759854+\\
&a_2^\dagger a_2^\dagger a_2 a_2 \cdot 0.348825752245232+\\
&a_2^\dagger a_3^\dagger a_1 a_0 \cdot 0.09060523100759854+\\
&a_2^\dagger a_3^\dagger a_3 a_2 \cdot 0.348825752245232+\\
&a_3^\dagger a_0^\dagger a_0 a_3 \cdot 0.3318557006754069+\\
&a_3^\dagger a_0^\dagger a_2 a_1 \cdot 0.09060523100759854+\\
&a_3^\dagger a_1^\dagger a_1 a_3 \cdot 0.3318557006754069+\\
&a_3^\dagger a_1^\dagger a_3 a_1 \cdot 0.09060523100759854+\\
&a_3^\dagger a_2^\dagger a_0 a_1 \cdot 0.09060523100759854+\\
&a_3^\dagger a_2^\dagger a_2 a_3 \cdot 0.348825752245232+\\
&a_3^\dagger a_3^\dagger a_1 a_1 \cdot 0.09060523100759854+\\
&a_3^\dagger a_3^\dagger a_3 a_3 \cdot 0.348825752245232
&&
\end{flalign*}

Once we apply the Jordan-Wigner transformation \ref{def:jw_transform_paulis} to obtain a Hamiltonian acting on qubits, we obtain:

\begin{flalign*}
-&0.09706626816762856 +\\
-&0.04530261550379927 \cdot X_0 X_1 Y_2 Y_3 +\\
&0.04530261550379927 \cdot X_0 Y_1 Y_2 X_3 +\\
&0.04530261550379927 \cdot Y_0 X_1 X_2 Y_3 +\\
-&0.04530261550379927 \cdot Y_0 Y_1 X_2 X_3 +\\
&0.17141282644776898 \cdot Z_0 +\\
&0.1686889817036120 \cdot Z_0 Z_1 +\\
&0.12062523483390417 \cdot Z_0 Z_2 +\\
&0.16592785033770344 \cdot Z_0 Z_3 +\\
&0.17141282644776903 \cdot Z_1 +\\
&0.16592785033770344 \cdot Z_1 Z_2 +\\
&0.12062523483390417 \cdot Z_1 Z_3 +\\
-&0.22343153690813572 \cdot Z_2 +\\
&0.174412876122616 \cdot Z_2 Z_3 +\\
-&0.22343153690813572 \cdot Z_3
&&
\end{flalign*}

%% file: Chapters/appendix3.tex
\pagestyle{plain}
\chapter{ADAPT-VQE Gradient Formula}
\label{ap:adaptgradient}

This appendix will show how the gradient formula from \gls{ADAPT2}-\gls{VQE} can be derived. 

First of all, it is important to remember that the pool operators $\hat{A}_i^\dagger$ are anti-hermitian:

\[\hat{A}_i^\dagger=-\hat{A}_i\]

This condition forces the unitarity of their parameterized and exponentiated versions that will actually appear in the ansatz. Since the variational parameters $\theta_i$ are real, we can see that

\[\left(e^{\theta_i\hat{A}_i}\right)^\dagger=e^{-\theta_i\hat{A}_i}\]

We want to calculate the derivative of the energy with respect to each possible variational parameter $\theta_i$, as an indicator of how much the corresponding operator will impact the energy. For the $i$th operator, this derivative can be calculated as

\[\pdv{}{\theta_i}\bra{\psi^{(n)}}e^{-\theta_i\hat{A}_i}\hat{H}e^{\theta_i\hat{A}_i}\ket{\psi^{(n)}}\]

\[= \bra{\psi^{(n)}}e^{-\theta_i\hat{A}_i}(-\hat{A}_i)\hat{H}e^{\theta_i\hat{A}_i}\ket{\psi^{(n)}} 
+ \bra{\psi^{(n)}}e^{-\theta_i\hat{A}_i}\hat{H}A_ie^{\theta_i\hat{A}_i}\ket{\psi^{(n)}}\]

\[=\bra{\psi^{(n)}}e^{-\theta_i\hat{A}_i}[\hat{H},\hat{A}_i]e^{\theta_i\hat{A}_i}\ket{\psi^{(n)}}\]

All operators will be initialized with null coefficients: when a new operator $e^{\theta_i\hat{A}_i}$ is added to the ansatz, the corresponding variational parameter $\theta_i$ is set to zero. Thus, the relevant derivative is that calculated at this point, and we obtain \ref{eq:adapt_gradient} as desired.

\begin{equation}
    \pdv{E^{(n)}}{\theta_i}\Bigg|_{\theta_i=0} = \bra{\psi^{(n)}}\left[\hat{H},\hat{A}_i\right]\ket{\psi^{(n)}}
    \label{eq:adapt_gradient_ap}
\end{equation}

The expectation value of the commutator can be measured efficiently on the quantum computer.

Given that $\hat{H}$ is hermitian and $\hat{A}_i$ is anti-hermitian, we can also write the commutator in \ref{eq:adapt_gradient} as

\[\bra{\psi^{(n)}}[\hat{H},\hat{A}_i]\ket{\psi^{(n)}} = \bra{\psi^{(n)}}\hat{H}\hat{A}_i\ket{\psi^{(n)}}-\bra{\psi^{(n)}}\hat{A}_i\hat{H}\ket{\psi^{(n)}}\]

\[=\bra{\psi^{(n)}}\hat{H}\hat{A}_i\ket{\psi^{(n)}}-\left(\bra{\psi^{(n)}}\hat{H}(-\hat{A}_i)\ket{\psi^{(n)}}\right)^\dagger\]

\[=\bra{\psi^{(n)}}\hat{H}\hat{A}_i\ket{\psi^{(n)}}+\left(\bra{\psi^{(n)}}\hat{H}\hat{A}_i\ket{\psi^{(n)}}\right)^\dagger = 2Re\bra{\psi^{(n)}}\hat{H}\hat{A}_i\ket{\psi^{(n)}}\]

In the sampling noise free, matrix algebra based simulations, formula \ref{eq:adapt_gradient_simulations} was used instead of \ref{eq:adapt_gradient}.

\begin{equation}
    \pdv{E^{(n)}}{\theta_i}\Bigg|_{\theta_i=0} = 2Re\bra{\psi^{(n)}}\hat{H}\hat{A}_i\ket{\psi^{(n)}}
    \label{eq:adapt_gradient_simulations}
\end{equation}

%% file: thesis.bbl
\begin{thebibliography}{10}
\bibitem{Grimsley2020}
Is the trotterized {UCCSD} ansatz chemically well-defined?
\end{thebibliography}
